# A new non-extensive equation of state for the fluid phases of argon including the metastable states, from the melting line to 2300 K and 50 GPa

F. Aitken[1], A. Denat and F. Volino

Univ. Grenoble Alpes, CNRS, Grenoble INP, G2Elab, F-38000 Grenoble, France

**Abstract:** A new equation of state for argon has been developed in view to extend the range of validity of the equation of state previously proposed by Tegeler *et al.* and to obtain a better physical description of the experimental thermodynamic data for the whole fluid region (single-phase, metastable and saturation states). As proposed by Tegeler *et al.*, this equation is also based on a functional form of the residual part of the reduced Helmholtz free energy. However in this work, the fundamental equation for the Helmholtz free energy has been derived from the measured quantities $C_V(\rho,T)$ and $P(\rho,T)$. The empirical description of the isochoric heat capacity $C_V(\rho,T)$ is based on an original empirical description containing explicitly the metastable states. The thermodynamic properties (internal energy, entropy, free energy) are then obtained by combining integration of $C_V(\rho,T)$. The arbitrary functions introduced by the integration process have been deduced from a comparison between calculated and experimental pressure $P(\rho,T)$ data. The new formulation is valid for the whole fluid region from the melting line to 2300 K and for pressures up to 50 GPa. It also predicts existence of a maximum of the isochoric heat capacity $C_V$ along isochors as experimentally observed in several other fluids. For many applications, an approximate form of the equation of state for the liquid phase may be sufficient. A Tait-Tammann equation is therefore proposed between the triple point temperature and 148 K.

[1] Corresponding author: frederic.aitken@g2elab.grenoble-inp.fr.

# List of Symbols

Symbol description

| | |
|---|---|
| $c$ | Sound speed |
| $C_P$ | Isobaric heat capacity |
| $C_V$ | Isochoric heat capacity |
| $F$ | Helmholtz free energy |
| $G$ | Gibbs energy |
| $H$ | Enthalpy |
| $i,j,k$ | Serial numbers |
| $M$ | Molar mass |
| $\mathcal{N}_a$ | Avogadro number |
| $P$ | Pressure |
| $R_\mathrm{A}$ | Specific gas constant |
| $S$ | Entropy |
| $T$ | Thermodynamic temperature |
| $U$ | Internal energy |
| $V$ | Specific volume ($V = 1/\rho$) |
| $x$ | Dimensionless parameter ($x = T_\mathrm{div}/T$) |
| $y$ | Dimensionless parameter ($y = x^{-1} = T/T_\mathrm{div}$) |
| $Z$ | Compressibility factor |

Greek

| | |
|---|---|
| $\partial$ | Partial differential |
| $\delta_\mathrm{T}$ | Isothermal throttling coefficient |
| $\rho$ | Density |
| $\Gamma$ | Incomplete gamma function |

Superscripts

| | |
|---|---|
| o | Ideal-gas property |
| r, * | Residual terms |
| ~ | Dimensionless quantity |
| ∧ | Dimensionless quantity (for energy only) using $\frac{3}{2}R_\mathrm{A}\,T_c$ as a reference |

Subscripts

| | |
|---|---|
| c | At the critical point |
| calc | Calculated |
| exp | Experimental |
| sat | Denotes states at saturation |
| sp | Denotes spinodal states |
| t | At the triple point |
| σl | Saturated liquid state |
| σv | Saturated vapor state |
| 0 | Terms that do not contribute to $C_V$ |

## Physical Constants for Argon

$M$ Molar mass

$M = 39.948$ g mol$^{-1}$

$R$ Universal gas constant

$R = 8.314\ 51$ J mol$^{-1}$ K$^{-1}$

$R_A$ Specific gas constant

$R_A = 0.208\ 133\ 3$ kJ kg$^{-1}$ K$^{-1}$

$T_c$ Critical temperature

$T_c = 150.687$ K

$P_c$ Critical pressure

$P_c = 4.863$ MPa

$\rho_c$ Critical density

$\rho_c = 0.535\ 599$ g cm$^{-3}$

$T_t$ Triple-point temperature

$T_t = 83.805\ 8$ K

$P_t$ Triple-point pressure

$P_t = 68.891$ kPa

$\rho_{t,Gaz}$ Triple-point gas density

$\rho_{t,Gaz} = 0.004\ 0546$ g cm$^{-3}$

$\rho_{t,Liq}$ Triple-point liquid density

$\rho_{t,Liq} = 1.416\ 80$ g cm$^{-3}$

$\rho_{t,Sol}$ Triple-point solid density

$\rho_{t,Sol} = 1.623\ 9$ g cm$^{-3}$

## 1 Introduction

Argon is a noble gas and, on earth, its isotopic composition is 99.6% $^{40}$Ar, 0.34% $^{36}$Ar and 0.06% $^{38}$Ar. Argon is very stable and chemically inert under most conditions. Due to those properties and its low cost, argon is largely used in scientific and industrial applications. For instance in high-temperature industrial processes, an argon atmosphere can prevent material burning, material oxidation, material defects during growing of crystals, etc. Due to its molecular simplicity (monoatomic, quasi-spherical geometry), argon is also considered as a reference fluid with well-known properties, i.e.: its triple point temperature (83.8058 K) is a defining fixed point in the International Temperature Scale of 1990 (Ref. 1). Its simple fluid characteristics allow, for example, to understand the fundamental mechanisms of interaction between ions and neutral species and thus gain a deeper insight into ion transport regimes (e.g. Ref. 2). The widespread use of argon requires an accurate knowledge of its thermodynamic properties in the largest possible temperature and pressure ranges, i.e. covering both stable and metastable states. Numerous empirical equations of state can be found in the literature, but most of them cover only small parts of the fluid region. For example, Shamsundar *et al.* (Ref. 3) have shown that the development of cubic-like equations of state provides very accurate thermodynamic properties of liquids on the coexistence curve and in the metastable (superheated) state. However, this approach has flaws on the vapor side. A very detailed overview of argon's experimental thermodynamics and the most important equations of state published prior to 1999 can be found in Ref. 4, which we won't go into here. In Ref. 4, Tegeler *et al.* also describe a new equation of state for argon which covers a very wide range of the fluid phase and which will serve as the reference equation of state for this study.

The development of the equation of state generally starts by an empirical description of the Helmholtz free energy $F$ (i.e. an arbitrary set of mathematical functions form is *a priori* chosen) with two independent variables, density $\rho$ and temperature $T$. All thermodynamic properties of a pure substance can then be obtained by combining derivatives of $F(\rho,T)$. The dimensionless Helmholtz free energy $\tilde{a} = F/(R_\mathrm{A} T)$ is commonly split into a part $\tilde{a}^\circ(\rho,T)$ which represents the properties of the ideal gas at given temperature and density, and a residual part $\tilde{a}^\mathrm{r}(\rho,T)$ which takes into account the dense fluid behavior. While statistical thermodynamics can predict the behavior of fluids in the ideal-gas state with high accuracy, no physically founded equation is known which describes accurately the actual thermodynamic behavior in the whole fluid region. Thus, an equation for the residual fluid behavior, in this case for the residual part of the Helmholtz free energy $\tilde{a}^\mathrm{r}$, has to be determined in an empirical way. However, as the Helmholtz free energy is not accessible to direct measurements, a suitable mathematical structure and some fitted coefficients have to be determined from properties for which experimental data are available. Hence all the physical properties are contained in the mathematical form given to the Helmholtz free energy.

In the wide-range equation of state for argon developed by Tegeler *et al.* (Ref. 4), the residual part of the Helmholtz free energy $\tilde{a}^\mathrm{r}(\rho,T)$ contains polynomial terms, Gaussian terms and exponential terms which results in a total of 41 coefficients (named $n_i$ in Ref. 4) which represent the number of mathematical distinct entities (each mathematical entity containing itself several adjustable parameters). This equation of state is valid for the fluid region delimited by

$$83.8058 \text{ K} < T < 700 \text{ K},$$

and

$$0 \text{ MPa} < P < 1000 \text{ MPa}.$$

The large number (~120) of adjustable parameters of the equation of state of Tegeler *et al.* (see table 30 in Ref. 4) are determined by a sophisticated fitting technique which is a powerful mathematical tool and a practical way for representing data sets (by assigning weights to each of them subjectively). This technique provides an easily practical overall numerical representation of the data but it also allows completing the representation of measurable quantities in areas where no measurements have been made. However, passing in a set of data points does not mean that the obtained variations have a physical meaning or that the physical ideas underlying mathematical representation are unique. For example, it can be noticed the following drawbacks of the equation of state of Tegeler et *al*. (Ref. 4):

1. Extrapolation of the equation for the isochoric heat capacity in regions of high or low density and high temperature is non physical.
2. The extrapolation of polynomial developments does not generally give valid results; indeed, polynomial development is very sensitive (i.e. instable) with respect to the values of its coefficients and these coefficients cannot support to be truncated, even slightly. So all the coefficients $n_i$ of Tegeler's *et al*. (Ref. 4) model have 14 digits, thus the coefficient have no physical sense.
3. The model applies for the pure fluid phases and cannot in its actual form take into account particular properties inside the coexistence liquid-vapor region. Moreover, the model gives negative values of $C_V$ on some isotherms inside the coexistence liquid-vapor region ($C_V < 0$ is never observed for classical thermodynamic systems). This implies for example some non-physical variations of the liquid spinodal curve.

The aim of this paper is not to increase the precision of the equation of state of Tegeler *et al*. (Ref. 4) in its own domain of validity but to develop a new equation of state based on different physical ideas that can fill the drawbacks previously expressed in order to obtain a more physical description of the experimental thermodynamic data of argon in a broader temperature and pressure ranges. In the classical approach, the ideal part of the free energy is generally determined from the well-known properties of the ideal-gas heat capacities. We propose to extend the classical approach also to the residual part, therefore the proposed new equation of state is based on an original empirical description of the isochoric heat capacity $C_V(\rho,T)$ containing explicitly the metastable states. Then, the thermodynamics properties (internal energy, entropy, free energy) are obtained by combining integration of functions involving $C_V(\rho,T)$. For instance, internal energy $U$ can be deduced from $U(\rho,T) = \int C_V(\rho,T)dT + U_0(\rho) + \text{constant}$, where $U_0(\rho)$ is an arbitrary function of density. In this way, possible data noise are smoothed. However, an integration process introduces arbitrary functions (e.g. $U_0(\rho)$). These functions can be deduced from a comparison between calculated and experimental data. The pressure equation of state $P(\rho,T)$ was chosen because it is the largest available data set. The set of experimental data taken into account by the model of Tegeler *et al*. (Ref. 4) will be further extended with the inclusion of the L'Air Liquide database (Ref. 19), thus extending the temperature validity range of this new modeling compared to that of Tegeler *et al*. The interest of this new approach is that it can be easily extended to all other fluids that exhibit a first order transition with metastable states.

*In the following, the model of Tegeler et al. (Ref. 4) will be simply named the* ***TSW model***.

## 2 New equation of state for the isochoric heat capacity

As stated previously, the present approach starts with the empirical description of a chosen thermodynamic quantity. It was chosen to describe an experimentally measured quantity, which is not the case for the Helmholtz free energy. The quantity which has the simplest mathematical and physical comprehensive variation is the isochoric heat capacity $C_V$ as function of density $\rho$ and temperature $T$. Starting with this quantity, we therefore lose the advantage of the description provided by the Helmholtz free energy from which all other thermodynamic quantities can be obtained by derivation but, it allows to introduce more easily new physical bases, in particular the non-extensivity, and simplicity is enhanced. Indeed, the number of coefficients $\alpha_i$ (without $\alpha_{\mathrm{crit,b}}$) for the description of $C_V$ is 11 (see Table 1) and it will be shown in the next section that the number of coefficients $\alpha_i$ for the description of the Helmholtz free energy is only **26** compared to **41** for the TSW model.

After choosing the thermodynamic quantity to be described, one must find a mathematical structure for its representation. A virial like development is an easy and widespread approximation. The problem of polynomial terms is that they introduce very small oscillations that are not physical. To avoid such effects, it is assumed that the description must not contain any form of polynomial expression. Thus, the description is made in terms of power laws and exponentials with density dependent exponents. We shall see that with such a description, we get among the parameters of the model a dozen of different characteristic densities instead of only $\rho_{\mathrm{c}}$ (which is consistent with the fact that argon does not follow the law of Corresponding States), it was therefore chosen to express all the equations of states in a dimensionless form according to the variables $\rho$ and $T$ which lead to simpler expressions than if one had considered the dimensionless variables $\rho/\rho_c$ and $T/T_c$. In addition, the most suitable units for density $\rho$ and temperature $T$ have been chosen as g/cm$^3$ and Kelvin, respectively.

As for the Helmholtz free energy, the isochoric heat capacity is split into a part $C_V^{\mathrm{o}}$ which represents the properties of the ideal gas and a part $C_V^{\mathrm{r}}$ which takes into account the residual fluid behavior at given $T$ and $\rho$. Note here that the ideal part of the free energy is in fact determined from known properties of $C_V^{\mathrm{o}}$ in the classical approach. Since argon is monoatomic only the translational contribution to the ideal-gas heat capacity $C_{P,tr}^{\mathrm{o}} = \frac{5}{2}R_{\mathrm{A}}$ has to be taken into account, so it is deduced from the Mayer's law that: $C_V^{\mathrm{o}} = \frac{3}{2}R_{\mathrm{A}}$. In dimensionless form, the isochoric heat capacity is written:

$$\tilde{c}_V(\rho,T) = \frac{C_V(\rho,T)}{R_{\mathrm{A}}} = \tilde{c}_V^{\mathrm{o}} + \tilde{c}_V^{\mathrm{r}}(\rho,T) = \frac{3}{2}\left(1 + \frac{2}{3}\tilde{c}_V^{r}(\rho,T)\right) \tag{1}$$

To take into account all fluid domain including the liquid-vapor coexistence region and the region around the critical point, $\tilde{c}_V^{\mathrm{r}}(\rho,T)$ must be split up into 3 terms - named regular, non-regular and critical - such that:

$$\tilde{c}_V^{\mathrm{r}}(\rho,T) = \tilde{c}_{V,\mathrm{reg}}^{r} + \tilde{c}_{V,\mathrm{nonreg}}^{r} + \tilde{c}_{V,\mathrm{crit}}^{r} \tag{2a}$$

with

$$
\left.
\begin{aligned}
\tilde{c}^{\mathrm{r}}_{V,\mathrm{reg}} &= n_{\mathrm{reg}}(\rho)\left\{1-\exp\left(-\left(\lambda\frac{T}{T_c}\right)^{1-(m(\rho)-1)}\right)\right\}\left(\frac{T}{T_c}\right)^{m(\rho)-1} \\
\tilde{c}^{r}_{V,\mathrm{nonreg}} &= n_{\mathrm{nonreg}}(\rho)\exp\left(-\left(\frac{T_{\mathrm{div}}(\rho)}{T}\right)^{-3/2}\right)\frac{1}{1-\dfrac{T_{\mathrm{div}}(\rho)}{T}} \\
\tilde{c}^{\mathrm{r}}_{\mathrm{V,crit}} &= n_{\mathrm{crit}}(\rho)\left(\frac{T_{\mathrm{div}}(\rho)}{T}\right)^{\varepsilon_{\mathrm{crit}}(\rho)}
\end{aligned}
\right\}
\quad \text{for } T \geq T_{\mathrm{div}} \tag{2b}
$$

where $\lambda$ = 6.8494 and $n_{\mathrm{reg}}(\rho)$, $m(\rho)$, $n_{\mathrm{nonreg}}(\rho)$, $T_{\mathrm{div}}(\rho)$, $n_{\mathrm{crit}}(\rho)$, $\varepsilon_{\mathrm{crit}}(\rho)$ are empirical functions determined from the best fitting of NIST (Ref. 5) and Ronchi (Ref. 6) data, whose expressions are given further on. In other words, it is assumed here that these two data sets are *a priori* consistent with each other.

An important feature of the Ronchi model is to predict the appearance of a maximum on the isochoric heat capacity $C_V$ along isochors. A maximum of $C_V$ along isochors has experimentally been observed in several fluids as for example in water. Consequently, the extrapolation of $C_V$ along isochors from a given model must show a maximum as predicted by the Ronchi model. This predicted behavior of $C_V$ along isochors constitutes the main interest of Ronchi model.

The Ronchi calculated data cover the largest available temperature range from 300 to 2300 K and the largest available pressure range from 9.9 to 47 058.9 MPa (420 data points). It is important to notice that they are consistent with many experimental data which were used to establish by Tegeler *et al*. (Ref. 4) but for which they assigned them to their groups 2-3. The data of Ronchi were assigned to group 2 by Tegeler *et al*. From the data of Ronchi, the highest available density is called $\rho_{\mathrm{max,Ronc}}$ and its value is given in Table 2.

By construction, a part of the Ronchi (Ref. 6) and NIST (Ref. 5) data overlap. For their common range of density values, it is observed in Fig. 1 that the deviation is always less than 2.5%, so that the data of Ronchi can be considered to be consistent with the data of NIST. *In the following, the term “NIST” will simply be used to refer to the data from Ref. 5 and the term “Ronchi” to refer to the data in Ref. 6.*

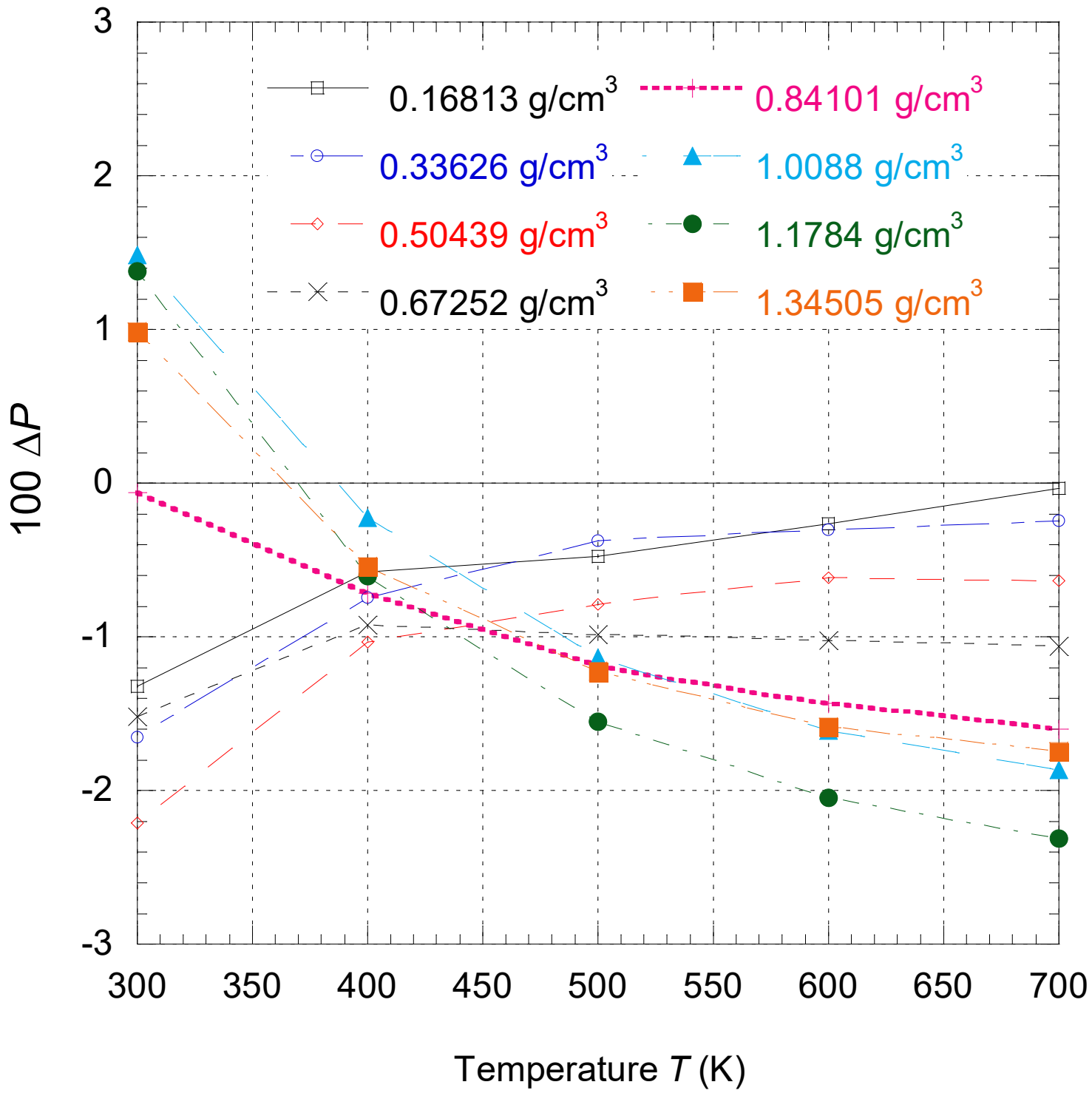

**Fig. 1:** Percentage deviations of pressure $\Delta P = (P_{\text{Ronchi}} - P_{\text{NIST}})/P_{\text{Ronchi}}$ on different isochors in the common region of data from NIST (Ref. 5) and from Ronchi (Ref. 6). The lines are eye guides.

The relations constituting Eq. (2) are therefore consistent with two sets of coherent data, but at this stage of the theoretical development, it must be strongly emphasized that these relations are only valid for temperatures $T \geq T_{\text{div}}(\rho)$, i.e. in particular for all states in the single phase region. In this way, $T_{\text{div}}(\rho)$ defines a divergence curve (i.e. it defines an asymptotic curve) and we will see later that it is related to the spinodal curve. We shall see in section 3.4 how this relation is transformed for $T < T_{\text{div}}(\rho)$ (i.e. for states inside the coexistence region).

It is very important to notice that with the present model, once chosen the mathematical form of the regular term, it is not possible to envisage any mathematical form for the two other residual terms. Indeed the two remaining terms must have a consistent mathematical form with that of the first one otherwise the amplitude terms $n_i$ become erratic functions of density and are no more smooth functions. The mathematical forms are certainly not unique but there are strong constraints on them. This is a fundamental difference with the classical fitting approach of the free energy function where there is no mathematical constraint between the different terms that are summed.

The 3 terms of the residual part of $C_V$ are now clarified.

- The first term $\tilde{c}^{\,\text{r}}_{V,\text{reg}}$ which is a simple power law, is called “**regular**”. It shows no singularity and can be calculated for temperature from $T_{\text{t}}$ (triple point temperature) up to infinity. When $T \rightarrow 0$ K following isochors, this term can be approximated by:

$$\tilde{c}^{\,\text{r}}_{V,\text{reg}}(\rho, T \rightarrow 0\text{K}) \cong \frac{3}{2} n_{\text{reg}}(\rho)\, \lambda \frac{T}{T_c} \qquad (3),$$

which tends towards zero as a linear law.

For $T >> T_c/\lambda$, $\tilde{c}^{\,r}_{V,\text{reg}}$ reduces to:

$$\tilde{c}^{\,r}_{V,\mathrm{reg}}(\rho, T >> T_c/\lambda) \cong \frac{3}{2} n_{\mathrm{reg}}(\rho) \left(\frac{T}{T_c}\right)^{m(\rho)-1} \tag{4}$$

The characteristic temperature $T_c/\lambda$ = 22.0 K is not the Debye temperature of argon which is equal to 85 K. This new characteristic temperature was chosen to minimize the relative error on $C_V$ on the saturated vapor pressure curve so that ***the term containing $\lambda T$ / $T_c$ becomes important only for temperatures smaller than the triple point temperature*** (i.e. for $T << T_t$ with $\lambda = 6.8494$).

- The second term $\tilde{c}^{\,r}_{V,\mathrm{nonreg}}$ called "**non regular**", presents an asymptote for $T = T_{\mathrm{div}}(\rho)$ (i.e. $C_V$ is infinite for this value of temperature). This term is only significant near the liquid-vapor coexistence region. We can also note that the divergence is weak.
- The third term $\tilde{c}^{\,r}_{V,\mathrm{crit}}$ is important only in a small region around the critical point. This term allows us to reproduce the very sharp evolution of $C_V$ very close to the critical point. It can be understood as the macroscopic contribution of the critical fluctuations. This term plays the same mathematical role as the contribution of the four last terms in the second derivative with temperature of the residual free energy in the TSW model.

We have pointed out that the regular term $\tilde{c}^{\,r}_{V,\mathrm{reg}}$ tends to zero when $T$ tends to zero. It will be shown in section 3.4 that for $T < T_{\mathrm{div}}$, the non-regular and critical terms have also a limit equal to zero when $T \to 0$ K. Hence, $\tilde{c}^{\,r}_{V}(\rho, T) \to 0$ if $T \to 0$; this result is in agreement with the third law of thermodynamics (Nernst-Planck assumption). Since $\tilde{c}_V = \tilde{c}^{\,0}_V + \tilde{c}^{\,r}_V$, this law imposes $\tilde{c}_V \to 0$ if $T \to 0$ and then $\tilde{c}^{\,o}_V \to 0$ if $T \to 0$. To reach this result, $\tilde{c}^{\,o}_V$ is rewritten on the following form:

$$\tilde{c}^{\,o}_V(T) = \frac{3}{2}\left(1 - \exp\left(-\lambda_0 \frac{T}{T_c}\right)\right) \tag{5}$$

where $\lambda_0 = 18.2121$.

In the expression of $\tilde{c}^{\,r}_V(\rho, T)$, all coefficients depend on density $\rho$ in the following way:

$$n_{\mathrm{reg}}(\rho) = \alpha_{\mathrm{reg},1}\left(\frac{\rho}{\rho + \rho_{\mathrm{t,Liq}}}\right)^{\varepsilon_{\mathrm{reg,1a}}} \exp\left(-\left(\frac{\rho}{\rho_{\mathrm{t,Gas}}}\right)^{\varepsilon_{\mathrm{reg,1b}}}\right) + \alpha_{\mathrm{reg},2}\left(\frac{\rho}{\rho_{\mathrm{t,Liq}}}\right)^{\varepsilon_{\mathrm{reg,2a}}}\left\{1 - \exp\left(-\left(\frac{\rho}{\rho_{\mathrm{reg,Ronc}}}\right)^{-\varepsilon_{\mathrm{reg,2b}}}\right)\right\} \tag{6}$$

$$m(\rho) = -\alpha_{\mathrm{m},1} + \alpha_{\mathrm{m},2}\exp\left(-\left(\frac{\rho}{\rho_{\mathrm{t,Liq}}}\right)^{3/2}\right) + \alpha_{\mathrm{m},3}\left(\frac{\rho}{\rho_c}\right)^{3/2}\exp\left(-\frac{\rho}{\rho_c}\right) - \alpha_{m,4}\ln\left(\frac{\rho}{\rho_c}\right) + m_{\mathrm{Ronc}}(\rho) \tag{7}$$

$$m_{\text{Ronc}}(\rho)=\begin{cases}\left[\alpha_{m,1}+\alpha_{m,4}\ln\left(\dfrac{\rho}{\rho_c}\right)\right]\left(1+\dfrac{\rho}{\rho_{\text{m,Ronc}}}\right)^{\varepsilon_{m,5a}}\exp\left(-\exp\left(\left(\dfrac{\rho_{\text{m,Ronc}}}{\rho}\right)^{\varepsilon_{m,5b}}\right)\right) & \text{for } \rho\geq\dfrac{M}{12.9}\text{ g/cm}^3\\ 0 \quad \text{otherwise} & \end{cases} \tag{8}$$

$$n_{\text{nonreg}}(\rho)=\begin{cases}\alpha_{\text{nonreg,1}}\left(\dfrac{\rho}{\rho_{\text{t,Gas}}}\right)^{\varepsilon_{\text{nonreg,1a}}}\exp\left(-\left(\dfrac{\rho_{\text{t,Liq}}}{\rho_{\text{t,Gas}}}\right)^{\varepsilon_{\text{nonreg,1b}}}\left(\dfrac{\rho}{\rho_{\text{t,Liq}}-\rho}\right)^{\varepsilon_{\text{nonreg,1b}}}\right) & \\ \quad+\alpha_{\text{nonreg,2}}\left(\dfrac{\rho}{\rho_{\text{t,Gas}}}\right)^{\varepsilon_{\text{nonreg,2a}}}\exp\left(-\left(\dfrac{\rho_{\text{t,Liq}}}{\rho_{\text{t,Gas}}}\right)^{\varepsilon_{\text{nonreg,2b}}}\left(\dfrac{\rho}{\rho_{\text{t,Liq}}-\rho}\right)^{\varepsilon_{\text{nonreg,2b}}}\right) & \text{for } \rho\leq\rho_{\text{t,Liq}}\\ 0 \quad \text{otherwise} & \end{cases} \tag{9}$$

$$T_{\text{div}}(\rho)=\alpha_{\text{div,1}}\left(\frac{\rho}{\rho_c}\right)^{\varepsilon_{\text{div,1a}}}\exp\left(-\left(\frac{\rho}{\rho_c}\right)^{\varepsilon_{\text{div,1b}}}\right)+\alpha_{\text{div,2}}\left(\frac{\rho}{\rho_{\text{t,Liq}}}\right)^{\varepsilon_{\text{div,2a}}}\exp\left(-\left(\frac{\rho}{\rho_{\text{t,Liq}}}\right)^{\varepsilon_{\text{div,2b}}}\right) \tag{10}$$

$$n_{\text{crit}}(\rho)=\alpha_{\text{crit,a}}\left(\frac{\rho}{\rho_c}\right)^{\varepsilon_{\text{crit,a}}}\exp\left(-\left(\left(\alpha_{\text{crit,b}}\frac{\rho-\rho_c}{\rho_c}\right)^2\right)^{\varepsilon_{\text{crit,b}}}\right) \tag{11}$$

$$\varepsilon_{\text{crit}}(\rho)=\varepsilon_{\text{crit,c}}+\varepsilon_{\text{crit,d}}\exp\left(-\left(\varepsilon_{\text{crit,e}}\frac{\rho-\rho_{\text{crit,a}}}{\rho_{\text{crit,a}}}\right)^2\right)+\varepsilon_{\text{crit,f}}\exp\left(-\left(\varepsilon_{\text{crit,g}}\frac{\rho-\rho_{\text{crit,b}}}{\rho_{\text{crit,b}}}\right)^2\right) \tag{12}$$

where $\varepsilon_i$ are exponents and $\alpha_i$ characteristic coefficients. Table 1 lists the values of these parameters.

Note that the function $m(\rho)$ is decomposed into two parts so that Eq. (8) can represent Ronchi data at very high density (i.e. for $\rho\geq\dfrac{M}{12.9}$ g/cm$^3$). Indeed, the variations imposed by Ronchi data are too complex to be taken into account by a single function.

Now, some explanations will be given on the properties of these coefficients. Most of them involve the three characteristic densities of argon:

- the density $\rho_{\text{t,Liq}}$ of liquid at the triple point,
- the density $\rho_{\text{t,Gas}}$ of gas at the triple point,
- the critical density $\rho_c$.

Moreover, two other characteristic densities, $\rho_{\text{reg,Ronc}}$ and $\rho_{\text{m,Ronc}}$, have to be added in view to correctly fit the data of Ronchi at very high densities. All values of these characteristic densities are given in Table 2.

Obviously, other mathematical forms for Eqs. (6) to (12) could be used, but the proposed equations are the simplest ones that have been found and that lead to an accurate fitting of the whole data set. The consequence of this representation will be seen in section 3.1.

The density dependence of the $n_i$ coefficients are shown on Fig. 2. Each coefficient is equal to zero when $\rho\to 0$ and $\rho\to\infty$ and get through a maximum in-between (remark: the maximum of $n_{\text{reg}}$ really occurs but outside the range of density shown on Fig. 2).

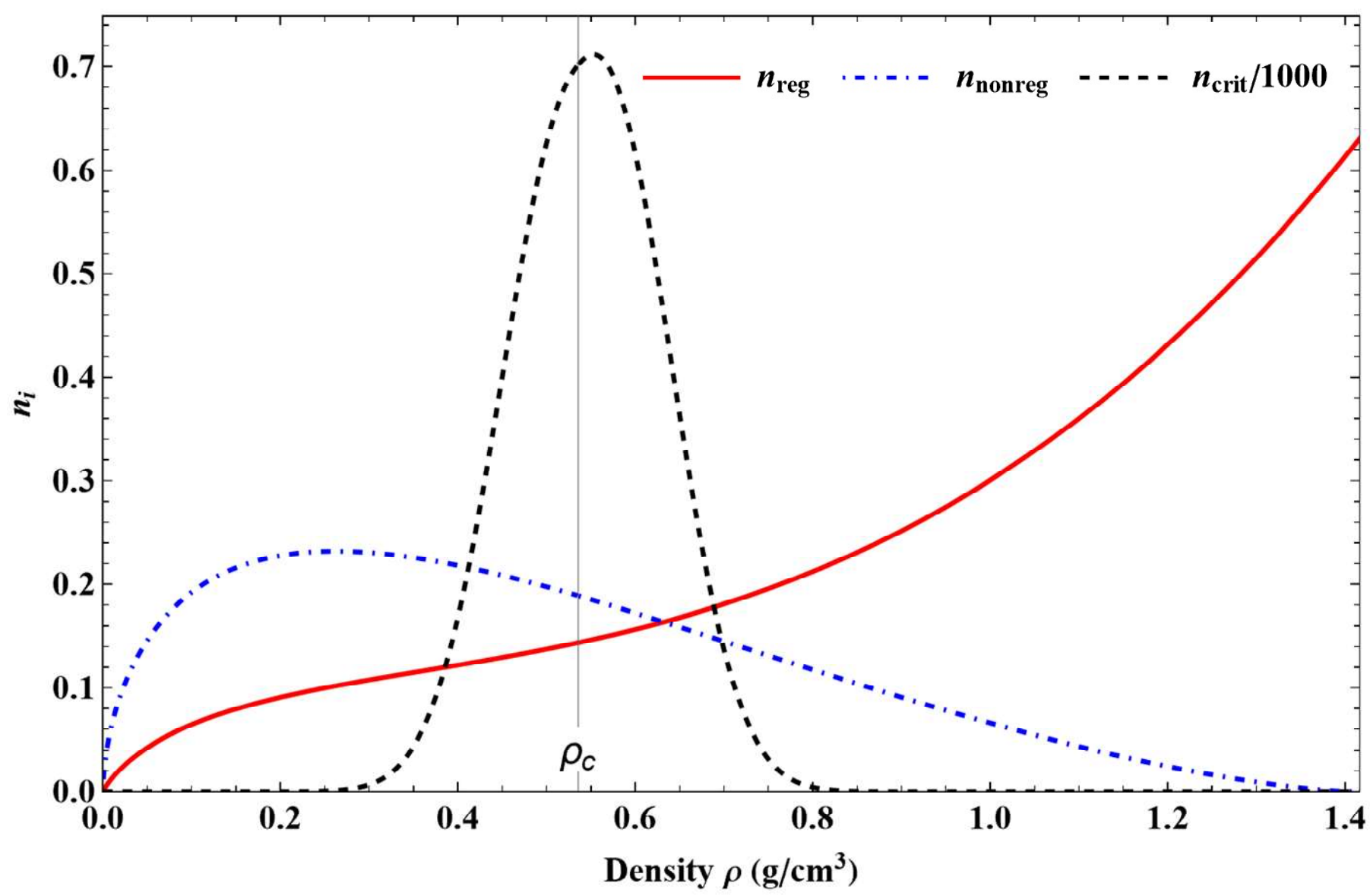

**Fig. 2:** Variations with density of the functions $n_{\mathrm{reg}}(\rho)$ (red curve), $n_{\mathrm{nonreg}}(\rho)$ (blue dot-dashed curve) and $n_{\mathrm{crit}}(\rho)$ (black dashed curve) between $\rho = 0$ and $\rho = \rho_{\mathrm{t,Liq}}$.

The density dependence of exponent $m$ is shown in Fig. 3. This coefficient is always strictly smaller than one and it tends to $-\infty$ when $\rho \to 0$ and $\rho \to \infty$. Then, there are two density values for which $m = 0$. This means that, for the region where $T >> T_c/\lambda$, $\tilde{c}^{\,\mathrm{r}}_{V,\mathrm{reg}}$ is always decreasing along isochors when the temperature is increasing.

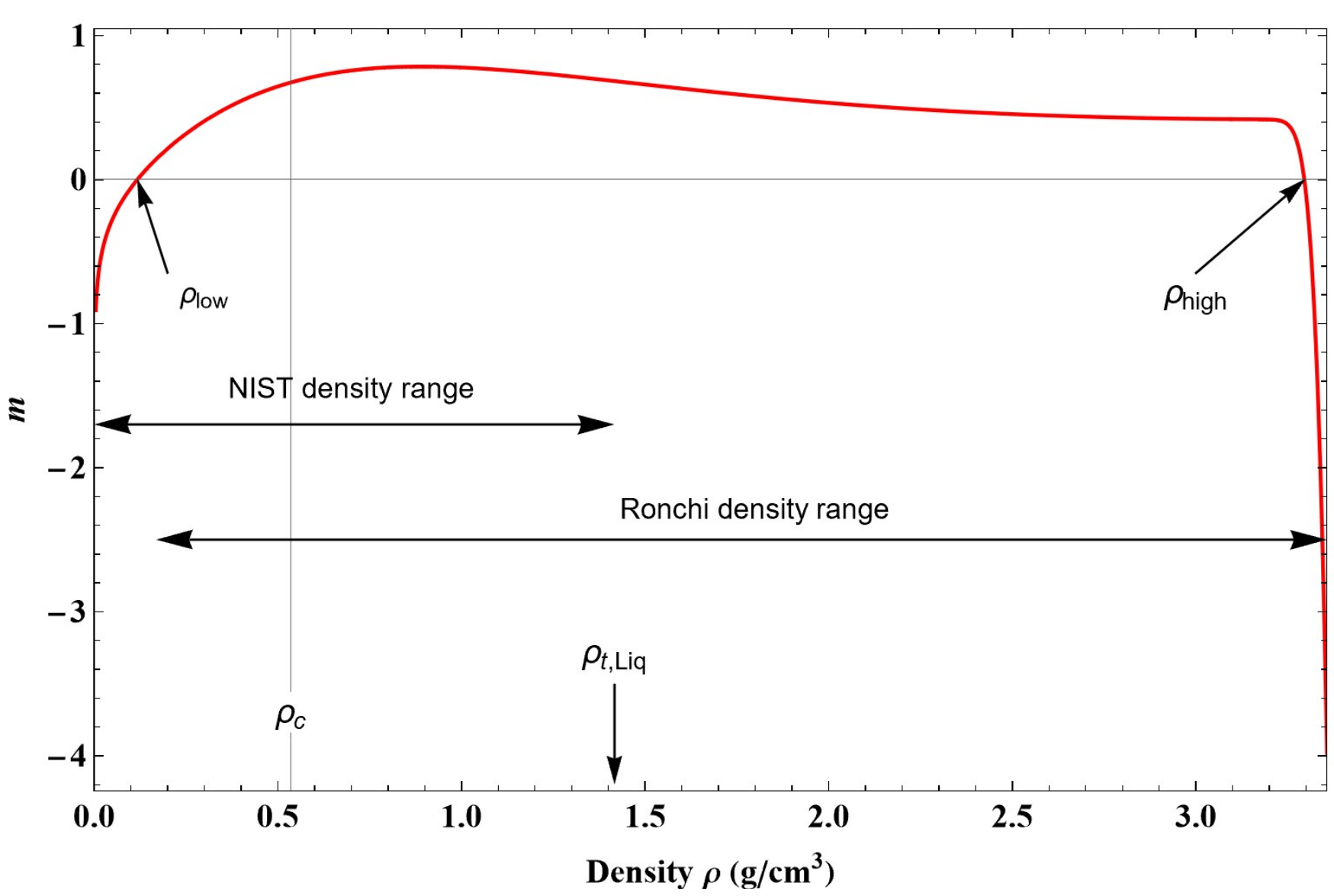

**Fig. 3:** Variation with density of the function $m(\rho)$ from $\rho_{\mathrm{t,Gas}}$ to $\rho_{\mathrm{max,Ronc}}$.

The characteristic temperature $T_{\mathrm{div}}$ as a function of density defines a curve $T_{\mathrm{div}}(\rho)$ which lies entirely inside the coexistence vapor-liquid region defined by $T_{\mathrm{sat}}(\rho)$ (see Fig. 4). For a first order phase transition the divergence of $C_V$ must occurs on the spinodal curve (i.e. loci of thermodynamic mechanical instability), corresponding to:

$$\left(\frac{\partial P}{\partial V}\right)_T = 0 \tag{13}$$

Because no experimental data of the spinodal curve can be found in all the density range from $\rho_{\mathrm{t,Gas}}$ to $\rho_{\mathrm{t,Liq}}$, $T_{\mathrm{div}}$ was only determined by fitting the data of $C_V$ from NIST. If this set of data is enough accurate and consistent with the $P\rho T$ data set, one should be able to identify $T_{\mathrm{div}}(\rho)$ as the spinodal temperature curve. We will discuss in more detail the results obtained for the spinodal states in section 4.3.3.

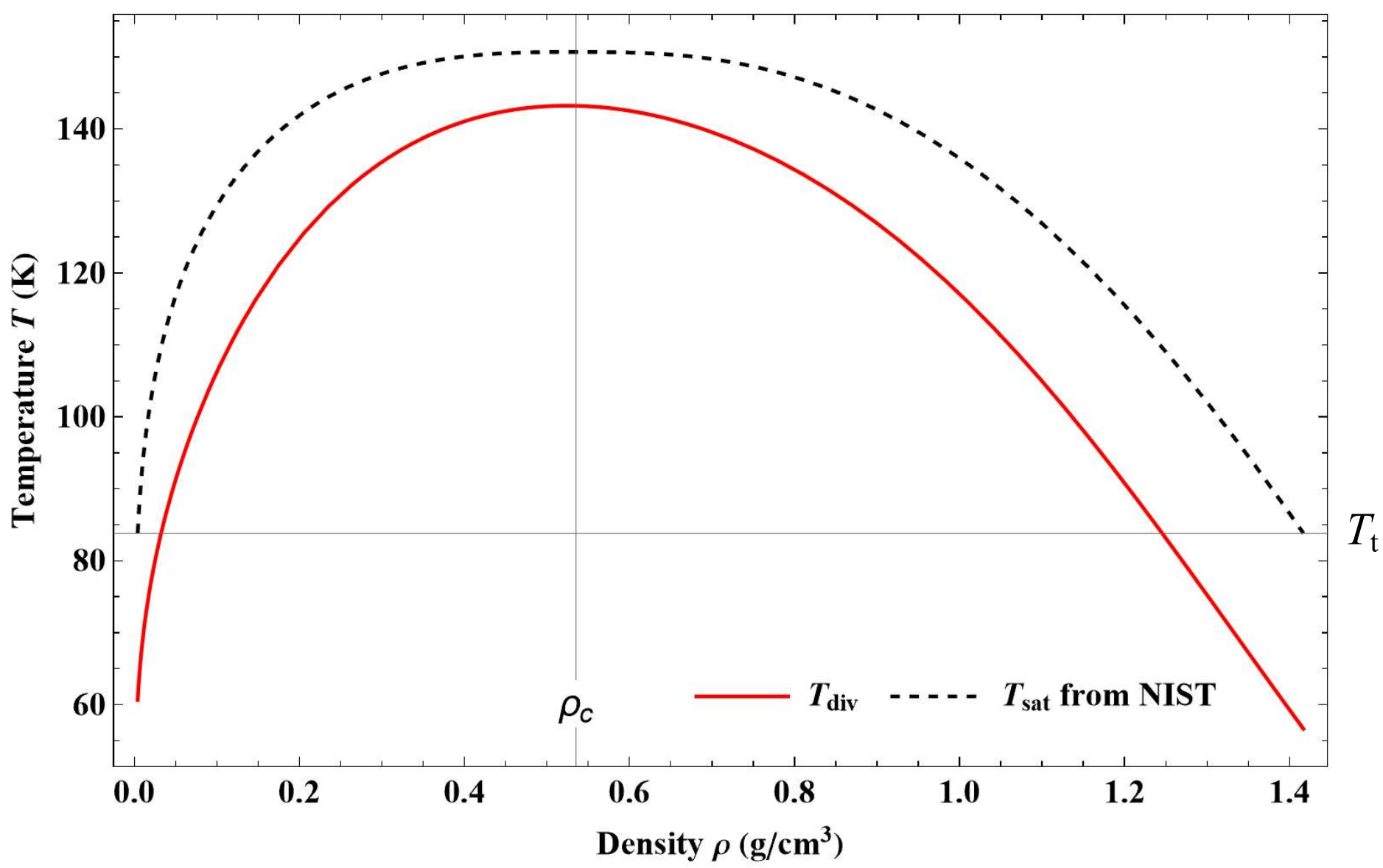

**Fig. 4:** Variations with density of the functions $T_{\mathrm{div}}(\rho)$ (red curve) and $T_{\mathrm{sat}}(\rho)$ (black dashed curve) from $\rho_{\mathrm{t,Gas}}$ to $\rho_{\mathrm{t,Liq}}$. The curve $T_{\mathrm{sat}}(\rho)$ is deduced from the data of NIST (Ref. 5).

From Eq. (2) it is also easy to see that the second thermodynamic instability (i.e. the thermal instability), defined by

$$C_V < 0 \tag{14}$$

will never occur in the present approach contrary to the TSW model.

Consequently, Eq. (2) being valid for $T > T_{\mathrm{div}}$, this relation can be used into the coexistence vapor-liquid region by crossing $T_{\mathrm{sat}}(\rho)$ till to approximately the spinodal curve. No trouble occurs as long as $T > T_{\mathrm{div}}$, though the model is based on a pure fluid description. The fact that there is no discontinuity of $C_V$ when crossing the coexistence curve (excepted at the critical point) is a characteristic of a first order transition. We shall see in section 3.4 how to treat the crossing of the divergence curve defined by $T_{\mathrm{div}}(\rho)$. Finally, it can be noticed that $T_{\mathrm{div}} = 0$ for $\rho = 0$ and $T_{\mathrm{div}} \to 0$ when $\rho \to \infty$, hence, $T_{\mathrm{div}}(\rho)$ shows the right density dependence which allows us to investigate the fluid properties from the gas phase up to the sublimation curve.

The flexibility of the present method is now illustrated on the equation of state for the isochoric heat capacity. So if one wants to represent the data from NIST instead of the data of Ronchi for higher densities than $\rho_{\mathrm{t,Liq}}$, it is only necessary to change the values of the couple ($\rho_{\text{reg,Ronc}}$, $\varepsilon_{\text{reg,2b}}$) and the mathematical form of the exponent function $m(\rho)$. In this case Eq. (7) must be replaced by the following function:

$$
\begin{aligned}
m_{\text{NIST}}(\rho) = {} & \alpha_{\text{m},1} - \alpha_{\text{m},2} \ln\left( \frac{\rho}{\rho_{\text{m},2}} + \frac{\rho_{\text{m},2}}{\rho} \right) + \alpha_{\text{m},3} \left( \frac{\rho}{\rho_{\text{m},3}} \right)^{\varepsilon_{\text{m},3\text{a}}} \exp\left( -\left( \frac{\rho}{\rho_{\text{m},3}} \right)^{\varepsilon_{\text{m},3\text{b}}} \right) \\
& + \alpha_{m,4} \exp\left( \left( \frac{\rho_{\text{m},4\text{b}}}{\rho_{\text{m},4\text{a}}} \right)^{2\varepsilon_{m,4}} \left( \frac{\rho_{\text{m},4\text{a}}}{\rho - 1} \right)^{2\varepsilon_{m,4}} \right) + m_{\text{Extrapol}}(\rho)
\end{aligned}
\qquad \text{(7bis)}
$$

with

$$
m_{\text{Extrapol}}(\rho) = \frac{\rho_{\text{t,Liq}}}{\rho} \left[ -\frac{\alpha_{\text{m},5}}{\varepsilon_{\text{m},5} - 1} \left( \frac{\rho}{\rho_{\text{m},4\text{a}}} \right)^{\varepsilon_{m,5}} + \alpha_{m,6} \left( \frac{\rho}{\rho_{\text{m},4\text{a}}} \right)^{\varepsilon_{m,6}} E_{\varepsilon_{m,6}}\left( \frac{\rho}{\rho_{\text{m},6}} \right) \right]
$$

where $\mathrm{E}_n(z) = \int_1^\infty \frac{e^{-zt}}{t^n} dt$ represents the exponential integral function. The corresponding parameters have the following values:

- Coefficients: $\alpha_{\text{m},1}$=0.48962315, $\alpha_{\text{m},2}$=0.24014465, $\alpha_{\text{m},3}$=1.0932969, $\alpha_{\text{m},4}$=0.08936644, $\alpha_{\text{m},5}$=67.4598, $\alpha_{\text{m},6}$=1331.29.
- Exponents: $\varepsilon_{\text{m},3\text{a}}$=1.56671, $\varepsilon_{\text{m},3\text{b}}$=0.930273, $\varepsilon_{\text{m},4}$=4.785, $\varepsilon_{\text{m},5}$=166.594, $\varepsilon_{\text{m},6}$=5.93118 and $\varepsilon_{\text{reg,2b}} = 5.248961$.
- Characteristic densities in g/cm$^3$: $\rho_{\text{m},2}$=1.35802, $\rho_{\text{m},3}$=0.449618, $\rho_{\text{m},4\text{a}}$=3.30149, $\rho_{\text{m},4\text{b}}$=4.05911, $\rho_{\text{m},6}$=24.5967 and the new value for $\rho_{\text{reg,Ronc}}$ is now equal to 2.22915.

It immediately follows that this new function needs more parameters than for Eq. (7) but the global shape of the function $m_{\text{NIST}}(\rho)$ is very similar to that of Eq. (7) except that this new function has a strong oscillation around the density value $\rho$ = 1.8 g/cm$^3$. This oscillation is needed for a good representation of the data but it is physically difficult to understand. Equation (7bis) is therefore of no practical use compared to Eq. (7), and will not be analyzed further in the following.

**Table 1.** Coefficients and exponents of Eqs. (6) to (12).

| $i$ | $\varepsilon_i$ | $\alpha_i$ |
|---|---|---|
| reg,1 | | 11.23233957 |
| reg,1a | 1.1178177 | |
| reg,1b | 0.23513928 | |
| reg,2 | | 0.53278931 |
| reg,2a | 2.9322362 | |
| reg,2b | 15.5957 | |
| m,1 | | 0.07079238 |
| m,2 | | 0.33623345 |
| m,3 | | 1.3019754 |
| m,4 | | -0.24008716 |
| m,5a | 14.4899 | |
| m,5b | 7.20862 | |
| nonreg,1 | | 0.089409 |
| nonreg,1a | 0.71915 | |
| nonreg,1b | 0.22569 | |
| nonreg,2 | | 0.015481 |
| nonreg,2a | 1.3401 | |
| nonreg,2b | 0.29485 | |
| div,1 | | 102.06515 |
| div,1a | 0.9218165 | |
| div,1b | 1.1328347 | |
| div,2 | | 120.40518 |
| div,2a | 0.12035802 | |
| div,2b | 4.424004 | |
| crit,a | 0.80803 | 701.52 |
| crit,b | 1.134 | 4.27385 |
| crit,c | 1.436786 | |
| crit,d | 123.1335 | |
| crit,e | 2.205614 | |
| crit,f | 26.32662 | |
| crit,g | 4.437711 | |

**Table 2.** Characteristic values of densities of argon and there corresponding molar volumes used in Eqs. (6) to (12).

| $i$ | $\rho_i$ (g/cm$^3$) | $V_i$ (cm$^3$/mole) |
|---|---|---|
| t,Gas | 0.0040546 | 9852.51318 |
| t,Liq | 1.41680 | 28.1959 |
| c | 0.53559 | 74.5857 |
| crit,a | 0.51182 | 78.0502 |
| crit,b | 0.73085 | 54.6589 |
| max,Ronc | 3.35697 | 11.9 |
| reg,Ronc | 3.53159 | 11.3116 |
| m,Ronc | 3.67875 | 10.8591 |
| u,1 | 6.61153 | 6.04217 |
| u,2 | 3.99925 | 9.98884 |
| u,3 | 3.90870 | 10.22026 |
| s,1 | 1.50915 | 26.47047 |
| s,4 | 1.18697 | 33.65528 |
| sRonc,1 | 3.28898 | 12.146 |
| sRonc,2 | 4.31602 | 9.25574 |

## 3 Thermodynamic properties derived from the isochoric heat capacity

Since the Helmholtz free energy versus density and temperature is one of the four basic forms of an equation of state, we focus here on the process for deducing its expression. For this purpose we use the thermodynamic relation:

$$C_V = \left(\frac{\partial U}{\partial T}\right)_V = T\left(\frac{\partial S}{\partial T}\right)_V = -T\left(\frac{\partial^2 F}{\partial T^2}\right)_V \tag{15}$$

Consequently, $F$ can be deduced from: (i) two successive integrations of $C_V$ or, (ii) a single integration of $C_V$ to calculate $U$ and $S$ and then use the thermodynamic relation:

$$F = U - TS \tag{16}$$

with $U(\rho,T) = \int C_V(\rho,T)dT + U_0(\rho) + \text{constant}$, $S(\rho,T) = \int \frac{C_V(\rho,T)}{T}dT + S_0(\rho) + \text{constant}$ and $C_V(\rho,T) = R_{\text{A}}\left(\tilde{c}_V^{\,o}(T) + \tilde{c}_V^{\,r}(\rho,T)\right)$ (given by Eq. 1, 2 and 5).

We chose the second approach accounting that the two integrations to find $U$ and $S$ induced the existence of two arbitrary functions, respectively $U_0(\rho)$ and $S_0(\rho)$, which are simpler to determine than directly finding the arbitrary function for $F$. The later simply writes $F_0(\rho,T) = U_0(\rho) - TS_0(\rho)$. It will be seen in section 3.1 how the two arbitrary functions $U_0(\rho)$ and $S_0(\rho)$ can be determined.

There is no difficulty to find a primitive of $\tilde{c}_V^{\,\text{o}}(T)$ for $U$ or $S$. For the residual part of $C_V$ (see Eq. 2), there is also no difficulty to find a primitive of $\tilde{c}_{V,\text{reg}}^{\,\text{r}}$ and $\tilde{c}_{V,\text{crit}}^{\,\text{r}}$. However, for $\tilde{c}_{V,\text{reg}}^{\,\text{r}}$ when $T \gg T_c/\lambda$, two expressions can be obtained for the primitive of $U$ depending on whether the value of $m(\rho)$ is zero or not, that is to say a power law if $m \neq 0$ and a logarithmic law if $m = 0$. It can be seen on Fig. 3 that there are two values of $\rho$ for which $m = 0$, namely for $\rho_{\text{low}} = 0.11726382$ g/cm$^3$ and $\rho_{\text{high}} = 3.29510771$ g/cm$^3$. To obtain a single expression uniformly valid, the primitive is written as follows:

$$\int \tilde{c}_{V,\text{reg}}^{\,\text{r}}(T \gg T_c/\lambda)dT = \frac{3}{2}n_{\text{reg}}(\rho)T_c\frac{(T/T_c)^{m(\rho)} - 1}{m(\rho)} \tag{17}$$

By using the Hospital's rule, it can be easily verified that: $\lim\limits_{m\to 0}\frac{(T/T_c)^{m(\rho)} - 1}{m(\rho)} = \ln\left(\frac{T}{T_c}\right)$, which corresponds to the right expression for the primitive when $m = 0$.

The same problem occurs for the primitive of $S$, but this time the expression depends whether $m = 1$ or not. For argon, the value $m = 1$ is never reached, but to maintain a general expression, we proceed in the same manner to determine the expression for the primitive of $S$.

For the integration of $C_V$, the only term for which it may be difficult to find a primitive is $\tilde{c}_{V,\text{nonreg}}^{\,\text{r}}$ (see Eq. 2). It could be integrated numerically but a reference state must be chosen; this will be done in section 3.4. To find a primitive it is also possible to perform a series expansion of the term $(1-x)^{-1}$ with $x = T_{\text{div}}/T$. Hence $\tilde{c}_{V,\text{nonreg}}^{\,\text{r}}$ can be written in the following form:

$$\tilde{c}^{\,\mathrm{r}}_{V,\mathrm{nonreg}}(\rho,x)=\frac{3}{2}n_{\mathrm{nonreg}}(\rho)\sum_{k=0}^{\infty}\exp\left(-x^{-3/2}\right)x^{k}\ ,\ \ k\in\mathbb{N} \tag{18}$$

A primitive for each term of the series can be obtained. For a practical calculation the series expansion must be truncated. The convergence is slower as $x$ approaches the unit value, and as a result the number of terms that must be considered increases. An empirical formula for calculating the number of terms required is given below so that the residual error due to truncation is less than 0.1% (except for $x > 0.99$ since the function weakly diverges as $T \rightarrow T_{\mathrm{div}}$)

$$k_{\max}(x)=1+\left\lfloor 400\frac{\exp\left[\exp\left(-8.52\left|1-x\right|\right)-1\right]}{1+16.6\left|1-x\right|}\right\rfloor \tag{19}$$

where $\lfloor\bullet\rfloor$ represents the integer part function.

Finally, the equations for $U$ and $S$ can be written in standard dimensionless form (i.e. an ideal gas part and a residual one) as:

$$\tilde{u}(\rho,T)=\frac{U(\rho,T)}{R_{\mathrm{A}}T}=\frac{3}{2}\left[1+\frac{T_c}{\lambda_0 T}\exp\left(-\frac{\lambda_0 T}{T_c}\right)\right]+\tilde{u}^{*}(\rho,T)+\tilde{u}_0(\rho) \tag{20}$$

with

$$\begin{aligned}\tilde{u}^{*}(\rho,T)=&\frac{3}{2}n_{\mathrm{reg}}(\rho)\frac{T_c}{T}\left\{\frac{\left(\dfrac{T}{T_c}\right)^{m(\rho)}-1}{m(\rho)}+\frac{\lambda^{-m(\rho)}}{2-m(\rho)}\Gamma\left(\frac{m(\rho)}{2-m(\rho)},\left(\frac{\lambda T}{T_c}\right)^{2-m(\rho)}\right)\right\}\\&-n_{\mathrm{nonreg}}(\rho)\frac{T_{\mathrm{div}}(\rho)}{T}\sum_{k=0}^{\infty}\Gamma\left(-\frac{2}{3}(k-1),\left(\frac{T_{\mathrm{div}}(\rho)}{T}\right)^{-3/2}\right)+\frac{3}{2}n_{\mathrm{crit}}(\rho)\frac{\left(T_{\mathrm{div}}(\rho)/T\right)^{\varepsilon_{\mathrm{crit}}(\rho)}}{1-\varepsilon_{\mathrm{crit}}(\rho)}\end{aligned} \tag{21}$$

and

$$\tilde{s}(\rho,T)=\frac{S(\rho,T)}{R_{\mathrm{A}}}=\frac{3}{2}\left[\ln(T)-\mathrm{Ei}\left(-\frac{\lambda_0 T}{T_c}\right)\right]+\tilde{s}^{*}(\rho,T)+\tilde{s}_0(\rho) \tag{22}$$

with

$$\begin{aligned}\tilde{s}^{*}(\rho,T)=&\frac{3}{2}n_{\mathrm{reg}}(\rho)\left\{\frac{\left(\dfrac{T}{T_c}\right)^{m(\rho)-1}-1}{m(\rho)-1}+\frac{\lambda^{1-m(\rho)}}{2-m(\rho)}\Gamma\left(\frac{m(\rho)-1}{2-m(\rho)},\left(\frac{\lambda T}{T_c}\right)^{2-m(\rho)}\right)\right\}\\&-n_{\mathrm{nonreg}}(\rho)\sum_{k=0}^{\infty}\Gamma\left(-\frac{2}{3}k,\left(\frac{T_{\mathrm{div}}(\rho)}{T}\right)^{-3/2}\right)-\frac{3}{2}n_{\mathrm{crit}}(\rho)\,\varepsilon_{\mathrm{crit}}(\rho)^{-1}\left(\frac{T_{\mathrm{div}}(\rho)}{T}\right)^{\varepsilon_{\mathrm{crit}}(\rho)}\end{aligned} \tag{23}$$

where $\Gamma(a,z)=\int_z^{\infty} t^{a-1}\exp(-t)dt$ represents the incomplete gamma function and $\mathrm{Ei}(z)=\int_{-z}^{\infty}\frac{\exp(-t)}{t}dt$ represents the exponential integral function.

In the ideal gas limit, the relations for the internal energy and entropy must be written respectively as:

$$U^{\circ}(T)=\frac{3}{2}R_{\mathrm{A}}T+U_0^{\circ} \tag{24}$$

$$S^{\circ}(\rho,T)=\frac{3}{2}R_{\mathrm{A}}\ln(T)-R_{\mathrm{A}}\ln(\rho)+S_0^{\circ} \tag{25}$$

where $U_0^{\circ}$ and $S_0^{\circ}$ are arbitrary constants. These formulas can be rewritten as follows (using Eq. (5) for $\tilde{c}_V^{\circ}$):

$$U^{\circ}(T)=\frac{3}{2}R_{\mathrm{A}}T\left[1+\frac{T_c}{\lambda_0 T}\exp\left(-\frac{\lambda_0 T}{T_c}\right)\right]+U_0^{\circ} \tag{26}$$

$$S^{\circ}(\rho,T)=\frac{3}{2}R_{\mathrm{A}}\left[\ln(T)-\mathrm{Ei}\left(-\frac{\lambda_0 T}{T_c}\right)\right]-R_{\mathrm{A}}\ln(\rho)+S_0^{\circ} \tag{27}$$

Equations (26) and (27) are used to express the Helmholtz free energy and its various derivatives as shown in Table 3.

The above expressions can now be used to rearrange Eqs. (20) and (22) in order to extract the residual part for the internal energy (i.e. $\tilde{u}(\rho,T)$ minus the ideal gas part) and for the entropy:

$$\tilde{u}^{\mathrm{r}}(\rho,T)=\tilde{u}^{*}(\rho,T)+\tilde{u}_0(\rho)-\varsigma_0\frac{T_c}{T} \quad \text{with} \quad \varsigma_0=\frac{U_0^{\circ}}{R_{\mathrm{A}}T_c} \tag{28}$$

$$\tilde{s}^{\mathrm{r}}(\rho,T)=\tilde{s}^{*}(\rho,T)+\tilde{s}_0(\rho)+\ln(\rho)-\omega_0 \quad \text{with} \quad \omega_0=\frac{S_0^{\circ}}{R_{\mathrm{A}}} \tag{29}$$

where $\varsigma_0$ and $\omega_0$ are two arbitrary constants. In view to fit NIST data, the constant values must be such that: $\varsigma_0=-0.00070133$ and $\omega_0=2.71428$.

**Table 3.** The ideal-gas part $\tilde{a}^\circ$ of the dimensionless Helmholtz free energy function and its derivatives.

$$\tilde{a}^\circ = \frac{3}{2}\left[1 - \ln(T) + \frac{T_c}{\lambda_0 T}\exp\left(-\frac{\lambda_0 T}{T_c}\right) + \mathrm{Ei}\left(-\frac{\lambda_0 T}{T_c}\right)\right] + \ln(\rho) + \varsigma_0 \frac{T_c}{T} - \omega_0$$

$$T_c\left(\frac{\partial \tilde{a}^\circ}{\partial T}\right)_\rho = -\frac{T_c}{T}\left[\frac{3}{2}\left(1 + \frac{T_c}{\lambda_0 T}\exp\left(-\frac{\lambda_0 T}{T_c}\right)\right) + \varsigma_0 \frac{T_c}{T}\right]$$

$$T_c^2\left(\frac{\partial^2 \tilde{a}^\circ}{\partial T^2}\right)_\rho = \frac{3}{2}\left(\frac{T_c}{T}\right)^2\left(1 + \frac{4}{3}\frac{T_c}{T}\varsigma_0\right) + \frac{3}{2}\left(\frac{T_c}{T}\right)^2\left(1 + 2\frac{T_c}{\lambda_0 T}\right)\exp\left(-\frac{\lambda_0 T}{T_c}\right)$$

$$\rho_c\left(\frac{\partial \tilde{a}^\circ}{\partial \rho}\right)_T = \frac{\rho_c}{\rho}$$

$$\rho_c^2\left(\frac{\partial^2 \tilde{a}^\circ}{\partial \rho^2}\right)_T = -\left(\frac{\rho_c}{\rho}\right)^2$$

$$\left(\frac{\partial^2 \tilde{a}^\circ}{\partial \rho \partial T}\right) = 0$$

### 3.1. Determination of the arbitrary functions for the internal energy and entropy

The two arbitrary functions $U_0(\rho)$ and $S_0(\rho)$ can be determined in two different ways. One way is to make the difference between previously published data of $U$ (or $S$) and the $U$ (or $S$) values calculated by the present modeling and then finding a function ($U_0(\rho)$ or $S_0(\rho)$) which best fits this difference. However, this way could be problematic as $U$ and $S$ are not measured quantities and depend on a chosen reference state. Another way is to use a new experimentally measured quantity namely pressure $P$, and by using the relation:

$$P = -\left(\frac{\partial F}{\partial V}\right)_T = \rho^2\left(\frac{\partial F}{\partial \rho}\right)_T \tag{30}$$

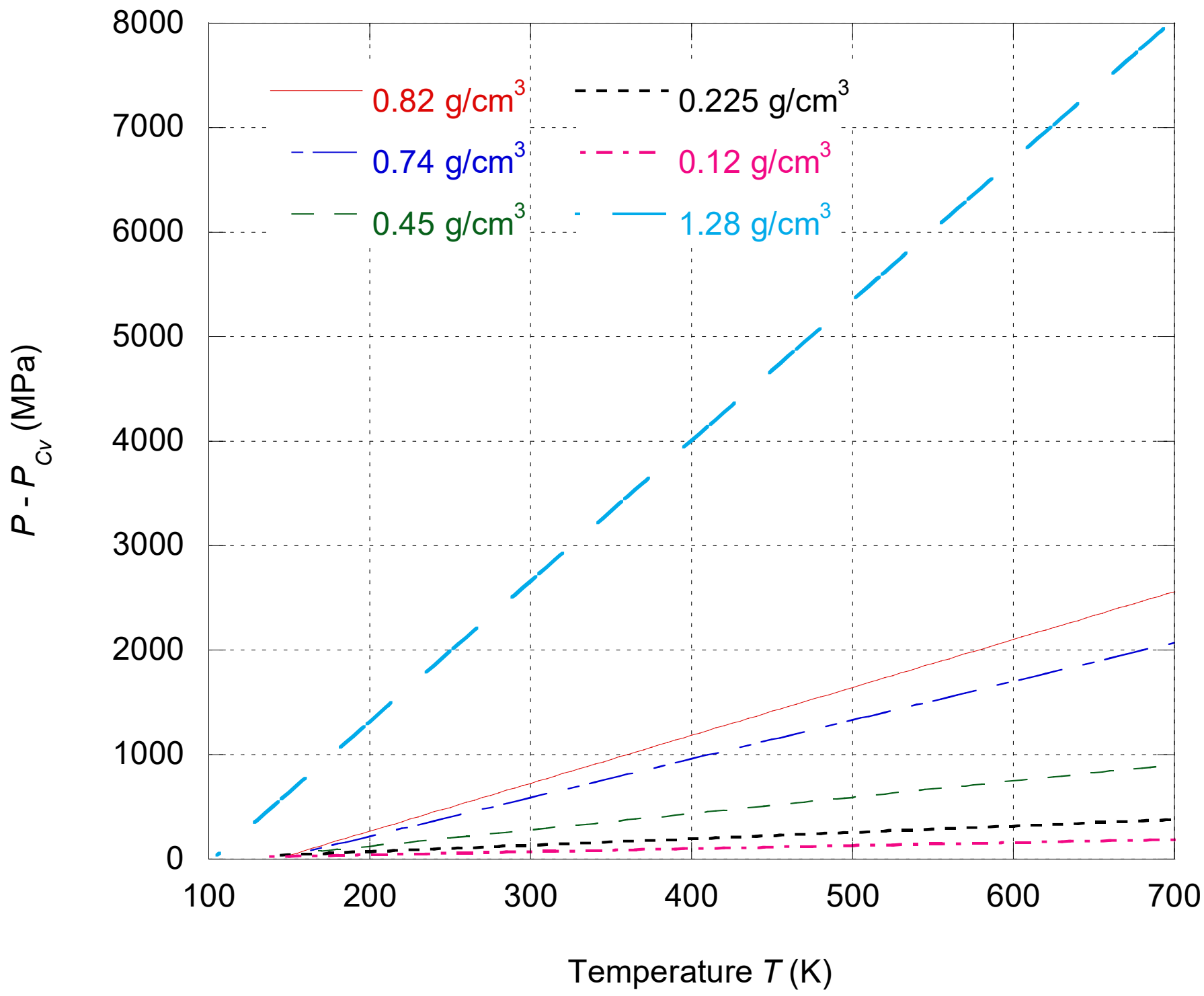

**Fig. 5:** Variation of $P - P_{C_V}$ as function of temperature for different isochors such that $\rho < \rho_{t,\mathrm{Liq}}$.

Along isochors, from Eqs. (15) and (16) it is deduced that $F = \int C_V dT + U_0 - T\left(\int \frac{C_V}{T} dT + S_0\right)$ and, its derivative versus $V$ (or $\rho$) gives: $P - P_{C_V} = U_0' - TS_0'$, with $P_{C_V} = \frac{\partial}{\partial V}\left(-\int C_V dT + T\int \frac{C_V}{T} dT\right)$, $U_0' = \frac{\partial}{\partial V}(U_0)$ and $S_0' = \frac{\partial}{\partial V}(S_0)$. Here $P_{C_V}$ is calculated from $C_V$ values given by Eq. (1). For a given isochor of density $\rho$, the difference $P - P_{C_V}$ must be a straight line (of slope $S_0'$ and ordinate at origin $U_0'$) if the $C_V$ values are well predicted by Eq. (1). This is effectively observed on Fig. 5 which displays $P - P_{C_V}$ versus $T$ on different isochors (i.e. the quasi-infinite curvature is a consequence of the extremely good representation of $C_V$ along isochors). The best and simplest functions that represent $U_0'(\rho)$ and $S_0'(\rho)$ are:

$$\hat{u}_0'(\rho) = \frac{U_0'(\rho)}{\frac{3}{2}R_{\mathrm{A}}T_c} = \frac{\alpha_{\mathrm{u},0}}{\rho_{\mathrm{t,Liq}}} - \frac{1}{\rho}\left(\frac{\rho_{\mathrm{t,Liq}}}{\rho}\right)\left[\alpha_{\mathrm{u},1}\exp\left(-\frac{\rho_{\mathrm{u},1}}{\rho}\right) - \alpha_{\mathrm{u},2}\left(\frac{\rho_{\mathrm{u},2}}{\rho_{\mathrm{t,Liq}}}\right)^{\varepsilon_{\mathrm{u},2}}\left(\frac{\rho}{\rho_{\mathrm{u},2}-\rho}\right)^{\varepsilon_{\mathrm{u},2}}\right.$$
$$\left. + \alpha_{\mathrm{u},3}\left(\frac{\rho_{\mathrm{u},3}}{\rho_{\mathrm{t,Liq}}}\right)^{\varepsilon_{\mathrm{u},3}}\left(\frac{\rho}{\rho_{\mathrm{u},3}-\rho}\right)^{\varepsilon_{\mathrm{u},3}} + \alpha_{\mathrm{u},4}\left(\frac{\rho}{\rho_{\mathrm{t,Liq}}}\right)^{\varepsilon_{\mathrm{u},4}} - \alpha_{\mathrm{u},5}\left(\frac{\rho}{\rho_{\mathrm{t,Liq}}}\right)^{3}\exp\left(-\left(\frac{\rho_{\mathrm{u},1}}{\rho}\right)^{2}\right) + \alpha_{\mathrm{u},6}\left(\frac{\rho}{\rho+\rho_{\mathrm{c}}}\right)^{\varepsilon_{\mathrm{u},6}}\right] \quad (31)$$

and

$$\tilde{s}_0'(\rho) = \frac{S_0'(\rho)}{R_{\mathrm{A}}} = -\frac{1}{\rho}\left[1 + \alpha_{\mathrm{s},1}\left(\frac{\rho}{\rho_{\mathrm{t,Liq}}}\right)^{\varepsilon_{\mathrm{s},1}-1}\left(\frac{\rho_{\mathrm{s},1}}{\rho+\rho_{\mathrm{s},1}}\right)^{\varepsilon_{\mathrm{s},1}} - \alpha_{\mathrm{s},2}\left(\frac{\rho}{\rho_{\mathrm{t,Liq}}}\right)^{\varepsilon_{\mathrm{s},2}-1}\ln\left(\frac{\rho}{\rho_{\mathrm{t,Liq}}}\right) + \alpha_{\mathrm{s},3}\left(\frac{\rho}{\rho_{\mathrm{t,Liq}}}\right)^{\varepsilon_{\mathrm{s},3}-1}\ln\left(\frac{\rho}{\rho_{\mathrm{t,Liq}}}\right)\right.$$
$$\left. + \alpha_{\mathrm{s},4}\left(\frac{\rho}{\rho_{\mathrm{t,Liq}}}\right)^{\varepsilon_{\mathrm{s},4}-1}\exp\left(-\frac{\rho}{\rho_{\mathrm{s},4}}\right) + \alpha_{\mathrm{s},5}\left(\frac{\rho}{\rho_{\mathrm{t,Liq}}}\right)\exp\left(-\left(\frac{\rho}{\rho_{\mathrm{s},1}}\right)^{\varepsilon_{\mathrm{s},5}}\right) + \alpha_{\mathrm{s},6}\left(\frac{\rho}{\rho_{\mathrm{t,Liq}}}\right)\exp\left(-\left(\frac{\rho-\rho_{\mathrm{s},4}}{\rho_c}\right)^{2}\right)\right] \quad (32)$$
$$+ \tilde{s}'_{0,\mathrm{Ronc}}(\rho)$$

with

$$\tilde{s}'_{0,\mathrm{Ronc}}(\rho) = \frac{1}{\rho_{\mathrm{sRonc},1}}\left[\alpha_{\mathrm{sRonc},1}\left(\frac{\rho}{\rho_{\mathrm{sRonc},1}}\right)^{\varepsilon_{\mathrm{sRonc},1}} - \alpha_{\mathrm{sRonc},2}\left(\frac{\rho}{\rho_{\mathrm{sRonc},1}}\right)\exp\left(-\frac{\rho_{\mathrm{sRonc},2}}{\rho}\right)\right]\left[1-\exp\left(-\left(\frac{\rho}{\rho_{\mathrm{c}}}\right)^{2}\right)\right] \quad (33)$$

Before continuing, it is worth noting that $\lim_{\rho\to 0}\frac{\rho U_0'(\rho)}{\frac{3}{2}R_{\mathrm{A}}T_c} = 0$ and $\lim_{\rho\to 0}\frac{\rho S_0'(\rho)}{R_{\mathrm{A}}} = -1$.

A primitive of expressions (31) and (32) leads to the functions $U_0(\rho)$ and $S_0(\rho)$ such that:

$$\hat{u}_0(\rho) = \frac{U_0(\rho)}{\frac{3}{2}R_{\mathrm{A}}T_c} = \alpha_{\mathrm{u},0}\frac{\rho}{\rho_{\mathrm{t,Liq}}} - \frac{\alpha_{\mathrm{u},1}}{\varepsilon_{\mathrm{u},1}-1}\left(\frac{\rho_{\mathrm{t,Liq}}}{\rho_{\mathrm{u},1}}\right)\exp\left(-\frac{\rho_{\mathrm{u},1}}{\rho}\right) + \frac{\alpha_{\mathrm{u},2}}{\varepsilon_{\mathrm{u},2}-1}\left(\frac{\rho_{\mathrm{u},2}}{\rho_{\mathrm{t,Liq}}}\right)^{\varepsilon_{\mathrm{u},2}-1}\left(\frac{\rho}{\rho_{\mathrm{u},2}-\rho}\right)^{\varepsilon_{\mathrm{u},2}-1}$$
$$- \frac{\alpha_{\mathrm{u},3}}{\varepsilon_{\mathrm{u},3}-1}\left(\frac{\rho_{\mathrm{u},3}}{\rho_{\mathrm{t,Liq}}}\right)^{\varepsilon_{\mathrm{u},3}-1}\left(\frac{\rho}{\rho_{\mathrm{u},3}-\rho}\right)^{\varepsilon_{\mathrm{u},3}-1} - \frac{\alpha_{\mathrm{u},4}}{\varepsilon_{\mathrm{u},4}-1}\left(\frac{\rho}{\rho_{\mathrm{t,Liq}}}\right)^{\varepsilon_{\mathrm{u},4}-1} \quad (34)$$
$$+ \frac{1}{2}\alpha_{\mathrm{u},5}\left(\frac{\rho}{\rho_{\mathrm{t,Liq}}}\right)^{2}\left[\exp\left(-\left(\frac{\rho_{\mathrm{u},1}}{\rho}\right)^{2}\right) - \left(\frac{\rho_{\mathrm{u},1}}{\rho}\right)^{2}\Gamma\left(0,\left(\frac{\rho_{\mathrm{u},1}}{\rho}\right)^{2}\right)\right]$$
$$- \frac{\alpha_{\mathrm{u},6}}{\varepsilon_{\mathrm{u},6}-1}\left(\frac{\rho_{\mathrm{t,Liq}}}{\rho_{\mathrm{c}}}\right)\left(\frac{\rho}{\rho+\rho_{\mathrm{c}}}\right)^{\varepsilon_{\mathrm{u},6}-1}$$

and

$$\tilde{s}_0(\rho) = \frac{S_0(\rho)}{R_{\mathrm{A}}} = -\ln\left(\frac{\rho}{\rho_{\mathrm{t,Liq}}}\right) - \frac{\alpha_{\mathrm{s},1}}{\varepsilon_{\mathrm{s},1}-1}\left(\frac{\rho}{\rho_{\mathrm{t,Liq}}}\right)^{\varepsilon_{\mathrm{s},1}-1}\left(\frac{\rho_{\mathrm{s},1}}{\rho+\rho_{\mathrm{s},1}}\right)^{\varepsilon_{\mathrm{s},1}-1}$$
$$+\frac{\alpha_{\mathrm{s},2}}{\left(\varepsilon_{\mathrm{s},2}-1\right)^2}\left(\frac{\rho}{\rho_{\mathrm{t,Liq}}}\right)^{\varepsilon_{\mathrm{s},2}-1}\left[-1+\left(\varepsilon_{\mathrm{s},2}-1\right)\ln\left(\frac{\rho}{\rho_{\mathrm{t,Liq}}}\right)\right] \tag{35}$$
$$-\frac{\alpha_{\mathrm{s},3}}{\left(\varepsilon_{\mathrm{s},3}-1\right)^2}\left(\frac{\rho}{\rho_{\mathrm{t,Liq}}}\right)^{\varepsilon_{\mathrm{s},3}-1}\left[-1+\left(\varepsilon_{\mathrm{s},3}-1\right)\ln\left(\frac{\rho}{\rho_{\mathrm{t,Liq}}}\right)\right]-\alpha_{\mathrm{s},4}\left(\frac{\rho}{\rho_{\mathrm{t,Liq}}}\right)^{\varepsilon_{\mathrm{s},4}-1}\mathrm{E}_{2-\varepsilon_{\mathrm{s},4}}\left(\frac{\rho}{\rho_{\mathrm{s},4}}\right)$$
$$+\frac{\alpha_{\mathrm{s},5}}{\varepsilon_{\mathrm{s},5}}\left(\frac{\rho}{\rho_{\mathrm{t,Liq}}}\right)\mathrm{E}_{\frac{\varepsilon_{\mathrm{s},4}-1}{\varepsilon_{\mathrm{s},4}}}\left(\left(\frac{\rho}{\rho_{\mathrm{s},1}}\right)^{\varepsilon_{\mathrm{s},5}}\right)-\frac{\sqrt{\pi}}{2}\alpha_{\mathrm{s},6}\left(\frac{\rho_{\mathrm{c}}}{\rho_{\mathrm{t,Liq}}}\right)\mathrm{erf}\left(\frac{\rho-\rho_{\mathrm{s},4}}{\rho_{\mathrm{c}}}\right)+\tilde{s}_{0,\mathrm{Ronc}}(\rho)$$

with

$$\tilde{s}_{0,\mathrm{Ronc}}(\rho) = \frac{\alpha_{\mathrm{sRonc},1}}{2\left(\varepsilon_{\mathrm{sRonc},1}+1\right)}\left(\frac{\rho}{\rho_{\mathrm{sRonc},1}}\right)^{1+\varepsilon_{\mathrm{sRonc},1}}\left[2+\left(1+\varepsilon_{\mathrm{sRonc},1}\right)\mathrm{E}_{\frac{1-\varepsilon_{\mathrm{sRonc},1}}{2}}\left(\left(\frac{\rho}{\rho_{\mathrm{c}}}\right)^2\right)\right]$$
$$-\frac{\alpha_{\mathrm{sRonc},2}}{\rho_{\mathrm{sRonc},1}}\int_0^{\rho}\left(\frac{t}{\rho_{\mathrm{sRonc},1}}\right)\exp\left(-\frac{\rho_{\mathrm{sRonc},2}}{\mathrm{t}}\right)\left(1-\exp\left(-\left(\frac{t}{\rho_c}\right)^2\right)\right)dt \tag{36}$$

where $\mathrm{E}_n(z) = \int_1^{\infty}\frac{e^{-zt}}{t^n}dt$ represents the exponential integral function and erf($x$) represents the error function.

The coefficient and exponent values appearing in these equations are given in Table 4. The density dependence of the terms $\tilde{u}_0$ and $\tilde{s}_0$ are shown on Fig. 6.

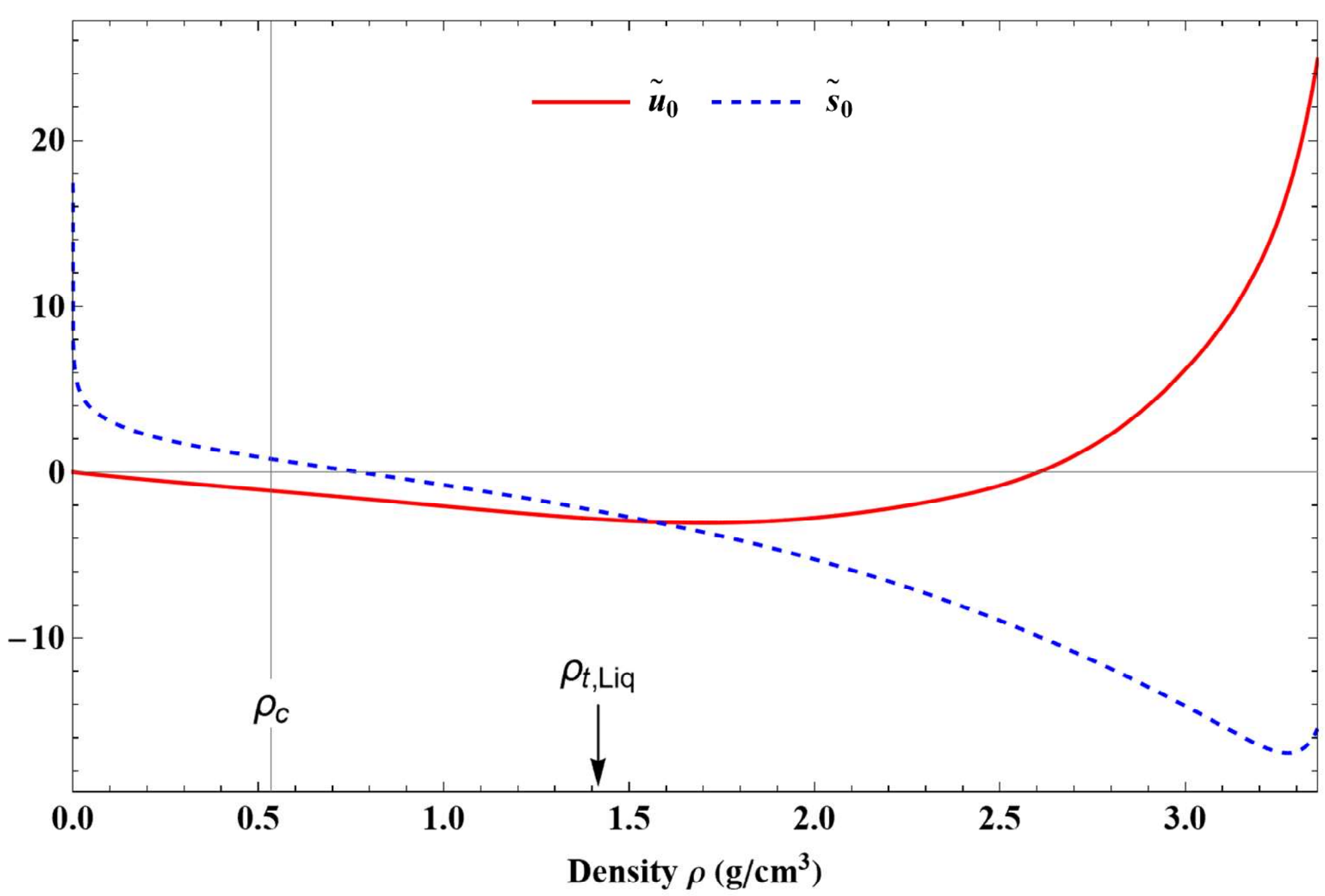

**Fig. 6:** Variations with density of the functions $\tilde{u}_0(\rho)$ (red curve) and $\tilde{s}_0(\rho)$ (blue dashed curve) between $\rho = 0$ and $\rho = \rho_{\mathrm{max,Ronc}}$.

It is important to remember that the two above primitives depend on an arbitrary constant $\alpha_{u0}$ and $\alpha_{s0}$ respectively. Moreover, it must be ensured that the dimensionless form for internal energy is such that:

$$\tilde{u}_0(\rho)=\frac{T_c}{T}\left[\frac{3}{2}\hat{u}_0(\rho)+\alpha_{u0}\right] \tag{37}$$

and for the entropy:

$$\tilde{s}_0^*(\rho)=\tilde{s}_0(\rho)+\ln(\rho) \tag{38}$$

From Eqs. (20) and (22) and Eqs. (34) to (38), the expression for the Helmholtz free energy can be easily deduced. In dimensionless form, this one writes:

$$\tilde{a}(\rho,T)=\frac{F}{R_\mathrm{A}T}=\frac{U}{R_\mathrm{A}T}-\frac{S}{R_\mathrm{A}}=\tilde{u}(\rho,T)-\tilde{s}(\rho,T)=\tilde{a}^\circ(\rho,T)+\tilde{a}^\mathrm{r}(\rho,T) \tag{39}$$

where $\tilde{a}^\circ(\rho,T)$ is given in Table 3 and

$$\tilde{a}^\mathrm{r}(\rho,T)=\underbrace{\tilde{a}^\mathrm{r}_\mathrm{reg}(\rho,T)+\tilde{a}^\mathrm{r}_\mathrm{nonreg}(\rho,T)+\tilde{a}^\mathrm{r}_\mathrm{crit}(\rho,T)}_{\tilde{u}^*-\tilde{s}^*}+\tilde{a}^\mathrm{r}_0(\rho,T) \tag{40}$$

with

$$\tilde{a}^\mathrm{r}_0(\rho,T)=\tilde{u}_0(\rho,T)-\left[\tilde{s}_0^*(\rho)-\omega_0\right]-\varsigma_0\frac{T_c}{T}=\frac{T_c}{T}\left[\frac{3}{2}\hat{u}_0(\rho)+\alpha_{u0}-\varsigma_0\right]-\left[\tilde{s}_0^*(\rho)+\alpha_{s0}-\omega_0\right] \tag{41}$$

For the sake of simplicity, the two constants $\alpha_{u0}$ and $\alpha_{s0}$ can be chosen such that $\alpha_{u0}=\varsigma_0$ and $\alpha_{s0}=\omega_0$. It follows that:

$$\tilde{a}^\mathrm{r}_0(\rho,T)=\frac{3}{2}\frac{T_c}{T}\hat{u}_0(\rho)-\tilde{s}_0^*(\rho) \tag{42}$$

and

$$\tilde{u}^\mathrm{r}(\rho,T)=\tilde{u}^*(\rho,T)+\frac{3}{2}\frac{T_c}{T}\hat{u}_0(\rho) \tag{43}$$

$$\tilde{s}^\mathrm{r}(\rho,T)=\tilde{s}^*(\rho,T)+\tilde{s}_0^*(\rho) \tag{44}$$

It appears that if one wants to represent for higher densities than $\rho_\mathrm{t,Liq}$ the data from NIST instead of the data of Ronchi, it is only necessary to change the mathematical form of Eqs. (31), (32) and (33). For example, the new function for $\hat{u}_0'(\rho)$ can be written with the same mathematical terms as in Eq. (31) but without the two last terms and with different values of the parameters. Therefore, it can be understood that the global shape of the new functions $\hat{u}_0'(\rho)$ and $\tilde{s}_0'(\rho)$ will have very similar variations. This remark shows also that the two data sets discussed above for high densities can be represented by only small variations of the shape of the two derivative functions $\hat{u}_0'(\rho)$ and $\tilde{s}_0'(\rho)$. But once these two functions are determined all the thermodynamic equations of state are known.

Table 5 summarizes the various terms making up the residual part of the Helmholtz free energy.

**Table 4.** Coefficients and exponents for $\hat{u}_0$ and $\tilde{s}_0$.

| $i$ | $\varepsilon_i$ | $\alpha_i$ |
|---|---|---|
| u,0 | | 16.86969325 |
| u,1 | | 71.08169282 |
| u,2 | 2.57795090 | 12.16671437 |
| u,3 | 2.01916041 | 22.41395798 |
| u,4 | 12.94678106 | 0.13352634 |
| u,5 | | 16612.44198645 |
| u,6 | 1.78738624 | 0.04855950 |
| s,1 | 2.23951150 | 11.75732913 |
| s,2 | 3.18259094 | 9.91697667 |
| s,3 | 2.71140252 | 12.27973100 |
| s,4 | 1.55994791 | 0.04075918 |
| s,5 | 21.47158258 | 0.31499626 |
| s,6 | | 0.46391511 |
| sRonc,1 | 62.32164244 | 57.01690712 |
| sRonc,2 | | 187.65045674 |

**Table 5.** Mathematical expressions of the dimensionless terms in the residual part $\tilde{a}^{\mathrm{r}}$ of the Helmholtz free energy for $T \geq T_{\mathrm{div}}$.

$$\tilde{a}^{\mathrm{r}}_{\mathrm{reg}} = \frac{3}{2} n_{\mathrm{reg}} \frac{T_c}{T} \left\{ \frac{\left(\frac{T}{T_c}\right)^m - 1}{m} + \frac{\lambda^{-m}}{2-m} \Gamma\left( \frac{m}{2-m}, \left(\frac{\lambda T}{T_c}\right)^{2-m} \right) \right\}$$

$$- \frac{3}{2} n_{\mathrm{reg}} \left\{ \frac{\left(\frac{T}{T_c}\right)^{m-1} - 1}{m-1} + \frac{\lambda^{1-m}}{2-m} \Gamma\left( \frac{m-1}{2-m}, \left(\frac{\lambda T}{T_c}\right)^{2-m} \right) \right\}$$

$$\tilde{a}^{\mathrm{r}}_{\mathrm{nonreg}} = n_{\mathrm{nonreg}} \sum_{k=0}^{\infty} \left\{ \Gamma\left( -\frac{2}{3}k, \left(\frac{T_{div}}{T}\right)^{-3/2} \right) - \frac{T_{div}}{T} \Gamma\left( -\frac{2}{3}(k-1), \left(\frac{T_{div}}{T}\right)^{-3/2} \right) \right\}$$

$$\tilde{a}^{\mathrm{r}}_{\mathrm{crit}} = \frac{3}{2} n_{\mathrm{crit}} \frac{\left(T_{\mathrm{div}}/T\right)^{\varepsilon_{\mathrm{crit}}}}{\left(1-\varepsilon_{crit}\right)\varepsilon_{crit}}$$

$$\tilde{a}^{\mathrm{r}}_{0} = \frac{3}{2} \frac{T_c}{T} \hat{u}_0(\rho) - \left\{ \tilde{s}_0(\rho) + \ln(\rho) \right\}$$

### 3.2. Analytic expression of the thermal equation of state for $T \geq T_{\mathrm{div}}$

The thermal state equation $P = P(\rho,T)$ which is a fundamental equation to calculate the basic thermal properties of argon, can easily be established using Eqs. (30) and (39) for free energy. The free energy is made up of four terms coming from the residual part of the Helmholtz free energy and a term that represents the behavior of the ideal gas:

$$P(\rho,T) = P_{\mathrm{reg}}(\rho,T) + P_{\mathrm{nonreg}}(\rho,T) + P_{\mathrm{crit}}(\rho,T) + P_0(\rho,T) + \rho R_{\mathrm{A}} T \tag{45}$$

or in dimensionless form

$$Z = \frac{P}{\rho R_{\mathrm{A}} T} = \underbrace{\frac{P_{\mathrm{reg}}(\rho,T)}{\rho R_{\mathrm{A}} T}}_{Z_{\mathrm{reg}}} + \underbrace{\frac{P_{\mathrm{nonreg}}(\rho,T)}{\rho R_{\mathrm{A}} T}}_{Z_{\mathrm{nonreg}}} + \underbrace{\frac{P_{\mathrm{crit}}(\rho,T)}{\rho R_{\mathrm{A}} T}}_{Z_{\mathrm{crit}}} + \underbrace{\frac{P_0(\rho,T)}{\rho R_{\mathrm{A}} T}}_{Z_0} + 1 \tag{46}$$

with

$$Z_{\mathrm{reg}} = \rho\left(\frac{\partial \tilde{a}^{\mathrm{r}}_{\mathrm{reg}}}{\partial \rho}\right)_T$$

$$\begin{aligned} Z_{\mathrm{nonreg}} = \rho\left(\frac{\partial \tilde{a}^{\mathrm{r}}_{\mathrm{nonreg}}}{\partial \rho}\right)_T &= -\rho\, n'_{\mathrm{nonreg}}(\rho) \sum_{k=0}^{\infty}\left[x\,\Gamma\left(-\frac{2}{3}(k-1), x^{-3/2}\right) - \Gamma\left(-\frac{2}{3}k, x^{-3/2}\right)\right] \\ &\quad - n_{\mathrm{nonreg}}(\rho)\frac{\rho\, T'_{\mathrm{div}}(\rho)}{T}\sum_{k=0}^{\infty}\Gamma\left(-\frac{2}{3}(k-1), x^{-3/2}\right) \end{aligned} \tag{47}$$

$$Z_0 = \rho\left(\frac{\partial \tilde{a}^{\mathrm{r}}_0}{\partial \rho}\right)_T + 1 = \rho\left[\frac{3}{2}\frac{T_c}{T}\hat{u}'_0(\rho) - \tilde{s}'_0(\rho)\right] \tag{48}$$

$$\begin{aligned} Z_{\mathrm{crit}} = \rho\left(\frac{\partial \tilde{a}^{\mathrm{r}}_{\mathrm{crit}}}{\partial \rho}\right)_T &= \frac{3}{2}\rho\, n'_{\mathrm{crit}}(\rho)\frac{\left(T_{\mathrm{div}}(\rho)/T\right)^{\varepsilon_{\mathrm{crit}}(\rho)}}{\varepsilon_{\mathrm{crit}}(\rho)\left(1-\varepsilon_{\mathrm{crit}}(\rho)\right)} \\ &\quad + \frac{3}{2}\rho\, n_{\mathrm{crit}}(\rho)\left\{\frac{2\varepsilon_{\mathrm{crit}}(\rho)-1}{\varepsilon_{\mathrm{crit}}(\rho)^2\left(\varepsilon_{\mathrm{crit}}(\rho)-1\right)^2}\varepsilon'_{\mathrm{crit}}(\rho)\left(\frac{T_{\mathrm{div}}(\rho)}{T}\right)^{\varepsilon_{\mathrm{crit}}(\rho)}\right. \\ &\quad \left. - \frac{\left(T_{\mathrm{div}}(\rho)/T\right)^{\varepsilon_{\mathrm{crit}}(\rho)}}{\varepsilon_{\mathrm{crit}}(\rho)\left(\varepsilon_{\mathrm{crit}}(\rho)-1\right)}\left[\varepsilon'_{\mathrm{crit}}(\rho)\ln\left(\frac{T_{\mathrm{div}}(\rho)}{T}\right) + \varepsilon_{\mathrm{crit}}(\rho)\frac{T'_{\mathrm{div}}(\rho)}{T_{\mathrm{div}}(\rho)}\right]\right\} \end{aligned} \tag{49}$$

It is recalled that $x = T_{\mathrm{div}}/T$ in the expression of $Z_{\mathrm{nonreg}}$. $Z_{\mathrm{reg}}(\rho,T)$ displaying too many terms, its expression is given in Appendix A. The expression of the first derivatives of Eqs. (6) to (12) are listed in Appendix B. From the expression of these factors, it is easy to see that $Z_{\mathrm{reg}} = Z_{\mathrm{nonreg}} = Z_{\mathrm{crit}} = 0$ and $Z_0 = 1$ for $\rho \to 0$ and, therefore, $Z \to 1$ for any temperature when density tends to zero. In a certain range of temperature, isotherms intersect the line $Z =$ 1 for $\rho$ values that are not identically zero. As thermodynamic quantities corresponding to $Z$ = 1 are physically important and not easy to find in the literature, they are listed in Table 6.

### 3.3. The liquid-vapor coexistence curve

At a given temperature $T$, vapor pressure and densities of the coexisting phases can be determined by simultaneous resolution of the equations:

$$\frac{P_{\text{sat}}}{\rho_{\sigma l} R_{\text{A}} T} = 1 + \rho_{\sigma l} \left( \frac{\partial \tilde{a}^{\text{r}}_{\sigma l}}{\partial \rho_{\sigma l}} \right)_T = Z(\rho_{\sigma l}, T) \tag{50}$$

$$\frac{P_{\text{sat}}}{\rho_{\sigma v} R_{\text{A}} T} = 1 + \rho_{\sigma v} \left( \frac{\partial \tilde{a}^{\text{r}}_{\sigma v}}{\partial \rho_{\sigma v}} \right)_T = Z(\rho_{\sigma v}, T) \tag{51}$$

$$\frac{P_{\text{sat}}}{R_{\text{A}} T} \left( \frac{1}{\rho_{\sigma v}} - \frac{1}{\rho_{\sigma l}} \right) - \ln\left( \frac{\rho_{\sigma l}}{\rho_{\sigma v}} \right) = \tilde{a}^{\text{r}}(\rho_{\sigma l}, T) - \tilde{a}^{\text{r}}(\rho_{\sigma v}, T) \tag{52}$$

where the indices $\sigma l$ and $\sigma v$ represent the liquid and the vapor coexistence states, respectively.

These equations represent the phase equilibrium conditions, i.e., the equality of pressure, temperature and specific Gibbs energy (Maxwell criterion) in the coexisting phases. The calculated values on the liquid-vapor coexistence curve (vapor pressure, saturated liquid density, saturated vapor density, etc.) are given in Table 12. Approximate formulas for representing the pressure and densities of liquid and vapor as a function of temperature along the coexistence curve are given in Appendix C.

**Table 6.** Thermodynamic properties corresponding to Z = 1 (i.e. ideal curve) deduced from Eq. (46).

| $T/T_c$ | $P/P_c$ | $\rho/\rho_c$ | $C_V/R_A$ | $C_P/R_A$ | $c\Big/\sqrt{\dfrac{R_A T}{M}}$ |
|---|---|---|---|---|---|
| 0.56408 | 5.3565 | 2.7491 | 2.7545 | 5.1425 | 7.1030 |
| 0.62500 | 5.7641 | 2.6699 | 2.6239 | 5.0445 | 6.5008 |
| 0.69444 | 6.1935 | 2.5819 | 2.4987 | 4.9633 | 5.9013 |
| 0.76389 | 6.5850 | 2.4956 | 2.3942 | 4.9091 | 5.3765 |
| 0.83333 | 6.9372 | 2.4100 | 2.3054 | 4.8869 | 4.9093 |
| 0.90278 | 7.2480 | 2.3242 | 2.2290 | 4.8871 | 4.4926 |
| 0.97222 | 7.5154 | 2.2379 | 2.1626 | 4.8993 | 4.1212 |
| 1.0417 | 7.7380 | 2.1505 | 2.1044 | 4.9136 | 3.7909 |
| 1.1111 | 7.9149 | 2.0622 | 2.0532 | 4.9213 | 3.4978 |
| 1.1806 | 8.0462 | 1.9731 | 2.0078 | 4.9165 | 3.2384 |
| 1.2500 | 8.1324 | 1.8835 | 1.9673 | 4.8967 | 3.0090 |
| 1.3194 | 8.1744 | 1.7935 | 1.9309 | 4.8621 | 2.8062 |
| 1.3889 | 8.1729 | 1.7036 | 1.8979 | 4.8148 | 2.6267 |
| 1.4583 | 8.1284 | 1.6136 | 1.8676 | 4.7565 | 2.4674 |
| 1.5278 | 8.0406 | 1.5236 | 1.8395 | 4.6881 | 2.3258 |
| 1.5972 | 7.9091 | 1.4335 | 1.8133 | 4.6093 | 2.1994 |
| 1.6667 | 7.7330 | 1.3432 | 1.7887 | 4.5188 | 2.0865 |
| 1.7361 | 7.5118 | 1.2526 | 1.7653 | 4.4151 | 1.9854 |
| 1.8056 | 7.2453 | 1.1617 | 1.7429 | 4.2976 | 1.8947 |
| 1.8750 | 6.9343 | 1.0707 | 1.7214 | 4.1672 | 1.8133 |
| 1.9444 | 6.5801 | 0.97968 | 1.7005 | 4.0261 | 1.7402 |
| 2.0139 | 6.1845 | 0.88903 | 1.6803 | 3.8775 | 1.6743 |
| 2.0833 | 5.7492 | 0.79890 | 1.6605 | 3.7247 | 1.6150 |
| 2.1528 | 5.2754 | 0.70942 | 1.6412 | 3.5712 | 1.5617 |
| 2.2222 | 4.7635 | 0.62056 | 1.6223 | 3.4196 | 1.5137 |
| 2.2917 | 4.2127 | 0.53218 | 1.6038 | 3.2719 | 1.4707 |
| 2.3611 | 3.6210 | 0.44397 | 1.5857 | 3.1294 | 1.4322 |
| 2.4306 | 2.9847 | 0.35551 | 1.5682 | 2.9927 | 1.3977 |
| 2.5000 | 2.2996 | 0.26630 | 1.5511 | 2.8616 | 1.3667 |
| 2.5500 | 1.7737 | 0.20137 | 1.5389 | 2.7700 | 1.3462 |
| 2.6000 | 1.2211 | 0.13596 | 1.5266 | 2.6803 | 1.3270 |
| 2.6500 | 0.66245 | 0.072369 | 1.5144 | 2.5944 | 1.3095 |
| 2.6900 | 0.29167 | 0.031390 | 1.5061 | 2.5396 | 1.2987 |
| 2.7100 | 0.14787 | 0.015796 | 1.5029 | 2.5193 | 1.2947 |
| 2.7125 | 0.11575 | 0.012354 | 1.5022 | 2.5149 | 1.2939 |

## 3.4. Thermodynamic state inside the liquid-vapor coexistence curve for $T < T_{\text{div}}(\rho)$

The thermodynamic properties of argon have been calculated from the isochoric heat capacity equation $C_V(\rho,T)$. However, the equation being only valid for $T \geq T_{\text{div}}$, a new equation, valid for $T < T_{\text{div}}$ (i.e. inside the coexistence liquid-vapor region), has to be established. This requires solving three mathematical problems.

- First, an expression of $C_V(\rho,T)$ for $T < T_{\text{div}}(\rho)$ has to be found.
- Secondly, to integrate $C_V$, it is required to remove the artificial divergence introduced with the term $\tilde{c}^{\,\text{r}}_{V,\text{nonreg}}$ in order to have a finite value of $C_V$ for $T = T_{\text{div}}(\rho)$.
- Finally, also for the integration of $C_V$, a reference state must be chosen.

The procedure used to develop the modified equation is now presented. It will be shown that this new formulation leads to a better description of the two-phase thermodynamic properties than the polynomial approach.

### *3.4.1. Expression of $C_V$*

The two terms in $C_V(\rho,T)$ creating difficulties are: $\tilde{c}^{\,\text{r}}_{V,\text{nonreg}}$ and $\tilde{c}^{\,\text{r}}_{V,\text{crit}}$. For $T < T_{\text{div}}$, $\tilde{c}^{\,\text{r}}_{V,\text{nonreg}}$ becomes negative which has no physical meaning; indeed, the thermodynamic thermal stability has always to be satisfied. And the term $\tilde{c}^{\,\text{r}}_{V,\text{crit}}$, for $T < T_{\text{div}}$, diverges when $T \to 0$, which has also no physical meaning. The easiest way to solve these problems is to take a symmetric function by changing the variable $T_{\text{div}}/T$ into $T/T_{\text{div}}$, hence it is obtained:

$$\tilde{c}^{\,\text{r}}_{\substack{V,\text{nonreg}\\ \text{inside}}}\left(\rho, T < T_{\text{div}}\right) = \frac{3}{2} n_{\text{nonreg}}(\rho)\, \exp\left[-\left(\frac{T}{T_{\text{div}}(\rho)}\right)^{-3/2}\right] \frac{1}{1-\dfrac{T}{T_{\text{div}}(\rho)}} \tag{53}$$

$$\tilde{c}^{\,\text{r}}_{\substack{V,\text{crit}\\ \text{inside}}}\left(\rho, T < T_{\text{div}}\right) = \frac{3}{2} n_{\text{crit}}(\rho) \left(\frac{T}{T_{\text{div}}(\rho)}\right)^{\varepsilon_{\text{crit}}(\rho)} \tag{54}$$

However, a problem remains as the two equations ($\tilde{c}^{\,\text{r}}_{V,\text{nonreg}}$ and $\tilde{c}^{\,\text{r}}_{\substack{V,\text{nonreg}\\ \text{inside}}}$) become infinite for $T = T_{\text{div}}$. In fact, this is the consequence of the extensive nature of $C_V$. Therefore, this divergence can be removed by introducing explicitly into the equations for $C_V$ a finite number $N_V$ of particles. $N_V$ have to be the largest possible without to be infinite (which is the condition for an extensive property). Then, as these equations must converge for $T = T_{\text{div}}$, the terms $\dfrac{1}{1-\dfrac{T}{T_{\text{div}}}}$ and $\dfrac{1}{1-\dfrac{T_{\text{div}}}{T}}$ have to be corrected so that the two equations must tend to the same finite value for $T = T_{\text{div}}$. The following functions have the required properties:

- $\dfrac{1}{1-\dfrac{T}{T_{\text{div}}}}$ is replaced by $\dfrac{1-N_V^{-\left(1-\frac{T}{T_{\text{div}}}\right)}}{1-\dfrac{T}{T_{\text{div}}}}$, so $\tilde{c}^{\,\text{r}}_{V,\text{nonreg}}$ becomes now

$$\tilde{c}^{\,\mathrm{r}}_{\substack{V,\mathrm{nonreg}\\ \mathrm{outside}}} = \frac{3}{2} n_{\mathrm{nonreg}}(\rho)\, \exp\left[-\left(\frac{T_{\mathrm{div}}(\rho)}{T}\right)^{-3/2}\right] \frac{1-N_V^{-\left(1-\frac{T}{T_{\mathrm{div}}}\right)}}{1-\dfrac{T}{T_{\mathrm{div}}}} \tag{55}$$

- and $\dfrac{1}{1-\dfrac{T_{\mathrm{div}}}{T}}$ is replaced by $\dfrac{1-N_V^{-\left(1-\frac{T_{\mathrm{div}}}{T}\right)}}{1-\dfrac{T_{\mathrm{div}}}{T}}$, so $\tilde{c}^{\,\mathrm{r}}_{\substack{V,\mathrm{nonreg}\\ \mathrm{inside}}}$ becomes now

$$\tilde{c}^{\,\mathrm{r}}_{\substack{V,\mathrm{nonreg}\\ \mathrm{inside}}} = \frac{3}{2} n_{\mathrm{nonreg}}(\rho)\, \exp\left[-\left(\frac{T}{T_{\mathrm{div}}(\rho)}\right)^{-3/2}\right] \frac{1-N_V^{-\left(1-\frac{T_{\mathrm{div}}}{T}\right)}}{1-\dfrac{T_{\mathrm{div}}}{T}} \tag{56}$$

The two corrections tends to $\ln(N_V)$ when $T \to T_{\mathrm{div}}$. $N_V$ may be thought as a quantity representing the number of particles in the volume $V$ for a given experiment, so it can be written as:

$$N_V = fmol\ \mathcal{N}_a \frac{\rho}{\rho_c} \tag{57}$$

where $fmol = 10^{20}$ is an arbitrary constant required to remove the divergence. This means that near the transition, $C_V$ and its related quantities are no longer extensive quantities. This is not surprising since sample size effects are known to exist around the phase transition. Thus, the divergence occurs only for an infinite number of particles. This non-extensive contribution introduced in Eqs. (55) and (56) has also been used to revisit liquid physics in order to explain rheological behavior under a wide variety of thermodynamic and mechanical conditions (Refs. 7 to 10).

Outside the coexistence liquid-vapor region, the percentage deviation between $\tilde{c}^{\,\mathrm{r}}_{V,\mathrm{nonreg}}$ calculated by Eq. (18) and $\tilde{c}^{\,\mathrm{r}}_{V,\mathrm{nonreg}}$ calculated by Eq. (55) is shown on Fig. 7. It is observed that the difference is only significant in the close vicinity of the critical point.

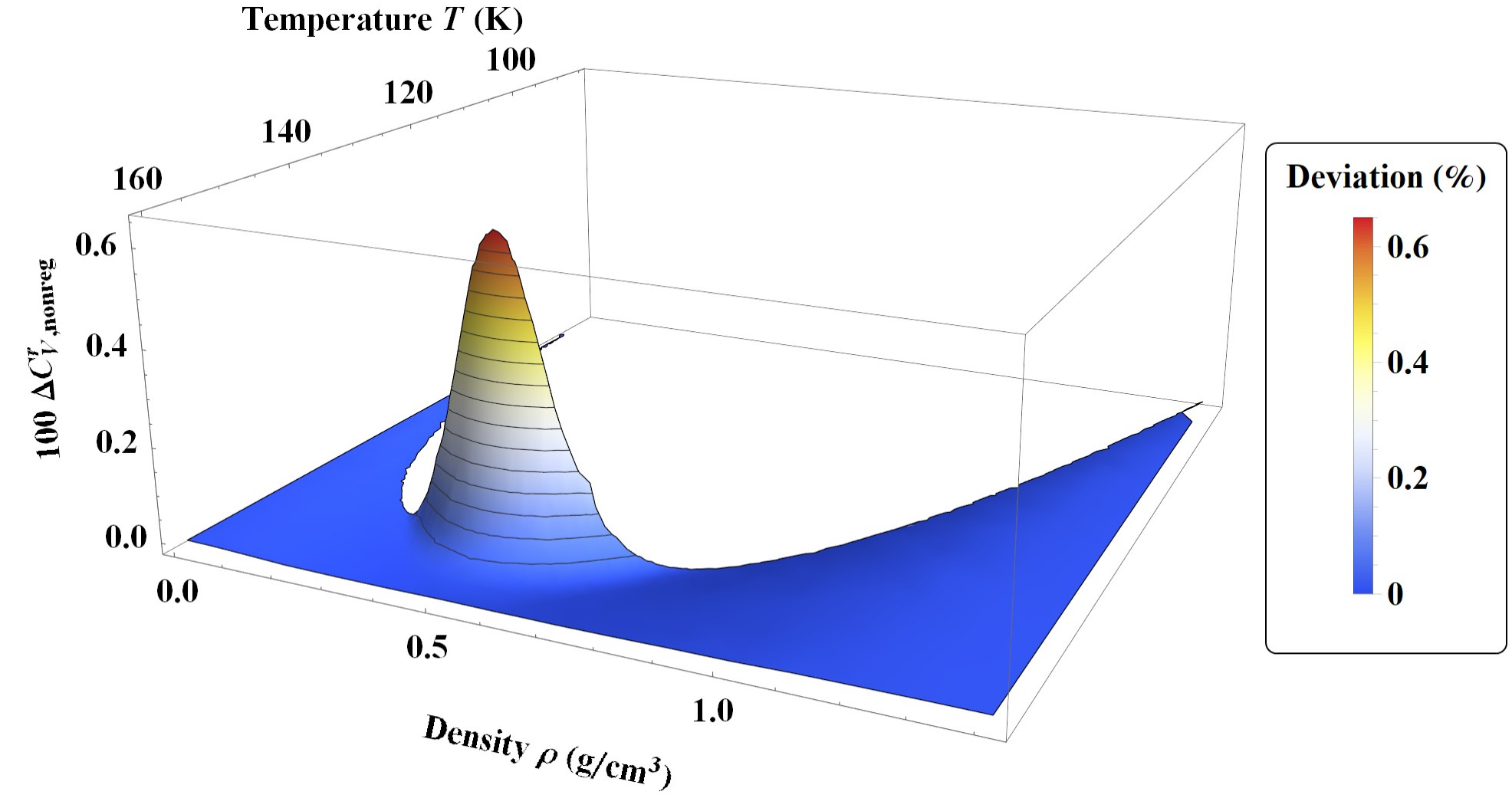

**Fig. 7:** Percentage deviations of the non-regular term of the isochoric heat capacity

$$\Delta C^r_{V,\text{nonreg}} = \left( \underset{\text{Eq.(18)}}{C^r_{V,\text{nonreg}}} - \underset{\text{Eq.(55)}}{C^r_{V,\text{nonreg}}} \right) \Bigg/ \underset{\text{Eq.(18)}}{C^r_{V,\text{nonreg}}}$$

for the single phase region in the temperature range from $T_\text{t}$ to 160 K and the density range from $\rho_\text{t,Gas}$ to $\rho_\text{t,Liq}$.

Fig. 8 shows the behavior of $C_V$ on two isotherms that are crossing the coexistence phase: one can observe that, on both isotherms, the new model gives always positive values of $C_V$ with maximum values as experimentally observed. It can be noticed that the TSW model leads to erroneous $C_V$ variations in this coexistence region.

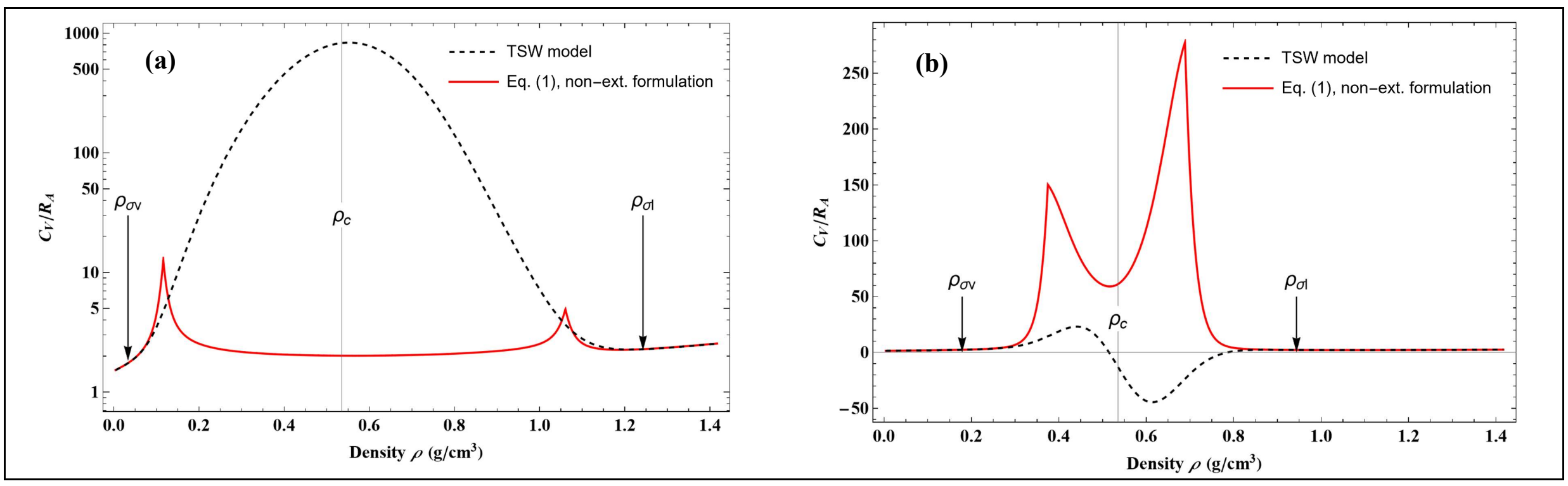

**Fig. 8:** Variations with density of the dimensionless isochoric heat capacity $C_V/R_\text{A}$ from $\rho_\text{t,Gas}$ to $\rho_\text{t,Liq}$ along two isotherms: (a) 110 K and (b) 140 K. The plotted curves correspond to values calculated from Eq. (1) with Eqs. (53) to (57) (red curves) and from the TSW model (black dashed curves).

### *3.4.2. Choice of a reference state*

The functions $U$, $S$ and $F$ are obtained by successive integrations of $C_V$ along isochors; this means a reference temperature is necessary. From Eqs. (55) and (56), it is obvious that the only state which is identical for all isochors is for $T$ infinite.

Due to the fact that, first, the development such as Eq. (18) for $\tilde{c}^\text{r}_{V,\text{nonreg}}$ becomes more complex with a much slower convergence of the series and, secondly, there are two expressions for this term inside the coexistence liquid-vapor region, it is preferable and easier to integrate numerically these terms. Thus the respective expressions for $\tilde{u}^\text{r}_\text{nonreg}$ and $\tilde{s}^\text{r}_\text{nonreg}$ are now:

$$\widetilde{u}^{\mathrm{r}}_{\mathrm{nonreg}}(\rho,T)=T^{-1}\int_{\infty}^{T}\widetilde{c}^{\mathrm{r}}_{V,\mathrm{nonreg}}(\rho,t)\,dt \tag{58}$$

$$\widetilde{s}^{\mathrm{r}}_{\mathrm{nonreg}}(\rho,T)=\int_{\infty}^{T}\widetilde{c}^{\mathrm{r}}_{V,\mathrm{nonreg}}(\rho,t)\,t^{-1}dt \tag{59}$$

with $\widetilde{c}^{\mathrm{r}}_{V,\mathrm{nonreg}}(\rho,T)=\begin{cases}\widetilde{c}^{\mathrm{r}}_{V,\substack{\mathrm{nonreg}\\ \mathrm{outside}}}(\rho,\mathrm{T}) & \text{if } T\geq T_{\mathrm{div}}\\ \widetilde{c}^{\mathrm{r}}_{V,\substack{\mathrm{nonreg}\\ \mathrm{inside}}}(\rho,\mathrm{T}) & \text{otherwise}\end{cases}$.

To complete the description, the expression for the non-regular compressibility factor is also given, i.e.:

$$Z_{\mathrm{nonreg}}=\rho\left(\frac{\partial\widetilde{a}^{\mathrm{r}}_{\mathrm{nonreg}}}{\partial\rho}\right)_{T}=\frac{\rho}{T}\int_{\infty}^{T}\left(1-\frac{T}{t}\right)\frac{\partial}{\partial\rho}\widetilde{c}^{\mathrm{r}}_{V,\mathrm{nonreg}}(\rho,t)\,dt \tag{60}$$

To calculate the partial derivative of $\widetilde{c}^{\mathrm{r}}_{V,\mathrm{nonreg}}$, $N_V$ must be considered constant because the experiments imagined here to measure the pressure are performed on a closed system. Thus, the expressions of the partial derivative of $\widetilde{c}^{\mathrm{r}}_{V,\mathrm{nonreg}}$ are:

$$\begin{aligned}\frac{\partial}{\partial\rho}\widetilde{c}^{\mathrm{r}}_{V,\substack{\mathrm{nonreg}\\ \mathrm{outside}}}(\rho,T)=\frac{3}{2}\,\frac{\exp\left(-\left(\frac{T_{\mathrm{div}}(\rho)}{T}\right)^{-3/2}\right)}{1-\frac{T_{\mathrm{div}}(\rho)}{T}}&\left\{n'_{\mathrm{nonreg}}(\rho)\left(1-N_V^{-1+\frac{T_{\mathrm{div}}(\rho)}{T}}\right)\right.\\ &\left.+\,n_{\mathrm{nonreg}}(\rho)\frac{T_{\mathrm{div}}(\rho)}{T}\left[\frac{1-N_V^{-1+\frac{T_{\mathrm{div}}(\rho)}{T}}}{1-\frac{T_{\mathrm{div}}(\rho)}{T}}-N_V^{-1+\frac{T_{\mathrm{div}}(\rho)}{T}}\ln(N_V)+\frac{3}{2}\left(\frac{T_{\mathrm{div}}(\rho)}{T}\right)^{-5/2}\left(1-N_V^{-1+\frac{T_{\mathrm{div}}(\rho)}{T}}\right)\right]\right\}\end{aligned} \tag{61}$$

$$\begin{aligned}\frac{\partial}{\partial\rho}\widetilde{c}^{\mathrm{r}}_{V,\substack{\mathrm{nonreg}\\ \mathrm{inside}}}(\rho,T)=\frac{3}{2}\,\frac{\exp\left(-\left(\frac{T}{T_{\mathrm{div}}(\rho)}\right)^{-3/2}\right)}{1-\frac{T}{T_{\mathrm{div}}(\rho)}}&\left\{n'_{\mathrm{nonreg}}(\rho)\left(1-N_V^{-1+\frac{T}{T_{\mathrm{div}}(\rho)}}\right)\right.\\ &\left.-\,n_{\mathrm{nonreg}}(\rho)\left(\frac{T}{T_{\mathrm{div}}(\rho)}\right)^{2}\frac{T'_{div}(\rho)}{T}\left[\frac{1-N_V^{-1+\frac{T}{T_{\mathrm{div}}(\rho)}}}{1-\frac{T}{T_{\mathrm{div}}(\rho)}}-N_V^{-1+\frac{T}{T_{\mathrm{div}}(\rho)}}\ln(N_V)+\frac{3}{2}\left(\frac{T}{T_{\mathrm{div}}(\rho)}\right)^{-5/2}\left(1-N_V^{-1+\frac{T}{T_{\mathrm{div}}(\rho)}}\right)\right]\right\}\end{aligned} \tag{62}$$

In the same manner as previously, one can deduce primitives for $U$ and $S$ corresponding to the term $\widetilde{c}^{\mathrm{r}}_{V,\substack{\mathrm{crit}\\ \mathrm{inside}}}$:

$$\widetilde{u}^{\mathrm{r}}_{\substack{\mathrm{crit}\\ \mathrm{inside}}}=\frac{3}{2}n_{\mathrm{crit}}(\rho)\left[\frac{\left(T/T_{\mathrm{div}}(\rho)\right)^{\varepsilon_{\mathrm{crit}}(\rho)}}{1+\varepsilon_{\mathrm{crit}}(\rho)}+\frac{2\varepsilon_{\mathrm{crit}}(\rho)}{1-\varepsilon_{\mathrm{crit}}(\rho)^{2}}\right] \tag{63}$$

$$\tilde{s}^{\mathrm{r}}_{\substack{\mathrm{crit}\\ \mathrm{inside}}} = \frac{3}{2} n_{\mathrm{crit}}(\rho)\,\varepsilon_{\mathrm{crit}}(\rho)^{-1}\left[\left(\frac{T}{T_{\mathrm{div}}(\rho)}\right)^{\varepsilon_{\mathrm{crit}}(\rho)} - 2\right] \tag{64}$$

Then it is deduced that:

$$\tilde{a}^{\mathrm{r}}_{\substack{\mathrm{crit}\\ \mathrm{inside}}}(\rho,T) = \frac{3}{2} n_{\mathrm{crit}}(\rho)\left[\frac{\left(T/T_{\mathrm{div}}(\rho)\right)^{\varepsilon_{\mathrm{crit}}(\rho)}}{\varepsilon_{\mathrm{crit}}(\rho)\left(1+\varepsilon_{\mathrm{crit}}(\rho)\right)} + \frac{2}{1-\varepsilon_{\mathrm{crit}}(\rho)^2}\right] \tag{65}$$

and

$$\begin{aligned}
\frac{P_{\substack{\mathrm{crit}\\ \mathrm{inside}}}(\rho,T)}{\rho R_{\mathrm{A}} T} = {} & \frac{3}{2}\rho\, n'_{\mathrm{crit}}(\rho)\left[\frac{\left(T/T_{\mathrm{div}}(\rho)\right)^{\varepsilon_{\mathrm{crit}}(\rho)}}{\varepsilon_{\mathrm{crit}}(\rho)\left(1+\varepsilon_{\mathrm{crit}}(\rho)\right)} + \frac{2}{1-\varepsilon_{\mathrm{crit}}(\rho)^2}\right] \\
& + \frac{3}{2}\rho\, n_{\mathrm{crit}}(\rho)\left\{-\frac{1+2\varepsilon_{\mathrm{crit}}(\rho)}{\varepsilon_{\mathrm{crit}}(\rho)^2\left(1+\varepsilon_{\mathrm{crit}}(\rho)\right)^2}\,\varepsilon'_{\mathrm{crit}}(\rho)\left(\frac{T}{T_{\mathrm{div}}(\rho)}\right)^{\varepsilon_{\mathrm{crit}}(\rho)} + \frac{4\varepsilon_{\mathrm{crit}}(\rho)\,\varepsilon'_{\mathrm{crit}}(\rho)}{\left(1-\varepsilon_{\mathrm{crit}}(\rho)^2\right)^2}\right. \\
& \left. + \frac{\left(T/T_{\mathrm{div}}(\rho)\right)^{\varepsilon_{\mathrm{crit}}(\rho)}}{\varepsilon_{\mathrm{crit}}(\rho)\left(1+\varepsilon_{\mathrm{crit}}(\rho)\right)}\left[\varepsilon'_{\mathrm{crit}}(\rho)\ln\left(\frac{T}{T_{\mathrm{div}}(\rho)}\right) - \varepsilon_{\mathrm{crit}}(\rho)\frac{T'_{\mathrm{div}}(\rho)}{T_{\mathrm{div}}(\rho)}\right]\right\}
\end{aligned} \tag{66}$$

We emphasize here that the choice of these expressions to describe the coexistence region has no effect on the properties of the pure fluid up to the saturation curve.

The present modeling with these new expressions of regular and non-critical terms is referred to as the "***non-extensive formulation***". The non-extensive residual part of the Helmholtz energy $\tilde{a}^{\mathrm{r}}(\rho,T)$ and its partial derivatives with temperature are given in Table 7. The first partial derivative with density can be easily deduced from the expression of the compressibility factor Z. Then the second partial derivatives $\dfrac{\partial^2 \tilde{a}^{\mathrm{r}}}{\partial\rho\partial T}$ and $\left.\dfrac{\partial^2 \tilde{a}^{\mathrm{r}}}{\partial\rho^2}\right|_T$ can be obtained from the first derivatives of the compressibility factor which are given in Table 8 and Table 9.

Table 26 of Ref. 4 summarizes how to calculate the thermodynamics properties from the empirical description of the Helmholtz free energy and its derivatives. In the present approach, the same thermodynamics properties are deduced from the isochoric heat capacity equation and the thermal equation of state, which are now two experimentally measured quantities. *A Mathematica application with the new equations of state corresponding to the non-extensive formulation can be freely downloaded by following* Ref. 11.

**Table 7.** Mathematical expressions of the dimensionless terms for the non-extensive residual part of the Helmholtz function and its partial derivatives with temperature.

$$\tilde{a}^{\mathrm{r}}_{\mathrm{reg}} = \frac{3}{2} n_{\mathrm{reg}} \frac{T_c}{T}\left\{\frac{1}{m}\left[\left(\frac{T}{T_c}\right)^{m} - 1\right] + \frac{\lambda^{-m}}{2-m}\Gamma\left(\frac{m}{2-m},\left(\frac{\lambda T}{T_c}\right)^{2-m}\right)\right\}$$

$$- \frac{3}{2} n_{\mathrm{reg}}\left\{\frac{1}{m-1}\left[\left(\frac{T}{T_c}\right)^{m-1} - 1\right] + \frac{\lambda^{1-m}}{2-m}\Gamma\left(\frac{m-1}{2-m},\left(\frac{\lambda T}{T_c}\right)^{2-m}\right)\right\}$$

$$T_c\left(\frac{\partial \tilde{a}^{\mathrm{r}}_{\mathrm{reg}}}{\partial T}\right)_{\rho} = \frac{3}{2} n_{\mathrm{reg}}\left(\frac{T_c}{T}\right)^{2} m^{-1}\left\{1 - \left(\frac{T}{T_c}\right)^{m} + \frac{\lambda^{-m} m}{m-2}\Gamma\left(\frac{m}{2-m},\left(\frac{\lambda T}{T_c}\right)^{2-m}\right)\right\}$$

$$T_c^2\left(\frac{\partial^2 \tilde{a}^{\mathrm{r}}_{\mathrm{reg}}}{\partial T^2}\right)_{\rho} = \frac{3}{2} n_{\mathrm{reg}}\left(\frac{T_c}{T}\right)^{3}\left\{\left(\frac{T}{T_c}\right)^{m}\left(\frac{2-m}{m} + \exp\left(-\left(\frac{\lambda T}{T_c}\right)^{2-m}\right)\right) - \frac{2}{m} + \frac{2\lambda^{-m}}{m-2}\Gamma\left(\frac{m}{2-m},\left(\frac{\lambda T}{T_c}\right)^{2-m}\right)\right\}$$

$$\tilde{a}^{\mathrm{r}}_{\mathrm{nonreg}} = \int_{\infty}^{T}\tilde{c}^{\mathrm{r}}_{V,\mathrm{nonreg}}\left(1 - \frac{T}{t}\right)\frac{dt}{T}$$

$$T_c\left(\frac{\partial \tilde{a}^{\mathrm{r}}_{\mathrm{nonreg}}}{\partial T}\right)_{\rho} = -\frac{T_c}{T^2}\int_{\infty}^{T}\tilde{c}^{\mathrm{r}}_{V,\mathrm{nonreg}}\, dt\ ,\quad T_c^2\left(\frac{\partial^2 \tilde{a}^{\mathrm{r}}_{\mathrm{nonreg}}}{\partial T^2}\right)_{\rho} = -\left(\frac{T_c}{T}\right)^{2}\tilde{c}^{\mathrm{r}}_{V,\mathrm{nonreg}} + 2\frac{T_c^2}{T^3}\int_{\infty}^{T}\tilde{c}^{\mathrm{r}}_{V,\mathrm{nonreg}}\, dt$$

$$\tilde{a}^{\mathrm{r}}_{\mathrm{crit}} = \frac{3}{2} n_{\mathrm{crit}}\begin{cases}\dfrac{\left(T_{\mathrm{div}}/T\right)^{\varepsilon_{\mathrm{crit}}}}{\left(1-\varepsilon_{\mathrm{crit}}\right)\varepsilon_{\mathrm{crit}}} & \text{if } T \geq T_{\mathrm{div}}\\[2ex] \dfrac{\left(T/T_{\mathrm{div}}\right)^{\varepsilon_{\mathrm{crit}}}}{\left(1+\varepsilon_{\mathrm{crit}}\right)\varepsilon_{\mathrm{crit}}} + \dfrac{2}{1-\varepsilon_{\mathrm{crit}}^2} & \text{otherwise}\end{cases}$$

$$T_c\left(\frac{\partial \tilde{a}^{\mathrm{r}}_{\mathrm{crit}}}{\partial T}\right)_{\rho} = \frac{3}{2} n_{\mathrm{crit}}\frac{T_c}{T}\begin{cases}\dfrac{1}{\varepsilon_{\mathrm{crit}}-1}\left(\dfrac{T_{\mathrm{div}}}{T}\right)^{\varepsilon_{\mathrm{crit}}} & \text{if } T \geq T_{\mathrm{div}}\\[2ex] \dfrac{1}{1+\varepsilon_{\mathrm{crit}}}\left(\dfrac{T}{T_{\mathrm{div}}}\right)^{\varepsilon_{\mathrm{crit}}} & \text{otherwise}\end{cases}$$

$$T_c^2\left(\frac{\partial^2 \tilde{a}^{\mathrm{r}}_{\mathrm{crit}}}{\partial T^2}\right)_{\rho} = \frac{3}{2} n_{\mathrm{crit}}\left(\frac{T_c}{T}\right)^{2}\begin{cases}\dfrac{1+\varepsilon_{\mathrm{crit}}}{\varepsilon_{\mathrm{crit}}-1}\left(\dfrac{T_{\mathrm{div}}}{T}\right)^{\varepsilon_{\mathrm{crit}}} & \text{if } T \geq T_{\mathrm{div}}\\[2ex] \dfrac{\varepsilon_{\mathrm{crit}}-1}{1+\varepsilon_{\mathrm{crit}}}\left(\dfrac{T}{T_{\mathrm{div}}}\right)^{\varepsilon_{\mathrm{crit}}} & \text{otherwise}\end{cases}$$

$$\tilde{a}^{\mathrm{r}}_{0} = \frac{3}{2}\frac{T_c}{T}\hat{u}_0(\rho) - \left\{\tilde{s}_0(\rho) + \ln(\rho)\right\}$$

$$T_c\left(\frac{\partial \tilde{a}^{\mathrm{r}}_{0}}{\partial T}\right)_{\rho} = -\frac{3}{2}\left(\frac{T_c}{T}\right)^{2}\hat{u}_0(\rho)$$

$$T_c^2\left(\frac{\partial^2 \tilde{a}^{\mathrm{r}}_{0}}{\partial T^2}\right)_{\rho} = 3\left(\frac{T_c}{T}\right)^{3}\hat{u}_0(\rho)$$

**Table 8.** Mathematical expressions of the first partial derivatives with temperature of the compressibility factor Z. All these derivatives are in $K^{-1}$.

---

$$\left(\frac{\partial Z_{\text{reg}}}{\partial T}\right)_{\rho} \quad \text{is given in Appendix A}$$

$$\left(\frac{\partial Z_{\text{nonreg}}}{\partial T}\right)_{\rho} = -\frac{\rho}{T^{2}}\int_{\infty}^{T}\left.\frac{\partial \tilde{c}_{V,\text{nonreg}}^{\,\text{r}}}{\partial \rho}\right|_{T} dt$$

$$\left(\frac{\partial Z_{\text{crit}}}{\partial T}\right)_{\rho} = \frac{3}{2}\frac{\rho}{T} n'_{\text{crit}} \begin{cases} \dfrac{1}{\varepsilon_{\text{crit}}-1}\left(\dfrac{T_{\text{div}}}{T}\right)^{\varepsilon_{\text{crit}}} & \text{if } T \geq T_{\text{div}} \\ \dfrac{1}{1+\varepsilon_{\text{crit}}}\left(\dfrac{T}{T_{\text{div}}}\right)^{\varepsilon_{\text{crit}}} & \text{otherwise} \end{cases}$$

$$+\frac{3}{2}\rho\, n_{\text{crit}} \begin{cases} \dfrac{\left(T_{\text{div}}/T\right)^{\varepsilon_{\text{crit}}-1}}{T^{2}\left(\varepsilon_{\text{crit}}-1\right)^{2}}\left\{\varepsilon_{\text{crit}}\left(\varepsilon_{\text{crit}}-1\right)T'_{\text{div}} + T_{\text{div}}\varepsilon'_{\text{crit}}\left(\left(\varepsilon_{\text{crit}}-1\right)\ln\left(\dfrac{T_{\text{div}}}{T}\right)\right)-1\right\} & \text{if } T \geq T_{\text{div}} \\ \dfrac{\left(T/T_{\text{div}}\right)^{\varepsilon_{\text{crit}}-1}}{T_{\text{div}}^{2}\left(1+\varepsilon_{\text{crit}}\right)^{2}}\left\{\varepsilon_{\text{crit}}\left(1+\varepsilon_{\text{crit}}\right)T'_{\text{div}} - T_{\text{div}}\varepsilon'_{\text{crit}}\left(\left(1+\varepsilon_{\text{crit}}\right)\ln\left(\dfrac{T}{T_{\text{div}}}\right)\right)-1\right\} & \text{otherwise} \end{cases}$$

$$\left(\frac{\partial Z_{0}}{\partial T}\right)_{\rho} = -\frac{3}{2}\frac{\rho}{T}\frac{T_{c}}{T}\hat{u}'_{0}(\rho)$$

---

**Table 9.** Mathematical expressions of the first partial derivatives with density of the compressibility factor Z for $T \geq T_{\mathrm{div}}$. All these derivatives are in $\mathrm{cm}^3/\mathrm{g}$.

$\left(\dfrac{\partial Z_{\mathrm{reg}}}{\partial\rho}\right)_T$ is given in Appendix A

$$\left(\frac{\partial Z_{\mathrm{nonreg}}}{\partial\rho}\right)_T = \frac{1}{T}\int_{\infty}^{T}\left(1-\frac{T}{t}\right)\left.\frac{\partial\tilde{c}_{V,\mathrm{nonreg}}^{\,\mathrm{r}}}{\partial\rho}\right|_T dt + \frac{\rho}{T}\int_{\infty}^{T}\left(1-\frac{T}{t}\right)\left.\frac{\partial^2\tilde{c}_{V,\mathrm{nonreg}}^{\,\mathrm{r}}}{\partial\rho^2}\right|_T dt$$

with

$$\left.\frac{\partial^2\tilde{c}_{V,\mathrm{nonreg}}^{\,\mathrm{r}}}{\partial\rho^2}\right|_T = -\frac{3}{2}\frac{T'_{\mathrm{div}}}{\left(T-T_{\mathrm{div}}\right)}\left[n'_{\mathrm{nonreg}} + \frac{T'_{\mathrm{div}}}{\left(T-T_{\mathrm{div}}\right)}n_{\mathrm{nonreg}} + \frac{1}{2}\frac{T''_{\mathrm{div}}}{T'_{\mathrm{div}}}n_{\mathrm{nonreg}}\right]\exp\left(-\left(\frac{T}{T_{\mathrm{div}}}\right)^{3/2}\right)$$

$$\times\left\{3\left(\frac{T}{T_{\mathrm{div}}}\right)^{5/2}\left(-1+N_V^{\;-1+\frac{T_{\mathrm{div}}}{T}}\right)+2N_V^{\;-1+\frac{T_{\mathrm{div}}}{T}}\ln\left(N_V\right)\right\}$$

$$-\frac{3}{8}\frac{n_{\mathrm{nonreg}}\,T'^{2}_{\mathrm{div}}}{T_{\mathrm{div}}\left(T-T_{\mathrm{div}}\right)}\exp\left(-\left(\frac{T}{T_{\mathrm{div}}}\right)^{3/2}\right)\left(\frac{T}{T_{\mathrm{div}}}\right)^{5/2}\left\{3\left(3\left(\frac{T}{T_{\mathrm{div}}}\right)^{3/2}-5\right)\left(-1+N_V^{\;-1+\frac{T_{\mathrm{div}}}{T}}\right)\right.$$

$$\left.+4N_V^{\;-1+\frac{T_{\mathrm{div}}}{T}}\ln\left(N_V\right)\frac{T_{div}}{T}\left(3+\ln\left(N_V\right)\left(\frac{T_{\mathrm{div}}}{T}\right)^{5/2}\right)\right\}$$

$$-\frac{3}{2}\frac{T}{\left(T-T_{\mathrm{div}}\right)^3}\exp\left(-\left(\frac{T}{T_{\mathrm{div}}}\right)^{3/2}\right)\left(-1+N_V^{\;-1+\frac{T_{\mathrm{div}}}{T}}\right)\left\{2\left(T-T_{\mathrm{div}}\right)T'_{\mathrm{div}}\,n'_{\mathrm{nonreg}}+\left(T-T_{\mathrm{div}}\right)^2 n''_{\mathrm{nonreg}}\right.$$

$$\left.+\left(2T'^{2}_{div}+\left(T-T_{\mathrm{div}}\right)T''_{\mathrm{div}}\right)n_{\mathrm{nonreg}}\right\}$$

$$\left(\frac{\partial Z_{\mathrm{crit}}}{\partial\rho}\right)_T = \frac{3}{2}\left(n'_{\mathrm{crit}}+\rho\,n''_{\mathrm{crit}}\right)\frac{\left(T_{\mathrm{div}}/T\right)^{\varepsilon_{\mathrm{crit}}}}{\left(1-\varepsilon_{\mathrm{crit}}\right)\varepsilon_{\mathrm{crit}}}+\frac{3}{2}\left(n_{\mathrm{crit}}+2\rho\,n'_{\mathrm{crit}}\right)\frac{\left(T_{\mathrm{div}}/T\right)^{\varepsilon_{\mathrm{crit}}}}{\left(\varepsilon_{\mathrm{crit}}-1\right)}\left\{\frac{2\varepsilon_{\mathrm{crit}}-1}{\left(\varepsilon_{\mathrm{crit}}-1\right)\varepsilon_{\mathrm{crit}}}\frac{\varepsilon'_{\mathrm{crit}}}{\varepsilon_{\mathrm{crit}}}-\frac{T'_{\mathrm{div}}}{T_{\mathrm{div}}}-\ln\left(\frac{T_{\mathrm{div}}}{T}\right)\frac{\varepsilon'_{\mathrm{crit}}}{\varepsilon_{\mathrm{crit}}}\right\}$$

$$-\frac{3}{2}\rho\,n_{\mathrm{crit}}\frac{\left(T_{\mathrm{div}}/T\right)^{\varepsilon_{\mathrm{crit}}}}{\left(\varepsilon_{\mathrm{crit}}-1\right)^2}\left\{\left(\varepsilon_{\mathrm{crit}}-1\right)^2\left(\frac{T'_{\mathrm{div}}}{T_{\mathrm{div}}}\right)^2+\frac{2\varepsilon'^{2}_{\mathrm{crit}}}{\left(\varepsilon_{\mathrm{crit}}-1\right)\varepsilon^{3}_{\mathrm{crit}}}+\frac{T'_{\mathrm{div}}}{T_{\mathrm{div}}}\left[\left(\varepsilon_{\mathrm{crit}}-1\right)\frac{T''_{\mathrm{div}}}{T'_{\mathrm{div}}}+2\varepsilon'_{\mathrm{crit}}\left(-1+\ln\left(\left(\frac{T_{\mathrm{div}}}{T}\right)^{\varepsilon_{\mathrm{crit}}-1}\right)\right)\right]\right.$$

$$\left.+\frac{1}{\varepsilon^{2}_{\mathrm{crit}}}\left[\varepsilon''_{\mathrm{crit}}\left(1-\varepsilon_{\mathrm{crit}}\left(2+\ln\left(\frac{T_{\mathrm{div}}}{T}\right)\right)+\varepsilon^{2}_{\mathrm{crit}}\ln\left(\frac{T_{\mathrm{div}}}{T}\right)\right)+6\,\varepsilon'^{2}_{\mathrm{crit}}+\varepsilon'^{2}_{\mathrm{crit}}\ln\left(\frac{T_{\mathrm{div}}}{T}\right)\left(2-\varepsilon_{\mathrm{crit}}\left(4+\ln\left(\frac{T_{\mathrm{div}}}{T}\right)\right)+\varepsilon^{2}_{\mathrm{crit}}\ln\left(\frac{T_{\mathrm{div}}}{T}\right)\right)\right]\right\}$$

$$\left(\frac{\partial Z_0}{\partial\rho}\right)_T = \frac{3}{2}\frac{T_c}{T}\left[\hat{u}'_0(\rho)+\rho\,\hat{u}''_0(\rho)\right]-\left[\tilde{s}'_0(\rho)+\rho\,\tilde{s}''_0(\rho)\right]$$

## 4 Comparison of the new equation of state with experimental data and the TSW model

In this section, the quality of the new equation of state in its **non-extensive formulation** (see Table 3 and Table 7) is analyzed in comparison with selected experimental and theoretical data. Most figures also show a comparison with the values calculated using the so-called reference equation of state established by Tegeler *et al.* (Ref. 4) which has been called here the TSW model.

### 4.1. Melting phase transition

In the TSW model, their Eq. (2.7) gives the melting-pressure variation (see Ref. 4). However, they discard arbitrarily some data sets for example the data of Zha *et al.* (Ref. 12). It is clear that these data are scattered but, as they are obtained at high temperature and pressure, it should be interesting to use them. New and more accurate data from Datchi *et al.* (Ref. 13) are almost in the same range of temperature of those of Zha *et al.* (Ref. 12) and are consistent with these data. It is possible to have a complete view of the melting line for the range of temperature corresponding to the Ronchi's data set by adding the data of Jephcoat *et al*. (Ref. 14).

Thus, using a two parameters Simon-Glatzel type function, it is possible to represent in a coherent manner and with continuity the data of Hardy *et al.* (Ref. 15), Zha *et al.* (Ref. 12), Datchi *et al*. (Ref. 13) and Jephcoat *et al*. (Ref. 14):

$$P_m - P_t = a_1\left[\left(\frac{T}{T_t}\right)^{a_2} - 1\right] \qquad (67)$$

with $a_1$ = 225.2858 MPa and $a_2$ = 1.5284. This equation is used in the following to represent the melting line.

It must be noticed that for determining the parameters in Eq. (67), the data from Bridgman (Ref. 16), Lahr *et al*. (Ref. 17), Crawford *et al*. (Ref. 18) and L'Air Liquide (Ref. 19) have also been used.

Fig. 9 compares some different data sets with values calculated from Eq. (67) (solid line), from Eq. (2) written by Datchi *et al*. (Ref. 13) (dashed curve) and from Eq. (1) written by Abramson (Ref. 20) (dot dashed curve). Equation (67) is very close to the function written by Datchi *et al*. (Ref. 13) and both equations are consistent with the data of Jephcoat *et al*. (Ref. 14). The main difference between Eq. (67) and Eq. (2) from Datchi *et al*. (Ref. 13) is at low temperature where this last one is very inaccurate and cannot be used when approaching the triple point. Equation (1) from Abramson is determined for the representation of its own data and it can be seen that the extrapolation of this function is not consistent with the data of Jephcoat *et al.* (Ref. 14). Also, Eq. (1) of Abramson (Ref. 20) is not very accurate at low temperature.

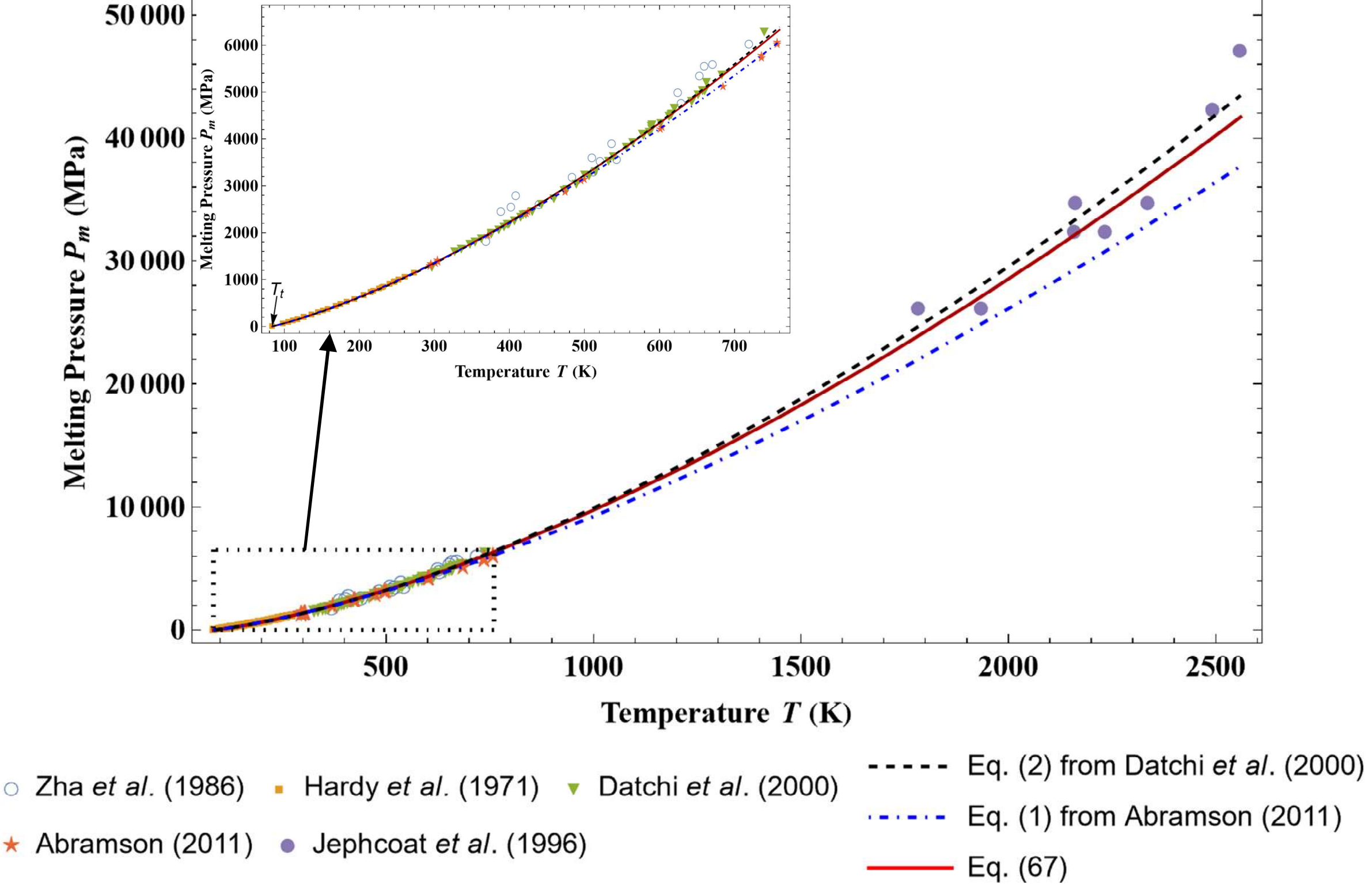

**Fig. 9:** Plot of the melting pressure data determined by Zha *et al*. (Ref. 12), Hardy *et al*. (Ref. 15), Datchi *et al*. (Ref. 13), Abramson (Ref. 20) and Jephcoat *et al*. (Ref. 14). The plotted curves correspond to values calculated from Eq. (67) (solid line), from Eq. (2) given by Datchi *et al*. (dashed curve) and from Eq. (1) given by Abramson (dot-dashed curve).

Some of the previous authors have also measured the liquid density on the melting line. But as it can be seen on Fig. 10, all the data sets have a large dispersion which makes difficult their representation. In particular the data of Lahr *et al*. (Ref. 17) at low temperature seem incompatible with the other data sets and at high temperature these data are incompatible with the data of Crawford *et al*. (Ref. 18). For these reasons, two equations are here proposed which give a greater importance at high temperature either the data from Lahr *et al*. (Ref. 17) either the data from Crawford *et al*. (Ref. 18):

$$T_{\mathrm{m,Low}}(\rho) = T_{\mathrm{t}} + 1015.4 \times \left( \frac{\rho}{\rho_{\mathrm{t,Liq}}} - 1 \right)^{1.843} + 250.46 \times \left( \frac{\rho}{\rho_{\mathrm{t,Liq}}} - 1 \right) \tag{67-Low}$$

$$T_{\mathrm{m,High}}(\rho) = T_{\mathrm{t}} + 677.25 \times \left( \frac{\rho}{\rho_{\mathrm{t,Liq}}} - 1 \right)^{1.236} + 94.955 \times \left[ 1 - \exp\left( - \left( \frac{\rho - \rho_{\mathrm{t,Liq}}}{0.25} \right)^{10.442} \right) \right] \tag{67-High}$$

These two empirical functions are represented on Fig. 10. As can be seen, for a liquid density smaller than 1.6 g/cm$^3$ (i.e. $\rho < \rho_{\mathrm{t,Sol}}$) the two functions are almost identical.

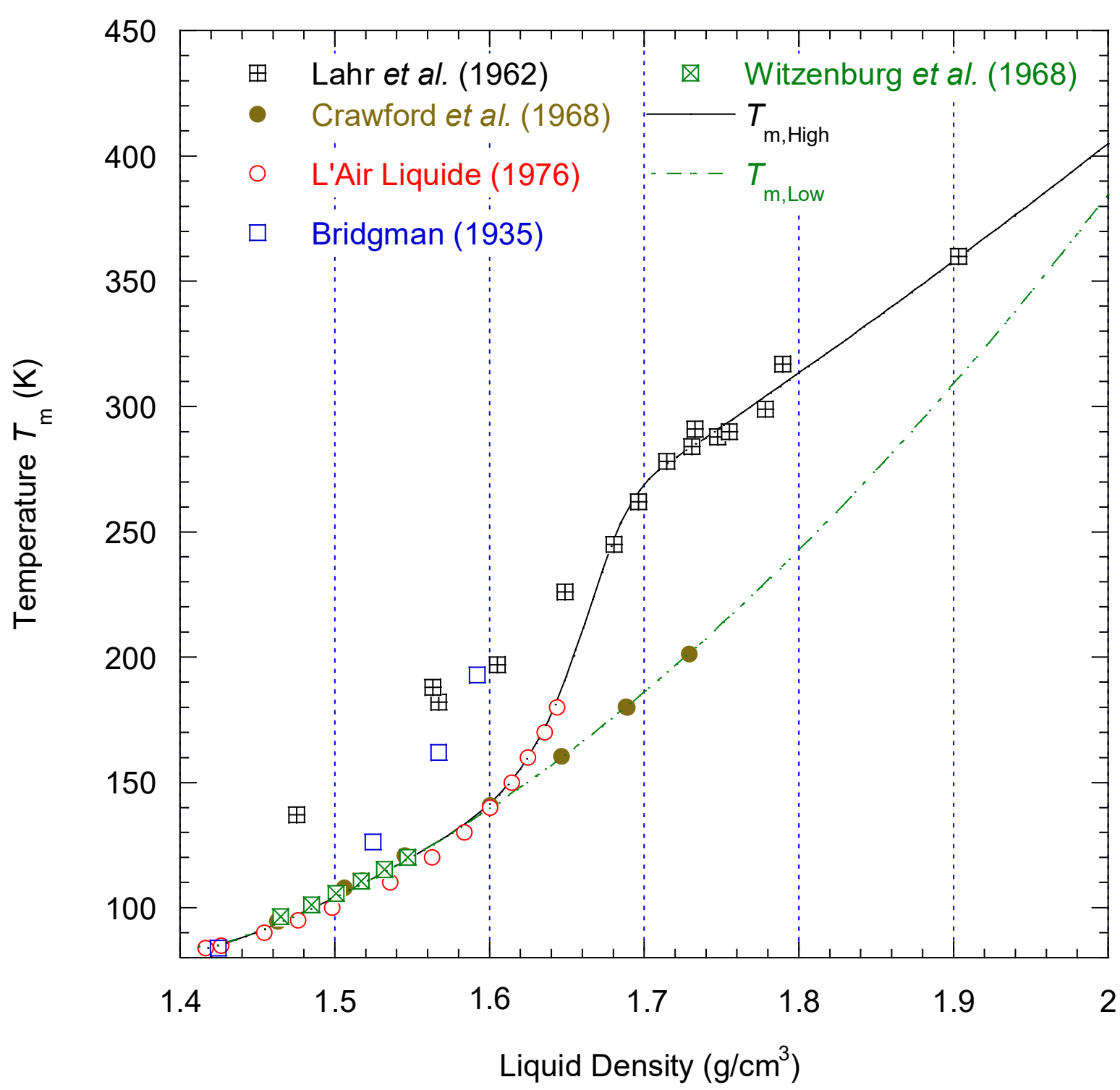

**Fig. 10:** Plot of the liquid density data on the melting line determined by Bridgman (Ref. 16), Lahr *et al*. (Ref. 17), Crawford *et al*. (Ref. 18), Witzenburg *et al*. (Ref. 21) and L'Air Liquide (Ref. 19). The plotted curves correspond to values calculated from the functions $T_{m,Low}(\rho)$ (i.e. Eq. (67-Low), dashed line) and $T_{m,High}(\rho)$ (i.e. Eq. (67-High), solid line).

## 4.2. Single phase region

### *4.2.1. Isochoric heat capacities*

The present modeling is mainly based on $C_V$'s data provided by NIST but since the data of NIST are identical, with some exceptions, to the numerical values deduced from Eq. (4.1) of the TSW model, a comparison between the results obtained with the two models is necessary. In the pressure-temperature region covered by NIST, the relative differences $\Delta C_V$ observed between the TSW model and the present one are less than the uncertainties given on Fig. 44 of Ref. 4. The most important relative difference is obtained in the vicinity of the critical point as shown in Fig. 11. It can also be noticed in Fig. 11, that outside the critical region, the error oscillates almost regularly with density (for all temperatures). These oscillations come from the different mathematical forms used in the two models. For the present modeling, a perfectly smooth monotonic function has been used for $C_V$ while the TSW model used a polynomial equation. This polynomial equation induced small oscillations on $C_V$ and these oscillations can be seen on $\Delta C_V$ (see Fig. 11). Such oscillations, more or less amplified, should also appear on other relative differences between thermodynamic quantities calculated from the two models.

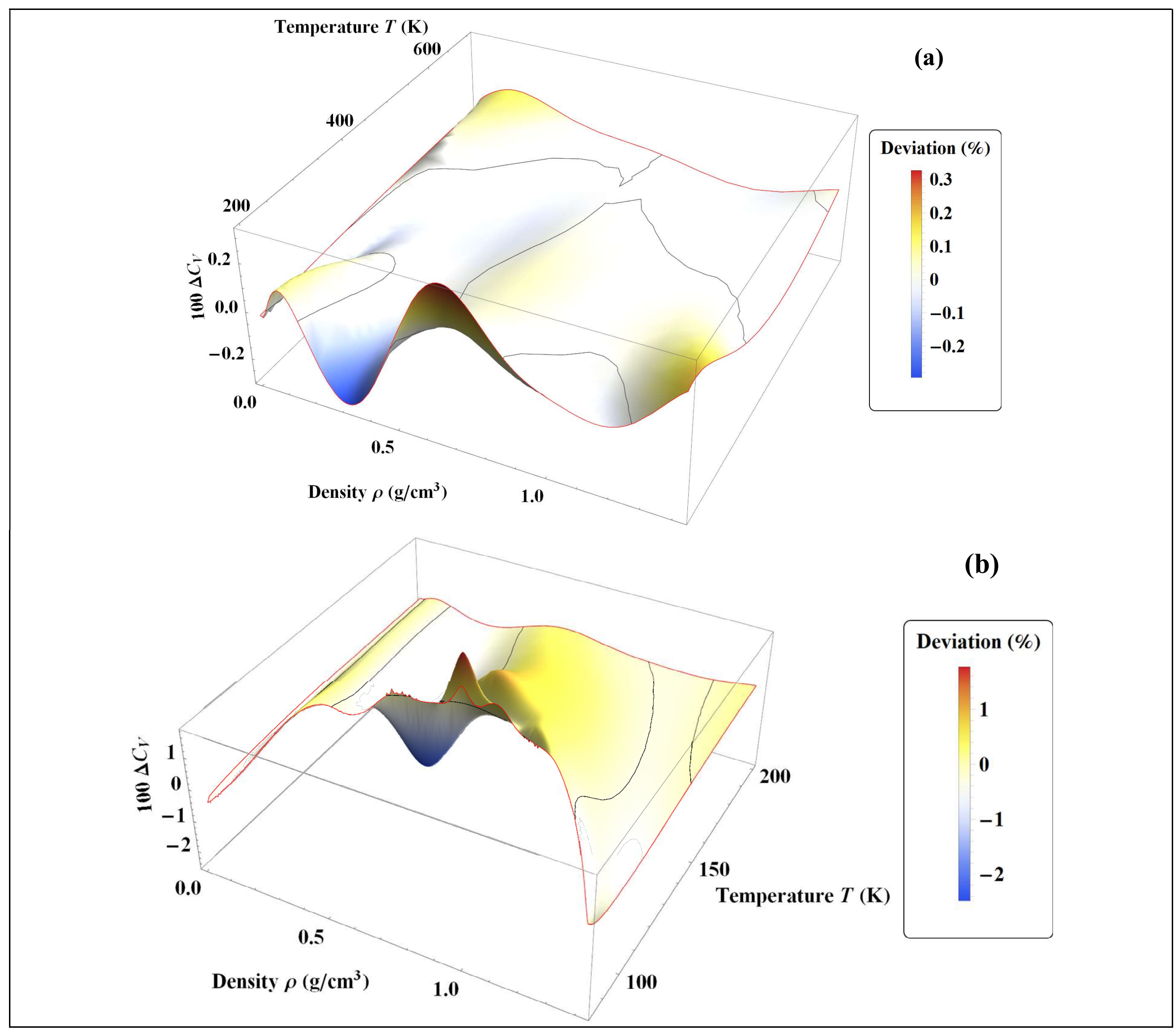

**Fig. 11:** Percentage deviations of isochoric heat capacity $\Delta C_V = \left( C_{V\,\text{TSW model}} - C_{V\,\text{Eq.(1)}} \right) \Big/ C_{V\,\text{TSW model}}$ between the TSW model and Eq. (1) in the density range between $\rho_{\text{t,Gas}}$ and $\rho_{\text{t,Liq}}$: (a) temperature range between 200 K and 700 K; (b) temperature range between $T_\text{t}$ and 200 K. The black lines correspond to values of $\Delta C_V$ equal to zero.

In the paper of Ronchi there are no $C_V$'s data, then, no direct comparison is possible with the present modeling. However, in the region covered by Ronchi, Vrabec *et al.* (Ref. 22) have calculated $C_V$'s data using a molecular dynamics calculation based on a (12,6) Lennard-Jones potential. Fig. 12 shows plots of the isochoric heat capacity on three high density isochors. The isochors with $\rho = 1.196$ g/cm$^3$ and $\rho = 1.393$ g/cm$^3$ are smaller than the density $\rho_{\text{t,Liq}}$ of the liquid at the triple point and hence are limited by the saturated liquid line at low temperatures. The isochor $\rho = 1.6$ g/cm$^3$ is limited by the solidification line. As can be seen, in the pressure-temperature region covered by NIST data, the difference between the present and the TSW model is insignificant. The difference becomes only significant for temperatures larger than 1000 K and for densities higher than $\rho_{\text{t,Liq}}$. For these conditions, the data of Vrabec *et al.* (Ref. 22) are better fitted with the present modeling than with the TSW model. This result was expected since the present model has been built to reproduce the data of Ronchi; data which are also based on a statistical model using a potential of type (12,7).

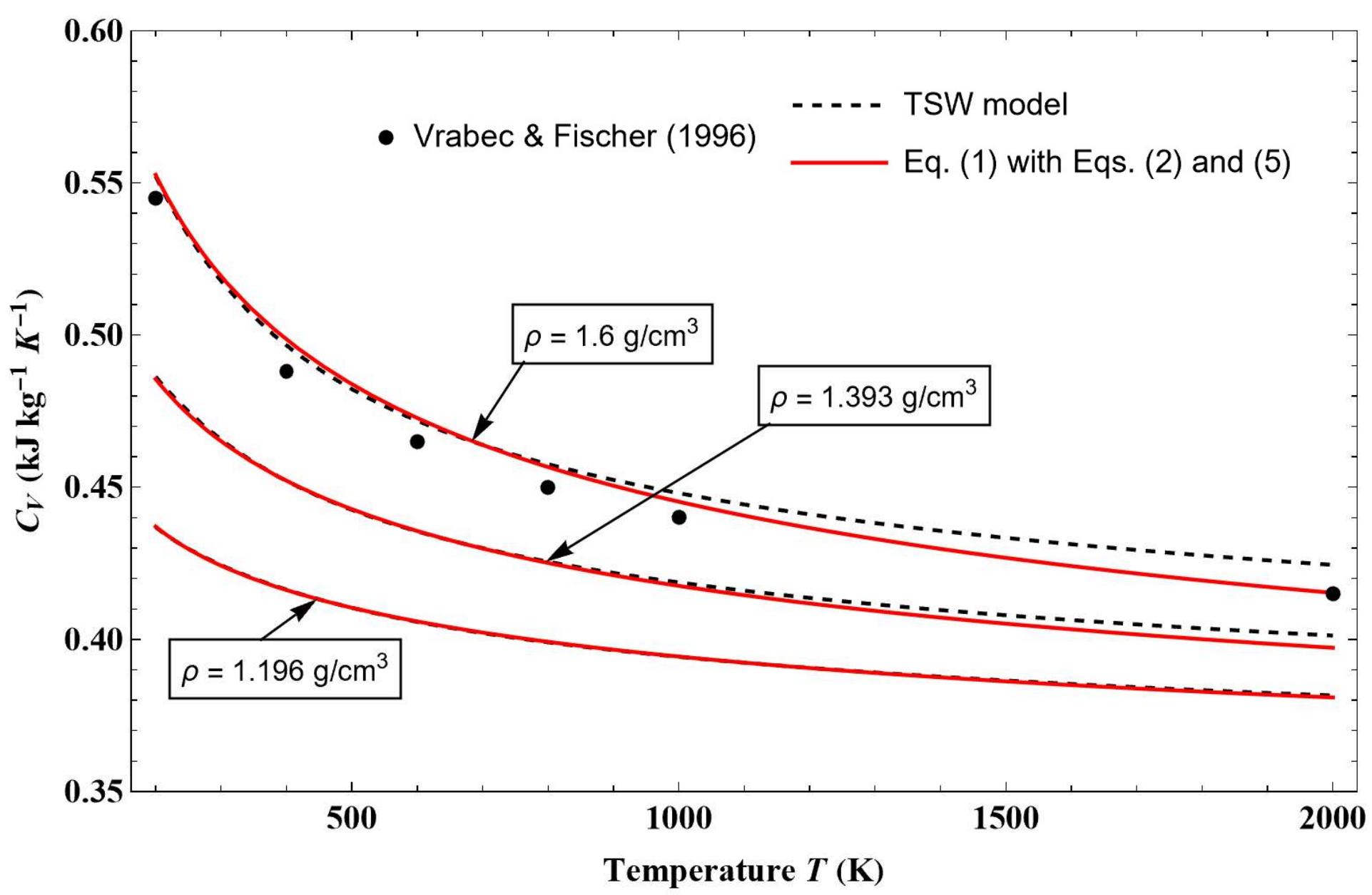

**Fig. 12:** Variations with temperature of the isochoric heat capacity along 3 isochors for high liquid densities (i.e. $\rho \sim \rho_{t,Liq}$). The black points correspond to the data of Vrabec *et al*. (Ref. 22). And the plotted curves to the values calculated from Eq. (1) (red curves) and from the TSW model (black dashed curves).

In the region covered by the data of Ronchi and not covered by the data from NIST, there exists the data of L'Air Liquide (685 data points, Ref. 19) that were not been considered in the TSW model. Fig. 13 shows plots of L'Air Liquide data (Ref. 19) on their highest isotherm at 1100 K and the corresponding calculated curves from the present and the TSW model. The maximum relative error is around 3% and, once again, these data are slightly better fitted with the present model than with the TSW model.

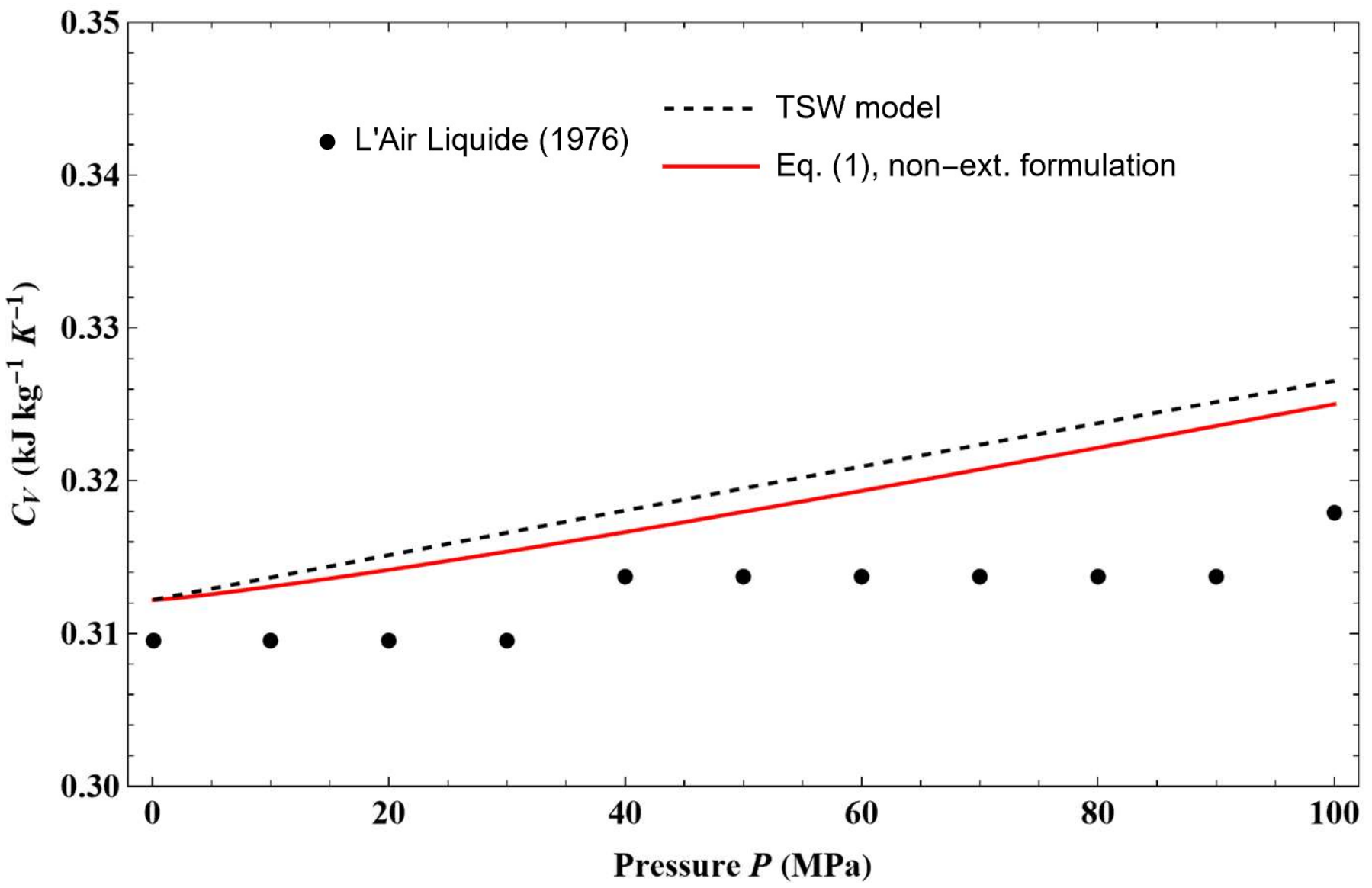

**Fig. 13:** Variation with pressure of the isochoric heat capacity along the isotherm at 1100 K: the data points are from L'Air Liquide (Ref. 19) and the plotted curves correspond to the values calculated from Eq. (1) (red curve) and from the TSW model (black dashed curve).

Even if the calculated values of Ronchi are not enough accurate, the isochors of $C_V$ show the right variation with a maximum when temperature tends to zero as expected for all liquids (Fig. 14). These maxima of $C_V$ are observed in water, and they should also exist in argon. The maximum is also well understood as an extension in the single phase of the same very sharp maximum which is observed in the region of vapor-liquid coexistence. On the contrary, with the TSW model, $C_V$ tends to infinity when temperature tends to zero (see Fig. 14), which is an improper variation.

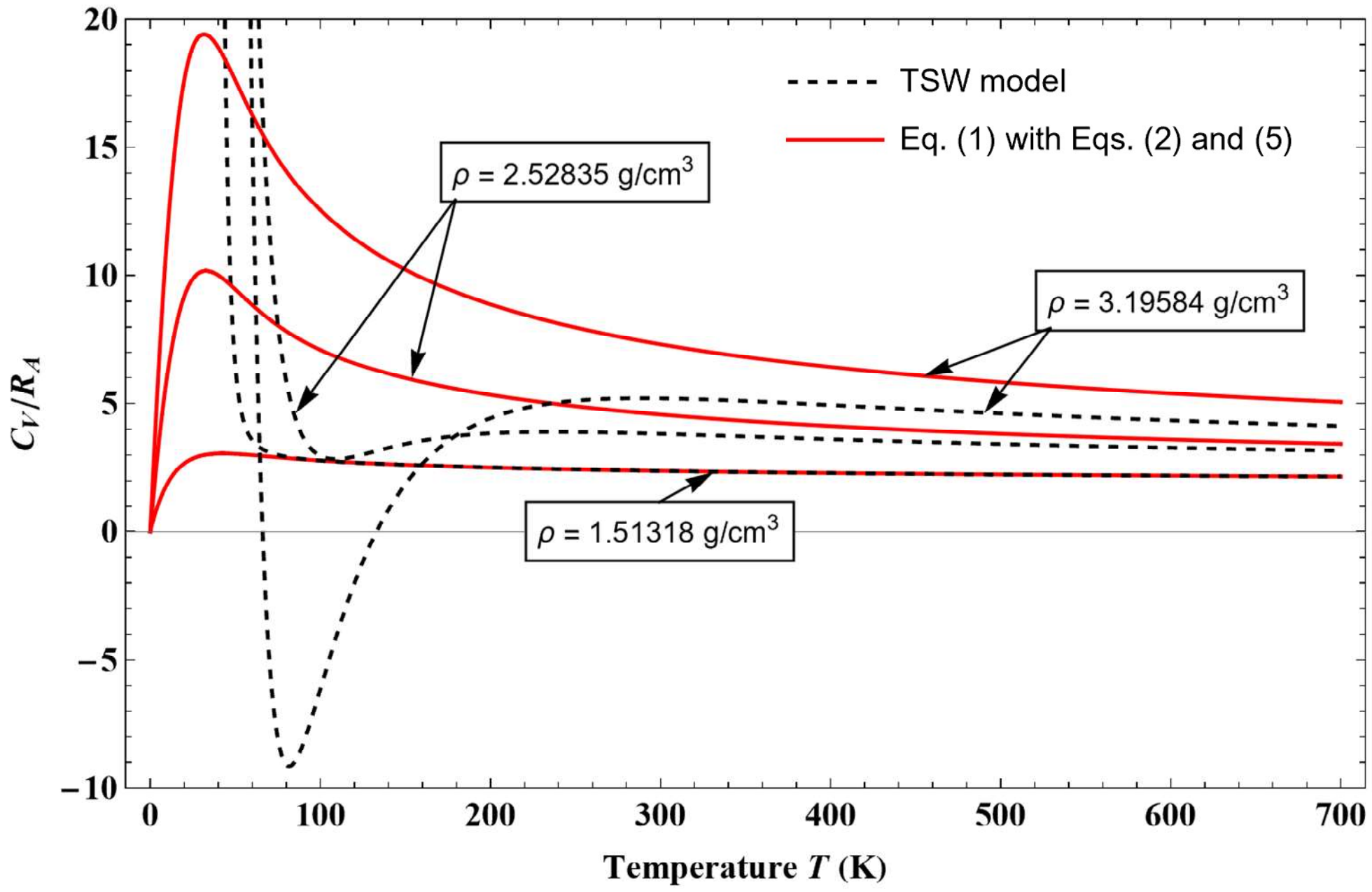

**Fig. 14:** Variations with temperature of the dimensionless **i**sochoric heat capacity along 3 isochors for very high liquid densities (i.e. $\rho > \rho_{t,Liq}$). The plotted curves correspond to the values calculated from Eq. (1) (red curves) and from the TSW model (black dashed curves).

### *4.2.2. Thermal properties*

As explained in section 3.1, the $P\rho T$ data from NIST and from Ronchi (420 data points, Ref. 6) were used to determine the arbitrary functions $U_0(\rho)$ and $S_0(\rho)$ of the internal energy and entropy respectively. Therefore, the equation of state $P(\rho,T)$ with the present approach depends on the accuracy obtained on the modeling of $U_0(\rho)$ and $S_0(\rho)$. Although the slopes of the straight lines $P - P_{C_V}$ are more accurately determined than the ordinates at the origin of these curves, Fig. 15 and Fig. 16 show that the average absolute errors obtained for $U(\rho,T)$ and $S(\rho,T)$ respectively, on all the isochors located in the region of pressure and temperature covered by NIST, are very small and of the same order of magnitude for a given temperature. Fig. 15 and Fig. 16 show in both cases an oscillation of the average error value that is nearly centered on zero. Errors bars represent the standard deviation of the absolute error and given the value of these standard deviations from the average one, it can be understood that the deviation on each isochor is nearly the same for all values of temperature. This indicates that the shape of the isochors as function of temperature is very well reproduced and the errors are due to small oscillations in the data arising from the mathematical form used in the TSW model. However, from the mathematical expressions we used, it is not possible to compensate for such oscillations.

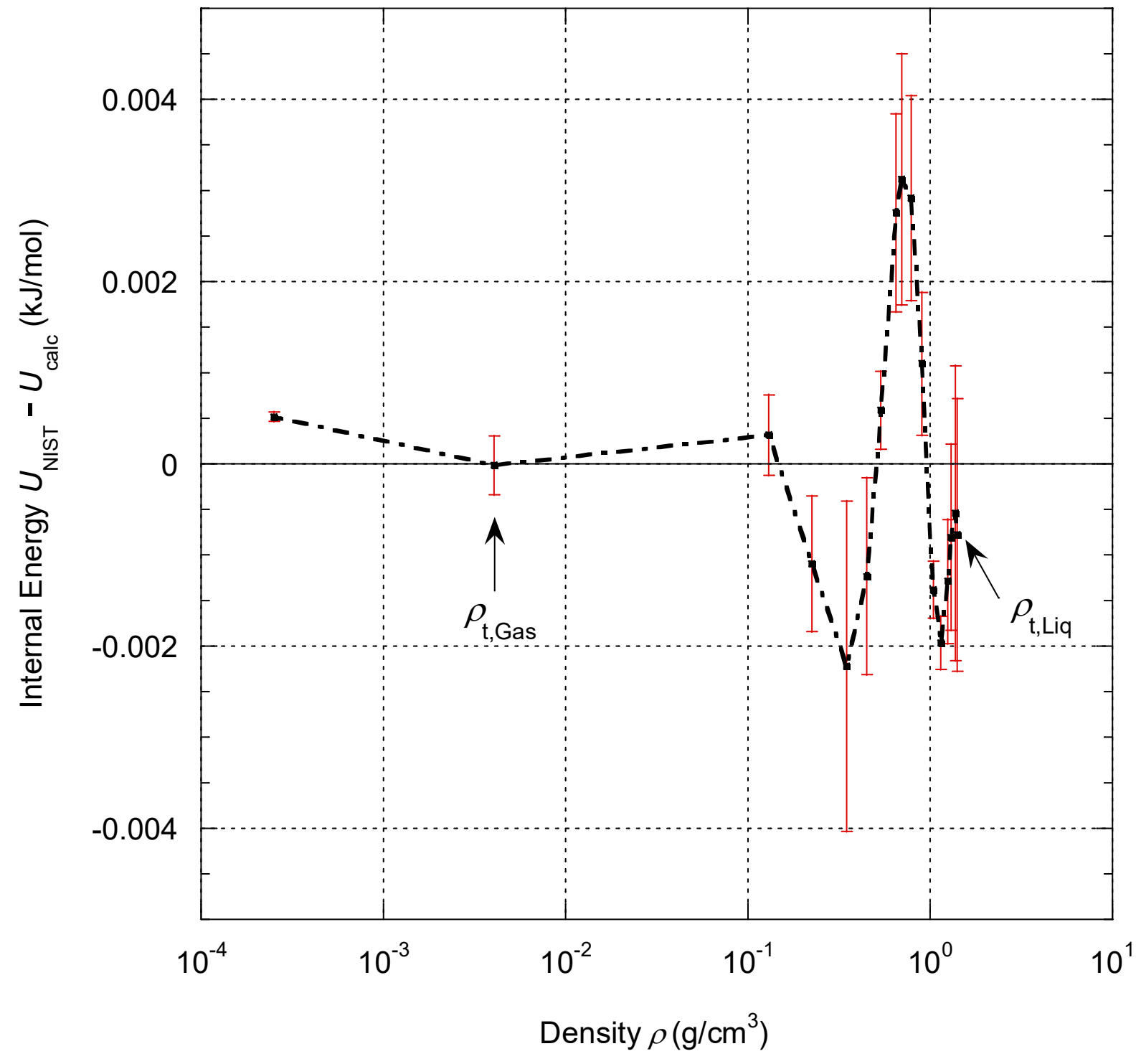

**Fig. 15:** Deviations of NIST internal energy data (Ref. 5) from Eq. (39) along isochors up to 700 K. The error bars correspond to standard deviations. The dashed lines are eye guides.

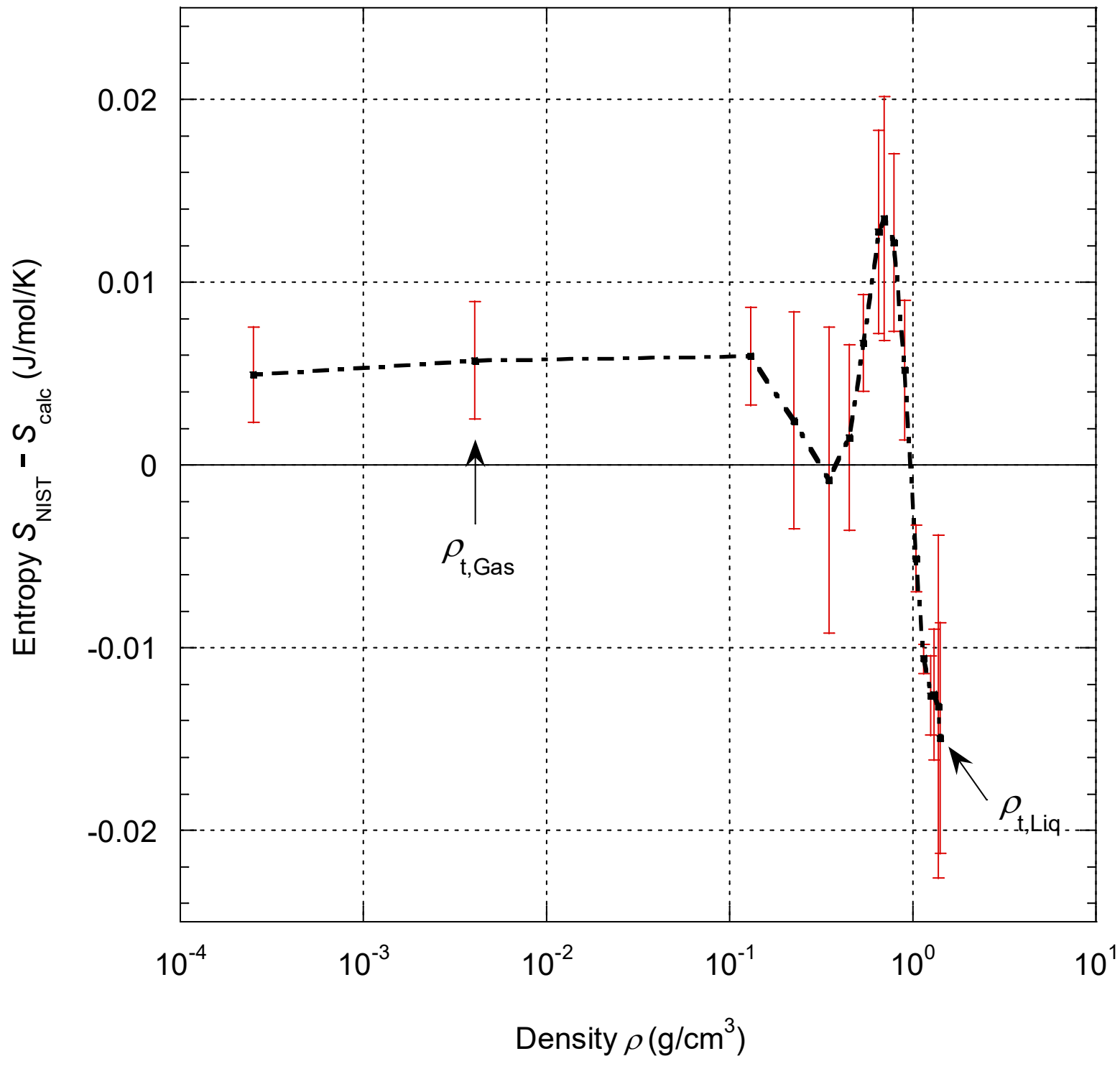

**Fig. 16:** Deviations of NIST entropy data (Ref. 5) from Eq. (39) along isochors up to 700 K. The error bars correspond to standard deviations. The dashed lines are eye guides.

Thus, for $T > T_c$ and for all densities in the range from $\rho_{t,Gas}$ to $\rho_{t,Liq}$, it is found that the relative error on pressure between the NIST data (or TSW model) and the data calculated by the present model shows a "beautiful" oscillation in density (i.e. along isotherms) between -0.2% to +0.4%. In the gas phase, the relative error remains well below 0.2%, value which is only reached on the coexistence curve and in the vicinity of $\rho$ = 0.3 g/cm$^3$. In the liquid phase, the relative error remains well below 0.5% except close to the coexistence line. These largest errors are due to the fact that in dense phase, small variations of density can lead to large variations of pressure. We will come back on this question in section 4.3.2 but it can be noticed that, in order to analyze clearly this problem, it is necessary to look at the inverse equation $\rho(P,T)$ obtained by inversion of Eq. (45).

Due to the strong nonlinearity of the equation $P(\rho,T)$, whatever the model (TSW, Ronchi or the present one), it is not possible to obtain an analytical form of the inverse equation $\rho(P,T)$, so a numerical method is used. The calculated data $\rho(P,T)$ from the different models are now compared. Fig. 17 shows the relative error on density between the data calculated by the present model and the NIST data. The same tolerance range (from ±0.03% to ±0.5% in density) as proposed on Fig. 42 of Ref. 4 was used. It is evident that Fig. 17 and Fig. 42 of Ref. 4 are comparable though the distribution of tolerance regions is different.

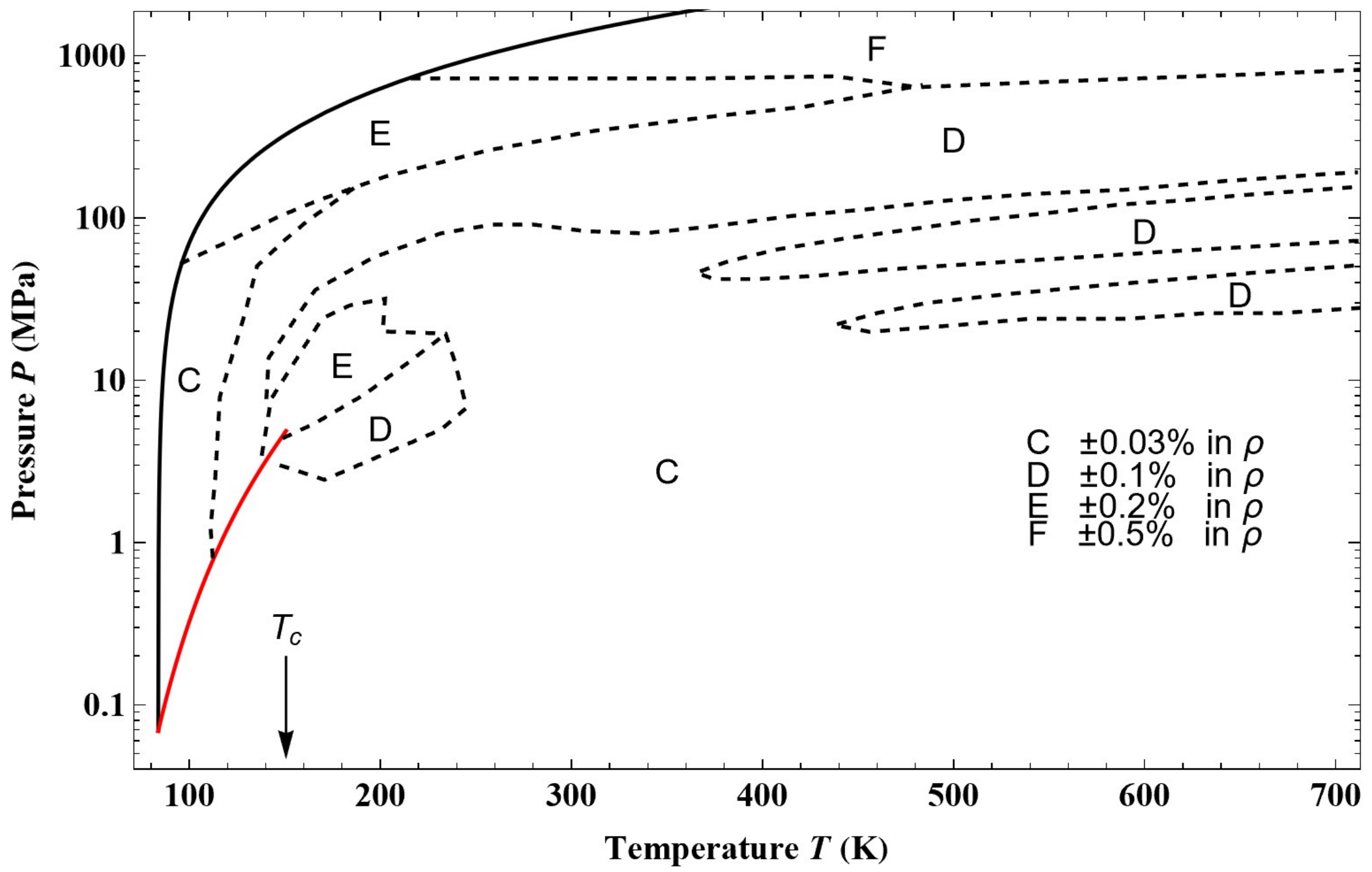

**Fig. 17:** Tolerance diagram for densities calculated from inversion of Eq. (45). The red curve corresponds to the saturated vapor pressure curve and the black one to the melting line.

For $P > P_c$ and for all temperatures corresponding to NIST data, it can be seen that the relative error on density and their oscillations (Fig. 17) of the present modeling are lower or close to the error obtained from the TSW model (see Fig. 42 of Ref. 4) except in the vicinity of the critical point. It can however be noticed that in this region, the TSW model shows an uncertainty on pressure, such uncertainty is obviously smaller than the uncertainty on density.

For $P < P_c$ and for all the gaseous phase, the relative error on density (in the range ±0.03 to ±0.1%) is close to the one given by the TSW model; globally, the error of the present model is in the range ±0.03%, i.e. category C in Ref. 4.

Before discussing these different tolerance diagrams, we will first look at the comparison of the calculated density data (from TSW model and the present one) with the data from L'Air Liquide (729 data points, Ref. 19). The accuracy claimed by L'Air Liquide on density measurements spread between ±0.1% and ±1.5% depending on the experimental method used.

Fig. 18 and Fig. 19 display the relative error on density as a function of temperature between calculated data and L'Air Liquide data (Ref. 19) along two isobars. The data along the isobar at 0.1 MPa are all in the gaseous phase while the data along the isobar at 100 MPa spread from the liquid phase to the supercritical one. Both relative errors show comparable variations with temperature. Fig. 18 shows that, using the TSW model, the relative errors are in agreement with the uncertainty obtained with the present model. The relative errors of the present modeling are slightly larger at low temperature but the error variation, in all the temperature range, is better centered on zero. This means that the shape of the isobars is better reproduced by the present modeling. On Fig. 19, it can be noticed that the relative errors using the TSW model agree again with the uncertainty obtained with the present model. The relative errors from the present model are almost everywhere slightly larger but remains in the tolerance range given by Tegeler *et al.*

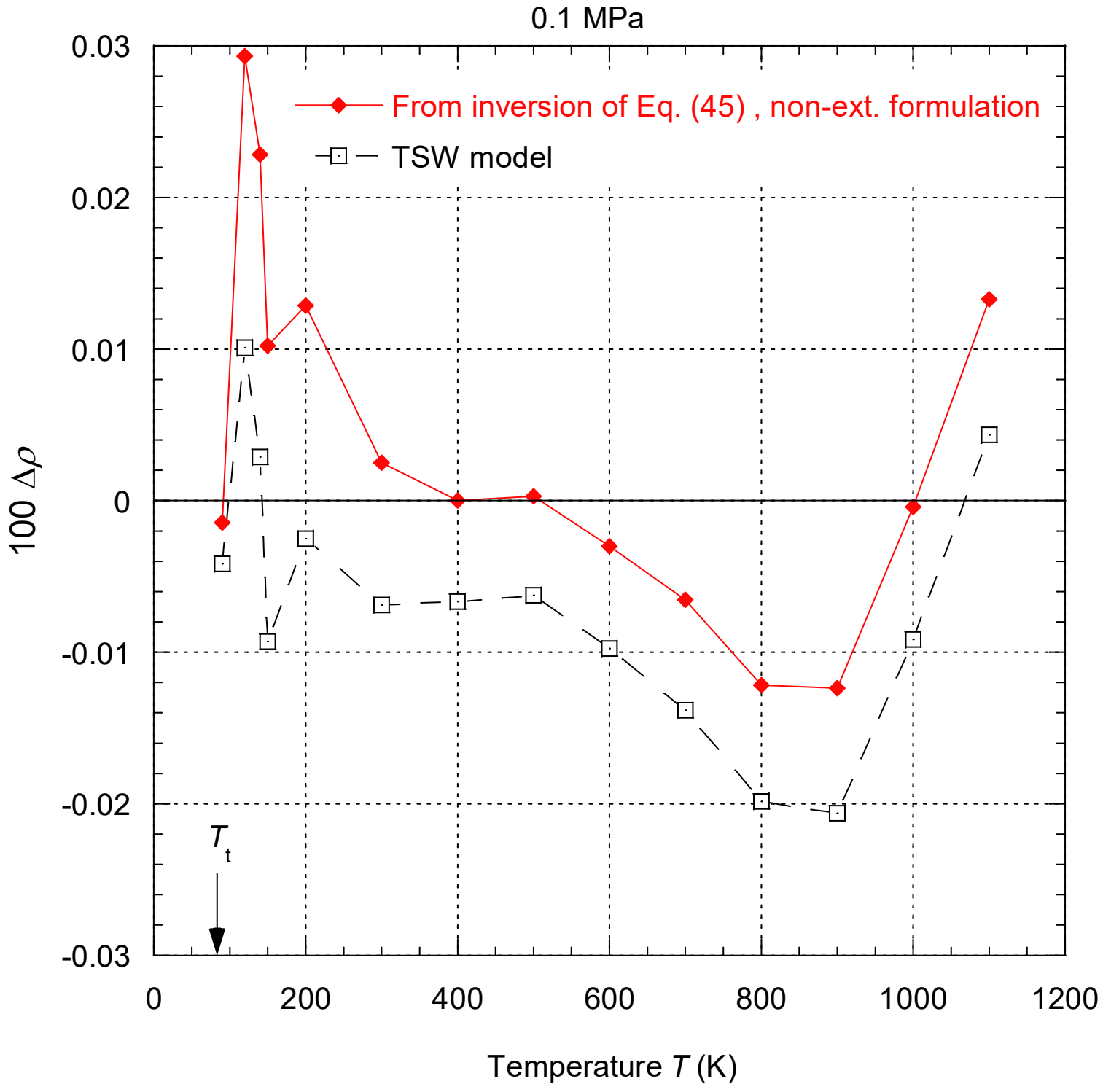

**Fig. 18:** Percentage deviations of density $\Delta\rho = \left(\rho_{\text{L'Air Liquide}} - \rho_{\text{calc}}\right) / \rho_{\text{L'Air Liquide}}$ on the isobar at 0.1 MPa between the data of L'Air Liquide (Ref. 19) and the inversion of Eq. (45) (red diamonds) or the TSW model (black open squares). The lines are eye guides.

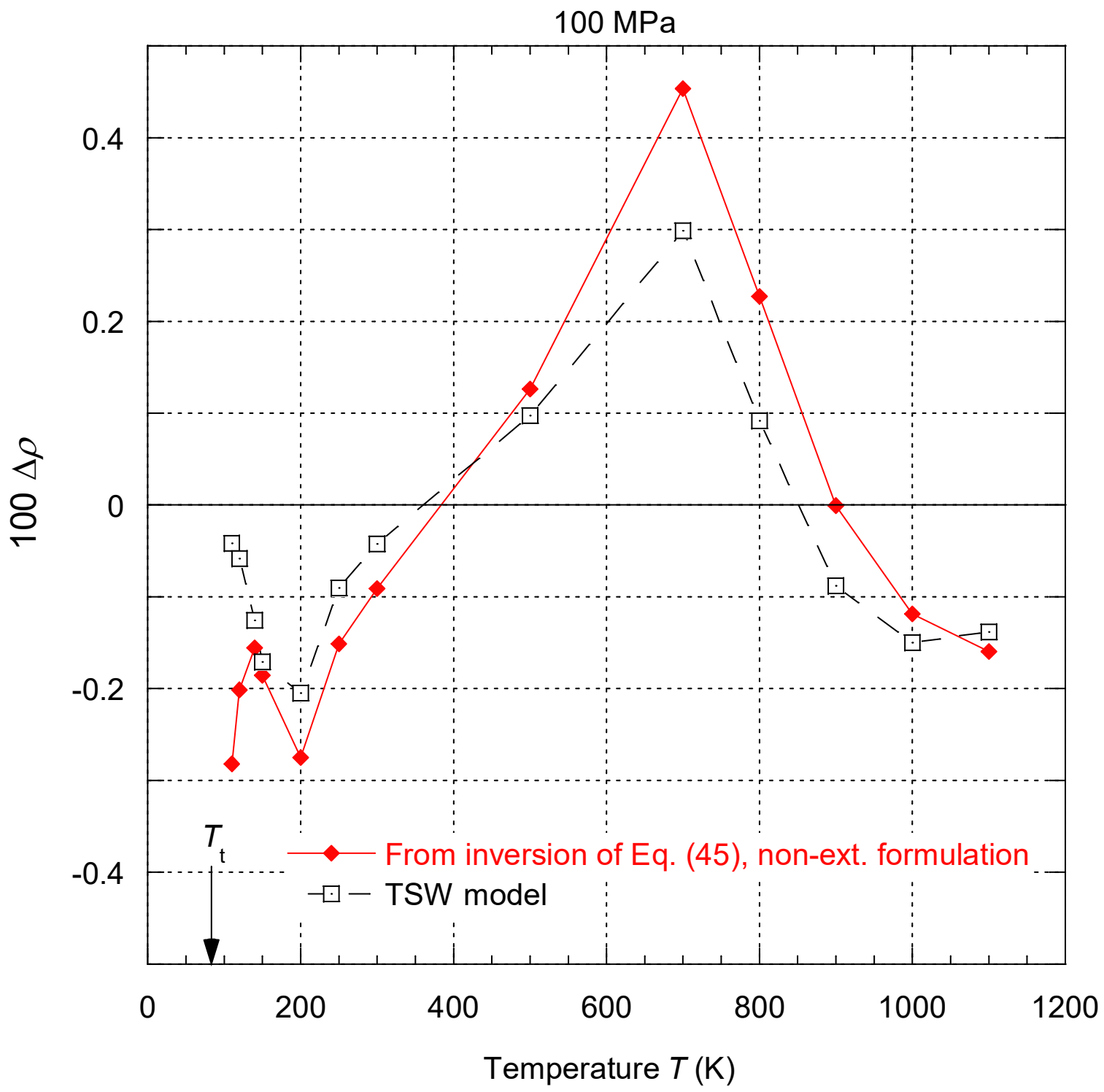

**Fig. 19:** Percentage deviations of density $\Delta\rho = \left(\rho_{\mathrm{L'Air\ Liquide}} - \rho_{\mathrm{calc}}\right) / \rho_{\mathrm{L'Air\ Liquide}}$ on the isobar at 100 MPa between the data of L'Air Liquide (Ref. 19) and the inversion of Eq. (45) (red diamonds) or the TSW model (black open squares). The lines are eye guides.

Some data from L'Air Liquide (Ref. 19) are outside the range of NIST data but are connected to data calculated from the model of Ronchi. Then, such data from L'Air Liquide can be compared to the calculated data from the two models (TSW and the present one). Fig. 20 shows plots of the relative errors on density between calculated and L'Air Liquide data, as a function of pressure on the highest isotherm at 1100 K. The maximum relative error is around 0.3% and the two models lead to a similar variation with temperature. The error variation is slightly better centered on zero using the present model than the TSW one.

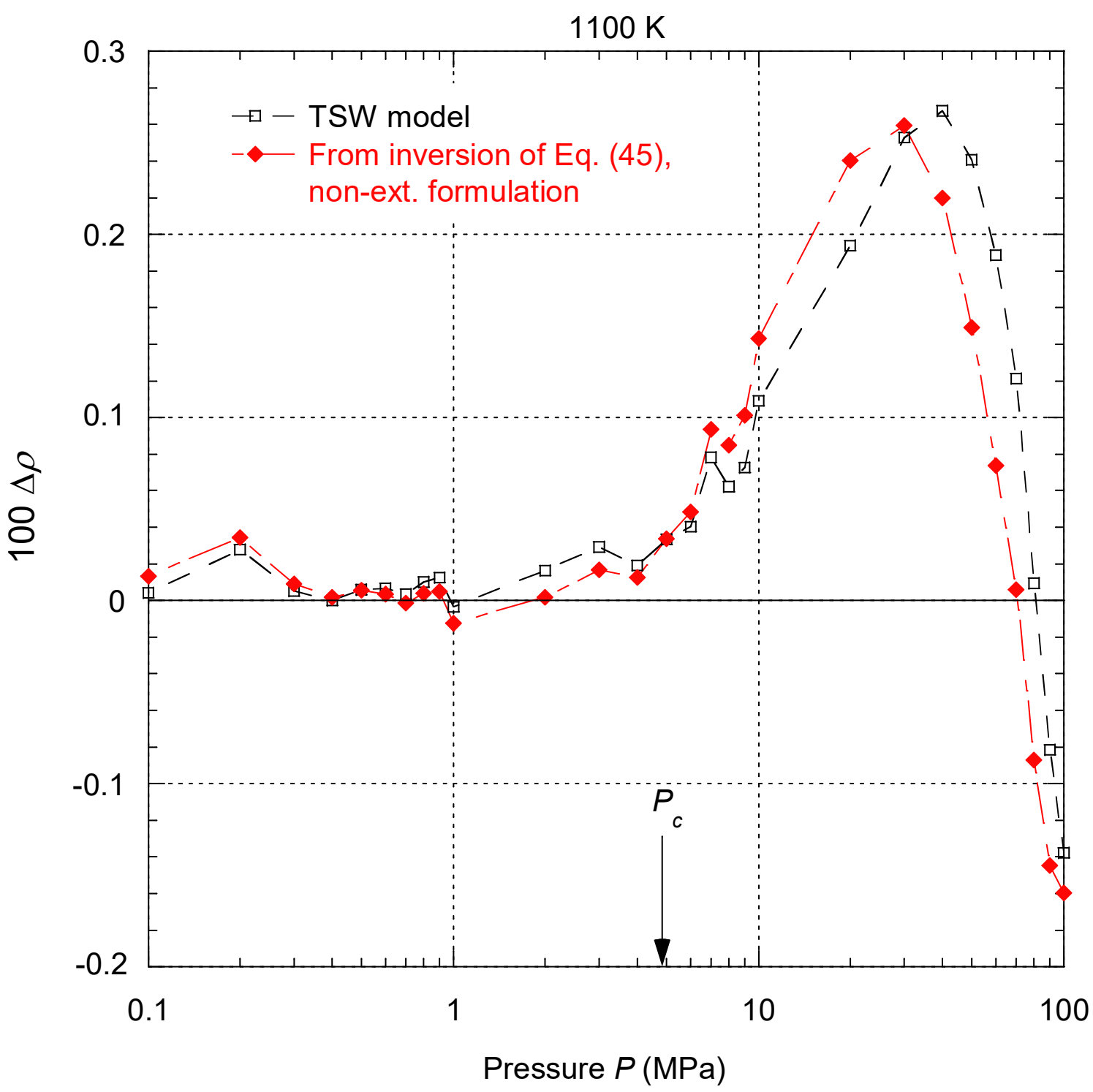

**Fig. 20:** Percentage deviations of density $\Delta\rho = \left(\rho_{\text{L'Air Liquide}} - \rho_{\text{calc}}\right) / \rho_{\text{L'Air Liquide}}$ on the isotherm at 1100 K between the data of L'Air Liquide (Ref. 19) and the inversion of Eq. (45) (red diamonds) or the TSW model (black open squares). The lines are eye guides.

It finally appears that the present model can ultimately better reproduce the thermal properties in the gas phase than in the liquid phase. This is consistent with the fact that the state equations for $U$ and $S$ are better reproduced at low densities than at high densities. So if one wants to better reproduce the data in the liquid phase, it is necessary to increase the accuracy on these two functions going towards high densities. It can however be noticed that the relative errors on pressure and density as defined by NIST remain very comparable for the two models, with a few exceptions.

In the region covered by the calculated data of Ronchi, it is only possible to use the thermal equations of state $P(\rho,T)$ to compare data. Fig. 21 shows the relative error on pressure versus temperature for different isochors. In the region of density covered by NIST data, the relative error from the present model below 700 K is similar to the relative error deduced from the TSW model (see Fig. 1). Above 700 K, the maximum of the error is -2.5% on the isochor $\rho = 1.1784$ g/cm$^3$ and, above 1000 K, the relative error on all the isochors decreases towards zero. Therefore, up to 2300 K, the overall error using the present model does not exceed the error obtained in the region covered by NIST data. Outside the region of density covered by NIST, Fig. 22 shows that the relative error corresponding to the present model is in the range ±5%, except at low temperature on the two isochors $\rho = 1.84944$ g/cm$^3$ and $\rho = 2.01758$ g/cm$^3$. For these isochors, the relative error can be reduced to zero by decreasing the density value corresponding to these isochors by about 0.6%. The uncertainty of ±5% corresponds to the uncertainty claimed by Ronchi between his model and the many experimental data he used. If the Ronchi's data are compared with the extrapolation of the TSW model, then Fig. 23 shows that the relative error increase with increasing the isochor's density and reach the value of 60% on the highest density isochor. This result was already mentioned in Ref. 4.

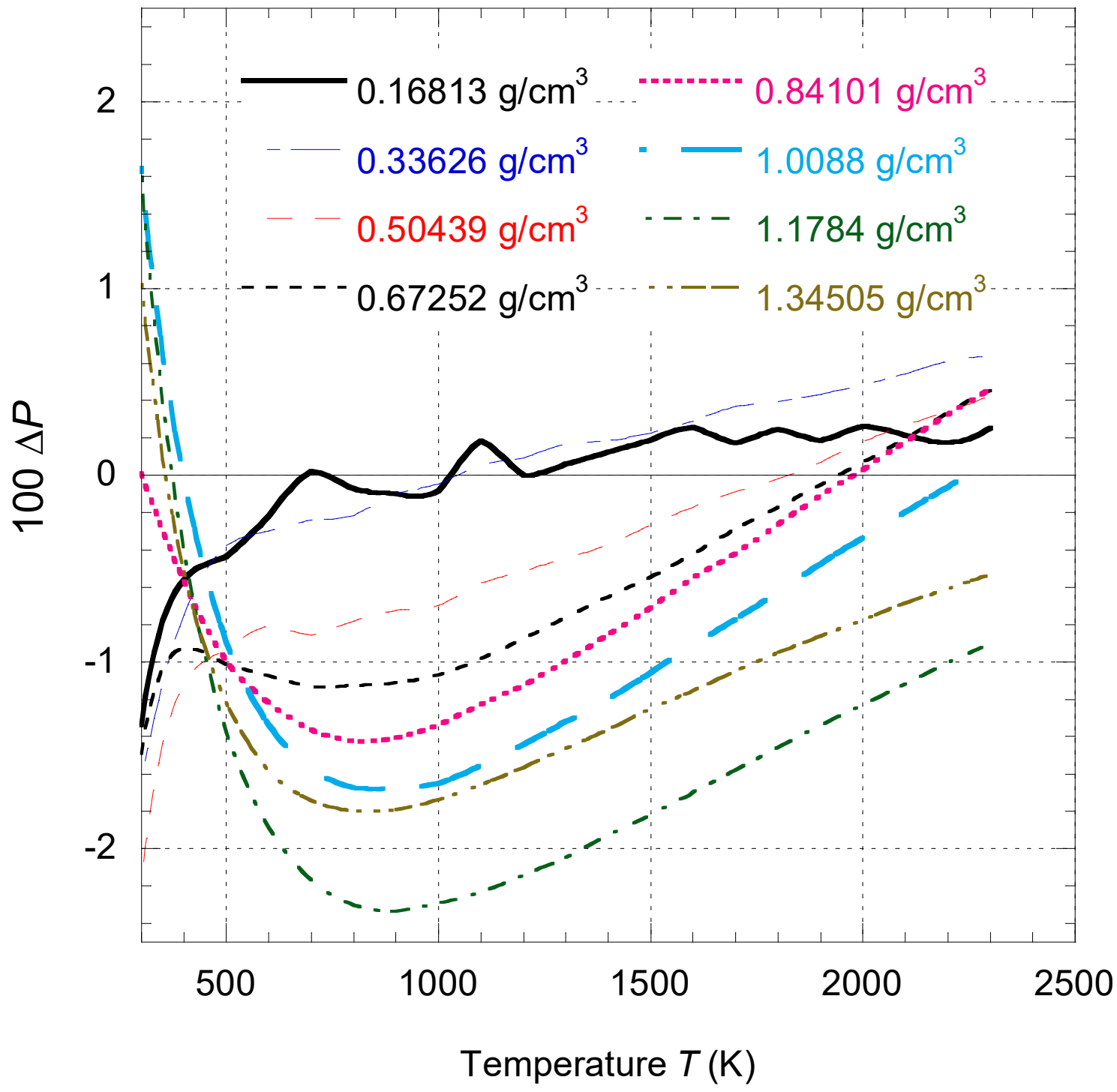

**Fig. 21:** Percentage deviations of pressure $\Delta P = \left(P_{\text{Ronchi}} - P_{\text{calc}}\right)/P_{\text{Ronchi}}$ between the data of Ronchi (Ref. 6) and Eq. (45) along isochors for densities less than $\rho_{\text{t,Liq}}$.

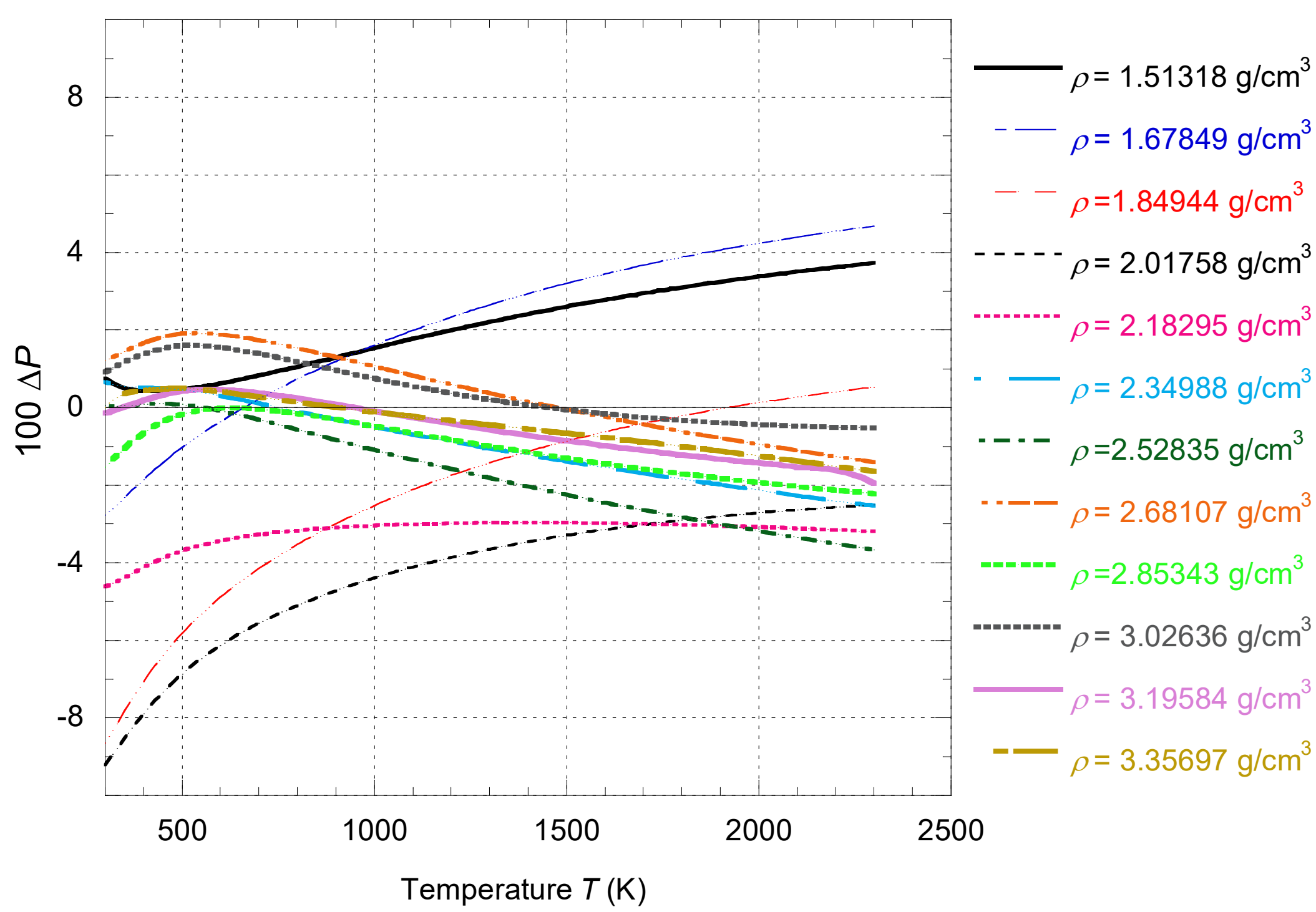

**Fig. 22:** Percentage deviations of pressure $\Delta P = \left(P_{\text{Ronchi}} - P_{\text{calc}}\right)/P_{\text{Ronchi}}$ between the data of Ronchi (Ref. 6) and Eq. (45) along isochors for densities greater than $\rho_{\text{t,Liq}}$.

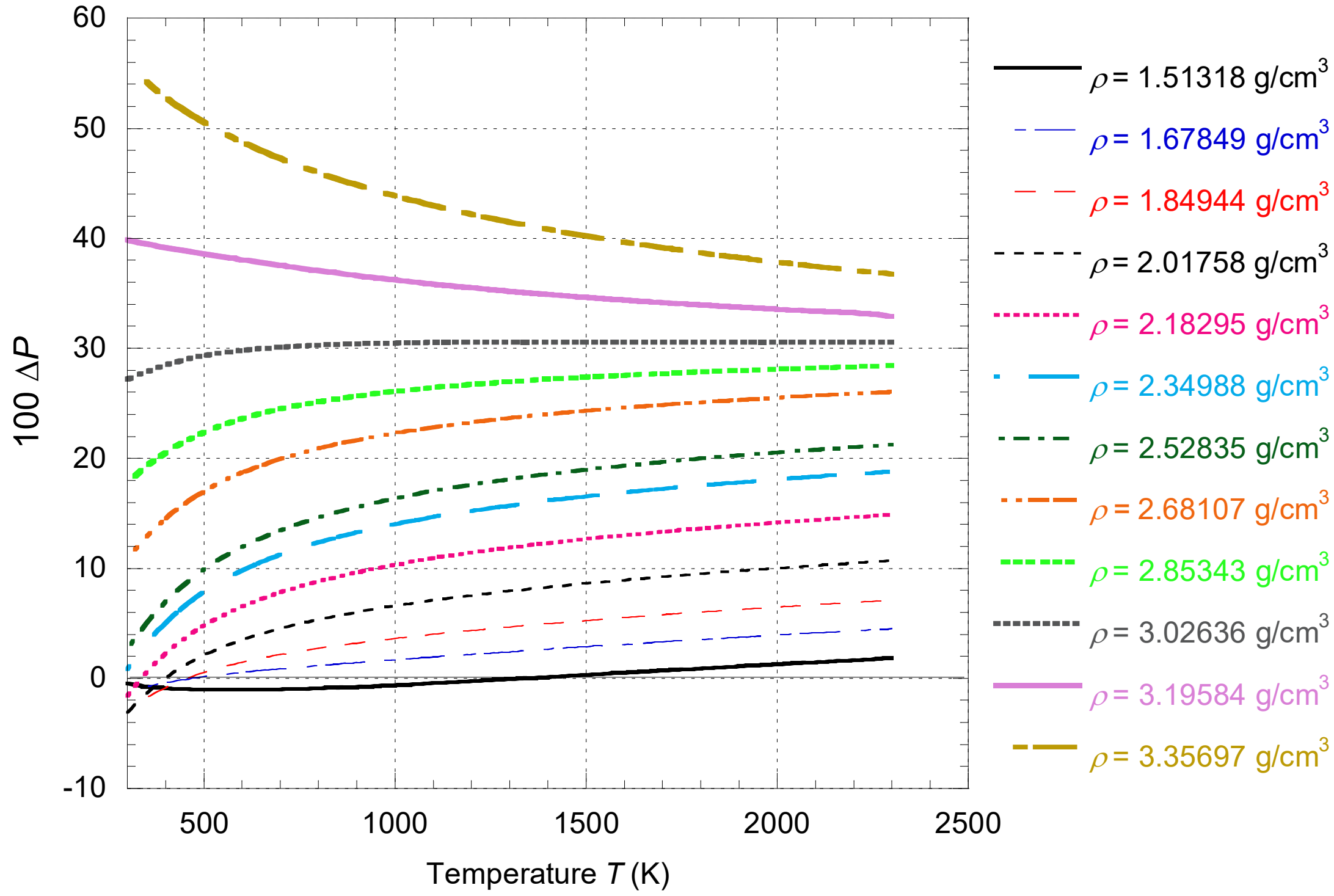

**Fig. 23:** Percentage deviations of pressure $\Delta P = \left(P_{\text{Ronchi}} - P_{\text{TSW model}}\right)/P_{\text{Ronchi}}$ between the data of Ronchi (Ref. 6) and the TSW model along isochors for densities greater than $\rho_{\text{t,Liq}}$.

New experimental $P\rho T$ data in the supercritical phase at 300 K have been determined by Hanna *et al*. (Ref. 23). There results have been compared with the TSW model and are consistent with it. But due to the large error bars, the present model is also consistent with these new data.

Precise experimental $P\rho T$ data in the gaseous phase, in the range of temperature from 234 K to 505 K, have been made by McLinden (Ref. 24). These results have been compared with the values calculated from the TSW model and are consistent with them. In the pressure and temperature ranges covered by these data, the present model has the same precision as that of the TSW model hence these data are also consistent with the present model.

#### *4.2.3. Isobaric heat capacities, sound velocities and isothermal throttling coefficient*

As it can be observed from Table 26 of Ref. 4, the isobaric heat capacity $C_P(\rho,T)$, the speed of sound $c(\rho, T)$ and the isothermal throttling coefficient $\delta_T(\rho,T) = \left(\partial H/\partial P\right)_T$ are functions expressed with first and second derivatives of the Helmholtz free energy, therefore these quantities are more complex with respect to the quantities shown in the previous sections. Given that the present model is not built on free energy but on the equation of state of $C_V(\rho,T)$ and thermal state equation $P(\rho,T)$, it is preferable to express the three above quantities as:

$$C_P = C_V + TVK_T\left(\frac{\partial P}{\partial T}\right)_V^2 = C_V + TV\frac{\beta^2}{K_T} \tag{68}$$

$$c^2 = \frac{V}{K_T}\frac{C_P}{C_V} \tag{69}$$

$$\delta_T = V(1 - T\beta) \tag{70}$$

where $K_T = -\frac{1}{V}\left(\frac{\partial V}{\partial P}\right)_T$ represents the isothermal compressibility coefficient and $\beta = \frac{1}{V}\left(\frac{\partial V}{\partial T}\right)_P$ represents the isobaric coefficient of thermal expansion. These quantities include the derivatives of pressure along the two directions $\rho$ and $T$. The two quantities $c$ and $\delta_T$ are functions of $K_T$ and of the ratio $C_P/C_V$ . So the errors on these two last quantities will reflect in a different way the errors on the state equations for pressure and for the isochoric heat capacity.

Since $C_P$ diverges at the critical point it is only possible to compare the two models (TSW and the present one) outside the region of coexistence. Fig. 24 shows the relative error on $C_P$ between the TSW model and the present one. The relative error is everywhere inside the uncertainty given on Fig. 44 of Ref. 4. In particular, Fig. 24 shows that, for most of the states, the relative error of the present model oscillates globally, without going into the details, between ±0.5%, except for high density states and states in the vicinity of the critical point.

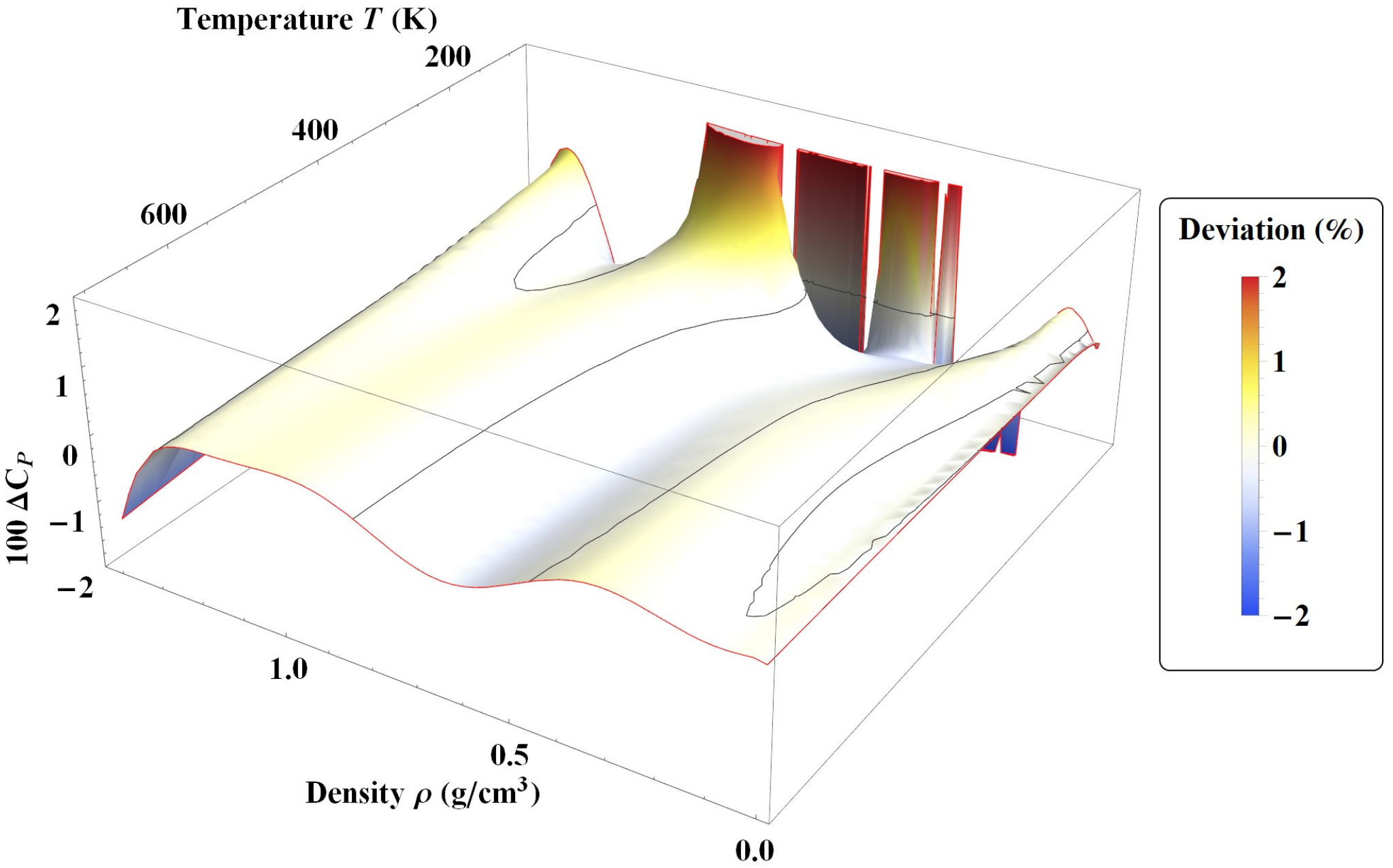

**Fig. 24:** Percentage deviations of isobaric heat capacity $\Delta C_P = \left(C_{P\,\text{TSW model}} - C_{P\,\text{Eq.(68)}}\right) / C_{P\,\text{TSW model}}$ between the TSW model and Eq. (68) in the range of density from $\rho_{t,\text{Gas}}$ to $\rho_{t,\text{Liq}}$. The black lines correspond to values of $\Delta C_P$ equal to zero.

It is again interesting to compare the results of the two models with the data from L'Air Liquide (729 data points, Ref. 19). Fig. 25 displays the data values of L'Air Liquide (Ref. 19) on two isotherms at 700 K and 1100 K (i.e. the highest isotherm) and, the corresponding calculated curves from the present and the TSW model. As for the $C_V$ data, the present model shows a closer fitting of the data from L'Air Liquide (Ref. 19) than the TSW model. The highest relative error (about 1%) is obtained for the isotherm at 1100K using the TSW model.

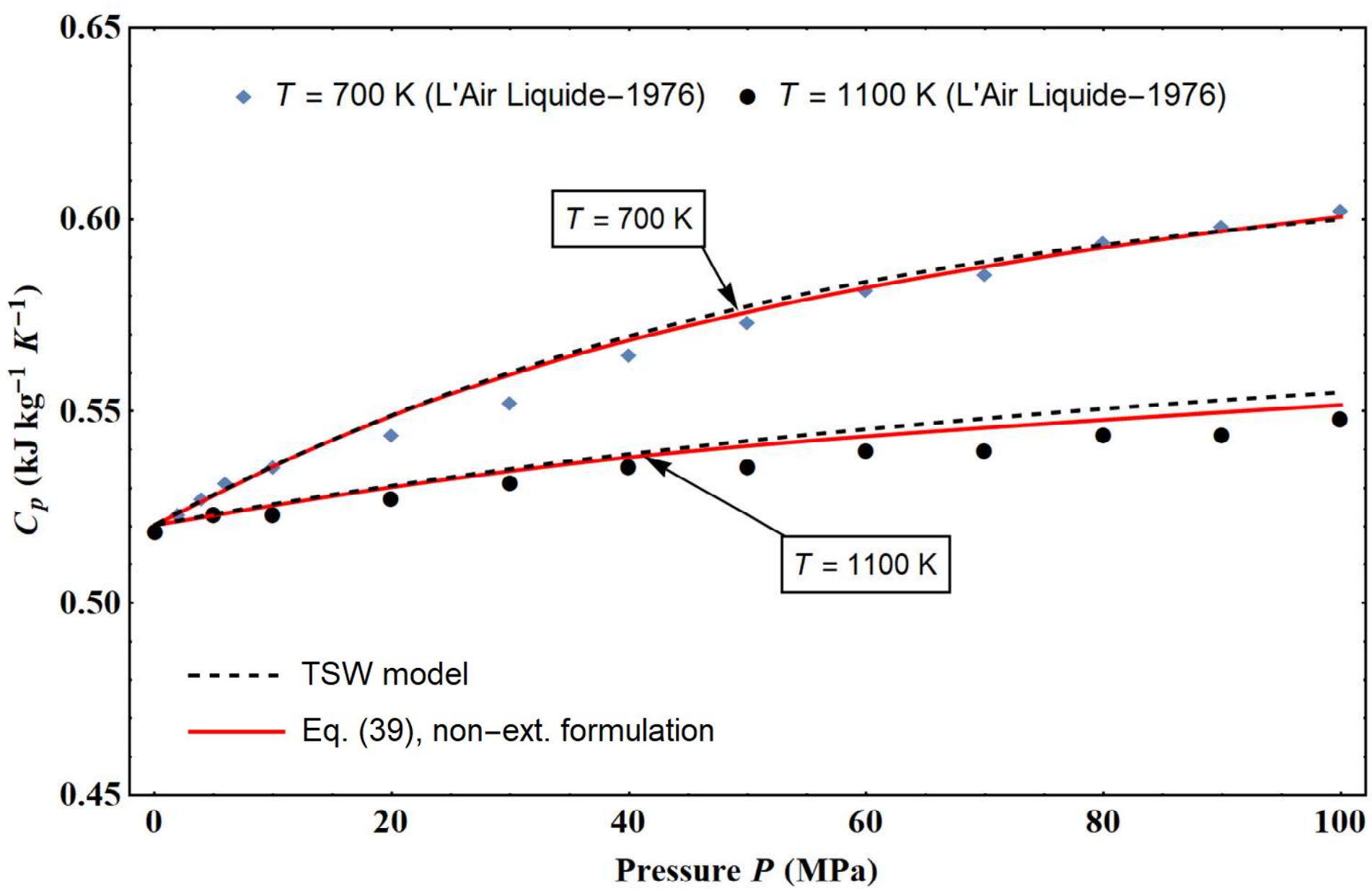

**Fig. 25:** Variations with pressure of the isobaric heat capacity $C_P$ along the two isotherms at 700 K and 1100 K: the blue diamonds and black points correspond to data from L'Air Liquide (Ref. 19) and the plotted curves correspond to the values calculated from Eq. (39) (red curves) and from the TSW model (black dashed curves).

The sound velocity $c$ does not diverge at the critical point, but exhibits a very pronounced minimum. However in the present model, $c$ is expressed on the basis of $C_P$ which diverges itself (Eq. (69)). Then for numerical reasons, we will compare the data calculated from the two models with the exception of the data on the respective curves of coexistence. The relative error on $c$ (see Fig. 26) between the TSW model and the present one shows a very similar variation with $\rho$ and $T$ that the one displayed by $C_P$ on Fig. 24. In the largest part of the ($\rho$, $T$) diagram, the relative error oscillates globally between ±0.5%, except for high density states and near the critical point where the error reaches 2%. On Fig. 43 of Ref. 4, the tolerance diagram for $c$ shows similar uncertainties that were obtained on Fig. 26, however some regions of their diagram present lower uncertainties (±0.02% and ±0.1%).

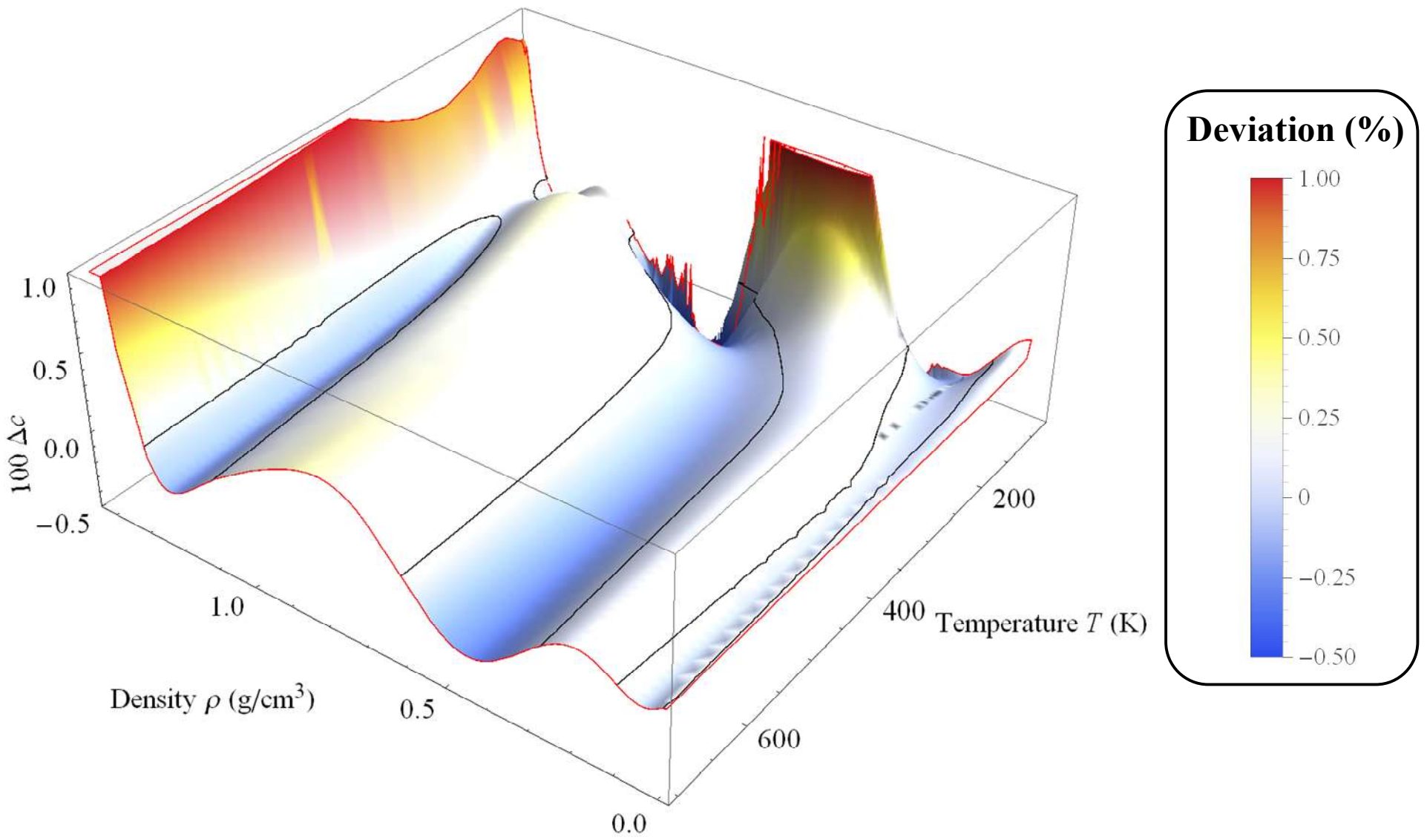

**Fig. 26:** Percentage deviations of sound speed $\Delta c = \left(c_{\text{TSW model}} - c_{\text{Eq.(69)}}\right)/c_{\text{TSW model}}$ between the TSW model and Eq. (69) in the range of density from $\rho_{\text{t,Gas}}$ to $\rho_{\text{t,Liq}}$. The black lines correspond to values of $\Delta c$ equal to zero.

If one compares the calculated data using the two models with the data of L'Air Liquide (296 data points, Ref. 19), it can be observed on Fig. 27 and Fig. 28 that, although the corresponding errors in the present model are sometimes higher than those of the TSW model, they are generally better centered on zero. This means that the isobars and isotherms variations are better predicted using the present model. Unfortunately, there are no data of sound speed from L'Air Liquide in the range 700 to 1100 K.

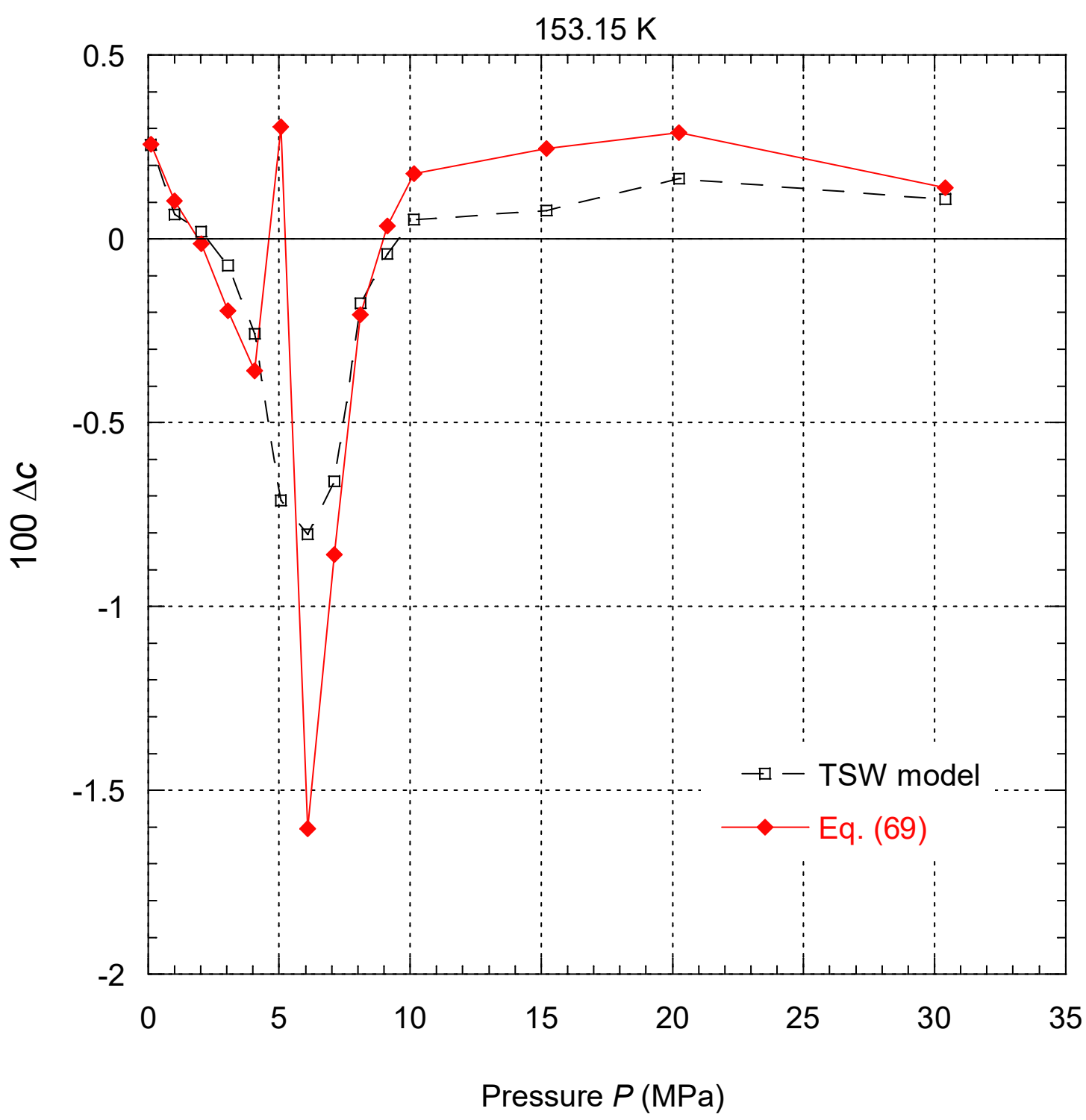

**Fig. 27:** Percentage deviations of sound speed $\Delta c = \left(c_{\text{L'Air Liquide}} - c_{\text{calc}}\right) / c_{\text{L'Air Liquide}}$ along the isotherm at 153.15 K between the data of L'Air Liquide (Ref. 19) and Eq. (69) (red diamonds) or the TSW model (black open squares). The lines are eye guides.

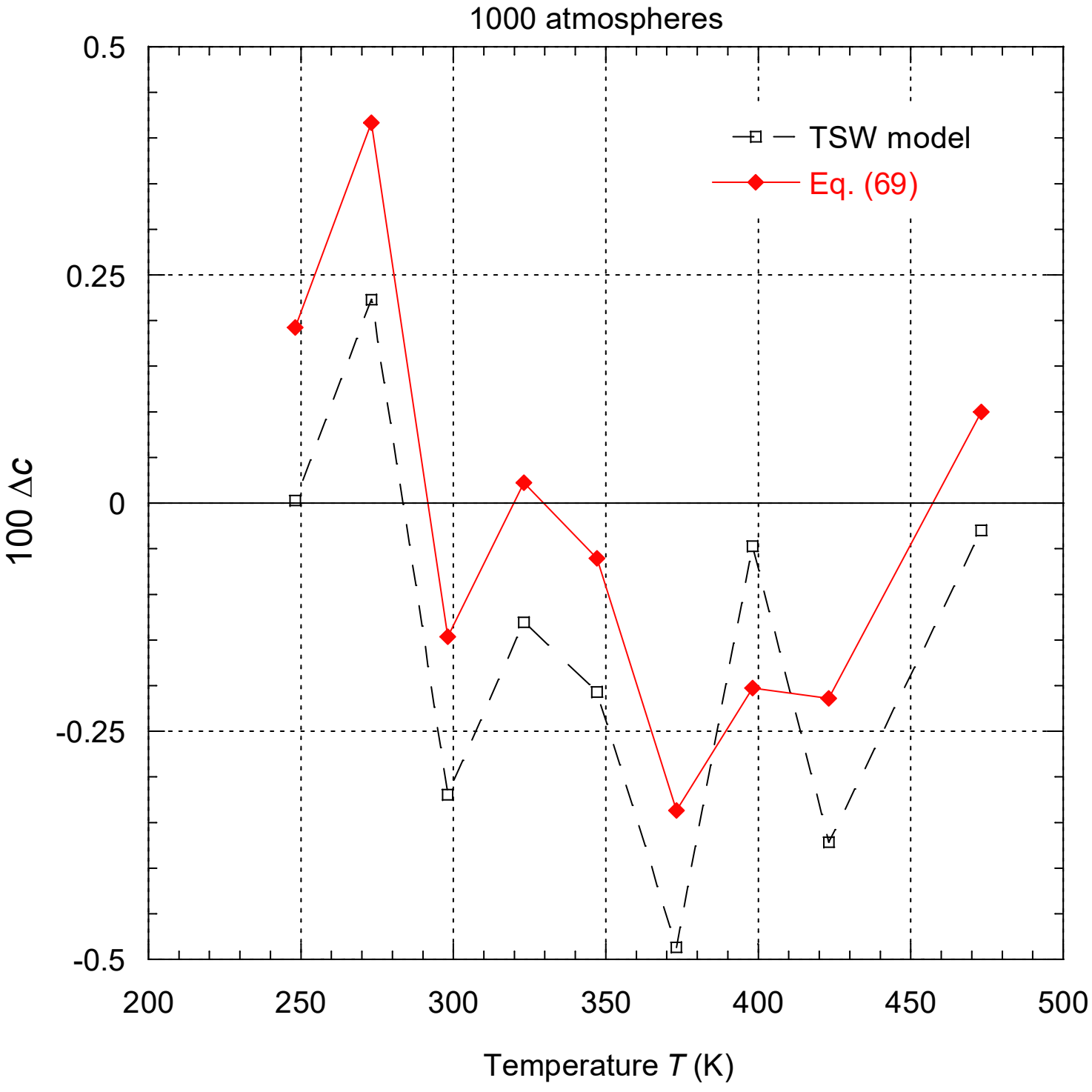

**Fig. 28:** Percentage deviations of sound speed $\Delta c = \left(c_{\text{L'Air Liquide}} - c_{\text{calc}}\right) / c_{\text{L'Air Liquide}}$ along the isobar at 1000 atmospheres between the data of L'Air Liquide (Ref. 19) and Eq. (69) (red diamonds) or the TSW model (black open squares). The lines are eye guides.

Equation (70) can be easily derived from the Gibbs-Helmholtz relations and its non-dimensional formulation $V^{-1}\delta_T$ reflects almost the behavior of thermal expansion coefficient. When this quantity is equal to zero, the fluid behaves as an ideal perfect gas. Due to the fact that zero is a possible value for this function; it is not possible to make a relative error analysis. Fig. 29 shows the absolute $\rho\delta_T$ vs. $P$ diagram for the same isotherms plotted on Fig. 33 of Ref. 4. It can be observed that the difference between the two models is very small and only more pronounced in the vicinity of the minimum of $\rho\delta_T$. On the isotherm at 162 K, the shape for the present modeling has a deeper well which seems slightly better in the light of data from Kim (Ref. 25).

Finally, it is important to note that, since the present model gives overall numerical results very close to those of the TSW model, both models have the same weakness for the representation of most of the experimental data very close to the critical point.

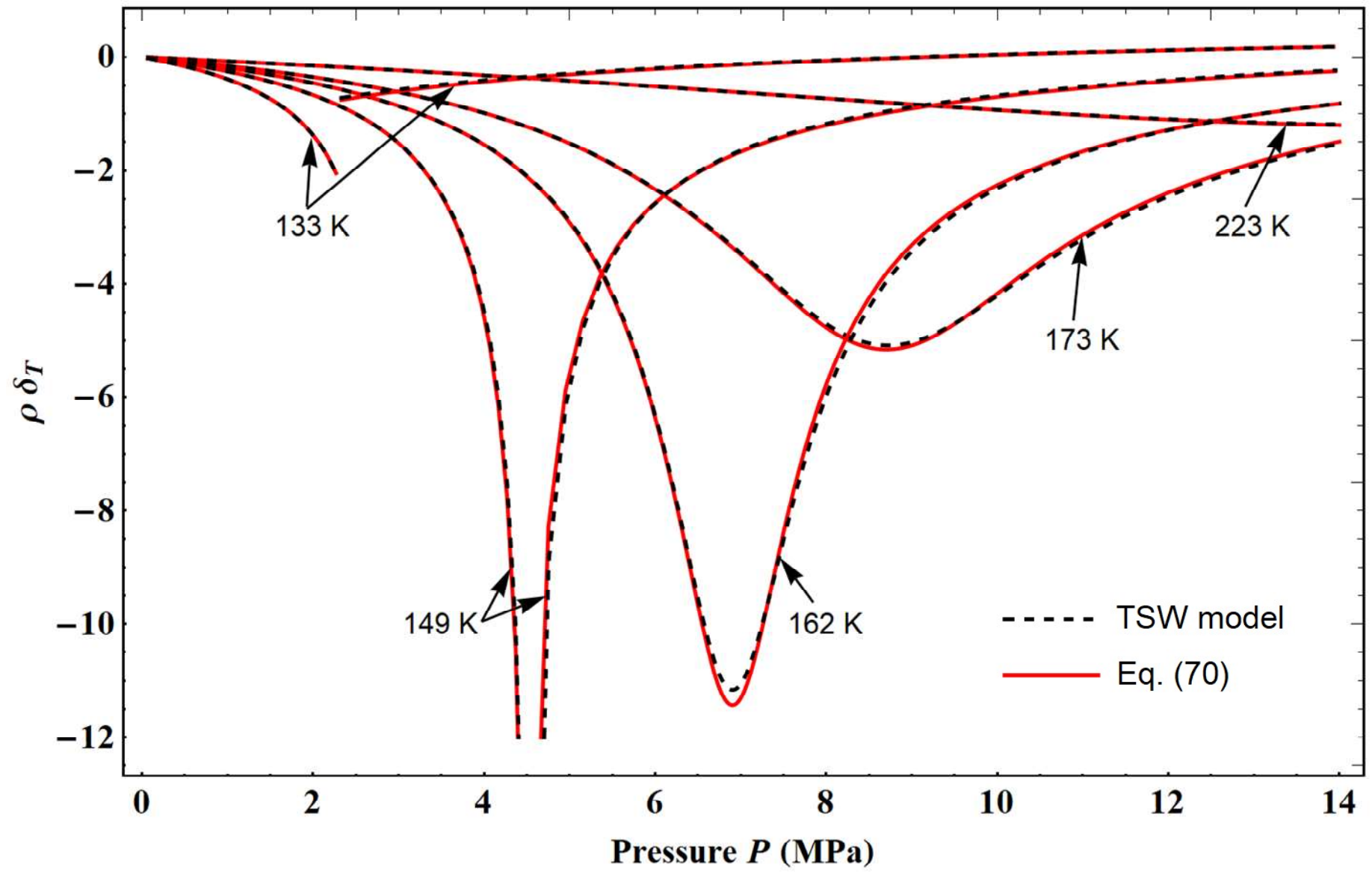

**Fig. 29:** Variations with pressure of the dimensionless isothermal throttling coefficient along 5 distinct isotherms. The plotted curves correspond to values calculated from Eq. (70) (red curves) and from the TSW model (black dashed curves).

### *4.2.4. The "ideal curves"*

Ideal curves are curves along which one property of a real fluid is equal to the corresponding property of the hypothetical ideal gas in the same state. The most important ideal curves can be obtained from the compressibility factor $Z$ and its first derivatives, i.e., the classical ideal curve ($Z$ = 1), the Boyle curve [$\left(\partial Z/\partial P\right)_T = 0$ or $\left(\partial Z/\partial V\right)_T = 0$], the Joule-Thomson inversion curve (or Charles curve) [$\left(\partial Z/\partial T\right)_P = 0$ or $\left(\partial Z/\partial V\right)_P = 0$], and the Amagat curve (or Joule curve) [$\left(\partial Z/\partial P\right)_\rho = 0$ or $\left(\partial Z/\partial T\right)_\rho = 0$]. For Argon, all ideal curves lie within the range covered by data from NIST and Ronchi, with the exception of the high-temperature part of the Amagat curve.

Fig. 30 shows the plot of the ideal curves calculated from Eq. (46) and its derivatives and from the TSW model. Inside the single phase domain where reliable data exists, both equations show the expected variations of ideal curves. The visible differences occur for the part of each curve corresponding to very low densities. This can be explained by the fact that, in the present model, the various thermodynamic functions are designed to converge to a physically admissible value when density tends towards zero. Another difference can be seen on the high-temperature Amagat curve, which is explained by a better representation of the Ronchi data than the TSW model (as shown in Fig. 22 and Fig. 23).

It thus appears that in the pressure and temperature range covered by the NIST and Ronchi data, the compressibility factor and its first derivatives are well represented by the present model.

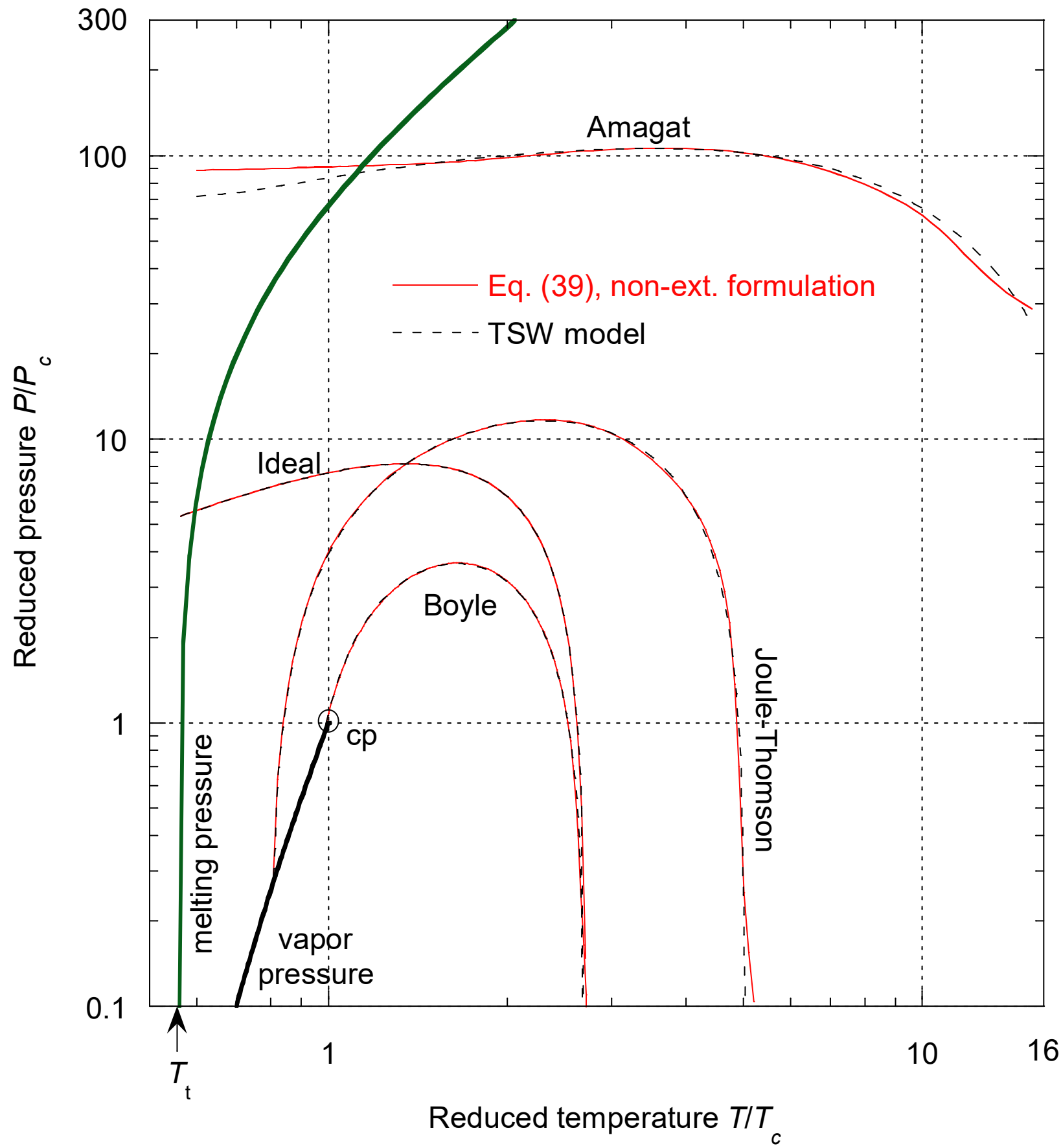

**Fig. 30:** Logarithmic plot of the so-called “ideal curves” calculated from Eq. (39) (red curves) and from the TSW model (black dashed curves) in the range of temperature covered by the data from NIST (Ref. 5) and from Ronchi (Ref. 6). The black curve represents the saturated vapor pressure curve, ending at the critical point cp. The dark green curve represents the melting curve calculated from Eq. (67).

#### *4.2.5. Extrapolation to high temperatures*

Tegeler *et al.* (Ref. 4) compared their model to data resulting from the shock wave experiments of van Thiel *et al.* (Ref. 26), Nellis *et al.* (Ref. 27) and Grigor’ev *et al.* (Ref. 28). The pressure and density are calculated from the Hugoniot relations by using experimental velocity measurements. All of these data are in the pressure range and density range of the Ronchi’s data but not in its temperature range. For example, all the data of Grigor’ev *et al.* (Ref. 28) correspond to a temperature range from 3700 K to 17000 K. In addition, for most of these experiments argon is ionized and the corresponding physics is clearly not included in the present approach but also it is not explicitly included in the TSW model.

For all data that are in the Ronchi’s domain, the Hugoniot curve determined with the present model is consistent with the data, as is the Hugoniot curve calculated by the TSW model. This can be easily understood because the Hugoniot states depend mainly on the behavior of the Poisson adiabatic curves which have close variations till to the melting line as can be seen on Fig. 31 for the two different initial states which corresponds to those of van Thiel *et al.* (Ref. 26).

From these results, we suggest not to extrapolate the present model outside the highest limit of the Ronchi’s domain.

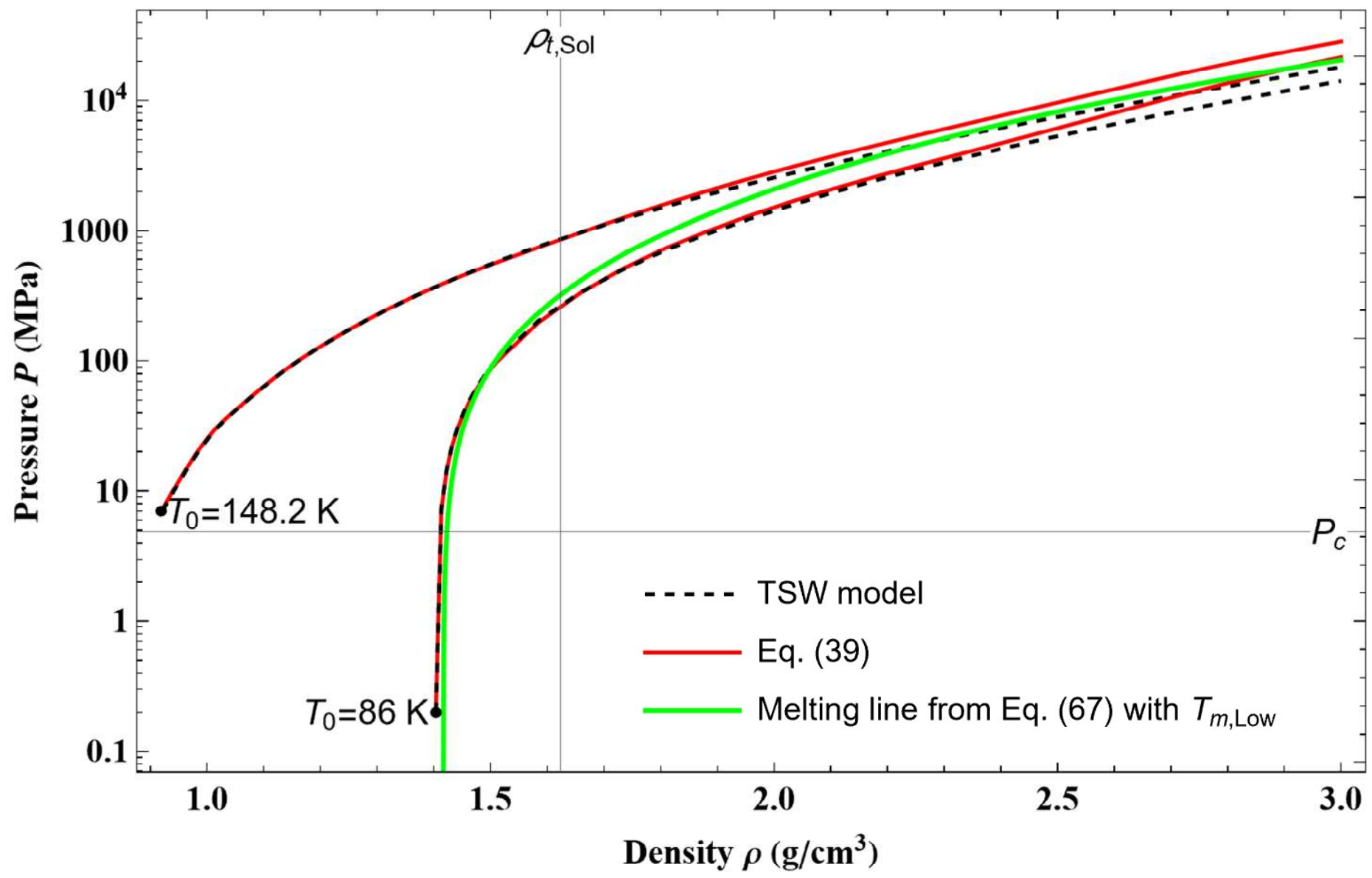

**Fig. 31:** Variations with density of the Poisson adiabatic curves calculated from Eq. (39) (red curves) and from the TSW model (black dashed curves) for the two different initial states of van Thiel *et al*. (Ref. 26). The density $\rho_{t,Sol}$ represents the triple point solid density.

## 4.3. Liquid-vapor phase boundary

### *4.3.1. Isochoric heat capacities*

Along the coexistence curve, the relative errors between the data from NIST and from the present modeling oscillate about ±0.45% (see Fig. 11). This result can be compared to the uncertainties given on Fig. 44 of Ref. 4. The relative error of the present model on the saturated liquid side is smaller than the one given in Fig. 44 of Ref. 4 (±0.45% instead of ±2%) and slightly larger on the saturated vapor side (±0.45% instead of ±0.3%).

Along this coexistence line, no NIST data are available in the density range 0.5 - 0.6 g/cm$^3$. However, this region that extends on both sides of the critical point is covered along some isochors that crossed the coexistence curve by the data of Voronel *et al.* (Refs. 29 and 30). Fig. 32 shows that the two models (TSW and present one) lead to similar variations with $T$ and the discrepancies with the data of Voronel *et al.* (Refs. 29 and 30) increase more and more as one approaches the coexistence curve. Inside the coexistence phase, the data of Voronel *et al.* (Refs. 29 and 30) show a peak in $C_V$ which is not symmetrical. Such $C_V$ variation is in any case impossible to reproduce using the TSW model (see Fig. 8). On the other hand, the present model could be modified to correctly describe such $C_V$ variation. Indeed, the parameter $T_{\text{div}}$ was inserted into the model (i.e. Eq. 10) to qualitatively describe the $C_V$ divergence inside the coexistence phase (see Fig. 8). It is however evident, from the data of Voronel *et al.* (Refs. 29 and 30), that the values defined by $T_{\text{div}}$ are not quantitatively suitable. The peak position of $C_V$ along the entire coexistence curve could be used to establish a new equation for $T_{\text{div}}$ leading to a reliable fitting of $C_V$ divergence inside the coexistence phase. The other side of $T_{\text{div}}$ (i.e. for $T < T_{\text{div}}$) could also be easily modeled without changing any properties for the single phase region. Unfortunately, the data of Voronel *et al.* (Refs. 29 and 30) which are limited to a very small density range are insufficient to be taken into account in view to improve the present model into the coexistence phase. It can also be noted that the data of Voronel *et al.* (Refs. 29 and 30) have been correctly modeled by Rizi *et al.* (Ref. 31) using the crossover model. However, this model, which contains coefficients among which a number are unknown, cannot be put in practice.

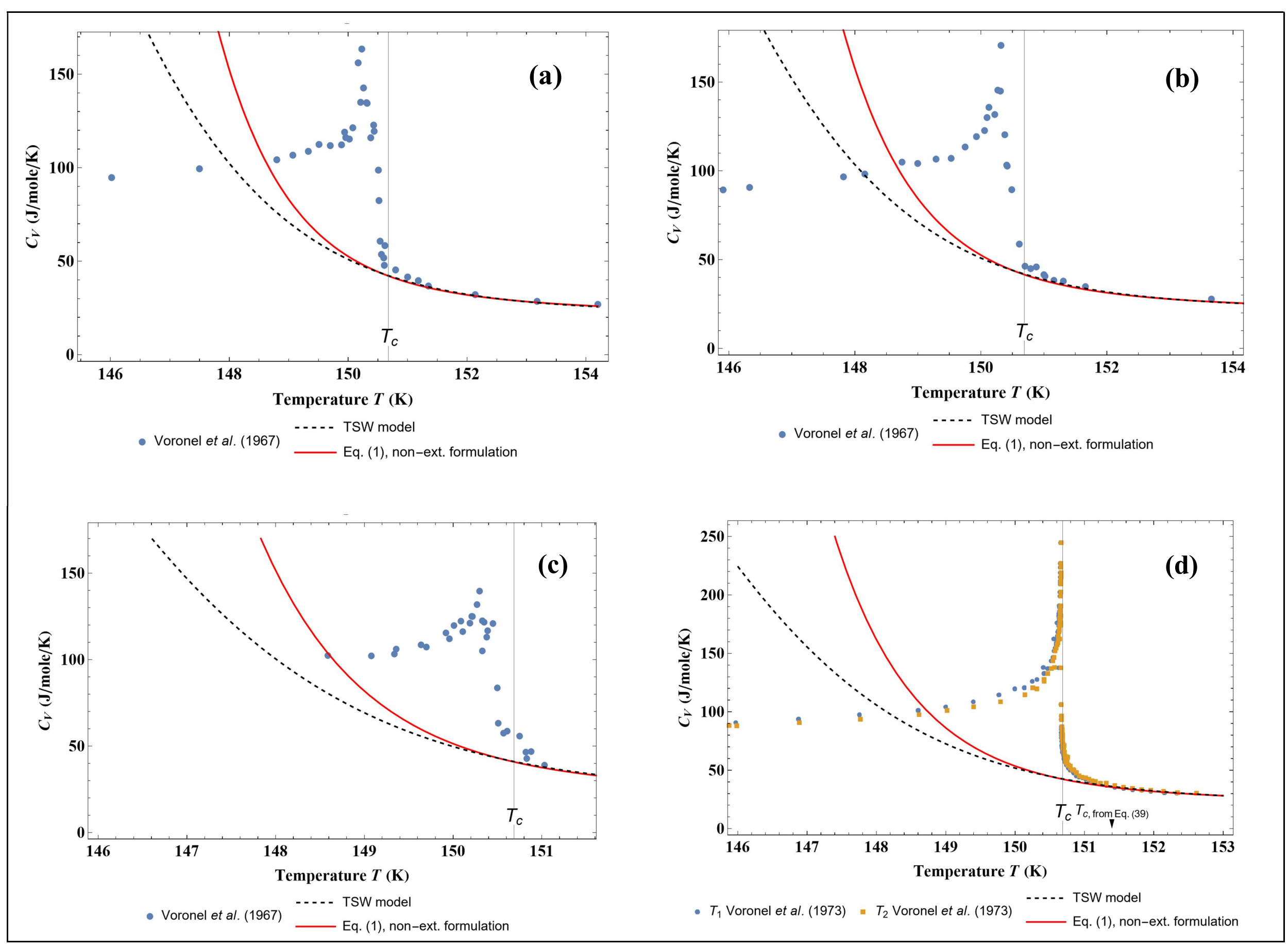

**Fig. 32:** Variations with temperature of the isochoric heat capacity $C_V$ in the vicinity of the critical point along 4 isochors: (a) 0.504 g/cm$^3$, (b) 0.549 g/cm$^3$, (c) 0.560 g/cm$^3$ and (d) 0.531 g/cm$^3$. The data points are from Voronel' *et al*. (Refs. 29 and 30) and the plotted curves correspond to values calculated from Eq. (1) (red curves) and from the TSW model (black dashed curves).

**Table 10.** Characteristic values of the coexistence line calculated from the thermal equation of state and using the NIST values.

| | Unit | NIST | TSW model | Non-extensive formulation, Eq. (45) |
|---|---|---|---|---|
| $P_c(\rho_c,T_c)$ | MPa | 4.863 | 4.86299 | 4.86298 |
| $\rho_c(P_c,T_c)$ | g/cm$^3$ | 0.535599 | 0.549928 | 0.535526 |
| $P_t(\rho_{t,Liq},T_t)$ | MPa | 0.068891 | 0.082671 | 0.0688907 |
| $\rho_{t,Liq}(P_t,T_t)$ | g/cm$^3$ | 1.4168 | 1.41676 | 1.41680 |
| $P_t(\rho_{tGas},T_t)$ | MPa | 0.068891 | 0.0688913 | 0.068891 |
| $\rho_{t,Gas}(P_t,T_t)$ | g/cm$^3$ | 0.0040546 | 0.00405458 | 0.00405460 |

**Table 11.** Characteristic values of the coexistence line calculated from the Maxwell relations, i.e. Eqs. (50) to (52).

| | NIST | TSW model | Non-extensive formulation |
|---|---|---|---|
| $T_c$ (K) | 150.687 | 150.687 | 151.396 |
| $P_c$ (MPa) | 4.863 | 4.86295 | 4.99684 |
| $\rho_c$ (g/cm$^3$) | 0.535599 | 0.533136 | 0.543786 |
| $T_t$ (K) | 83.8058 | 83.8058 | 83.8058 |
| $P_t$ (MPa) | 0.068891 | 0.0688908 | 0.0689657 |
| $\rho_{t,Liq}$ (g/cm$^3$) | 1.4168 | 1.41676 | 1.416802 |
| $\rho_{t,Gas}$ (g/cm$^3$) | 0.0040546 | 0.00405457 | 0.00405912 |

### *4.3.2. Thermal properties*

The coexistence phase is characterized by three specific points which are the saturated liquid triple point, the saturated vapor triple point and the critical point. These points correspond to different well-known thermodynamic states.

From the thermal equation of state $P(\rho,T)$ of the TSW model and the present one, the specific points have been calculated using for $\rho$ and $T$ the data of NIST. Table 10 shows that the calculated values of the three characteristic states using the present model are globally more accurate than the one calculated using the TSW model and, particularly for the triple point on the saturated liquid curve. As shown in Table 10, near the liquid saturated curve, even a tiny variation on density produces a large variation on the calculated pressure, i.e. the density values for these states must be extremely accurate. Thus, the TSW model gives an error of 0.003% on $\rho_{\mathrm{t,Liq}}(P_{\mathrm{t}},T_{\mathrm{t}})$ and this leads to an error of 20% on $P_{\mathrm{t}}(\rho_{\mathrm{t,Liq}},T_{\mathrm{t}})$.

At a given temperature, the vapor pressure and the densities of the coexisting phases ($\rho_{\sigma l}$ and $\rho_{\sigma v}$) can also be calculated from the Maxwell criterion of phase equilibrium conditions and, therefore, the characteristics values of the coexistence line (triple and critical points) as well. Table 11 shows these characteristics values calculated from the TSW model and the present model (i.e. Eqs. (50) to (52)). The data from NIST and those calculated with the TSW model are in good agreement, but this is not the case for the present model particularly for the critical point. However, it can be noticed that $T_{\mathrm{c}}$ and $T_{\mathrm{t}}$ are imposed values for the TSW model whereas only $T_{\mathrm{t}}$ is fixed in the present one. Then, the critical values ($P_{\mathrm{c}}$, $T_{\mathrm{c}}$, $\rho_{\mathrm{c}}$) are calculated ones in the present approach. Eq. (45) leads to better results for characteristics values than those calculated with the TSW model while, using the Maxwell equations, it is the opposite.

How to explain this result? The Maxwell equations represent the equality of pressure, temperature and specific Gibbs energy in the coexisting phases. The present approach built all the required thermodynamics quantities ($U$, $S$, $F$, etc.) from the empirical description of the experimental data of $C_V(\rho,T)$ and $P(\rho,T)$. The accuracy of these empirical equations to describe the experimental data has been shown to be of very high quality. Therefore, the discrepancy between the critical values ($P_{\mathrm{c}}$, $T_{\mathrm{c}}$, $\rho_{\mathrm{c}}$) calculated with the present model and those from NIST can be attributed to the inconsistency of the data in this critical state. The agreement between the calculated liquid triple point and the experimental one is better using the present model; this means that, on the liquid side, the present isochors network is slightly twisted compared to the TSW model network. This slight distortion of the isochors network on the liquid side has then a strong impact on the construction of the coexistence curve as already shown in Table 11. Indeed, considering that there is good agreement with the TSW model on the gas side but not as good on the liquid side, it is clear that the equilibrium conditions deduced from the Maxwell relations must be different. Table 12 gives a numerical summary of all the thermodynamic quantities on the saturation curve using the present non-extensive formulation.

Fig. 33 shows that the relative error between the TSW model and the present one for the saturated liquid density is less than ±0.2% in the range $T_{\mathrm{t}}$ to 139 K, which is within the uncertainty of the data selected by Tegeler *et al.* (see Fig. 5 of Ref. 4). Therefore, on the liquid side, the network of isochors induced by the present model below 139 K is clearly more realistic.

Also, Fig. 33 shows that the relative error between the TSW model and the present one for the saturated vapor density is within the uncertainty of the data selected by Tegeler *et al.* (see Fig. 6 of Ref. 4) in the range 100 K to 149.5 K. Below 100 K, the error of the present model is 3 to 4 times larger than the claimed experimental uncertainties of Gilgen *et al.* (Ref.

32). But, as mentioned by Tegeler *et al.*, the densities on the saturated vapor curve were extrapolated from measurement in the homogeneous region close to the phase boundary. Such density values are obviously depending on the method used for doing the extrapolation. Although the error of the present model in this region is relatively high, the calculated density values are compatible with the experimental data.

Finally, Fig. 33 also shows that the relative error between the TSW model and the present one for the saturated vapor pressure is within the uncertainty of the data selected by Tegeler *et al.* in the range 97 K to 144 K. Below 97 K, the error of the present model oscillates slightly around -0.1% which is within the uncertainty of the data assigned to Group 2 by Tegeler *et al.*, therefore, it can be said that these results are also in agreement with the experimental data.

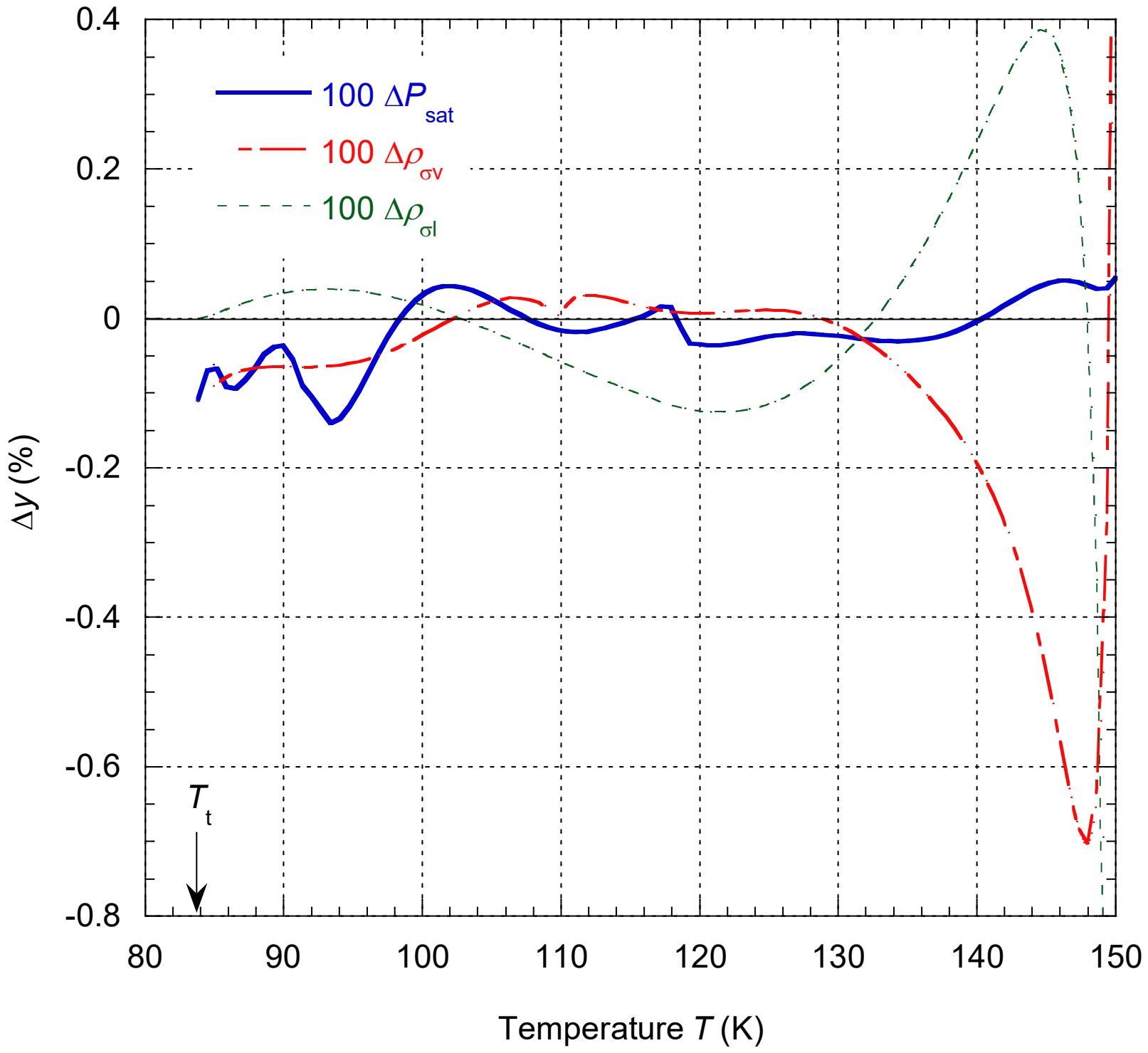

**Fig. 33:** Percentage deviations $\Delta y = \left(y_{\mathrm{NIST}} - y_{\mathrm{calc}}\right)/y_{\mathrm{NIST}}$ of the selected thermal data at saturation from values calculated from Eq. (39) in the range of temperature from $T_t$ to 149.5K.

However, in the vicinity of the critical state, the two models lead to very different values. This is due to the different approaches used for the two models. For the TSW model, the parameters of the critical point are imposed, whereas they are calculated in the present model. Apart from the numerical values, Fig. 34 shows that the shape of the saturated vapor pressure curve around the critical point depends on the model. The TSW model generates an extremely “flat” variation on a wide range of density around the critical point when the present model produces a more rounded variation in the same range of density. This last variation is closer of the experimental saturated vapor pressure curve from L’Air Liquide (Ref. 19) than the one given by the TSW model. The fact that the critical state is imposed in the TSW model seems to lead to a forced flattening-out of the saturated vapor pressure curve at the critical point.

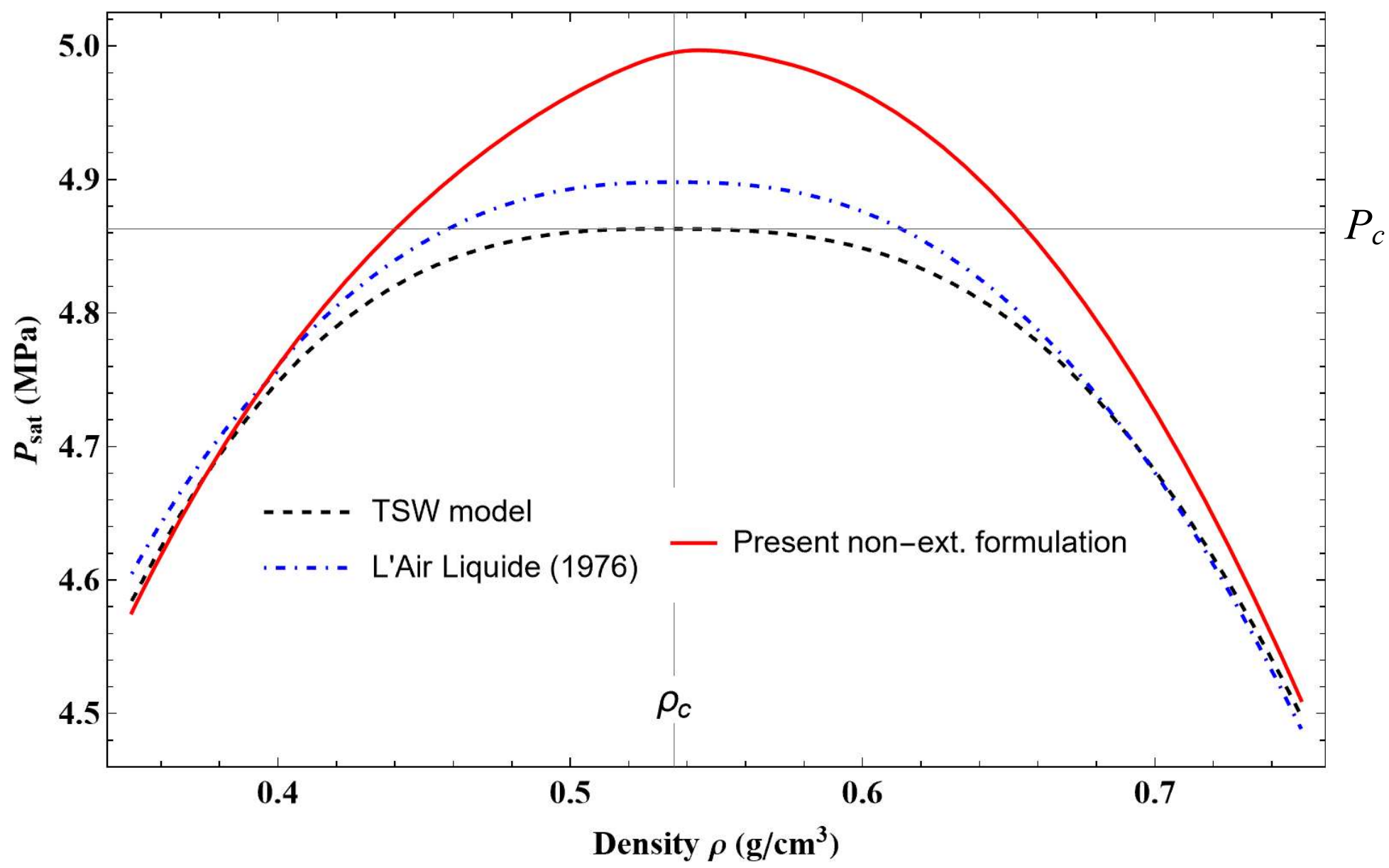

**Fig. 34:** Variations with density of the saturated pressure in the vicinity of the critical point (i.e. from 0.35 g/cm$^3$ to 0.75 g/cm$^3$). The plotted curves correspond to values calculated from Maxwell relations Eqs. (50) to (52) (red curve), from the TSW model (black dashed curve) and from the data of L'Air Liquide (blue dot-dashed curve, Ref. 19).

If we use $T_{\text{sat}}(\rho)$ on the phase boundary derived from NIST data and calculate the saturated pressure curve using Eq. (45), then Fig. 35 shows that the relative error on the saturated vapor pressure curve $P_{\text{sat}}(\rho)$ is less than ±0.2% below $\rho_c$. This uncertainty is compatible with the uncertainty of the data assigned to Group 2 by Tegeler *et al*. On the other hand, the relative error on the saturated liquid pressure curve $P_{\text{sat}}(\rho)$ in the range $\rho_c$ to 0.85 g/cm$^3$ is compatible with the uncertainty of the data assigned to Group 3 by Tegeler *et al*. From this, it can be concluded that the present thermal equation of state is probably not enough accurate in the range 145 K to $T_c$.

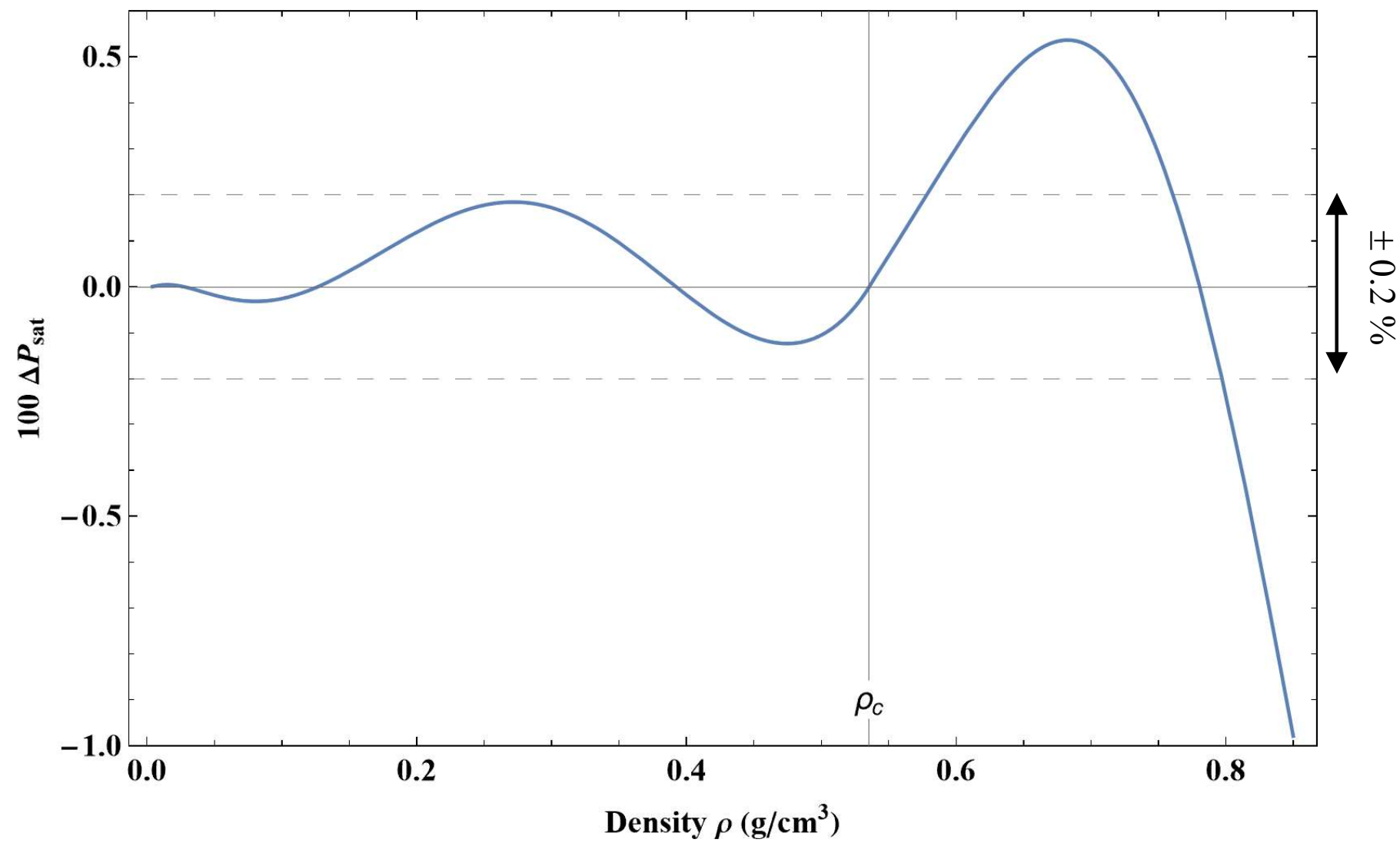

**Fig. 35:** Percentage deviations of saturated pressure $\Delta P_{\text{sat}} = \left(P_{\text{TSW mod.}}\left(T_{\text{sat}}(\rho),\rho\right) - P_{\text{calc}}\left(T_{\text{sat}}(\rho),\rho\right)\right)/P_{\text{TSW mod.}}\left(T_{\text{sat}}(\rho),\rho\right)$ between the TSW model and Eq. (45) in the range of density from $\rho_{\text{t,Gas}}$ to 0.85 g/cm$^3$. The curve $T_{\text{sat}}(\rho)$ used is an interpolation of the data from NIST (Ref. 5).

From the analysis of the latent heat of vaporization $L_v = H_{\sigma v} - H_{\sigma l}$, the effect of cumulative errors between the properties on the saturated vapor and saturated liquid sides can be determined. Fig. 36 shows that until 149.5 K, the relative error between the TSW model and ours is far insight of the experimental data uncertainties shown on Fig. 15 of Ref. 4 and, more particularly from $T_t$ to 134 K, the relative error is less than ±0.1%.

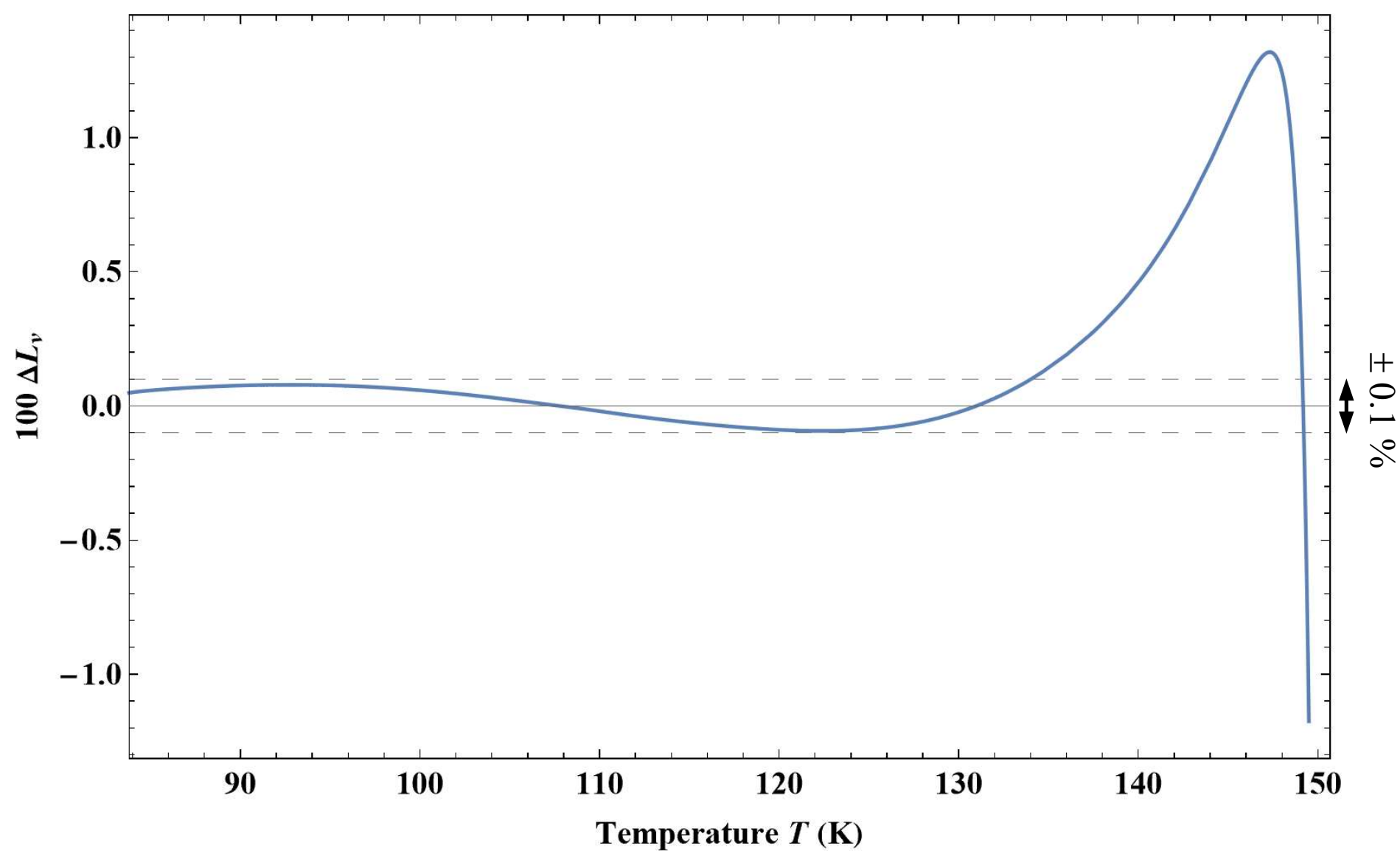

**Fig. 36:** Percentage deviations of latent heat of vaporization $\Delta L_v = \left(L_{v,\text{TSW model}} - L_{v,\text{Eq.}(39)}\right)/L_{v,\text{TSW model}}$ between the TSW model and Eq. (39) in the range of temperature from $T_t$ to 149.5 K.

Owing to the fact that the “ideal curves” are correctly described - in particular near the critical point- but that the saturated pressure is not correctly reproduced in the vicinity of the critical point, it can be concluded that the problem comes from the fact that the experimental data in this region are not correctly described and, moreover are not enough coherent between themselves. These arguments can be easily observed with the spinodal properties.

#### *4.3.3. The spinodal properties*

The spinodal properties correspond to the metastable states of the fluid system. The knowledge of these metastable states is important for industrial processes that are involving ever increasing heat fluxes and rapid transients but also for testing the validity of a new equation of state formulation.

Most of the available experimental data pertain to states much closer to the saturated liquid state than the spinodal limit except very close to the critical point. The experimental data of Voronel *et al*. (Refs. 29 and 30) crossed the spinodal limit in a very narrow range of density around the critical point and the divergence states are shown on Fig. 37. This figure also shows the liquid spinodal data points from Baĭdakov *et al*. (Ref. 33). These data where determined from experimental $P\rho T$ data combined with a simple theoretical equation of $C_V(\rho,T)$. So these data points are dependent of the theoretical variations of $C_V$ chosen by Baĭdakov *et al*. (Ref. 33). Fig. 37 shows that these data decrease rapidly as the density increase but they are compatible with the spinodal states determined from the present approach or from the TSW model, excepted for the TSW model around the density of 0.8 g/cm$^3$ where a strong unphysical hole appears due to uncontrollable strong oscillations of the polynomial terms.

Fig. 37 shows that globally the two spinodal curves determined from the present approach and the TSW model are very close except close the liquid triple point where our spinodal curve shows a strong decrease vs. density. This is due to the fact that in this region the $P\rho T$ data are not represented with enough accuracy by the TSW model. It has already be seen in previous sections that the present model have a better accuracy in this region, accuracy that was obtained by "twisting" the isochors network. The strong decrease of the spinodal curve near the liquid triple point is simply the results of this locally network deformation.

From Fig. 37, it is possible to compare our divergence curve of $C_V$ with the spinodal curve. If the data set of $C_V$ and of $P\rho T$ used for the theoretical developments were sufficiently coherent both curves would be identical. This is approximately true only on the liquid side for densities higher than 0.9 g/cm$^3$ and on the gaseous side in the range of density from 0.025 g/cm$^3$ to 0.15 g/cm$^3$. Elsewhere, this is not the case showing that the variations of $C_V$ close to the saturation curve are not correctly represented. Since high accuracy is obtained with the TSW model, it means that the variations of $C_V$ calculated from the TSW model have not the good shape. This shows that it is not enough to have great precision with *a priori* selected set of data to ensure a coherent representation. Local variations in some measured quantities have physical non-negligible importance.

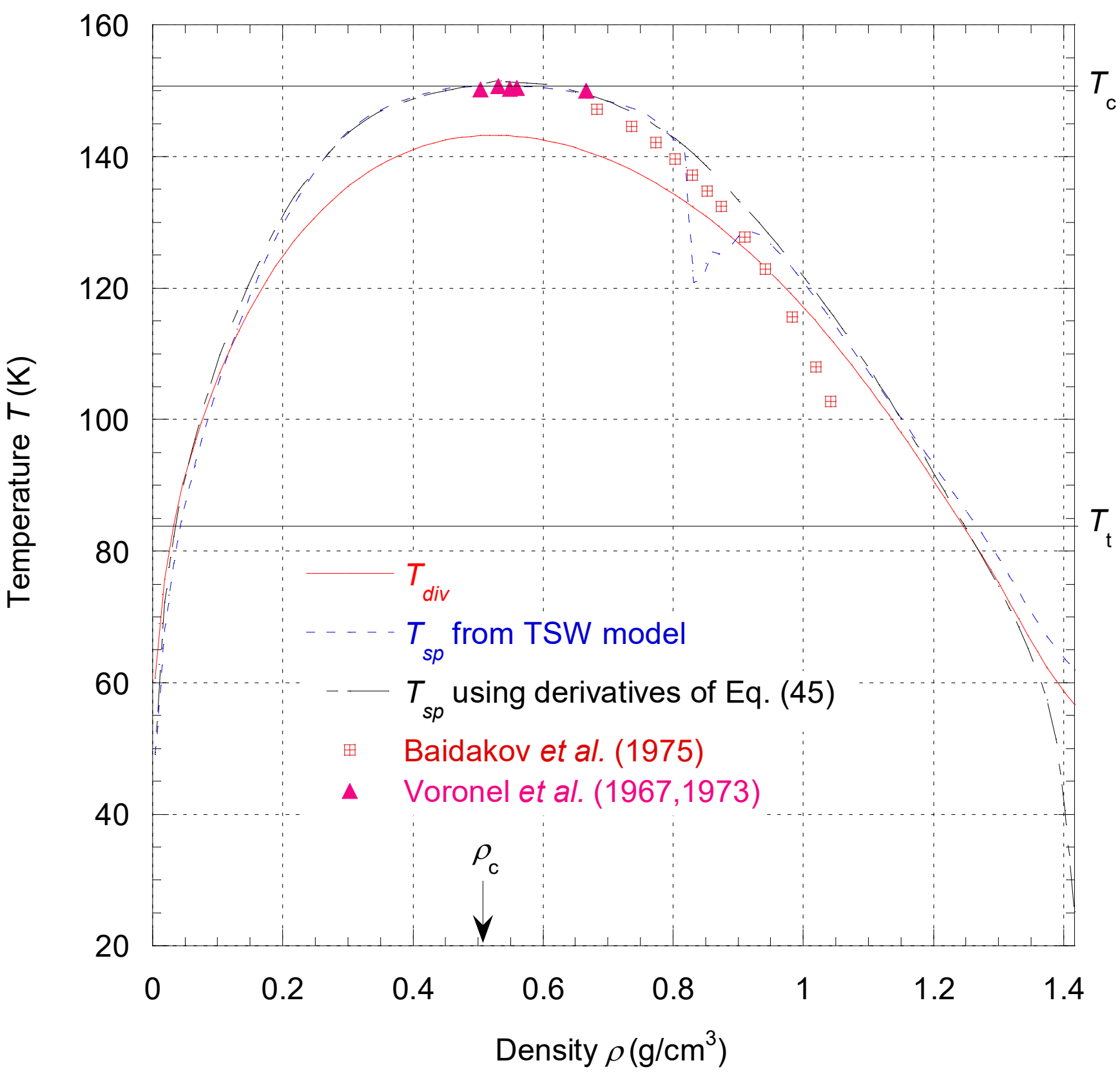

**Fig. 37:** Variations with density of the spinodal temperature. The plotted curves correspond to the divergence curve of $C_V$, the calculated values using derivation of Eq. (45) and the TSW model. The data points are from Voronel *et al*. (Refs. 29 and 30) and Baidakov *et al*. (Ref. 33).

**Table 12 :** Thermodynamic parameters of saturated argon. For each temperature, the first line corresponds to the liquid and the second line to the gas.

| Temperature (K) | Pressure (MPa) | Density (g cm$^{-3}$) | Enthalpy (kJ kg$^{-1}$) | Entropy (kJ kg$^{-1}$ K$^{-1}$) | $C_V$ (kJ kg$^{-1}$ K$^{-1}$) | $C_P$ (kJ kg$^{-1}$ K$^{-1}$) | $c$ (m s$^{-1}$) |
|---|---|---|---|---|---|---|---|
| 83.8058[a] | 0.068 891 | 1.41680 | -121.39 | 1.3297 | 0.548 64 | 1.11895 | 856.17 |
| | | 0.0040546 | 42.23 | 3.2824 | 0.326 40 | 0.55734 | 168.02 |
| 84 | 0.070 522 | 1.41561 | -121.17 | 1.3323 | 0.547 95 | 1.11968 | 854.73 |
| | | 0.0041430 | 42.31 | 3.2786 | 0.326 60 | 0.55794 | 168.17 |
| 86 | 0.088 193 | 1.40321 | -118.92 | 1.3586 | 0.540 97 | 1.12304 | 840.88 |
| | | 0.0050845 | 43.08 | 3.2425 | 0.328 55 | 0.56400 | 169.74 |
| 88 | 0.109 096 | 1.39072 | -116.67 | 1.3844 | 0.534 26 | 1.12305 | 828.22 |
| | | 0.0061790 | 43.82 | 3.2082 | 0.330 63 | 0.57071 | 171.24 |
| 90 | 0.133 597 | 1.37815 | -114.41 | 1.4095 | 0.527 82 | 1.12402 | 815.37 |
| | | 0.0074416 | 44.52 | 3.1755 | 0.332 84 | 0.57812 | 172.67 |
| 92 | 0.162 078 | 1.36549 | -112.15 | 1.4342 | 0.521 64 | 1.12613 | 802.28 |
| | | 0.0088882 | 45.18 | 3.1443 | 0.335 19 | 0.58629 | 174.02 |
| 94 | 0.194 930 | 1.35271 | -109.88 | 1.4583 | 0.515 69 | 1.12944 | 788.92 |
| | | 0.0105356 | 45.80 | 3.1145 | 0.337 68 | 0.59531 | 175.30 |
| 96 | 0.232 558 | 1.33979 | -107.60 | 1.4820 | 0.509 96 | 1.13400 | 775.28 |
| | | 0.0124010 | 46.37 | 3.0859 | 0.340 32 | 0.60523 | 176.51 |
| 98 | 0.275 373 | 1.32672 | -105.31 | 1.5053 | 0.504 45 | 1.13988 | 761.35 |
| | | 0.0145031 | 46.89 | 3.0584 | 0.343 11 | 0.61617 | 177.64 |
| 100 | 0.323 796 | 1.31346 | -103.00 | 1.5282 | 0.499 15 | 1.14718 | 747.12 |
| | | 0.0168613 | 47.36 | 3.0319 | 0.346 07 | 0.62824 | 178.69 |
| 102 | 0.378 255 | 1.29999 | -100.68 | 1.5508 | 0.494 05 | 1.15597 | 732.59 |
| | | 0.0194966 | 47.78 | 3.0063 | 0.349 19 | 0.64156 | 179.68 |
| 104 | 0.439 185 | 1.28630 | -98.34 | 1.5730 | 0.489 16 | 1.16640 | 717.75 |
| | | 0.0224312 | 48.14 | 2.9816 | 0.352 51 | 0.65630 | 180.58 |
| 106 | 0.507 023 | 1.27233 | -95.97 | 1.5951 | 0.484 46 | 1.17859 | 702.59 |
| | | 0.0256892 | 48.43 | 2.9575 | 0.356 03 | 0.67264 | 181.41 |
| 108 | 0.582 216 | 1.25808 | -93.58 | 1.6169 | 0.479 97 | 1.19271 | 687.10 |
| | | 0.0292969 | 48.67 | 2.9341 | 0.359 77 | 0.69081 | 182.17 |
| 110 | 0.665 211 | 1.24349 | -91.15 | 1.6385 | 0.475 68 | 1.20896 | 671.25 |
| | | 0.0332828 | 48.83 | 2.9112 | 0.363 75 | 0.71110 | 182.84 |
| 112 | 0.756 461 | 1.22855 | -88.70 | 1.6600 | 0.471 60 | 1.22758 | 655.04 |
| | | 0.0376785 | 48.92 | 2.8888 | 0.368 01 | 0.73383 | 183.44 |
| 114 | 0.856 424 | 1.21319 | -86.20 | 1.6814 | 0.467 75 | 1.24886 | 638.45 |
| | | 0.0425194 | 48.94 | 2.8669 | 0.372 57 | 0.75942 | 183.97 |
| 116 | 0.965 560 | 1.19739 | -83.66 | 1.7027 | 0.464 15 | 1.27315 | 621.45 |
| | | 0.0478449 | 48.86 | 2.8452 | 0.377 47 | 0.78839 | 184.41 |
| 118 | 1.0843 | 1.18108 | -81.07 | 1.7240 | 0.460 80 | 1.30090 | 604.01 |
| | | 0.0537001 | 48.70 | 2.8238 | 0.382 76 | 0.82138 | 184.77 |
| 120 | 1.2132 | 1.16422 | -78.43 | 1.7452 | 0.457 74 | 1.33265 | 586.10 |
| | | 0.0601365 | 48.44 | 2.8026 | 0.388 49 | 0.85921 | 185.05 |
| 122 | 1.3526 | 1.14673 | -75.72 | 1.7666 | 0.455 01 | 1.36910 | 567.70 |
| | | 0.0672141 | 48.08 | 2.7814 | 0.394 73 | 0.90296 | 185.24 |
| 124 | 1.5032 | 1.12853 | -72.95 | 1.7880 | 0.452 64 | 1.41114 | 548.75 |
| | | 0.0750035 | 47.60 | 2.7603 | 0.401 54 | 0.95400 | 185.33 |
| 126 | 1.6653 | 1.10955 | -70.11 | 1.8097 | 0.450 70 | 1.45995 | 529.20 |
| | | 0.0835889 | 46.99 | 2.7390 | 0.409 04 | 1.01421 | 185.34 |
| 128 | 1.8394 | 1.08966 | -67.17 | 1.8315 | 0.449 26 | 1.51708 | 509.01 |
| | | 0.0930725 | 46.24 | 2.7176 | 0.417 33 | 1.08612 | 185.24 |
| 130 | 2.0261 | 1.06875 | -64.14 | 1.8536 | 0.448 41 | 1.58468 | 488.11 |
| | | 0.103580 | 45.33 | 2.6958 | 0.426 58 | 1.17327 | 185.03 |
| 132 | 2.2260 | 1.04666 | -61.00 | 1.8762 | 0.448 30 | 1.66578 | 466.42 |
| | | 0.115272 | 44.23 | 2.6735 | 0.436 97 | 1.28078 | 184.69 |
| 134 | 2.4395 | 1.02319 | -57.73 | 1.8992 | 0.449 09 | 1.76485 | 443.85 |
| | | 0.128351 | 42.94 | 2.6505 | 0.448 77 | 1.41628 | 184.21 |
| 136 | 2.6672 | 0.998082 | -54.31 | 1.9229 | 0.451 02 | 1.88882 | 420.29 |
| | | 0.143090 | 41.39 | 2.6266 | 0.462 33 | 1.59167 | 183.57 |
| 138 | 2.9099 | 0.970997 | -50.71 | 1.9474 | 0.454 42 | 2.04902 | 395.56 |
| | | 0.159864 | 39.55 | 2.6015 | 0.478 18 | 1.82657 | 182.73 |
| 140 | 3.1682 | 0.941443 | -46.88 | 1.9730 | 0.459 84 | 2.26491 | 369.39 |
| | | 0.179210 | 37.35 | 2.5747 | 0.497 06 | 2.15574 | 181.63 |
| 142 | 3.4430 | 0.908698 | -42.77 | 2.0000 | 0.468 24 | 2.57120 | 341.35 |
| | | 0.201946 | 34.68 | 2.5455 | 0.520 30 | 2.64684 | 180.19 |
| 144 | 3.7351 | 0.871668 | -38.28 | 2.0292 | 0.481 83 | 3.03858 | 310.73 |
| | | 0.229420 | 31.38 | 2.5129 | 0.550 34 | 3.45008 | 178.23 |
| 146 | 4.0460 | 0.828488 | -33.24 | 2.0614 | 0.505 66 | 3.85590 | 276.57 |
| | | 0.264135 | 27.13 | 2.4749 | 0.592 79 | 4.97016 | 175.27 |
| 148 | 4.3773 | 0.775089 | -27.30 | 2.0989 | 0.550 48 | 5.71750 | 237.51 |
| | | 0.311718 | 21.29 | 2.4273 | 0.662 71 | 8.72009 | 169.94 |
| 150 | 4.7323 | 0.698343 | -19.28 | 2.1495 | 0.647 63 | 13.9176 | 190.54 |
| | | 0.390945 | 11.82 | 2.3569 | 0.809 39 | 26.4188 | 157.52 |
| 150.687[b] | 4.8607 | 0.656707 | -15.14 | 2.1758 | 0.715 68 | 29.7562 | 171.24 |
| | | 0.439127 | 6.406 | 2.3188 | 0.891 67 | 56.4503 | 149.66 |
| 151.396[c] | 4.99684 | 0.543786 | -4.184 | 2.2468 | 0.879 26 | 3580 | 146.56 |

[a] Triple-point temperature
[b] Critical temperature from NIST
[c] Critical temperature from Maxwell relations Eqs. (50) to (52)

## 5 Uncertainty of the new equation of state

Mainly guided by comparison with the TSW model, estimates for the uncertainty of calculated densities $\rho$, speeds of sound $c$, and isobaric heat capacities $C_P$ calculated from Eq. (39) have been made. These uncertainties are illustrated in the following tolerance diagrams, Fig. 17, Fig. 38 and Fig. 39. For all these tolerance diagrams, the variables are the pressure $P$ and the temperature $T$. Since the quantities $c$ and $C_P$ depend on $\rho$ and $T$, the pressure was converted to density by inversion of Eq. (45). In order to make an easier comparison with tolerance diagrams given in Ref. 4, we used the same tolerance ranges (±0.03 to ±5%) and identical notations (A, B, C, D, E, F) with their corresponding meanings.

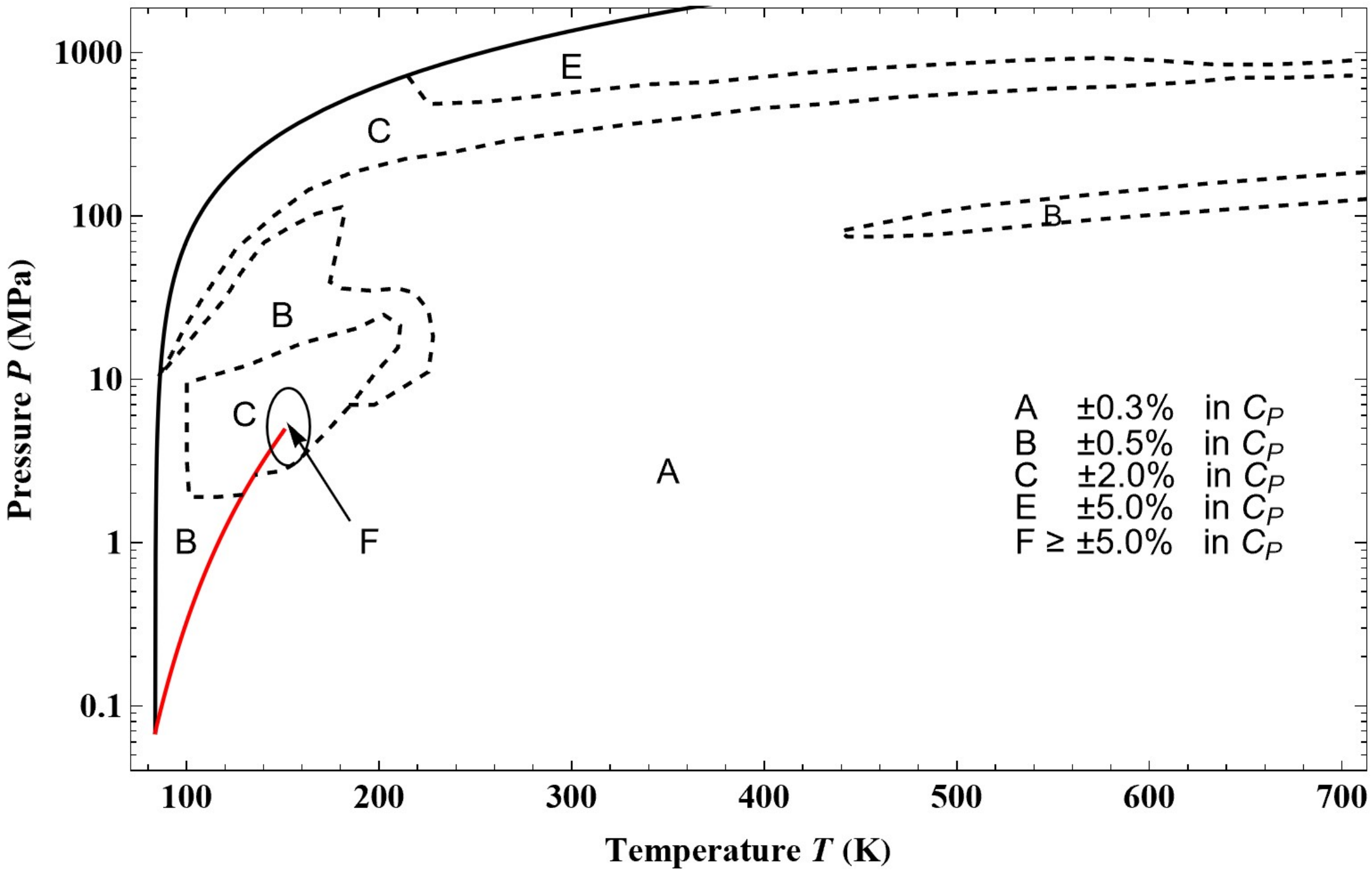

**Fig. 38:** Tolerance diagram for isobaric heat capacities calculated from Eq. (68) with the use of Eq. (45) for determining densities as function of pressure. The red curve corresponds to the saturated vapor pressure curve and the black one to the melting line.

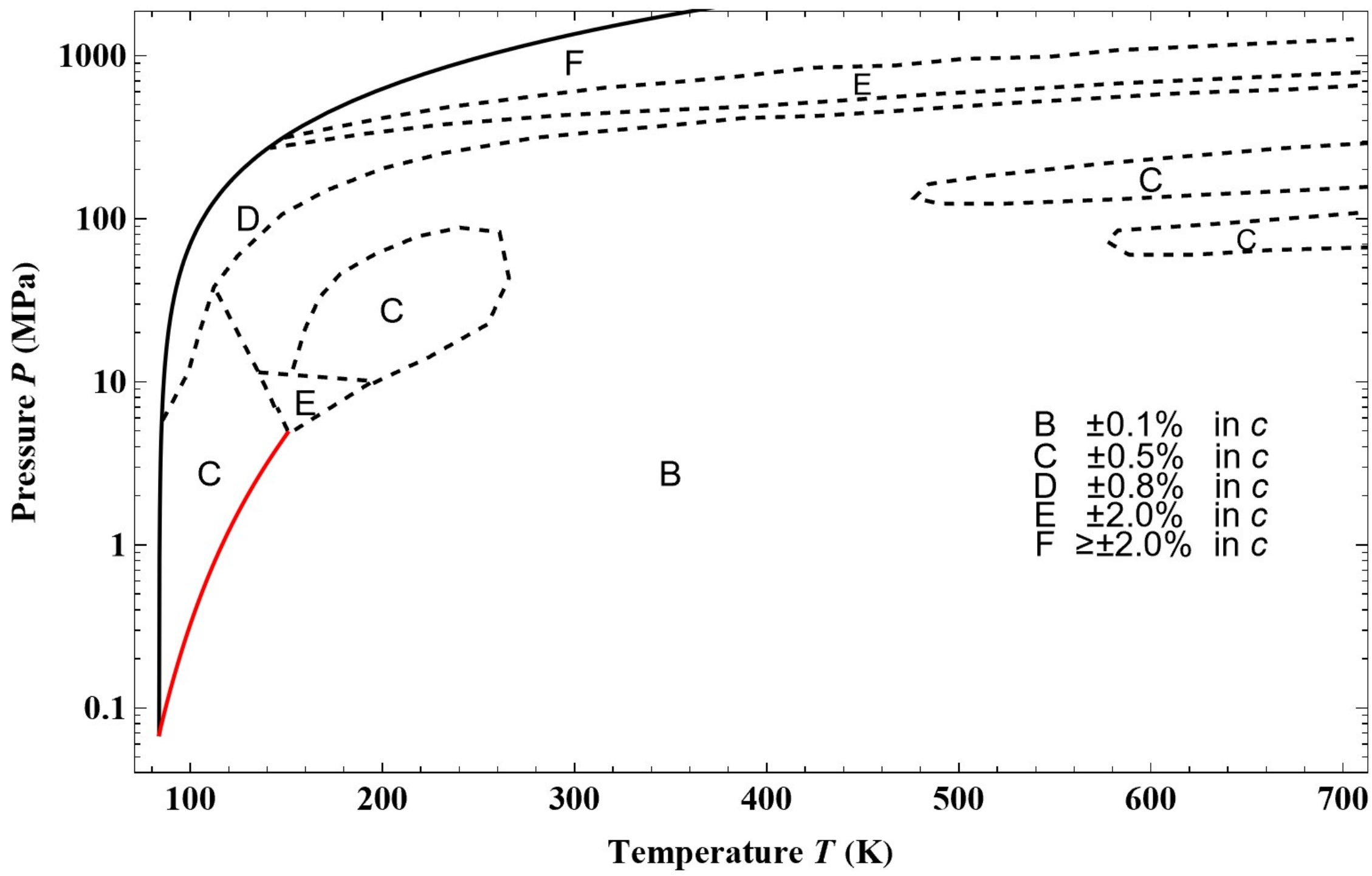

**Fig. 39:** Tolerance diagram for sound speed calculated from Eq. (69) with the use of Eq. (45) for determining densities as function of pressure. The red curve corresponds to the saturated vapor pressure curve and the black one to the melting line.

We do not plot a tolerance diagram for $C_V$ as it will not bring different information from Fig. 11. Moreover, the relative error on $C_V$ between the TSW model and the present one is everywhere far inside the errors shown on Fig. 44 of Ref. 4.

Comparisons with the data of L'Air Liquide (1710 data points, Ref. 19) allow completing the tolerance diagrams in temperature range from 700 K to 1100 K. These uncertainties for calculated densities $\rho$, isobaric heat capacities $C_P$ and isochoric heat capacities $C_V$ are illustrated in the tolerance diagrams, Fig. 40 to Fig. 42. Here again, in order to extend the comparison with the tolerance diagrams in Ref. 4, we have retained the same notations with their corresponding meanings. The new tolerance intervals are entered directly, without using new letters of the alphabet.

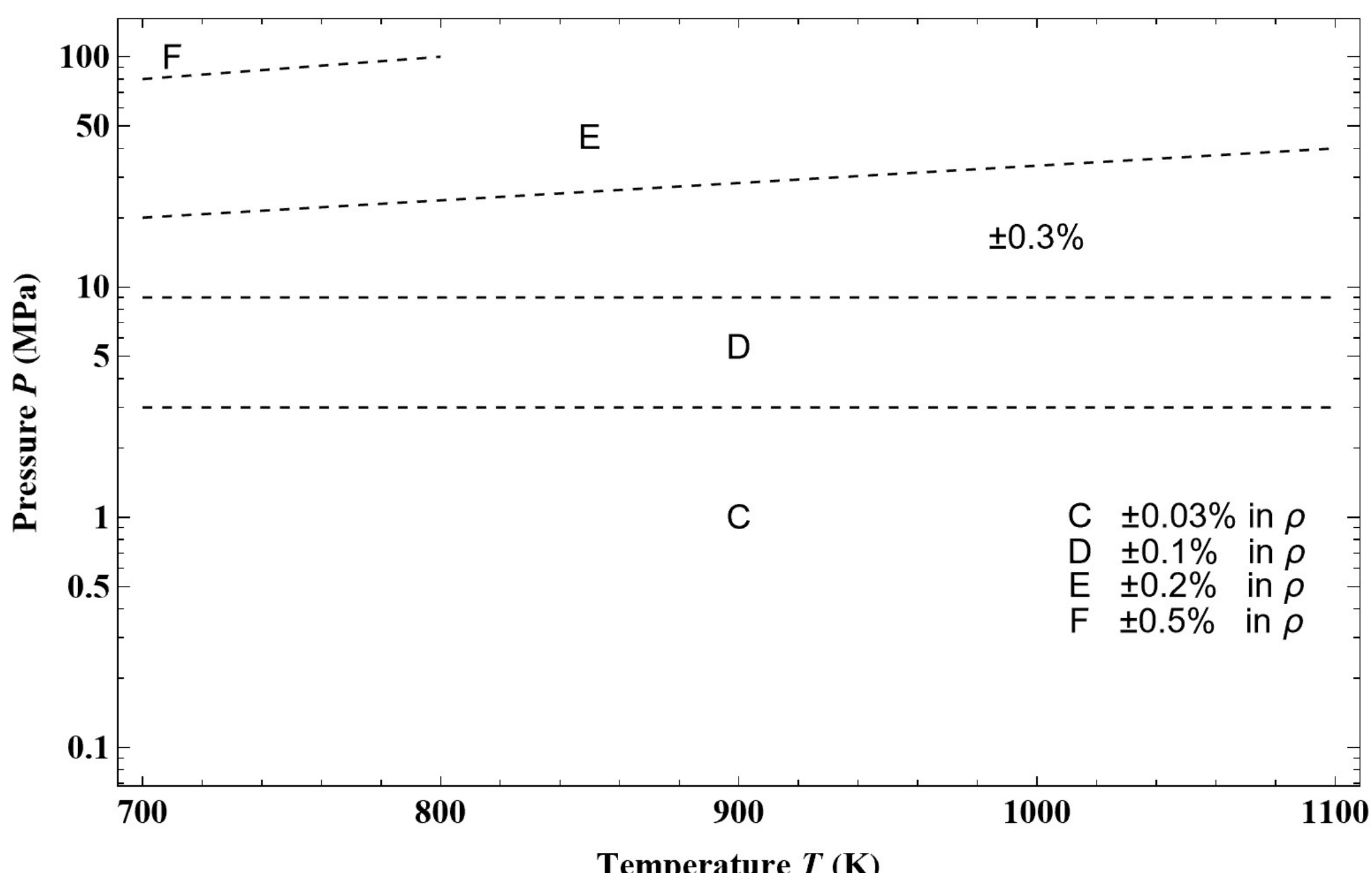

**Fig. 40:** Tolerance diagram for densities calculated from inversion of Eq. (45) in the range of temperature from 700 K to 1100 K.

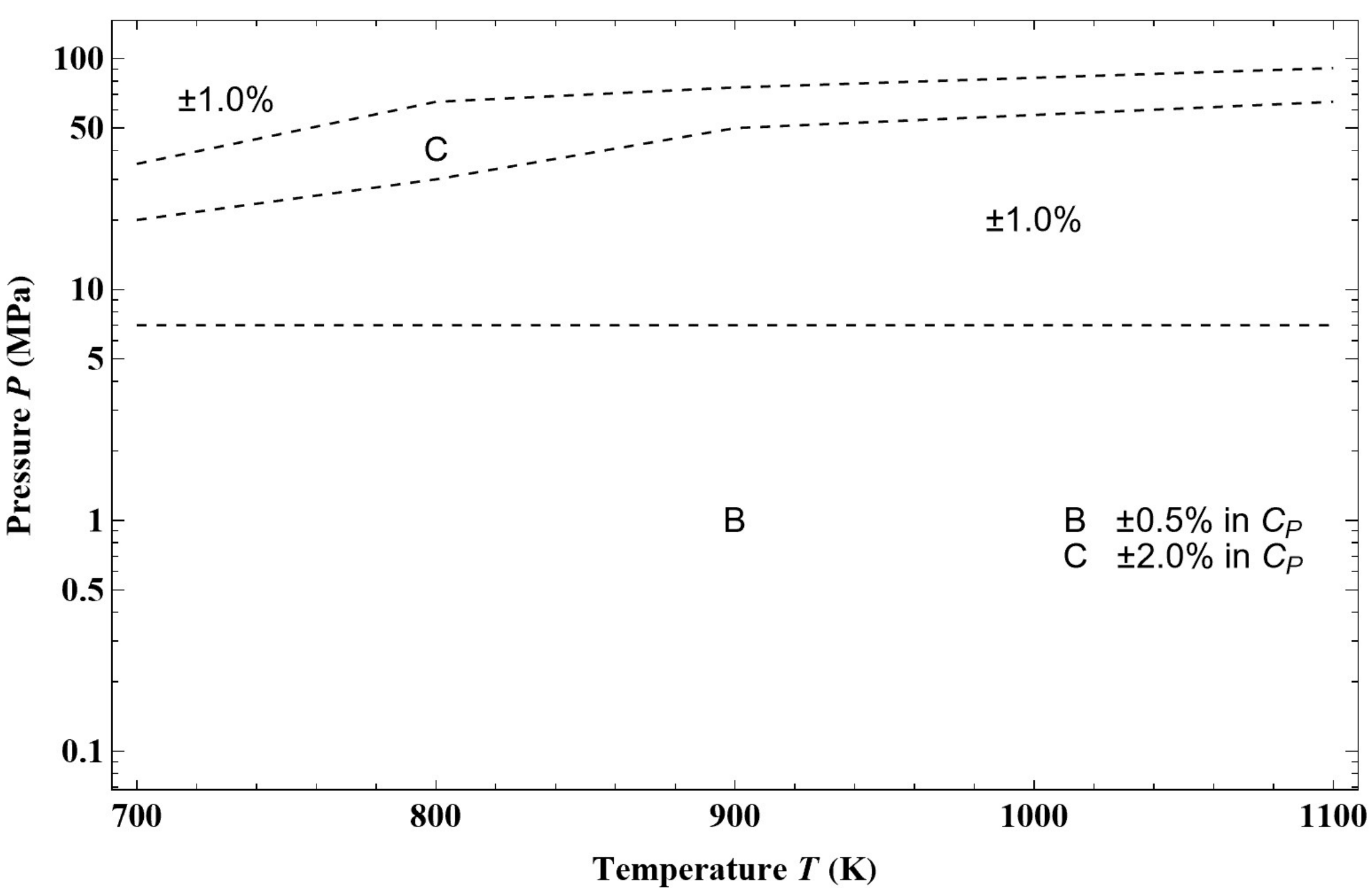

**Fig. 41:** Tolerance diagram for isobaric heat capacities calculated from Eq. (68) with the use of Eq. (45) for determining densities as function of pressure in the range of temperature from 700 K to 1100 K.

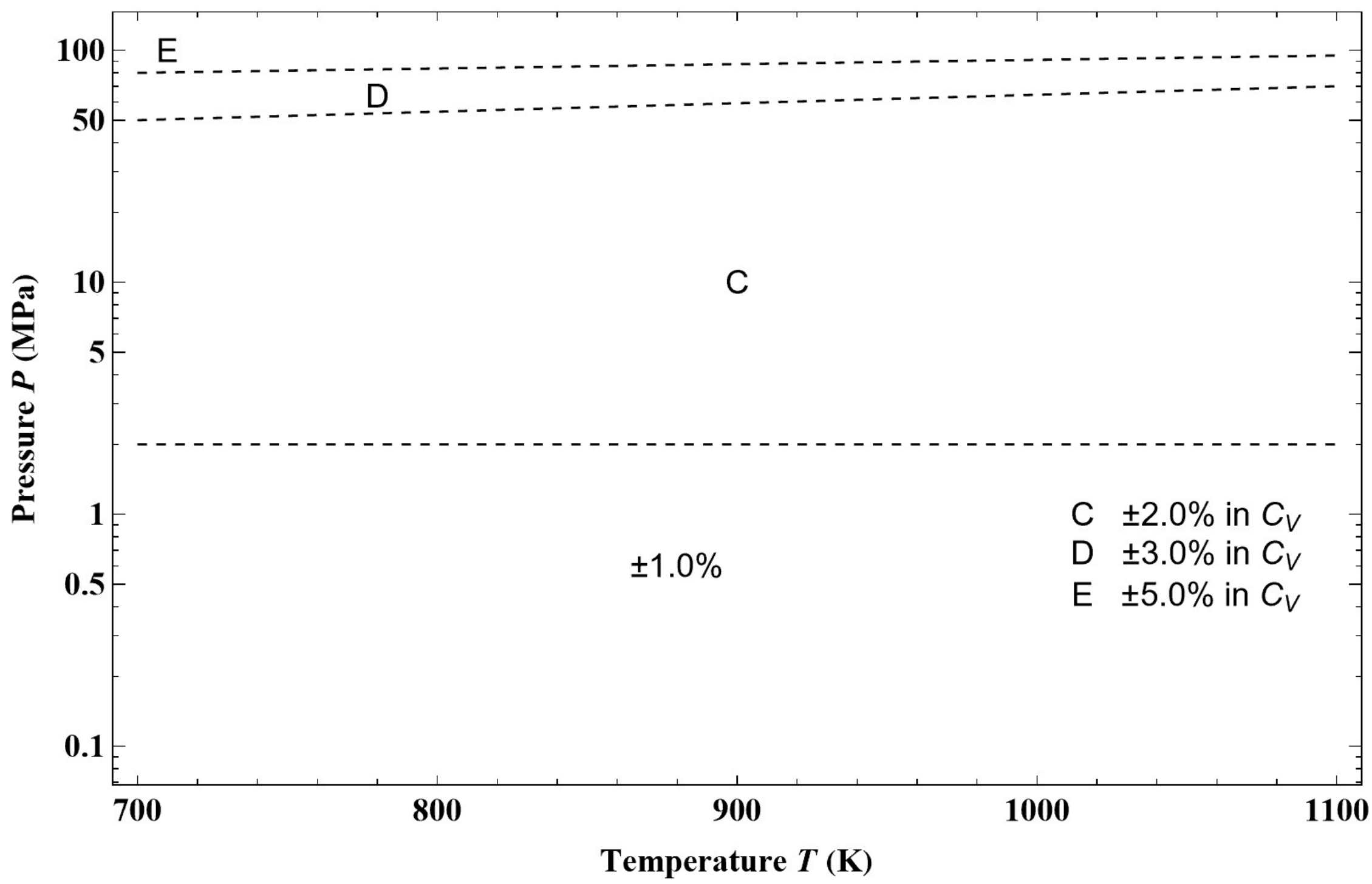

**Fig. 42:** Tolerance diagram for isochoric heat capacities calculated from Eq. (1) with the use of Eq. (45) for determining densities as function of pressure in the range of temperature from 700 K to 1100 K.

## 6 Tait’s equation of state to describe the liquid phase

As a result of Tait's work to understand and analyze the ocean temperature measurements (Ref. 34) from the first global oceanographic campaign of the H.M.S. Challenger (Ref. 35), it was shown that most liquids can be described empirically by a so-called Tait equation of state (Ref. 36). The most commonly used form of equation is the so-called Tait-Tammann equation defined from the isothermal mixed elasticity modulus in volume, which we will write here using the notations of Ref. 36 (see Chapter 3):

$$V(T,P) = V_{\sigma l}(T) - \tilde{J}(T)\ \ln\left(\frac{P+\tilde{\Pi}(T)}{P_{\text{sat}}(T)+\tilde{\Pi}(T)}\right) \qquad (71)$$

where $V$ represents the specific volume of the liquid, $V_{\sigma l}$ the specific volume of liquid along the coexistence curve deduced from Eq. (C2) and $P_{\text{sat}}$ is given by Eq. (C1). The two Tait-Tammann parameters $\tilde{\Pi}$ and $\tilde{J}$ have the following expression in the case of liquid argon:

$$\tilde{\Pi}(T) = -P_c + 7.5538\ (1-T_r)^{1.0728} \exp\left(-\left|\frac{T_r - 0.77453}{0.22816}\right|^{39}\right) \times \left(1+176\ \exp\left(-\left|\,T_r - 0.5614\,\right|^{1.1276}\right)^{2.3856}\right) \qquad (72)$$

$$\tilde{J}(T) = 1.7549\, T_r^6 \exp\left(-\left|\frac{T_r - 0.56277}{0.1147}\right|^{1.0685}\right) + 0.064031 \exp\left(-\left|\frac{T_r - 0.98098}{0.17611}\right|^{1.2217}\right) + 0.01 \exp\left(-\left|\frac{T_r - 0.577}{0.061826}\right|^{1.2}\right) \quad (73)$$

where $T_r = T/T_c$ with $T_c = T_{c,\text{non-ext. formulation}} = 151.396$ K and $P_c = P_{c,\text{non-ext. formulation}} = 49.9684$ bar. Equation (72) gives $\tilde{\Pi}$ in bar and Eq. (73) gives $\tilde{J}$ in cm$^3$/g.

It can be immediately noticed that Eqs. (72) and (73) verify the necessary conditions for the thermodynamic stability such that $\tilde{\Pi}(T_c) = -P_c$ and $\tilde{J} \geq 0$ (see section 3.3 of Ref. 36). Equation (72) implies that the corresponding temperature for which $\tilde{\Pi} = 0$ is for argon $T_\infty$ = 137.584 K. This temperature represents the transition between the liquid and gas-like behavior. Therefore, for temperatures $T_\infty < T < T_c$, Eq. (71) shows an asymptote that lies in the biphasic region for $T$ < 150.5 K but beyond that, it appears in the liquid phase. In other words, it is not advisable to use the Tait-Tammann approximation for temperatures above 150.5 K.

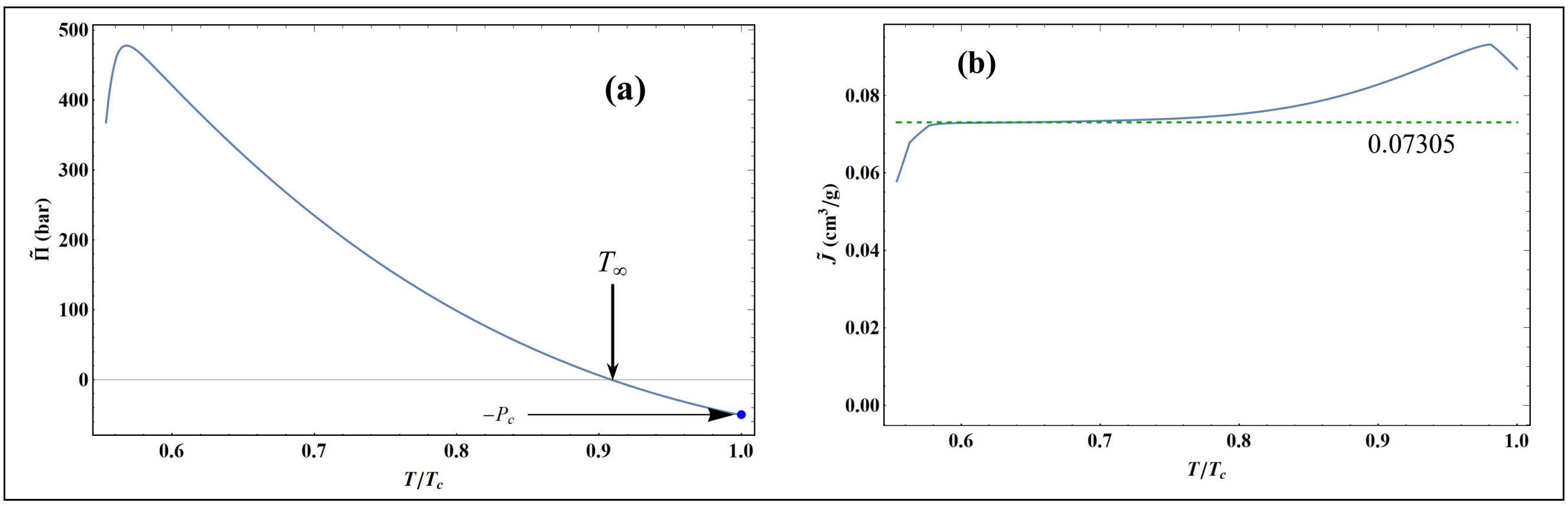

**Fig. 43:** Variations of Tait-Tammann parameters for liquid argon between $T_t$ and $T_c$ = 151.396 K. (a) Eq. (72); (b) Eq. (73). The green dotted curve represents the average value of $\tilde{J}$ between 90 and 110 K.

Fig. 43 shows that the variations of Tait-Tammann parameters for argon are very similar to those of water (see Figure 3.4 of Ref. 36) except for $\tilde{J}$ near the critical point, which here reaches a maximum. For many liquids it is generally assumed that $\tilde{J}$ is constant. Fig. 43b shows that this is approximately the case between 90 K and 110 K for liquid argon so that the average value is $\tilde{J}_{\text{average}} \approx 0.07305\, \text{cm}^3/\text{g}$. The closeness of the variations of the Tait-Tammann parameters to those of water still implies that the Ginell parameters (Ref. 37) will have the same variations along an isotherm or isobar as those shown in Figures 3.7 and 3.8 of Ref. 36 and consequently the same picture of the structure of liquid argon in terms of aggregates could be drawn.

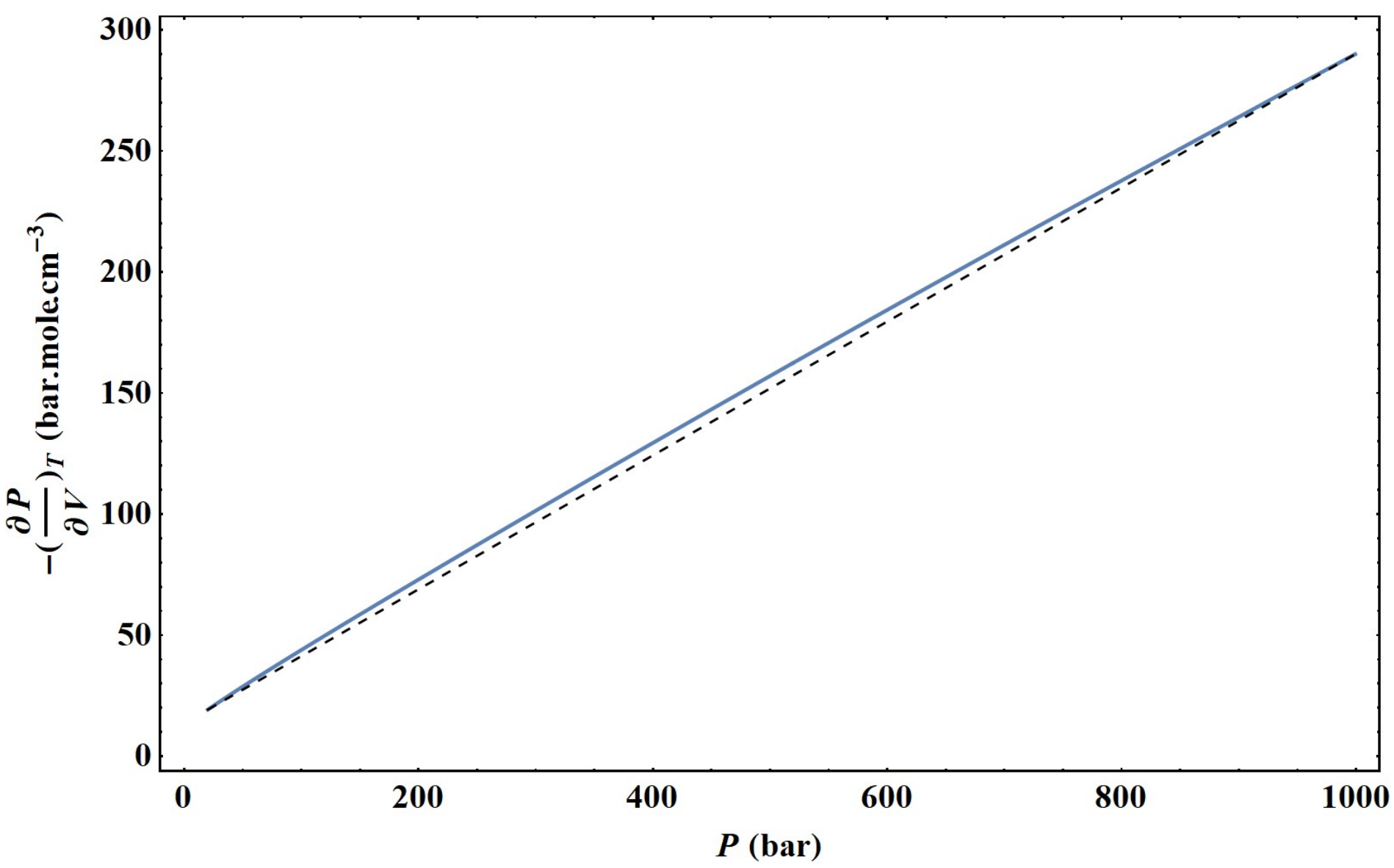

**Fig. 44:** Evolution of the isothermal mixed elasticity modulus along the isotherm at $T$ = 130 K, calculated from the present non-extensive formulation (i.e. blue curve). The black dotted line simply represents a straight line connecting the first and last points of the isotherm.

In Fig. 44, from an example on a given isotherm, it can be observed that the isothermal mixed elasticity modulus of liquid argon has an appreciable curvature over a pressure range of 1000 bar. Therefore, the Tait-Tammann equation over this pressure range can only be a rough approximation. If one wants to obtain a representation that fits into the tolerance diagram of Fig. 17, the pressure range must be reduced. In the case of liquid argon, the variation of the isothermal mixed elasticity modulus as a function of pressure can be considered linear for a pressure range globally below 200 bar. Above 200 bar, it was shown in chapter 4 of Ref. 36, that it is the adiabatic modulus of elasticity that satisfies the linearity with pressure, and therefore, a modified Tait equation must be used to describe the liquid and supercritical states of argon.

Fig. 45a shows the deviation obtained between the Tait-Tammann equation of state given by Eqs. (71) to (73) and the present non-extensive formulation given by the inversion of Eq. (45). It can be observed that the deviation increases essentially for temperatures above 140 K, so it diverges from the prescription given by the tolerance diagram in Fig. 17 which provides the tolerance ±0.2 % for subregion E. Fig. 45b shows that to achieve the tolerance of ±0.2%, the pressure range must be reduced between $P_{sat}$ and 100 bar. However, the tolerance corresponding to subregion E is still slightly exceeded between 146 K and 148 K.

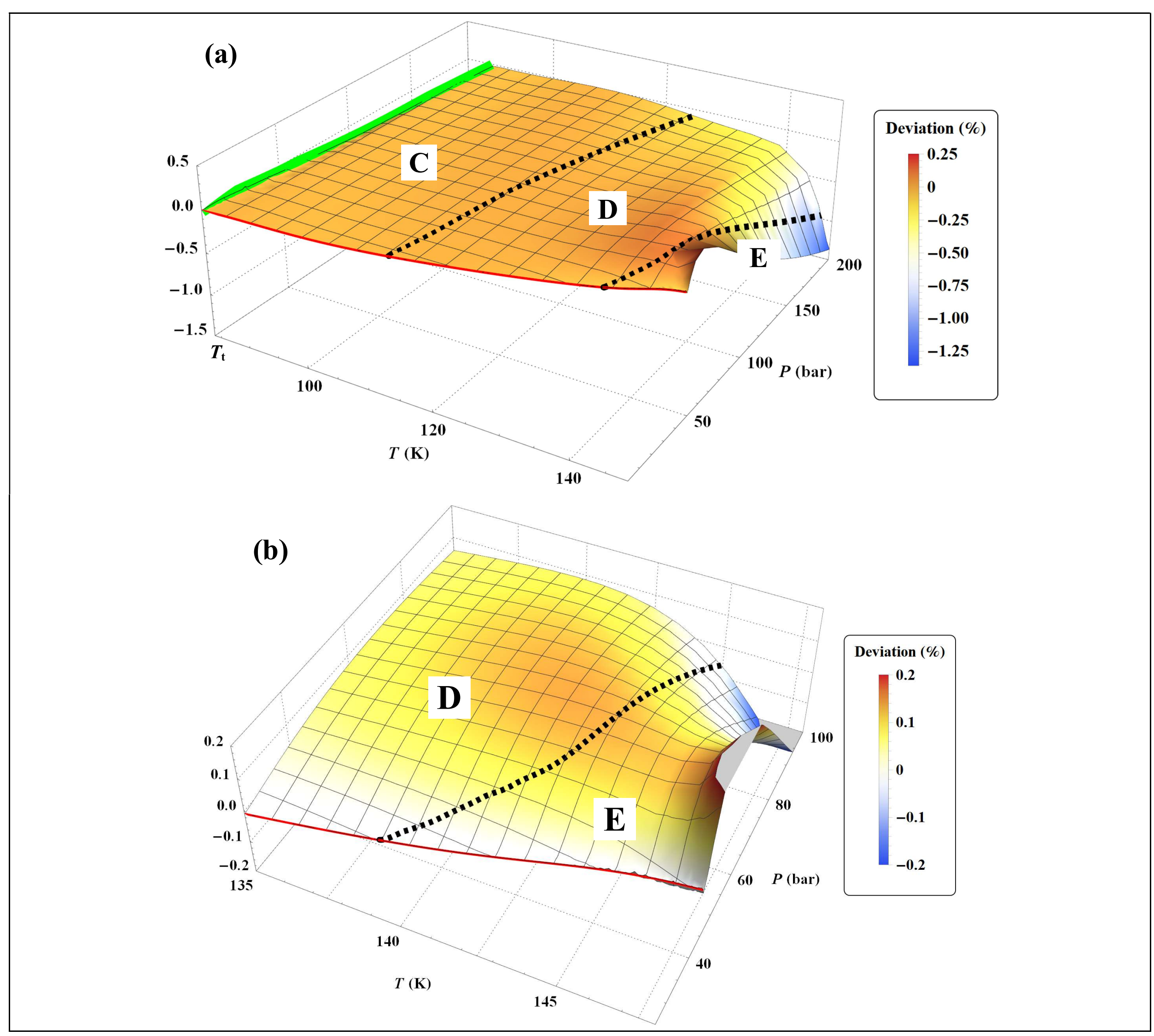

**Fig. 45:** Percentage deviations of the specific volume $100\left(V_{\substack{\text{non-ext.}\\ \text{formulation}}} - V_{\text{Eq.}(71)}\right) \Big/ V_{\substack{\text{non-ext.}\\ \text{formulation}}}$ between the Tait-Tammann equation of state Eq. (71) and the present non-extensive formulation from the inversion of Eq. (45). The thick green curve represents the melting line while the red curve represents the saturated vapor pressure curve. The black dashed curves are those in Fig. 17 that represent the separation between the subregions C, D and E in the liquid phase. (a) pressure range from $P_{sat}$ to 200 bar and temperature range from $T_t$ to 148 K; (b) pressure range from $P_{sat}$ to 100 bar and temperature range from 135 K to 148 K.

Fig. 45a shows a larger negative deviation for pressures above 100 bar and temperatures above 130 K. In other words, it seems that the tolerance diagram prescription of ±0.1% for subregion D is not held at high pressures. This is confirmed by Fig. 46a, which shows that the tolerance of ±0.1% for subregion D is slightly exceeded for temperatures above 130 K and pressures above 160 bar. On the other hand, Fig. 46b shows that the tolerance diagram prescription of ±0.03% for subregion C is well achieved for the pressure range from $P_{sat}$ to 200 bar. Moreover, Fig. 47 shows that the application of Tait-Tammann equation of state up to 300 bar in subregion C leads to a deviation of ±0.05% between $T_t$ and 110 K, and then becomes very slightly higher between 110 K and 115 K for pressures above 260 bar. Even if the deviation is slightly larger than the tolerance in Fig. 17, the use of the Tait-Tammann equation of state between $T_t$ and 115 K can be considered a “good” approximation between $P_{sat}$ and 300 bar.

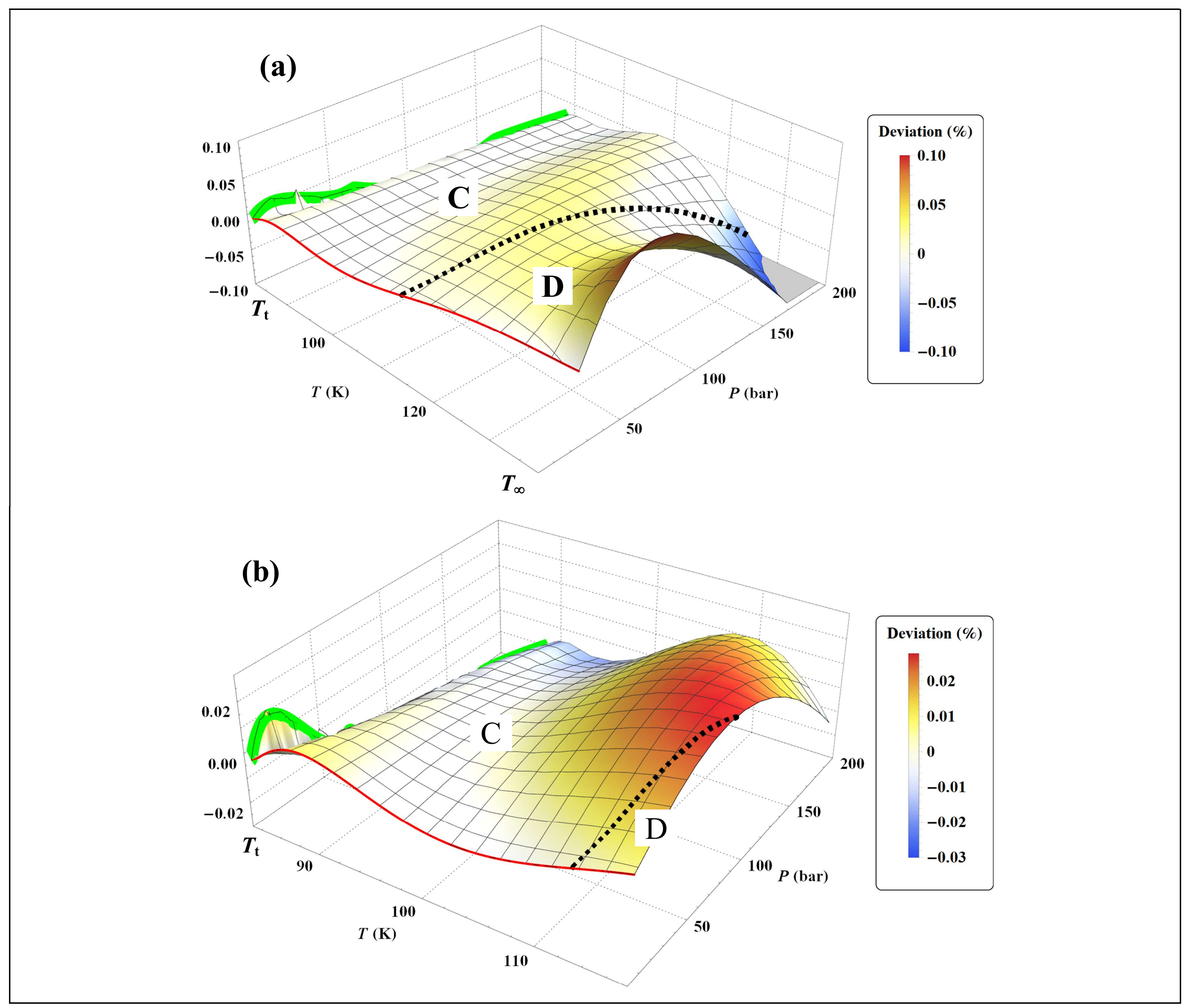

**Fig. 46:** Percentage deviations of the specific volume $100\left(V_{\substack{\text{non-ext.}\\ \text{formulation}}} - V_{\text{Eq.}(71)}\right)\Big/V_{\substack{\text{non-ext.}\\ \text{formulation}}}$ between the Tait-Tammann equation of state Eq. (71) and the present non-extensive formulation from the inversion of Eq. (45), in the pressure range from $P_{sat}$ to 200 bar. The thick green curve represents the melting line while the red curve represents the saturated vapor pressure curve. The black dashed curves are those in Fig. 17 that represent the separation between the subregions C and D in the liquid phase. (a) temperature range $T_t$ to $T_\infty$; (b) temperature range from $T_t$ to 115 K .

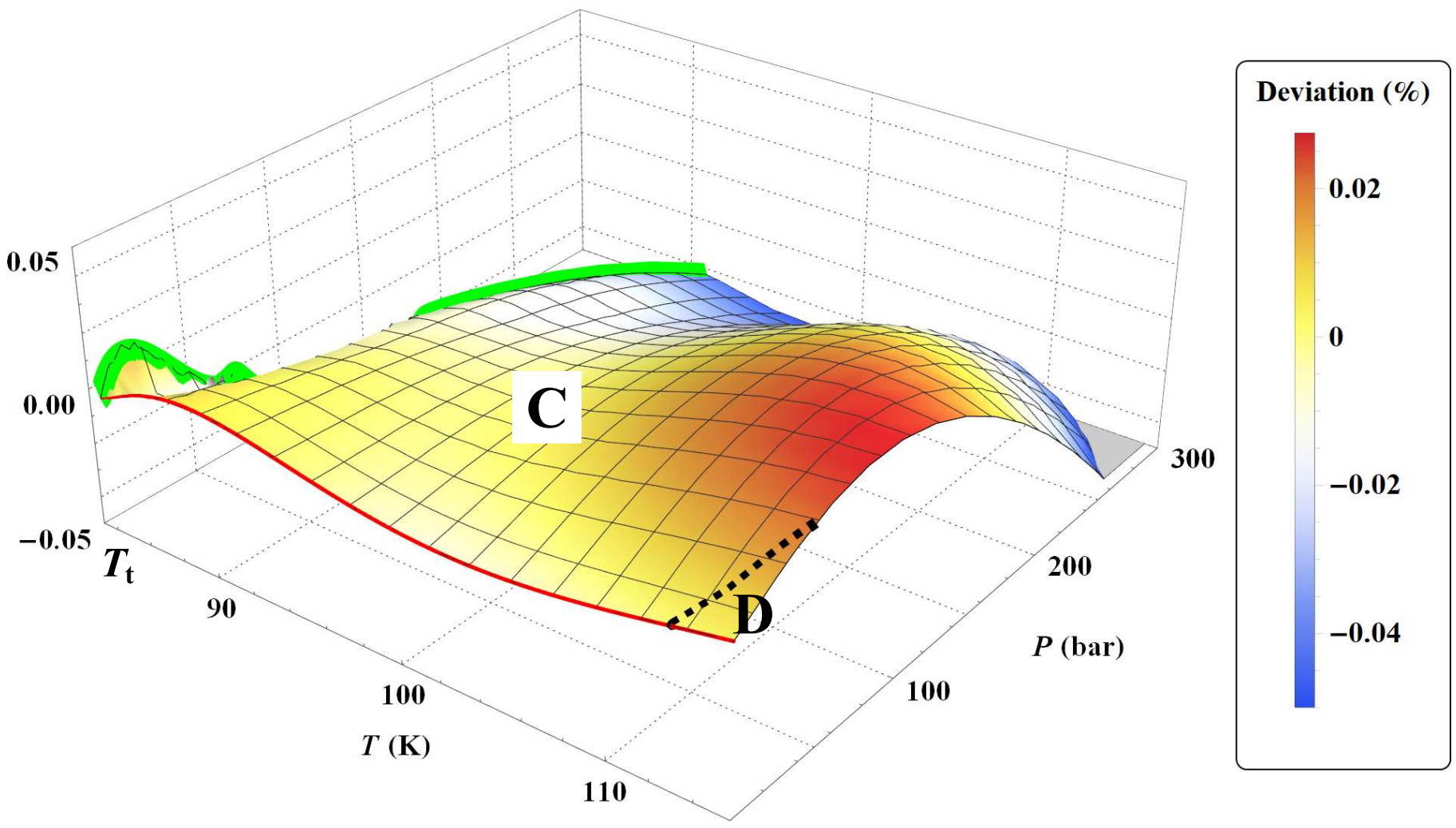

**Fig. 47:** Percentage deviations of the specific volume $100\left(V_{\substack{\text{non-ext.}\\ \text{formulation}}} - V_{\text{Eq.}(71)}\right) \Big/ V_{\substack{\text{non-ext.}\\ \text{formulation}}}$ between the Tait-Tammann equation of state Eq. (71) and the present non-extensive formulation from the inversion of Eq. (45), in the pressure range from $P_{\text{sat}}$ to 300 bar and temperature range from $T_{\text{t}}$ to 115 K. The thick green curve represents the melting line while the red curve represents the saturated vapor pressure curve. The black dashed curves are those in Fig. 17 that represent the separation between the subregions C and D in the liquid phase.

Jain *et al.* (Ref. 38) also developed a Tait-Tammann equation to describe the specific volume of liquid argon between 90 K and 148 K for a pressure range of 300 bar. It is therefore instructive to compare the evolution of the Tait-Tammann parameters with Eqs. (71) to (73). The form proposed by Jain *et al.* is such that their reference volume is no longer that of the liquid on the coexistence curve $V_{\sigma l}$ but the volume corresponding to zero pressure, noted $V_0$. Under this condition the pressure $P_{\text{sat}}$ is now replaced by the zero value in Eq. (71). With the notations of Jain *et al.*, $\widetilde{J}$ is equivalent to $cV_0$ where $c$ is a constant and $\widetilde{\Pi}$ to $B$.

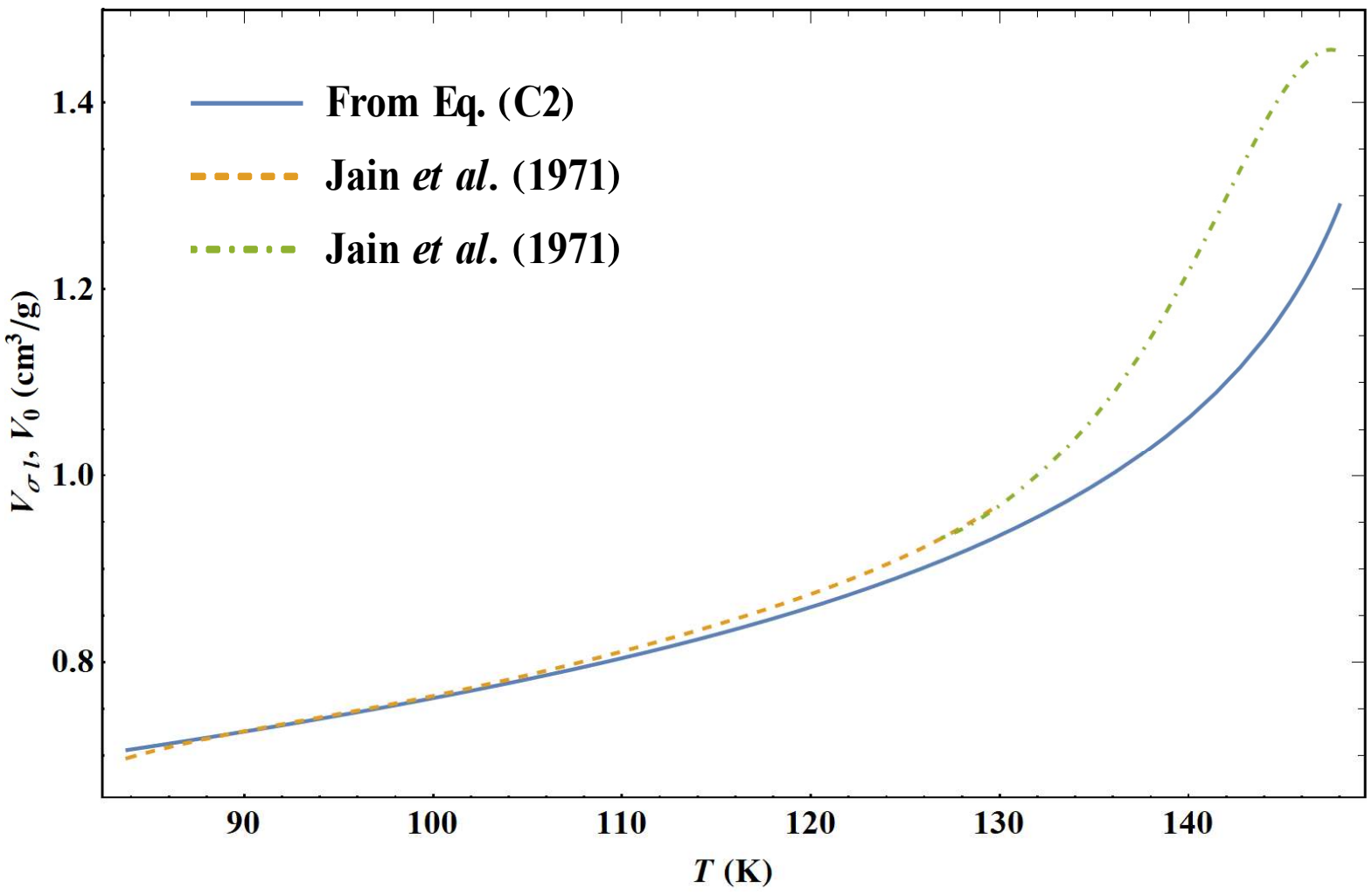

**Fig. 48:** Evolution of the reference volume for the Tait-Tammann equation of state as a function of temperature in the range from $T_{\text{t}}$ to 148 K: the dashed curves correspond to the two determinations of Jain *et al.* (Ref. 38) and the solid blue curve to $V_{\sigma l}$ deduced from Eq. (C2).

To cover the entire temperature range between 90 K and 148 K, Jain *et al.* (Ref. 38) performed a piecewise determination of their equation of state as they explained:

> "in two overlapping temperature ranges (i) 90 K to 130 K and (ii) 127 K to 148 K. […] The differences in the calculated values of $V_0$, and $B$ in these overlapping ranges for both the liquids are less than 1 in 4000."

Fig. 48 shows the comparison between the two reference specific volumes of the Tait-Tammann equation of state for the two determinations of Jain *et al.* (Ref. 38) and Eqs. (71)-(73). It can be observed that the specific volume at zero pressure $V_0$ is smaller and then becomes the same as the liquid coexistence specific volume when the temperature is lower than 100 K, which is not physically acceptable. However, the saturated vapor pressure $P_{sat}$ is not negligible for temperatures between 90 K and 100 K, therefore the equation of state of Jain *et al.* is expected to deviate strongly from the TSW model or the present non-extensive formulation in the vicinity of the saturated vapor pressure curve. However, this is a voluntary choice made by Jain *et al.* as they explained it below:

> "At sufficiently low saturation pressures, the observed volume can be taken equal to $V_0$, and the problem of finding a suitable value of $B$ to represent the experimental results presents little difficulty."

Fig. 49 shows significant differences between the parameters $B$ and $\tilde{\Pi}$, and $cV_0$ and $\tilde{J}$. However, the variations show some similarities like the maximum of $cV_0$. Indeed, Fig. 49b shows that Eq. (73) also has a maximum but for a slightly higher temperature than for Jain *et al.* (Ref. 38).

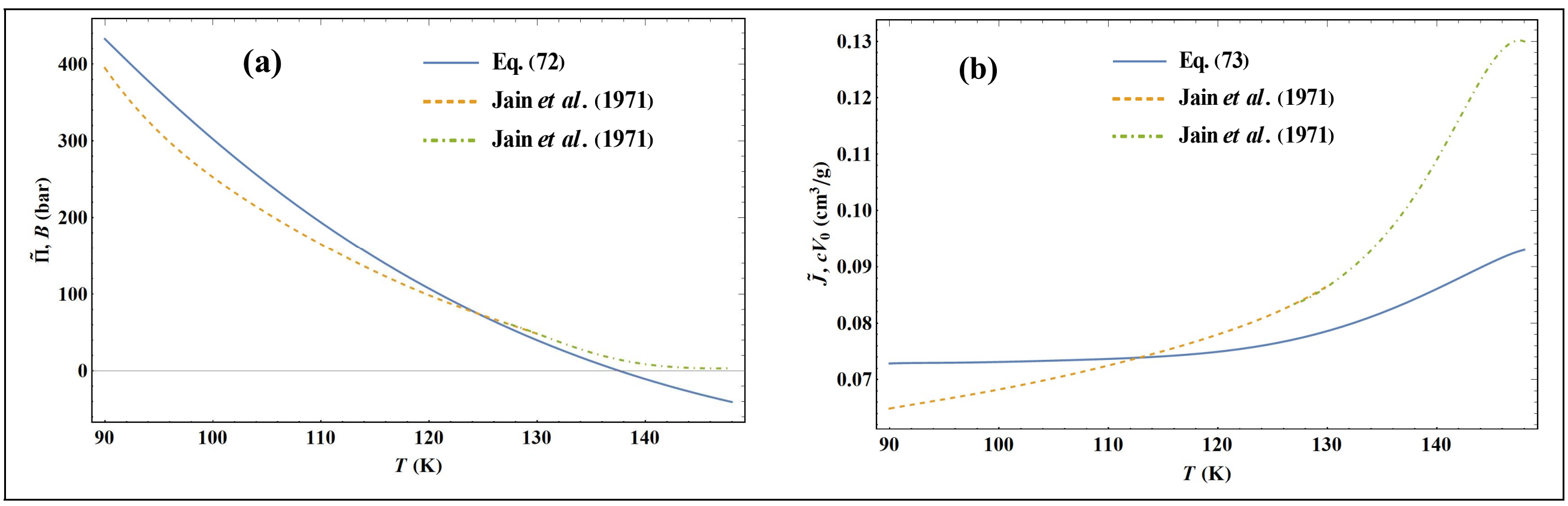

**Fig. 49:** Variations of Tait-Tammann parameters for liquid argon between 90 K and 148 K. (a) $B$ from the two determinations of Jain *et al.* (Ref. 38) and $\tilde{\Pi}$ from Eq. (72); (b) $cV_0$ from the two determinations of Jain *et al.* (Ref. 38) and $\tilde{J}$ from Eq. (73).

Jain *et al.* (Ref. 38) developed their Tait-Tammann equation of state to represent the experimental data of van Itterbeek *et al.* (11 isotherms that range from 90.15 K to 148.25 K, TABLE I of Ref. 39). These experimental data were assigned by Tegeler *et al.* (Ref. 4) to their group 3 and therefore do not fit into the tolerance diagram of Fig. 17. Thus, the equation of state of Jain *et al.* is expected to represent the raw data of van Itterbeek *et al.* much more accurately than the non-extensive formulation of the present model or the Tait-Tammann equation of state, i.e. Eqs. (71) to (73).

Fig. 50 shows the comparison of the different models to represent the raw van Itterbeek *et al.* data corresponding to subregion C in Fig. 17. It is first observed that Eq. (71) and the non-extensive formulation of the present model are not much different according to the deviation in Fig. 47. Then, it appears that the deviation of the model of Jain *et al.* (Ref. 38) is comparable to the other models for pressures above 100 bar, which is quite surprising. It is even more surprising to note that the model of Jain *et al.* appears entirely shifted by -0.15 % on the lowest isotherm at 90.15 K. However, the model of Jain *et al.* better represents data that are close to the saturated vapor pressure curve. This is consistent with the fact that these data were not considered by Tegeler *et al.* to determine the liquid density on the coexistence curve.

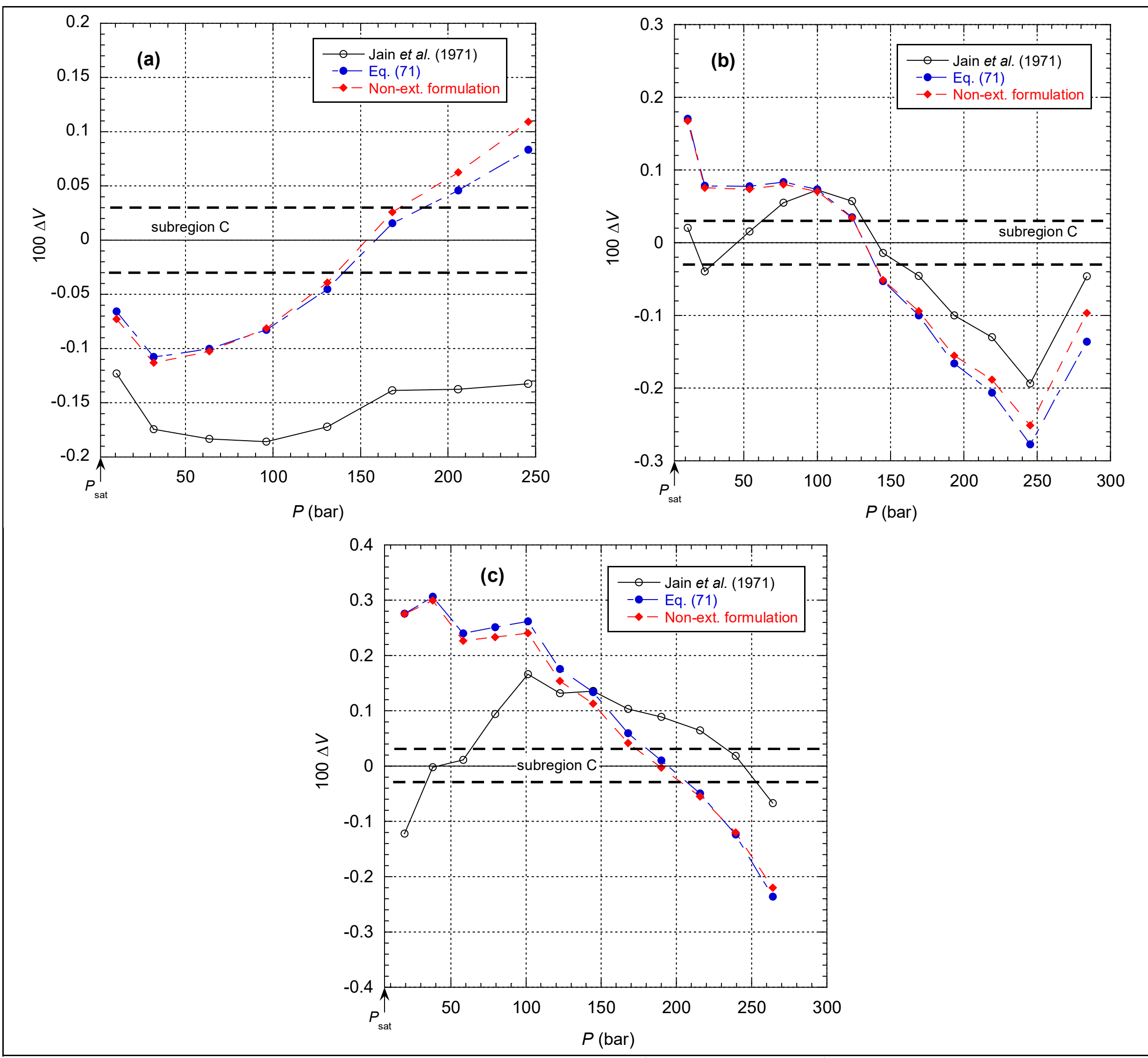

**Fig. 50:** Percentage deviations of the specific volume $\Delta V = \left(V_{\text{van Itterbeek}} - V_{\text{calc}}\right)/V_{\text{van Itterbeek}}$ between the raw data of van Itterbeek *et al.* (TABLE I of Ref. 39) and the different models for the three isotherms corresponding to subregion C in Fig. 17: (a) 90.15 K; (b) 96.99 K; (c) 108.18 K.

Fig. 51 shows now the comparison of the different models for the isotherms corresponding to subregion D in Fig. 17. It is first observed that there is a larger deviation between the non-extensive formulation and Eq. (71) beyond 150 bar, which is consistent with Fig. 46a. Then, it can be seen that the model of Jain *et al.* (Ref. 38) again has difficulties for reproducing the data near the saturated vapor pressure curve. This can only be explained by an incorrect extrapolation to determine their function $V_0(T)$, if we refer to their explanation:

> "However, at high saturation pressures both $V_0$, and $B$ have to be suitably chosen. The trial value of $V_0$ is first obtained by extrapolation of the $V_0$, against $T$ graph from the low temperature side."

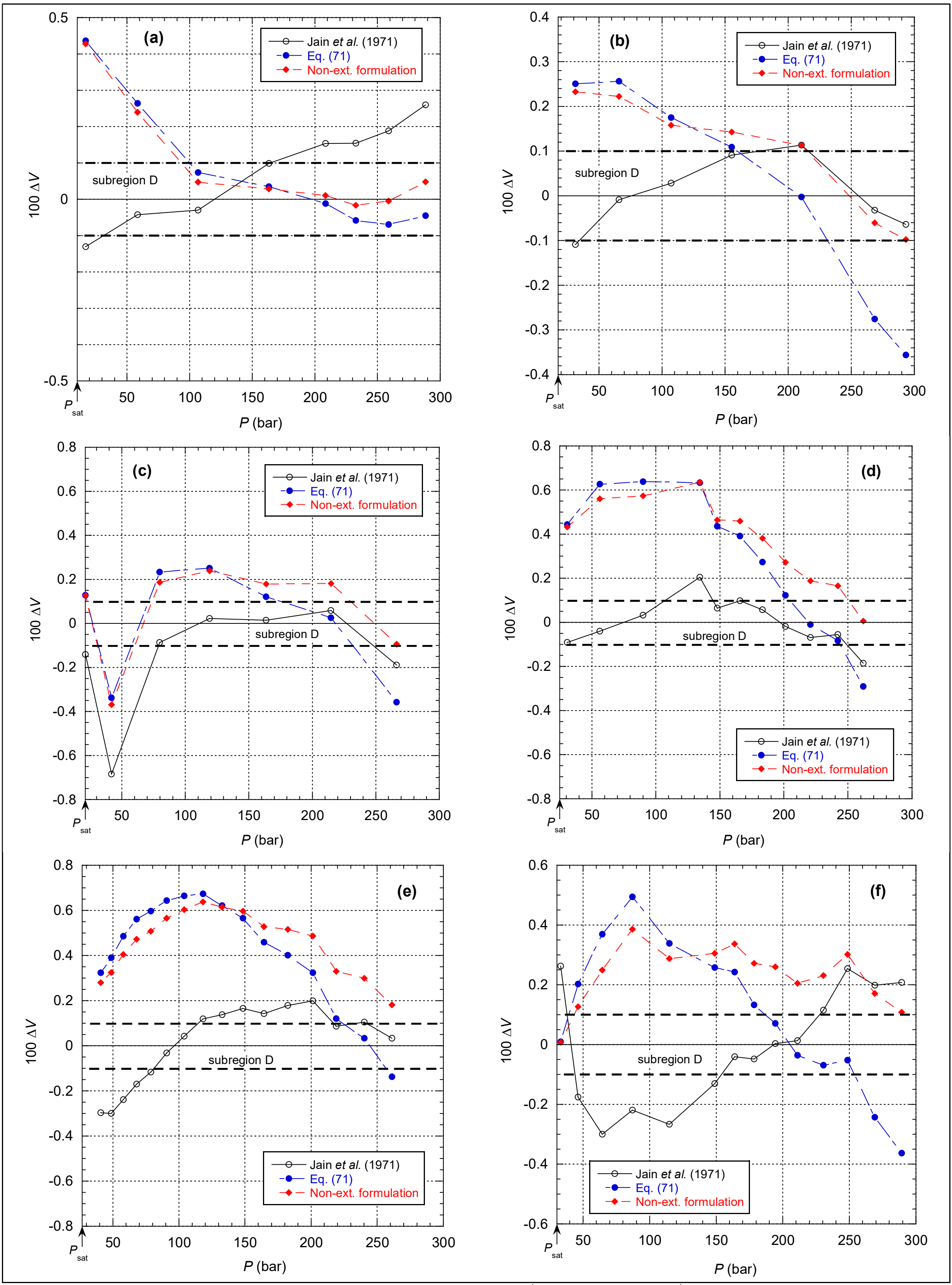

**Fig. 51:** Percentage deviations of the specific volume $\Delta V = \left(V_{\text{van Itterbeek}} - V_{\text{calc}}\right)/V_{\text{van Itterbeek}}$ between the raw data of van Itterbeek *et al.* (TABLE I of Ref. 39) and the different models for the six isotherms corresponding to subregion D in Fig. 17: (a) 117.10 K; (b) 127.05 K; (c) 130.85 K; (d) 134.40 K; (e) 136.02 K; (f) 138.98 K.

Fig. 52 shows now the comparison of the different models for the isotherms corresponding to subregion E in Fig. 17. It is observed here that the deviation of the Jain *et al.* model from the data is greatest in the vicinity of the saturated vapor pressure curve and the deviation is systematically in the same direction. On the other hand, this model allows data to be taken into account with the tolerance prescribed for the subregion E for pressures higher than 120 bar. It can be seen that the opposite process occurs to a smaller extent for Eq. (71). This is due to the curvature of the mixed elastic modulus becoming stronger and stronger and therefore the approximation of the Tait-Tammann equation only makes sense for a smaller and smaller range of pressures: Eq. (71) allows a satisfactory representation of data up to 150 bar while the model of Jain *et al.* is able to reproduce the data between 150 and 300 bar. But the two descriptions cannot be connected except by making the Tait-Tammann parameters depend on the pressure, but under these conditions, it is preferable to use the non-extensive formulation.

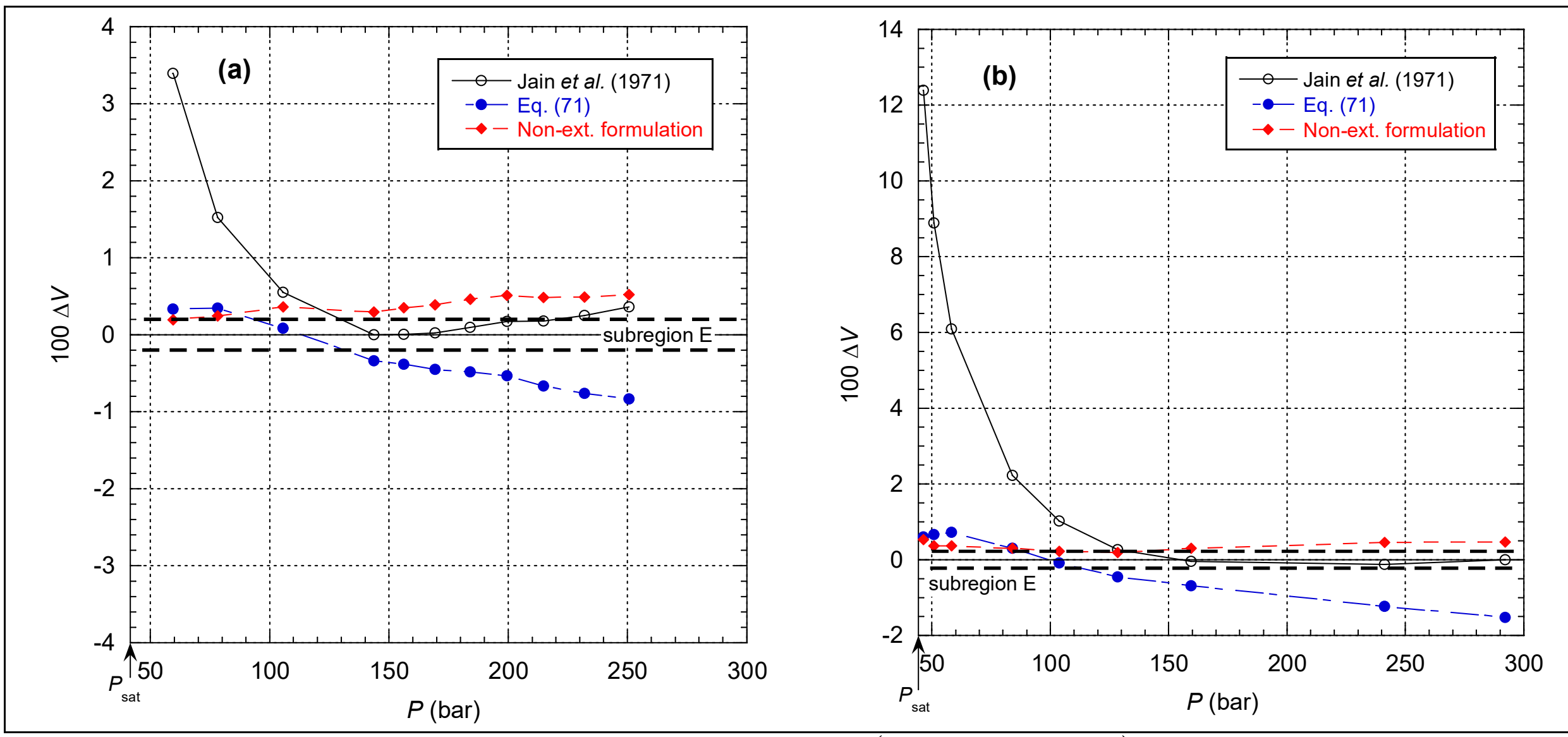

**Fig. 52:** Percentage deviations of the specific volume $\Delta V = \left(V_{\text{van Itterbeek}} - V_{\text{calc}}\right)/V_{\text{van Itterbeek}}$ between the raw data of van Itterbeek *et al.* (TABLE I of Ref. 39) and the different models for the two isotherms corresponding to subregion E in Fig. 17: (a) 146.63 K; (b) 148.25 K.

In this section, it has been shown that the Tait-Tammann equation of state can be a simple alternative to describe the specific volume of liquid argon between $T_t$ and 148 K, for pressures varying between $P_{sat}$ and 300 bar in subregion C, then between $P_{sat}$ and 200 bar in subregion D and finally between $P_{sat}$ and 100 bar for subregion E. These pressure ranges are sufficient for a very large number of applications.

## 7 Conclusion

A new equation of state for argon has been developed, which can be written in the form of a fundamental equation explicit in the reduced Helmholtz free energy. This equation has been derived from the measured quantities $C_V(\rho,T)$ and $P(\rho,T)$. It is valid for the whole fluid region (single-phase and coexistence states) from the melting line to 2300 K and for pressures up to 50 GPa. The formulation is based on data from NIST (or equivalently on the calculated values from the TSW model) and calculated values from the model of Ronchi (Ref. 6).

This new approach using mainly power laws with density dependent exponents involves much less coefficients than the TSW model and, more, eliminates the very small oscillations introduced by a polynomial description. This leads to a more physical description of the thermodynamic properties. On the other hand, the cut in the number of terms and parameters does not modify in an appreciable way the uncertainties of the calculated data (as shown in the different diagrams of tolerance). However, in an unexpected way, the present approach which generates more regular and monotonous expressions raises greater difficulty for the reversal of certain equations of state due to a highly nonlinear behavior of these expressions.

The new equation of state also shows a more physical behavior along isochors when $T$ tends to zero for the basic properties such as the isochoric heat capacity and the compressibility factor. It also shows a more reasonable behavior for the crossing of the coexistence phase. However, it does not describe correctly the properties in the vicinity of the critical point, in the same way as the TSW model does not properly describe the properties in the vicinity of the critical point with the exception of the saturation curve. However, variations of the isochoric heat capacity in the coexistence phase with the present model show peaks that are qualitatively in agreement with experimental observations, unlike the TSW model which produces unphysical variations. Comparison of the present model with the data of L'Air Liquide (Ref. 19), which had not previously been taken into account, shows that this model is consistent with these data up to 1100 K and 100 MPa, which allows, regardless of the data of Ronchi (Ref. 6), to extend the range of NIST data (Ref. 5).

In section 6 and Appendix C, simple expressions are also provided to describe the specific volume of the liquid states of argon between $T_t$ and 148 K, in the form of a Tait-Tammann equation of state, and some properties of the liquid-vapor coexistence curve. These approximate formulas can advantageously replace the complex non-extensive formulation of the present model for a large number of applications.

The non-extensive approach developed here shows that metastable states are by construction included as an extension of the single-phase isochoric heat capacity modeling. As $C_V$ data are generally known for the vast majority of fluids, this new approach can be easily extended to all of them.

## 8 ACKNOWLEDGEMENTS

The authors are grateful to J.-L. Garden (Institut Néel laboratory, France) for his valuable support.

This work benefited from the support of the project ZEROUATE under Grant ANR-19-CE24-0013 operated by the French National Research Agency (ANR).

## 10 APPENDIX A : Expression of the regular term of pressure $P_{\mathrm{reg}}$

The regular term of pressure is formed by the difference of two terms which come respectively from the derivative of the energy and entropy, such that:

$$P_{\mathrm{reg}}(\rho,T) = P_{\mathrm{Ureg}}(\rho,T) - P_{\mathrm{Sreg}}(\rho,T) \tag{A1}$$

and

$$Z_{\mathrm{reg}}(\rho,T) = \frac{P_{\mathrm{reg}}(\rho,T)}{\rho R_{\mathrm{A}} T} = Z_{\mathrm{Ureg}}(\rho,T) - Z_{\mathrm{Sreg}}(\rho,T) \tag{A2}$$

with

$$
\begin{aligned}
Z_{\mathrm{Ureg}} &= \frac{P_{\mathrm{Ureg}}(\rho,T)}{\rho R_{\mathrm{A}} T} = \rho \left( \frac{\partial \tilde{u}_{\mathrm{reg}}}{\partial \rho} \right)_T = \frac{3}{2} \rho n'_{\mathrm{reg}}(\rho) \frac{T_c}{T} \left[ \frac{(T/T_c)^{m(\rho)} - 1}{m(\rho)} + \frac{\lambda^{-m(\rho)}}{2-m(\rho)} \Gamma\left( \frac{m(\rho)}{2-m(\rho)}, \left( \frac{\lambda T}{T_c} \right)^{2-m(\rho)} \right) \right] \\
&+ \frac{3}{2} n_{\mathrm{reg}}(\rho) \frac{T_c}{T} \rho m'(\rho) \left\{ \frac{1-(T/T_c)^{m(\rho)}}{m(\rho)^2} + \frac{(T/T_c)^{m(\rho)} \ln(T/T_c)}{m(\rho)} + \frac{\lambda^{-m(\rho)}}{2-m(\rho)} \left( \frac{\lambda T}{T_c} \right)^{m(\rho)} \ln\left( \frac{\lambda T}{T_c} \right) \exp\left( -\left( \frac{\lambda T}{T_c} \right)^{\frac{5}{2}-m(\rho)} \right) \right\} \\
&+ \frac{3}{2} n_{\mathrm{reg}}(\rho) \frac{T_c}{T} \rho m'(\rho) \frac{\lambda^{-m(\rho)}}{[2-m(\rho)]^2} \Gamma\left( \frac{m(\rho)}{2-m(\rho)}, \left( \frac{\lambda T}{T_c} \right)^{2-m(\rho)} \right) \left\{ 1 - \ln\left( \lambda^{2-m(\rho)} \right) + \frac{2}{2-m(\rho)} \ln\left( \left( \frac{\lambda T}{T_c} \right)^{2-m(\rho)} \right) \right\} \\
&+ 3 n_{\mathrm{reg}}(\rho) \frac{T_c}{T} \rho m'(\rho) \frac{\lambda^{-m(\rho)}}{[2-m(\rho)]^3} G_{2\,3}^{3\,0}\left( \left( \frac{\lambda T}{T_c} \right)^{2-m(\rho)} \middle| \begin{matrix} 1,\ 1 \\ 0,\ 0,\ \frac{m(\rho)}{2-m(\rho)} \end{matrix} \right)
\end{aligned}
\tag{A3}
$$

and

$$
\begin{aligned}
Z_{\mathrm{Sreg}} &= \frac{P_{\mathrm{Sreg}}(\rho,T)}{\rho R_{\mathrm{A}} T} = \rho \left( \frac{\partial \tilde{s}_{\mathrm{reg}}}{\partial \rho} \right)_T = \frac{3}{2} \rho n'_{\mathrm{reg}}(\rho) \left[ \frac{(T/T_c)^{m(\rho)-1} - 1}{m(\rho)-1} + \frac{\lambda^{1-m(\rho)}}{2-m(\rho)} \Gamma\left( \frac{m(\rho)-1}{2-m(\rho)}, \left( \frac{\lambda T}{T_c} \right)^{2-m(\rho)} \right) \right] \\
&+ \frac{3}{2} n_{\mathrm{reg}}(\rho) \rho m'(\rho) \left\{ \frac{1-(T/T_c)^{m(\rho)-1}}{[m(\rho)-1]^2} + \frac{(T/T_c)^{m(\rho)-1} \ln(T/T_c)}{m(\rho)-1} + \frac{\lambda^{1-m(\rho)}}{2-m(\rho)} \left( \frac{\lambda T}{T_c} \right)^{m(\rho)-1} \ln\left( \frac{\lambda T}{T_c} \right) \exp\left( -\left( \frac{\lambda T}{T_c} \right)^{2-m(\rho)} \right) \right\} \\
&+ \frac{3}{2} n_{\mathrm{reg}}(\rho) \rho m'(\rho) \frac{\lambda^{1-m(\rho)}}{[2-m(\rho)]^2} \Gamma\left( \frac{m(\rho)-1}{2-m(\rho)}, \left( \frac{\lambda T}{T_c} \right)^{2-m(\rho)} \right) \left\{ 1 - \ln\left( \lambda^{2-m(\rho)} \right) + \frac{1}{2-m(\rho)} \ln\left( \left( \frac{\lambda T}{T_c} \right)^{2-m(\rho)} \right) \right\} \\
&+ \frac{3}{2} n_{\mathrm{reg}}(\rho) \rho m'(\rho) \frac{\lambda^{1-m(\rho)}}{[2-m(\rho)]^3} G_{2\,3}^{3\,0}\left( \left( \frac{\lambda T}{T_c} \right)^{2-m(\rho)} \middle| \begin{matrix} 1,\ 1 \\ 0,\ 0,\ \frac{m(\rho)-1}{2-m(\rho)} \end{matrix} \right)
\end{aligned}
\tag{A4}
$$

where $G_{p\,q}^{m\,n}\left( z \middle| \begin{matrix} a_1, \cdots, a_p \\ b_1, \ldots, b_q \end{matrix} \right)$ represents the Meijer G function. It is worth noting that the Meijer functions in $Z_{\mathrm{Ureg}}$ and $Z_{\mathrm{Sreg}}$ are equal to zero when $T \geq T_{\mathrm{t}}$ whatever the value of density.
For calculating some thermodynamics parameters, the first partial derivatives of this pressure term are needed.

The first partial derivatives of $Z_{\mathrm{Ureg}}$ and $Z_{\mathrm{Sreg}}$ with temperature are written below:

$$
\begin{aligned}
\left(\frac{\partial Z_{\text{Ureg}}}{\partial T}\right)_{\rho} &= \frac{3}{2}\frac{T_c}{T}\frac{\rho}{T}n'_{\text{reg}}\,\lambda^{-m}\left[\left(\frac{\lambda T}{T_c}\right)^{m}\left(\frac{m-1}{m}-\exp\left(-\left(\frac{\lambda T}{T_c}\right)^{2-m}\right)\right)+\frac{\lambda^{m}}{m}+\frac{1}{m-2}\Gamma\left(\frac{m}{2-m},\left(\frac{\lambda T}{T_c}\right)^{2-m}\right)\right]\\
&+\frac{3}{2}\frac{T_c}{T}\frac{\rho}{T}n_{\text{reg}}\,\frac{m'}{m^{2}}\left\{-1+\left(\frac{T}{T_c}\right)^{m}\left[1+m(m-1)\ln\left(\frac{T}{T_c}\right)\right]\right\}\\
&-3\frac{T_c}{T}\frac{\rho}{T}n_{\text{reg}}\,m'\frac{\lambda^{-m}}{(2-m)^{3}}\Gamma\left(\frac{m}{2-m},\left(\frac{\lambda T}{T_c}\right)^{2-m}\right)\left\{\ln\left(\left(\frac{\lambda T}{T_c}\right)^{2-m(\rho)}\right)+\frac{1}{2}\left(1+\ln\left(\lambda^{2-m}\right)\right)(m-2)\right\}\\
&+\frac{3}{2}\frac{T_c}{T}\frac{\rho}{T}n_{\text{reg}}\,m'\frac{\lambda^{-m}}{(2-m)}\left(\frac{\lambda T}{T_c}\right)^{2}\exp\left(-\left(\frac{\lambda T}{T_c}\right)^{2-m}\right)\\
&\qquad\times\left\{\left(\frac{\lambda T}{T_c}\right)^{m-2}\left[\ln\left(\lambda^{2-m}\right)-\ln\left(\left(\frac{\lambda T}{T_c}\right)^{2}\right)\right]-\ln\left(\left(\frac{\lambda T}{T_c}\right)^{2-m}\right)\left(1+\left(\frac{\lambda T}{T_c}\right)^{m-2}\frac{m-1}{m-2}\right)\right\}\\
&+3\frac{T_c}{T}\frac{\rho}{T}n_{\text{reg}}\,m'\frac{\lambda^{-m}}{(2-m)^{3}}\left\{(m-2)G^{2\,0}_{1\,2}\left(\left(\frac{\lambda T}{T_c}\right)^{2-m}\,\middle|\,\begin{matrix}1\\ 0,\ \frac{m}{2-m}\end{matrix}\right)-G^{3\,0}_{2\,3}\left(\left(\frac{\lambda T}{T_c}\right)^{2-m}\,\middle|\,\begin{matrix}1,\ 1\\ 0,\ 0,\ \frac{m}{2-m}\end{matrix}\right)\right\}
\end{aligned}
\qquad\text{(A5)}
$$

and

$$
\begin{aligned}
\left(\frac{\partial Z_{\text{Sreg}}}{\partial T}\right)_{\rho} &= \frac{3}{2}\frac{\rho}{T}n'_{\text{reg}}\left(\frac{T}{T_c}\right)^{m-1}\left[1-\exp\left(-\left(\frac{\lambda T}{T_c}\right)^{2-m}\right)\right]+\frac{3}{2}\frac{\rho}{T}n_{\text{reg}}\,m'\left(\frac{T}{T_c}\right)^{m-1}\ln\left(\frac{T}{T_c}\right)\\
&+\frac{3}{2}\frac{\rho}{T}n_{\text{reg}}\,m'\frac{\lambda^{1-m}}{(m-2)^{2}}\left\{\Gamma\left(\frac{1-m}{m-2},\left(\frac{\lambda T}{T_c}\right)^{2-m}\right)-G^{2\,0}_{1\,2}\left(\left(\frac{\lambda T}{T_c}\right)^{2-m}\,\middle|\,\begin{matrix}1\\ 0,\ \frac{1-m}{m-2}\end{matrix}\right)\right\}\\
&+\frac{3}{2}\frac{\rho}{T}n_{\text{reg}}\,m'\frac{\lambda^{1-m}}{(m-2)}\left(\frac{\lambda T}{T_c}\right)^{2-m}\exp\left(-\left(\frac{\lambda T}{T_c}\right)^{2-m}\right)\left\{\left(1+\left(\frac{\lambda T}{T_c}\right)^{2-m}\right)\ln\left(\left(\frac{\lambda T}{T_c}\right)^{2-m}\right)-\ln\left(\lambda^{2-m}\right)\right\}
\end{aligned}
\qquad\text{(A6)}
$$

The first partial derivatives of $Z_{\text{Ureg}}$ and $Z_{\text{Sreg}}$ with density are written hereafter:

$$
\begin{aligned}
\left(\frac{\partial Z_{\mathrm{Ureg}}}{\partial \rho}\right)_T &= \frac{3}{2}\frac{T_c}{T}\left(n'_{\mathrm{reg}} + \rho\, n''_{\mathrm{reg}}\right)\left[\frac{\lambda^{-m}}{2-m}\Gamma\left(\frac{m}{2-m},\left(\frac{\lambda T}{T_c}\right)^{2-m}\right) - m^{-1}\left(1-\left(\frac{T}{T_c}\right)^{m}\right)\right] \\
&+ \frac{3}{2}\frac{T_c}{T}\rho\, n_{\mathrm{reg}}\frac{m'^2}{m^3}\left\{-2+2\left(\frac{T}{T_c}\right)^{m}\left[1-\ln\left(\left(\frac{T}{T_c}\right)^{m}\right)+\frac{1}{2}\ln\left(\left(\frac{T}{T_c}\right)^{m}\right)^{2}\right]\right\} \\
&+ \frac{3}{2}\frac{T_c}{T}\rho\, n'_{\mathrm{reg}}\, m'\frac{\lambda^{-m}}{(m-2)^2}\Gamma\left(\frac{m}{2-m},\left(\frac{\lambda T}{T_c}\right)^{2-m}\right)\left(1+\ln\left(\lambda^{m-2}\right)\right) \\
&+ \frac{3}{2}\frac{T_c}{T}\left(n_{\mathrm{reg}}\, m' + \rho\, n_{\mathrm{reg}}\, m'' + 2\rho\, n'_{\mathrm{reg}}\, m'\right)\left[m^{-2}\left(1-\left(\frac{T}{T_c}\right)^{m}\right) + m^{-1}\left(\frac{T}{T_c}\right)^{m}\ln\left(\frac{T}{T_c}\right)\right] \\
&- \frac{3}{2}\frac{T_c}{T}\left(n_{\mathrm{reg}}\, m' + \rho\, n_{\mathrm{reg}}\, m'' + \rho\, n'_{\mathrm{reg}}\, m'\right)\frac{\lambda^{-m}}{(m-2)^3}\left\{\Gamma\left(\frac{m}{2-m},\left(\frac{\lambda T}{T_c}\right)^{2-m}\right)\left[2\ln\left(\left(\frac{\lambda T}{T_c}\right)^{2-m}\right)-(m-2)\left(1+\ln\left(\lambda^{m-2}\right)\right)\right]\right. \\
&\qquad\left. + (m-2)^2\left(\frac{\lambda T}{T_c}\right)^{m}\ln\left(\frac{\lambda T}{T_c}\right)\exp\left(-\left(\frac{\lambda T}{T_c}\right)^{2-m}\right) + 2G^{3\,0}_{2\,3}\left(\left(\frac{\lambda T}{T_c}\right)^{2-m}\middle|\begin{matrix}1, & 1\\ 0, & 0, & \frac{m}{2-m}\end{matrix}\right)\right\} \\
&+ \frac{3}{2}\frac{T_c}{T}\rho\, m'\left[(2\text{ - m})n'_{\mathrm{reg}} + n_{\mathrm{reg}}m'\left(1+2\ln\left(\frac{T}{T_c}\right)+\ln\left(\lambda^{m}\right)\right)\right]\frac{\lambda^{-m}}{(m-2)^4} \\
&\qquad\times\left\{(2-m)\left(\frac{\lambda T}{T_c}\right)^{m}\exp\left(-\left(\frac{\lambda T}{T_c}\right)^{2-m}\right)\ln\left(\left(\frac{\lambda T}{T_c}\right)^{2-m}\right)\right. \\
&\qquad\left. + 2\Gamma\left(\frac{m}{2-m},\left(\frac{\lambda T}{T_c}\right)^{2-m}\right)\ln\left(\left(\frac{\lambda T}{T_c}\right)^{2-m}\right) + 2G^{3\,0}_{2\,3}\left(\left(\frac{\lambda T}{T_c}\right)^{2-m}\middle|\begin{matrix}1, & 1\\ 0, & 0, & \frac{m}{2-m}\end{matrix}\right)\right\} \\
&- \frac{3}{2}\frac{T_c}{T}\rho\, n_{\mathrm{reg}}\, m'^2\frac{\lambda^{-m}}{(m-2)^3}\left\{\Gamma\left(\frac{m}{2-m},\left(\frac{\lambda T}{T_c}\right)^{2-m}\right)\left[2+\left(2+\ln\left(\lambda^{m-2}\right)\right)\left(2\ln\left(\frac{T}{T_c}\right)+\ln\left(\lambda^{m}\right)\right)\right]\right. \\
&\qquad\left. + \left(\frac{\lambda T}{T_c}\right)^{m}\exp\left(-\left(\frac{\lambda T}{T_c}\right)^{2-m}\right)\ln\left(\left(\frac{\lambda T}{T_c}\right)^{2-m}\right)\left[1+\ln\left(\left(\frac{T}{T_c}\right)^{2-m}\right)+\left(\frac{\lambda T}{T_c}\right)^{2-m}\ln\left(\left(\frac{\lambda T}{T_c}\right)^{2-m}\right)\right]\right\} \\
&- 3\frac{T_c}{T}\rho\, n_{\mathrm{reg}}\, m'^2\frac{\lambda^{-m}}{(m-2)^5}\left\{(2-m)\left(\frac{\lambda T}{T_c}\right)^{m-2}\ln\left(\left(\frac{\lambda T}{T_c}\right)^{2-m}\right)G^{2\,0}_{1\,2}\left(\left(\frac{\lambda T}{T_c}\right)^{2-m}\middle|\begin{matrix} & 2\\ 1, & \frac{m}{2-m}\end{matrix}\right)\right. \\
&\qquad\left. + (2-m)\left(3+2\ln\left(\frac{T}{T_c}\right)+\ln\left(\lambda^{m}\right)\right)G^{3\,0}_{2\,3}\left(\left(\frac{\lambda T}{T_c}\right)^{2-m}\middle|\begin{matrix}1, & 1\\ 0, & 0, & \frac{m}{2-m}\end{matrix}\right) + 4G^{4\,0}_{3\,4}\left(\left(\frac{\lambda T}{T_c}\right)^{2-m}\middle|\begin{matrix}1, & 1, & 1\\ 0, & 0, & 0, & \frac{m}{2-m}\end{matrix}\right)\right\} \qquad \text{(A7)}
\end{aligned}
$$

and

$$
\begin{aligned}
\left(\frac{\partial Z_{\mathrm{Sreg}}}{\partial\rho}\right)_T = {} & \frac{3}{2}\left(n'_{\mathrm{reg}} + \rho\, n''_{\mathrm{reg}}\right)\left[\frac{\lambda^{1-m}}{2-m}\Gamma\left(\frac{1-m}{m-2},\left(\frac{\lambda T}{T_c}\right)^{2-m}\right) - (m-1)^{-1}\left(1-\left(\frac{T}{T_c}\right)^{m-1}\right)\right] \\
& + \frac{3}{2}\left(n_{\mathrm{reg}}\, m' + \rho\, n_{\mathrm{reg}}\, m'' + 2\rho\, n'_{\mathrm{reg}}\, m'\right)\frac{1}{(m-1)^2}\left[\left(1-\left(\frac{T}{T_c}\right)^{m-1}\right) + (m-1)\left(\frac{T}{T_c}\right)^{m-1}\ln\left(\frac{T}{T_c}\right)\right] \\
& + \frac{3}{2}\left(n_{\mathrm{reg}}\, m' + \rho\, n_{\mathrm{reg}}\, m'' + \rho\, n'_{\mathrm{reg}} m'\right)\frac{\lambda^{1-m}}{(2-m)^2}\left\{\Gamma\left(\frac{1-m}{m-2},\left(\frac{\lambda T}{T_c}\right)^{2-m}\right)\left[1+\ln\left(\frac{\lambda T}{T_c}\right) - \ln\left(\lambda^{2-m}\right)\right]\right. \\
& \qquad \left. + \left(\frac{\lambda T}{T_c}\right)^{m-1}\ln\left(\left(\frac{\lambda T}{T_c}\right)^{2-m}\right)\exp\left(-\left(\frac{\lambda T}{T_c}\right)^{2-m}\right)\right\} \\
& + \frac{3}{2}\rho\, n'_{\mathrm{reg}}\, m' \frac{\lambda^{1-m}}{(m-2)^2}\Gamma\left(\frac{1-m}{m-2},\left(\frac{\lambda T}{T_c}\right)^{2-m}\right)\left(1+\ln\left(\lambda^{m-2}\right)\right) \\
& - \frac{3}{2}\rho\, n_{\mathrm{reg}}\, m'^2 \frac{1}{(m-1)^3}\left[2\left(1-\left(\frac{T}{T_c}\right)^{m-1}\right) + 2\left(\frac{T}{T_c}\right)^{m-1}\ln\left(\left(\frac{T}{T_c}\right)^{m-1}\right) - \left(\frac{T}{T_c}\right)^{m-1}\ln\left(\left(\frac{T}{T_c}\right)^{m-1}\right)^2\right] \\
& - \frac{3}{2}\rho\, n_{\mathrm{reg}}\, m'^2 \frac{\lambda^{1-m}}{(m-2)^3}\left(3+\ln\left(\lambda^{m-2}\right)\right)\left\{\Gamma\left(\frac{1-m}{m-2},\left(\frac{\lambda T}{T_c}\right)^{2-m}\right)\left[1+\ln\left(\frac{\lambda T}{T_c}\right) - \ln\left(\lambda^{2-m}\right)\right]\right. \\
& \qquad \left. + \left(\frac{\lambda T}{T_c}\right)^{m-1}\ln\left(\left(\frac{\lambda T}{T_c}\right)^{2-m}\right)\exp\left(-\left(\frac{\lambda T}{T_c}\right)^{2-m}\right)\right\} \\
& - \frac{3}{2}\rho\, n_{\mathrm{reg}}\, m'^2 \frac{\lambda^{1-m}}{(2-m)^3}\left\{\Gamma\left(\frac{1-m}{m-2},\left(\frac{\lambda T}{T_c}\right)^{2-m}\right)\left[1+\ln\left(\frac{\lambda T}{T_c}\right) - 2\ln\left(\lambda^{2-m}\right)\right] - \left(\frac{\lambda T}{T_c}\right)\ln\left(\left(\frac{\lambda T}{T_c}\right)^{2-m}\right)^2 \exp\left(-\left(\frac{\lambda T}{T_c}\right)^{2-m}\right)\right. \\
& \qquad \left. + \left(\frac{\lambda T}{T_c}\right)^{m-1}\ln\left(\left(\frac{\lambda T}{T_c}\right)^{2-m}\right)\left(2-\ln\left(\left(\frac{\lambda T}{T_c}\right)^{2-m}\right)\right)\exp\left(-\left(\frac{\lambda T}{T_c}\right)^{2-m}\right) - \left(\frac{\lambda T}{T_c}\right)^{m-2}\ln\left(\frac{\lambda T}{T_c}\right) G^{\,2\,0}_{\,1\,2}\left(\left(\frac{\lambda T}{T_c}\right)^{2-m}\middle|\begin{matrix} & 2 \\ 1, & \frac{1}{2-m}\end{matrix}\right)\right\} \\
& + \frac{3}{2}\rho\, m'\left[(2\text{-}m)\, n'_{\mathrm{reg}} + n_{\mathrm{reg}} m'\left(1+\ln\left(\frac{\lambda T}{T_c}\right) - \ln\left(\lambda^{2-m}\right)\right)\right]\frac{\lambda^{1-m}}{(m-2)^4}\ln\left(\left(\frac{\lambda T}{T_c}\right)^{2-m}\right)\left\{(2-m)\left(\frac{\lambda T}{T_c}\right)^{m-1}\exp\left(-\left(\frac{\lambda T}{T_c}\right)^{2-m}\right)\right. \\
& \qquad \left. + \Gamma\left(\frac{1-m}{m-2},\left(\frac{\lambda T}{T_c}\right)^{2-m}\right)\right\} \\
& - 3\rho\, n_{\mathrm{reg}}\, m'^2 \frac{\lambda^{1-m}}{(m-2)^5}\left\{\left(\ln\left(\left(\frac{\lambda T}{T_c}\right)^{2-m}\right) + (2-m)\left(2-\ln\left(\lambda^{2-m}\right)\right)\right) G^{\,3\,0}_{\,2\,3}\left(\left(\frac{\lambda T}{T_c}\right)^{2-m}\middle|\begin{matrix} & 1, & 1 \\ 0, & 0, & \frac{1-m}{m-2}\end{matrix}\right)\right. \\
& \qquad \left. + G^{\,4\,0}_{\,3\,4}\left(\left(\frac{\lambda T}{T_c}\right)^{2-m}\middle|\begin{matrix} & 1, & 1, & 1 \\ 0, & 0, & 0, & \frac{1-m}{m-2}\end{matrix}\right)\right\} \\
& + \frac{3}{2}\frac{\lambda^{1-m}}{(2-m)^3}\left(n_{\mathrm{reg}}\, m' + \rho\, n_{\mathrm{reg}}\, m'' + 2\rho\, n'_{\mathrm{reg}}\, m'\right) G^{\,3\,0}_{\,2\,3}\left(\left(\frac{\lambda T}{T_c}\right)^{2-m}\middle|\begin{matrix} & 1, & 1, \\ 0, & 0, & \frac{1-m}{m-2}\end{matrix}\right)
\end{aligned}
\tag{A8}
$$

## 11 APPENDIX B : Expression of the first and second derivatives of the coefficients in the $C_V$ expression

In this appendix, the expressions of the first derivatives of coefficients that appear in the expression of $C_V$ (see Eq. (2)), and that are useful for calculating pressure, are given below.

$$\begin{aligned}\rho\, n'_{\mathrm{reg}}(\rho) = {} & \alpha_{\mathrm{reg},1}\left(\frac{\rho}{\rho+\rho_{\mathrm{t,Liq}}}\right)^{\varepsilon_{\mathrm{reg,1a}}} \exp\left(-\left(\frac{\rho}{\rho_{\mathrm{t,Gas}}}\right)^{\varepsilon_{\mathrm{reg,1b}}}\right)\left\{\varepsilon_{\mathrm{reg,1a}}\frac{\rho_{\mathrm{t,Liq}}}{\rho+\rho_{\mathrm{t,Liq}}} - \varepsilon_{\mathrm{reg,1b}}\left(\frac{\rho}{\rho_{t,Gas}}\right)^{\varepsilon_{\mathrm{reg,1b}}}\right\} \\ & -\alpha_{\mathrm{reg},2}\left(\frac{\rho}{\rho_{\mathrm{t,Liq}}}\right)^{\varepsilon_{\mathrm{reg,2a}}}\left\{-\varepsilon_{\mathrm{reg,2a}}\left(1-\exp\left(-\left(\frac{\rho}{\rho_{\mathrm{reg,Ronc}}}\right)^{-\varepsilon_{\mathrm{reg,2b}}}\right)\right)+\varepsilon_{\mathrm{reg,2b}}\left(\frac{\rho}{\rho_{\mathrm{reg,Ronc}}}\right)^{-\varepsilon_{\mathrm{reg,2b}}}\exp\left(-\left(\frac{\rho}{\rho_{\mathrm{reg,Ronc}}}\right)^{-\varepsilon_{\mathrm{reg,2b}}}\right)\right\}\end{aligned} \tag{B1}$$

$$\begin{aligned}\rho\, m'(\rho) = {} & -\left\{\alpha_{m,4} + \alpha_{m,3}\left(-\frac{3}{2}+\frac{\rho}{\rho_c}\right)\left(\frac{\rho}{\rho_c}\right)^{\frac{3}{2}}\exp\left(-\frac{\rho}{\rho_c}\right)+\frac{3}{2}\alpha_{m,2}\left(\frac{\rho}{\rho_{\mathrm{t,Liq}}}\right)^{\frac{3}{2}}\exp\left(-\left(\frac{\rho}{\rho_{\mathrm{t,Liq}}}\right)^{\frac{3}{2}}\right)\right\} \\ & + \rho\, m'_{\mathrm{Ronc}}(\rho)\end{aligned} \tag{B2}$$

$$\begin{aligned}\rho\, m'_{\mathrm{Ronc}}\left(\rho \ge \frac{M}{12.9}\,\mathrm{g/cm^3}\right) = {} & \frac{\rho}{\rho_{\mathrm{m,Ronc}}}\left(\frac{\rho+\rho_{\mathrm{m,Ronc}}}{\rho_{\mathrm{m,Ronc}}}\right)^{\varepsilon_{m,5a}-1}\exp\left(-\exp\left(\left(\frac{\rho_{\mathrm{m,Ronc}}}{\rho}\right)^{\varepsilon_{m,5b}}\right)\right) \\ & \times\left\{\alpha_{m,1}\,\varepsilon_{m,5a}+\alpha_{m,4}\left(1+\frac{\rho_{\mathrm{m,Ronc}}}{\rho}+\varepsilon_{m,5a}\ln\left(\frac{\rho}{\rho_c}\right)\right)+\varepsilon_{m,5b}\left(\frac{\rho+\rho_{\mathrm{m,Ronc}}}{\rho_{\mathrm{m,Ronc}}}\right)\left(\frac{\rho_{\mathrm{m,Ronc}}}{\rho}\right)^{\varepsilon_{m,5b}}\right. \\ & \left.\times\exp\left(\left(\frac{\rho_{\mathrm{m,Ronc}}}{\rho}\right)^{\varepsilon_{m,5b}}\right)\left(\alpha_{m,1}+\alpha_{m,4}\ln\left(\frac{\rho}{\rho_c}\right)\right)\right\}\end{aligned} \tag{B3}$$

$$\begin{aligned}\rho\, n'_{\mathrm{nonreg}}\left(\rho \le \rho_{\mathrm{t,Liq}}\right) = {} & \alpha_{\mathrm{nonreg},1}\left(\frac{\rho}{\rho_{\mathrm{t,Gas}}}\right)^{\varepsilon_{\mathrm{nonreg,1a}}}\left(\frac{\rho_{\mathrm{t,Liq}}}{\rho}\right)\exp\left(-\left(\frac{\rho_{\mathrm{t,Liq}}}{\rho_{\mathrm{t,Gas}}}\right)^{\varepsilon_{\mathrm{nonreg,1b}}}\left(\frac{\rho}{\rho_{\mathrm{t,Liq}}-\rho}\right)^{\varepsilon_{\mathrm{nonreg,1b}}}\right) \\ & \times\left\{\varepsilon_{\mathrm{nonreg,1a}}\frac{\rho}{\rho_{\mathrm{t,Liq}}}-\varepsilon_{\mathrm{nonreg,1b}}\left(\frac{\rho_{\mathrm{t,Liq}}}{\rho_{\mathrm{t,Gas}}}\right)^{\varepsilon_{\mathrm{nonreg,1b}}}\left(\frac{\rho}{\rho_{\mathrm{t,Liq}}-\rho}\right)^{1+\varepsilon_{\mathrm{nonreg,1b}}}\right\} \\ & +\alpha_{\mathrm{nonreg},2}\left(\frac{\rho}{\rho_{\mathrm{t,Gas}}}\right)^{\varepsilon_{\mathrm{nonreg,2a}}}\left(\frac{\rho_{\mathrm{t,Liq}}}{\rho}\right)\exp\left(-\left(\frac{\rho_{\mathrm{t,Liq}}}{\rho_{\mathrm{t,Gas}}}\right)^{\varepsilon_{\mathrm{nonreg,2b}}}\left(\frac{\rho}{\rho_{\mathrm{t,Liq}}-\rho}\right)^{\varepsilon_{\mathrm{nonreg,2b}}}\right) \\ & \times\left\{\varepsilon_{\mathrm{nonreg,2a}}\frac{\rho}{\rho_{\mathrm{t,Liq}}}-\varepsilon_{\mathrm{nonreg,2b}}\left(\frac{\rho_{\mathrm{t,Liq}}}{\rho_{\mathrm{t,Gas}}}\right)^{\varepsilon_{\mathrm{nonreg,2b}}}\left(\frac{\rho}{\rho_{\mathrm{t,Liq}}-\rho}\right)^{1+\varepsilon_{\mathrm{nonreg,2b}}}\right\}\end{aligned} \tag{B4}$$

$$\begin{aligned}\rho\, T'_{\text{div}}(\rho) = {} & \alpha_{\text{div},1}\left(\frac{\rho}{\rho_c}\right)^{\varepsilon_{\text{div},1a}} \exp\left(-\left(\frac{\rho}{\rho_c}\right)^{\varepsilon_{\text{div},1b}}\right)\left\{\varepsilon_{\text{div},1a} - \varepsilon_{\text{div},1b}\left(\frac{\rho}{\rho_c}\right)^{\varepsilon_{\text{div},1b}}\right\} \\ & + \alpha_{\text{div},2}\left(\frac{\rho}{\rho_{\text{t,Liq}}}\right)^{\varepsilon_{\text{div},2a}} \exp\left(-\left(\frac{\rho}{\rho_{\text{t,Liq}}}\right)^{\varepsilon_{\text{div},2b}}\right)\left\{\varepsilon_{\text{div},2a} - \varepsilon_{\text{div},2b}\left(\frac{\rho}{\rho_{\text{t,Liq}}}\right)^{\varepsilon_{\text{div},2b}}\right\}\end{aligned} \tag{B5}$$

$$\begin{aligned}\rho\, n'_{\text{crit}}(\rho) = {} & \alpha_{\text{crit,a}}\left(\frac{\rho}{\rho_c}\right)^{\varepsilon_{\text{crit,a}}} \times \exp\left[-\left(\left(\alpha_{\text{crit,b}}\frac{\rho-\rho_c}{\rho_c}\right)^2\right)^{\varepsilon_{\text{crit,b}}}\right] \\ & \times\left\{\varepsilon_{\text{crit,a}} - 2\,\varepsilon_{\text{crit,b}}\,\rho\left(\frac{\alpha_{\text{crit,b}}}{\rho_c}\right)^{2\,\varepsilon_{\text{crit,b}}}\left((\rho-\rho_c)^2\right)^{\varepsilon_{\text{crit,b}}-\frac{1}{2}}\,\text{sign}(\rho-\rho_c)\right\}\end{aligned} \tag{B6}$$

$$\begin{aligned}\rho_c\,\varepsilon'_{\text{crit}}(\rho) = {} & 2\varepsilon_{\text{crit,d}}\,\varepsilon^2_{\text{crit,e}}\,\frac{\rho_{\text{crit,a}}-\rho}{\rho_{\text{crit,a}}}\,\frac{\rho_c}{\rho_{\text{crit,a}}}\exp\left(-\left(\varepsilon_{\text{crit,e}}\frac{\rho-\rho_{\text{crit,a}}}{\rho_{\text{crit,a}}}\right)^2\right) \\ & + 2\varepsilon_{\text{crit,f}}\,\varepsilon^2_{\text{crit,g}}\,\frac{\rho_{\text{crit,b}}-\rho}{\rho_{\text{crit,b}}}\,\frac{\rho_c}{\rho_{\text{crit,b}}}\exp\left(-\left(\varepsilon_{\text{crit,g}}\frac{\rho-\rho_{\text{crit,b}}}{\rho_{\text{crit,b}}}\right)^2\right)\end{aligned} \tag{B7}$$

It is interesting to note that the limits of $\rho\, n'_{\text{crit}}(\rho)$, $\rho\, T'_{\text{div}}(\rho)$, $\rho\, n'_{\text{nonreg}}(\rho)$ and $\rho\, n'_{\text{reg}}(\rho)$ are equal to zero when $\rho \to 0$. Moreover, the limit of $\rho\, m'(\rho)$ is equal to $0.240087$ when $\rho \to 0$.

The expressions of the second derivatives of the coefficients which appear in the expression of $C_V$ and which are useful for calculating the compressibility factor are expressed below.

$$\begin{aligned}\rho^2\, n''_{\text{reg}}(\rho) = {} & \alpha_{\text{reg},1}\left(\frac{\rho}{\rho+\rho_{\text{t,Liq}}}\right)^{\varepsilon_{\text{reg},1a}} \exp\left(-\left(\frac{\rho}{\rho_{\text{t,Gas}}}\right)^{\varepsilon_{\text{reg},1b}}\right)\left\{\varepsilon^2_{\text{reg},1b}\left(\frac{\rho}{\rho_{\text{t,Gas}}}\right)^{\varepsilon_{\text{reg},1b}}\left(-1+\left(\frac{\rho}{\rho_{\text{t,Gas}}}\right)^{\varepsilon_{\text{reg},1b}}\right)\right. \\ & \left. + \varepsilon_{\text{reg},1a}\left(\frac{\rho_{\text{t,Liq}}}{\rho+\rho_{\text{t,Liq}}}\right)^2\left(-2\frac{\rho}{\rho_{\text{t,Liq}}}-1+\varepsilon_{\text{reg},1a}\right) + \varepsilon_{\text{reg},1b}\left(\frac{\rho}{\rho_{\text{t,Gas}}}\right)^{\varepsilon_{\text{reg},1b}}\left(1-2\,\varepsilon_{\text{reg},1a}\frac{\rho_{\text{t,Liq}}}{\rho+\rho_{\text{t,Liq}}}\right)\right\} \\ & + \alpha_{\text{reg},2}\,\varepsilon_{\text{reg},2b}\left(\frac{\rho}{\rho_{\text{t,Liq}}}\right)^{\varepsilon_{\text{reg},2a}}\left(\frac{\rho}{\rho_{\text{reg,Ronc}}}\right)^{-2\,\varepsilon_{\text{reg},2b}} \exp\left(-\left(\frac{\rho}{\rho_{\text{reg,Ronc}}}\right)^{-\varepsilon_{\text{reg},2b}}\right) \\ & \qquad \times\left\{-\varepsilon_{\text{reg},2b} + \left(1-2\,\varepsilon_{\text{reg},2a}+\varepsilon_{\text{reg},2b}\right)\left(\frac{\rho}{\rho_{\text{reg,Ronc}}}\right)^{\varepsilon_{\text{reg},2b}}\right\} \\ & + \alpha_{\text{reg},2}\,\varepsilon_{\text{reg},2a}\left(\varepsilon_{\text{reg},2a}-1\right)\left(\frac{\rho}{\rho_{\text{t,Liq}}}\right)^{\varepsilon_{\text{reg},2a}}\left(1-\exp\left(-\left(\frac{\rho}{\rho_{\text{reg,Ronc}}}\right)^{\varepsilon_{\text{reg},2b}}\right)\right)\end{aligned} \tag{B8}$$

$$\rho^2\, m''(\rho) = \alpha_{m,4} + \alpha_{m,3}\left(\frac{\rho}{\rho_c}\right)^{\frac{3}{2}} \exp\left(-\frac{\rho}{\rho_c}\right)\left(\frac{3}{4} - 3\frac{\rho}{\rho_c} + \left(\frac{\rho}{\rho_c}\right)^2\right)$$

$$+ \frac{3}{2}\alpha_{m,2}\left(\frac{\rho}{\rho_{\mathrm{t,Liq}}}\right)^{\frac{3}{2}} \exp\left(-\left(\frac{\rho}{\rho_{\mathrm{t,Liq}}}\right)^{\frac{3}{2}}\right)\left(1 + \frac{3}{2}\left(-1 + \left(\frac{\rho}{\rho_{\mathrm{t,Liq}}}\right)^{\frac{3}{2}}\right)\right) + \rho^2\, m''_{\mathrm{Ronc}}(\rho) \qquad \text{(B9)}$$

$$\rho^2\, m''_{\mathrm{Ronc}}\left(\rho \geq \frac{M}{12.9}\ \mathrm{g/cm^3}\right) = \left(\frac{\rho + \rho_{\mathrm{m,Ronc}}}{\rho_{\mathrm{m,Ronc}}}\right)^{\varepsilon_{m,5a}} \exp\left(-\exp\left(\left(\frac{\rho_{\mathrm{m,Ronc}}}{\rho}\right)^{\varepsilon_{m,5b}}\right)\right)$$

$$\times \Bigg\{ \alpha_{m,4}\left[-1 + 2\,\varepsilon_{m,5a}\,\frac{\rho}{\rho + \rho_{\mathrm{m,Ronc}}} + 2\,\varepsilon_{m,5b}\left(\frac{\rho_{\mathrm{m,Ronc}}}{\rho}\right)^{\varepsilon_{m,5b}} \exp\left(\left(\frac{\rho_{\mathrm{m,Ronc}}}{\rho}\right)^{\varepsilon_{m,5b}}\right)\right]$$

$$+ \left(\alpha_{m,1} + \alpha_{m,4}\ln\left(\frac{\rho}{\rho_c}\right)\right)\Bigg[\varepsilon_{m,5a}\left(\varepsilon_{m,5a} - 1\right)\left(\frac{\rho}{\rho + \rho_{\mathrm{m,Ronc}}}\right)^2 \qquad \text{(B10)}$$

$$- \varepsilon_{m,5b}^2\left(\frac{\rho_{\mathrm{m,Ronc}}}{\rho}\right)^{2\varepsilon_{m,5b}} \exp\left(\left(\frac{\rho_{\mathrm{m,Ronc}}}{\rho}\right)^{\varepsilon_{m,5b}}\right)\left(1 - \exp\left(\left(\frac{\rho_{\mathrm{m,Ronc}}}{\rho}\right)^{\varepsilon_{m,5b}}\right)\right)$$

$$- \varepsilon_{m,5b}\left(\frac{\rho_{\mathrm{m,Ronc}}}{\rho}\right)^{\varepsilon_{m,5b}} \exp\left(\left(\frac{\rho_{\mathrm{m,Ronc}}}{\rho}\right)^{\varepsilon_{m,5b}}\right)\left(1 + \varepsilon_{m,5b} - 2\,\varepsilon_{m,5a}\,\frac{\rho}{\rho + \rho_{\mathrm{m,Ronc}}}\right)\Bigg]\Bigg\}$$

$$\rho^2\, n''_{\mathrm{nonreg}}\left(\rho \leq \rho_{\mathrm{t,Liq}}\right) = \left(\frac{\rho_{\mathrm{t,Liq}}}{\rho - \rho_{\mathrm{t,Liq}}}\right)^2 \Bigg\{ \alpha_{\mathrm{nonreg,1}}\left(\frac{\rho}{\rho_{\mathrm{t,Gas}}}\right)^{\varepsilon_{\mathrm{nonreg,1a}}} \exp\left(-\left(\frac{\rho_{\mathrm{t,Liq}}}{\rho_{\mathrm{t,Gas}}}\right)^{\varepsilon_{\mathrm{nonreg,1b}}}\left(\frac{\rho}{\rho_{\mathrm{t,Liq}} - \rho}\right)^{\varepsilon_{\mathrm{nonreg,1b}}}\right)$$

$$\times\Bigg[\varepsilon_{\mathrm{nonreg,1a}}\left(\varepsilon_{\mathrm{nonreg,1a}} - 1\right)\left(\frac{\rho}{\rho_{\mathrm{t,Liq}}} - 1\right)^2 + \varepsilon_{\mathrm{nonreg,1b}}^2\left(\frac{\rho_{\mathrm{t,Liq}}}{\rho_{\mathrm{t,Gas}}}\right)^{2\varepsilon_{\mathrm{nonreg,1b}}}\left(\frac{\rho}{\rho_{\mathrm{t,Liq}} - \rho}\right)^{2\varepsilon_{\mathrm{nonreg,1b}}}$$

$$+ \varepsilon_{\mathrm{nonreg,1b}}\left(\frac{\rho_{\mathrm{t,Liq}}}{\rho_{\mathrm{t,Gas}}}\right)^{\varepsilon_{\mathrm{nonreg,1b}}}\left(\frac{\rho}{\rho_{\mathrm{t,Liq}} - \rho}\right)^{\varepsilon_{\mathrm{nonreg,1b}}}\left(2\left(\varepsilon_{\mathrm{nonreg,1a}} - 1\right)\frac{\rho}{\rho_{\mathrm{t,Liq}}} + 1 - 2\,\varepsilon_{\mathrm{nonreg,1a}} - \varepsilon_{\mathrm{nonreg,1b}}\right)\Bigg]$$

$$+ \alpha_{\mathrm{nonreg,2}}\left(\frac{\rho}{\rho_{\mathrm{t,Gas}}}\right)^{\varepsilon_{\mathrm{nonreg,2a}}} \exp\left(-\left(\frac{\rho_{\mathrm{t,Liq}}}{\rho_{\mathrm{t,Gas}}}\right)^{\varepsilon_{\mathrm{nonreg,2b}}}\left(\frac{\rho}{\rho_{\mathrm{t,Liq}} - \rho}\right)^{\varepsilon_{\mathrm{nonreg,2b}}}\right)$$

$$\times\Bigg[\varepsilon_{\mathrm{nonreg,2a}}\left(\varepsilon_{\mathrm{nonreg,2a}} - 1\right)\left(\frac{\rho}{\rho_{\mathrm{t,Liq}}} - 1\right)^2 + \varepsilon_{\mathrm{nonreg,2b}}^2\left(\frac{\rho_{\mathrm{t,Liq}}}{\rho_{\mathrm{t,Gas}}}\right)^{2\varepsilon_{\mathrm{nonreg,2b}}}\left(\frac{\rho}{\rho_{\mathrm{t,Liq}} - \rho}\right)^{2\varepsilon_{\mathrm{nonreg,2b}}}$$

$$+ \varepsilon_{\mathrm{nonreg,2b}}\left(\frac{\rho_{\mathrm{t,Liq}}}{\rho_{\mathrm{t,Gas}}}\right)^{\varepsilon_{\mathrm{nonreg,2b}}}\left(\frac{\rho}{\rho_{\mathrm{t,Liq}} - \rho}\right)^{\varepsilon_{\mathrm{nonreg,2b}}}\left(2\left(\varepsilon_{\mathrm{nonreg,2a}} - 1\right)\frac{\rho}{\rho_{\mathrm{t,Liq}}} + 1 - 2\,\varepsilon_{\mathrm{nonreg,2a}} - \varepsilon_{\mathrm{nonreg,2b}}\right)\Bigg]\Bigg\} \qquad \text{(B11)}$$

$$\begin{aligned}\rho^2\, T''_{\text{div}}(\rho) = {} & \alpha_{\text{div,1}} \left(\frac{\rho}{\rho_c}\right)^{\varepsilon_{\text{div,1a}}} \exp\left(-\left(\frac{\rho}{\rho_c}\right)^{\varepsilon_{\text{div,1b}}}\right) \\ & \times \left\{ \varepsilon_{\text{div,1a}}\left(\varepsilon_{\text{div,1a}}-1\right) - \varepsilon_{\text{div,1b}}\left(\varepsilon_{\text{div,1b}} + 2\,\varepsilon_{\text{div,1a}} - 1\right)\left(\frac{\rho}{\rho_c}\right)^{\varepsilon_{\text{div,1b}}} + \varepsilon^2_{\text{div,1b}}\left(\frac{\rho}{\rho_c}\right)^{2\,\varepsilon_{\text{div,1b}}} \right\} \\ & + \alpha_{\text{div,2}} \left(\frac{\rho}{\rho_{\text{t,Liq}}}\right)^{\varepsilon_{\text{div,2a}}} \exp\left(-\left(\frac{\rho}{\rho_{\text{t,Liq}}}\right)^{\varepsilon_{\text{div,2b}}}\right) \\ & \times \left\{ \varepsilon_{\text{div,2a}}\left(\varepsilon_{\text{div,2a}}-1\right) - \varepsilon_{\text{div,2b}}\left(\varepsilon_{\text{div,2b}} + 2\,\varepsilon_{\text{div,2a}} - 1\right)\left(\frac{\rho}{\rho_{\text{t,Liq}}}\right)^{\varepsilon_{\text{div,2b}}} + \varepsilon^2_{\text{div,2b}}\left(\frac{\rho}{\rho_{\text{t,Liq}}}\right)^{2\,\varepsilon_{\text{div,2b}}} \right\}\end{aligned} \tag{B12}$$

$$\begin{aligned}\rho^2\, n''_{\text{crit}}(\rho) = {} & \alpha_{\text{crit,a}} \left(\frac{\rho}{\rho-\rho_c}\right)^2 \left(\frac{\rho}{\rho_c}\right)^{\varepsilon_{\text{crit,a}}} \exp\left[-\left(\left(\alpha_{\text{crit,b}}\,\frac{\rho-\rho_c}{\rho_c}\right)^2\right)^{\varepsilon_{\text{crit,b}}}\right] \\ & \times \left\{ \varepsilon_{\text{crit,a}}\left(\varepsilon_{\text{crit,a}}-1\right)\left(\frac{\rho-\rho_c}{\rho}\right)^2 + 4\,\varepsilon^2_{\text{crit,b}}\left(\left(\alpha_{\text{crit,b}}\,\frac{\rho-\rho_c}{\rho_c}\right)^2\right)^{2\varepsilon_{\text{crit,b}}} \right. \\ & \left. - 2\varepsilon_{\text{crit,b}}\left(\left(\alpha_{\text{crit,b}}\,\frac{\rho-\rho_c}{\rho_c}\right)^2\right)^{\varepsilon_{\text{crit,b}}}\left(-1 + 2\,\varepsilon_{\text{crit,b}} + 2\,\varepsilon_{\text{crit,a}}\left(1-\frac{\rho_c}{\rho}\right)\right)\right\}\end{aligned} \tag{B13}$$

$$\begin{aligned}\rho_c^2\, \varepsilon''_{\text{crit}}(\rho) = {} & 2\,\varepsilon_{\text{crit,d}}\,\varepsilon^2_{\text{crit,e}}\left(\frac{\rho_c}{\rho_{\text{crit,a}}}\right)^2\left(2\,\varepsilon^2_{\text{crit,e}}\left(\frac{\rho-\rho_{\text{crit,a}}}{\rho_{\text{crit,a}}}\right)^2 - 1\right)\exp\left(-\left(\varepsilon_{\text{crit,e}}\,\frac{\rho-\rho_{\text{crit,a}}}{\rho_{\text{crit,a}}}\right)^2\right) \\ & + 2\,\varepsilon_{\text{crit,f}}\,\varepsilon^2_{\text{crit,g}}\left(\frac{\rho_c}{\rho_{\text{crit,b}}}\right)^2\left(2\,\varepsilon^2_{\text{crit,g}}\left(\frac{\rho-\rho_{\text{crit,b}}}{\rho_{\text{crit,b}}}\right)^2 - 1\right)\exp\left(-\left(\varepsilon_{\text{crit,g}}\,\frac{\rho-\rho_{\text{crit,b}}}{\rho_{\text{crit,b}}}\right)^2\right)\end{aligned} \tag{B14}$$

It is interesting to note that the limits of $\rho^2\, n''_{\text{crit}}(\rho)$, $\rho^2\, T''_{\text{div}}(\rho)$, $\rho^2\, n''_{\text{nonreg}}(\rho)$ and $\rho^2\, n''_{\text{reg}}(\rho)$ are also equal to zero when $\rho \rightarrow 0$ and the limit of $\rho^2\, m''(\rho) = -0.240087$ is same absolute value as that of the limit of $\rho\, m'(\rho)$ but with opposite sign.

## 12 APPENDIX C: Approximate formulas to describe some properties along the liquid-vapor coexistence curve

In this appendix, we propose simple formulas, valid between $T_t$ and $T_c$, to approximate the pressure and densities of liquid and vapor deduced from Maxwell's relations. Thus, a simple formula to describe the variation of pressure with temperature along the Saturation Vapor Pressure curve (SVP) can be written:

$$P_{\text{sat}}(T) = P_c\, \exp\left(\frac{-5.9887\,\theta + 1.7151\,\theta^{3/2} + 3.344\,\theta^2}{T_r^{1.7791}}\right) \tag{C1}$$

where $T_r = T/T_c$ and $\theta = 1 - T_r$ with $T_c = T_{c,\text{non-ext formulation}} = 151.396$ K and $P_c = P_{c,\text{non-ext formulation}} = 49.9684$ bar (see Table 11).

The variation of the liquid density with temperature along SVP can be described approximately by the following formula:

$$\rho_{\sigma l}(T) = \rho_c \frac{1 + 5.38842\left(1 - T_r^{8.23084}\right)^{0.296407} - (5.38842 - 1.02985)\left(1 - T_r^{2.54783}\right)^{0.280344}}{T_r^{3.19766}} \tag{C2}$$

with $\rho_c = \rho_{c,\text{non-ext formulation}} = 0.543786$ g/cm$^3$ (see Table 11).

The variation of the vapor density with temperature along SVP can be described approximately by the following formula:

$$\begin{aligned}\rho_{\sigma v}(T) = \rho_c \exp&\left(-\left[5.5815\left(1 - T_r^{3.5104}\right)^{0.49565} - (5.5815 + 0.209)\,\theta^{0.44766}\right]\frac{\exp\left(1.1761\,\theta^2\right)}{T_r^2}\right)\\ &\times \exp\left(-5.5815\,\theta^{1.1685}\exp\left(-\left|\frac{T_r - 0.99338}{0.019791}\right|\right)\right)\end{aligned} \tag{C3}$$

with $\rho_c = \rho_{c,\text{non-ext formulation}} = 0.543786$ g/cm$^3$ (see Table 11). By construction, all formulas cross exactly the critical point.

Fig. 53 shows the deviations obtained between the approximate formulas Eqs. (C1) to (C3) from the values calculated with Maxwell's equations for the non-extensive formulation of the present model. It can be observed first of all that the maximum deviations occur systematically in a neighborhood very close to the critical point.

Fig. 53a shows that Eq. (C1) reproduces the pressure very well since the deviation obtained is smaller than that of Fig. 33 which compares the TSW model and the present model. Therefore, it can be said that Eq. (C1) is both a good representation of the TSW model and the present model.

With respect to the deviation of the liquid density, Fig. 53b appears to be consistent with the tolerance diagram of Fig. 17 all along the coexistence curve up to the critical point. Again, the deviation is smaller than the corresponding one in Fig. 33; therefore, Eq. (C2) is a good representation of both the TSW model and the present model.

The greatest deviation is obtained for the vapor density, which does not conform to the tolerance diagram of Fig. 17. Numerically, the deviation is comparable to the corresponding one in Fig. 33. However, Fig. 53c shows that the deviation is globally well centered on zero indicating that the overall variation is correctly reproduced. This last remark is also valid for the deviations on Fig. 53a and b.

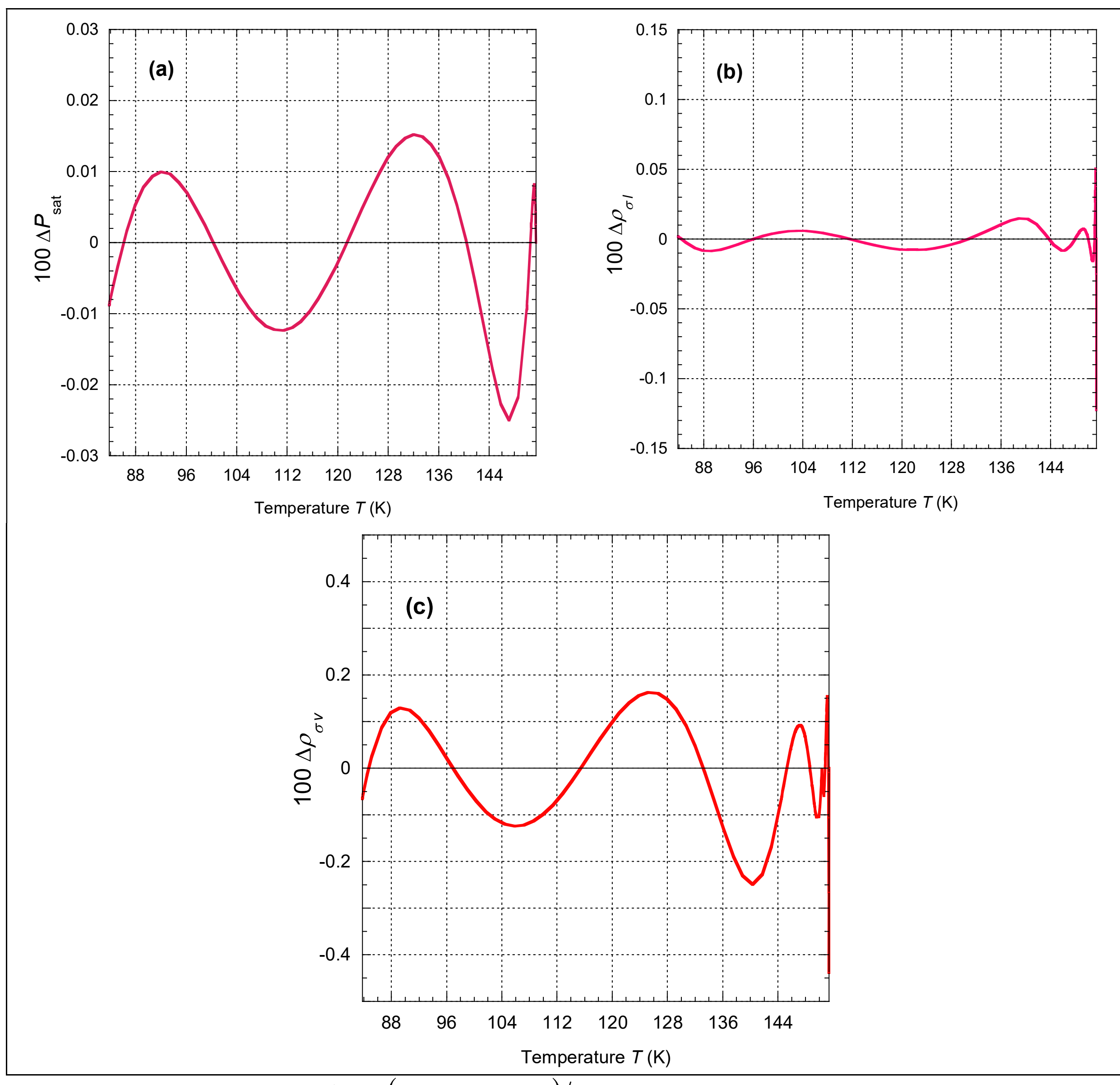

**Fig. 53:** Percentage deviations $\Delta y = \left(y_{\text{Maxwell}} - y_{\text{calc}}\right)/y_{\text{Maxwell}}$ of the approximate formulas Eqs. (C1) to (C3) from the values calculated from Maxwell's equations for the non-extensive formulation of the present model (i.e. Eqs. (50) to (52) with Eq. (39)), in the temperature range from 83.8058 K to 151.396 K.